\documentclass[useAMS,usenatbib]{mn2e}
\usepackage{natbib}
\usepackage{graphicx}
\usepackage[T1]{fontenc}
\usepackage{aecompl}
\usepackage{xr}
\input{psfig}
\def\ga{\mathrel{\mathchoice {\vcenter{\offinterlineskip\halign{\hfil
$\displaystyle##$\hfil\cr>\cr\sim\cr}}}
{\vcenter{\offinterlineskip\halign{\hfil$\textstyle##$\hfil\cr>\cr\sim\cr}}}
{\vcenter{\offinterlineskip\halign{\hfil$\scriptstyle##$\hfil\cr>\cr\sim\cr}}}
{\vcenter{\offinterlineskip\halign{\hfil$\scriptscriptstyle##$\hfil
\cr>\cr\sim\cr}}}}}
\def\la{\mathrel{\mathchoice {\vcenter{\offinterlineskip\halign{\hfil
$\displaystyle##$\hfil\cr<\cr\sim\cr}}}
{\vcenter{\offinterlineskip\halign{\hfil$\textstyle##$\hfil\cr<\cr\sim\cr}}}
{\vcenter{\offinterlineskip\halign{\hfil$\scriptstyle##$\hfil\cr<\cr\sim\cr}}}
{\vcenter{\offinterlineskip\halign{\hfil$\scriptscriptstyle##$\hfil
\cr<\cr\sim\cr}}}}}

\title{Vlasov-Poisson in 1D: waterbags}

\author[S. Colombi \& J. Touma]{St\'ephane
  Colombi$^1$\thanks{E-mail: colombi@iap.fr}  and Jihad Touma$^2$\thanks{E-mail: jt00@aub.edu.lb}\\
\\
$^1$Institut d'Astrophysique de Paris, CNRS UMR 7095 and UPMC, 98bis, bd Arago, F-75014 Paris, France\\
$^2$Department of Physics, American University of Beirut, PO Box 11-0236, Riad El-Solh, Beirut 11097 2020, Lebanon}
\begin{document}
\voffset -1cm
\date{\today}
\pagerange{\pageref{firstpage}--\pageref{lastpage}} \pubyear{2014}
\maketitle
\label{firstpage}
\begin{abstract}
We revisit in one dimension the waterbag method to solve numerically Vlasov-Poisson equations. In this approach, the phase-space distribution function $f(x,v)$ is initially sampled by an ensemble of patches, the waterbags, where $f$ is assumed to be constant. As a consequence of Liouville theorem it is only needed to follow the evolution of the border of these waterbags, which can be done by employing an orientated, self-adaptive polygon tracing isocontours of $f$. This method, which is entropy conserving in essence, is very accurate and can trace very well non linear instabilities as illustrated by specific examples. 

As an application of the method, we generate an ensemble of single waterbag simulations with decreasing thickness, to perform a convergence study to the cold case. Our measurements show that the system relaxes to a steady state where the gravitational potential profile is a power-law of slowly varying index $\beta$, with $\beta$ close to $3/2$ as found in the literature. However, detailed analysis of the properties of the gravitational potential shows that at the center, $\beta > 1.54$. Moreover, our measurements are consistent with the value $\beta=8/5=1.6$ that can be analytically derived by assuming that the average of the phase-space density per energy level obtained at crossing times is conserved during the mixing phase. These results are incompatible with the logarithmic slope of the projected density profile $\beta-2 \simeq -0.47$ obtained recently by \cite{Schulz2013} using a $N$-body technique. This sheds again strong doubts on the capability of $N$-body techniques to converge to the correct steady state expected in the continuous limit.
\end{abstract}

%========================
\section{Introduction}
%========================
The Vlasov-Poisson equations describe the evolution of the phase-space distribution function of a self-gravitating, collisionless system of particles in the fluid limit. In the proper units, they are given in one dimension by 
\begin{equation}
\frac{\partial f}{\partial t}+v \frac{\partial f}{\partial v} - \frac{\partial \phi}{\partial x} \frac{\partial f}{\partial v}=0,
\end{equation}
\begin{equation}
\frac{\partial^2 \phi}{\partial x^2} = 2 \rho(x,t),
\end{equation}
\begin{equation} 
\rho(x,t) \equiv  \int f(x,v',t) {\rm d}v',
\end{equation}
where $x$ is the position, $v$ the velocity, $t$ the time, $f(x,v,t)$ the phase-space density distribution function, $\phi(x,t)$ the gravitational potential and $\rho(x,t)$ the projected density. 

Resolving Vlasov-Poisson equations is very challenging from the analytical point of view. The long term nonlinear evolution of a system following these equations is indeed not yet fully understood, even in the simple one dimensional case.  In general, collisionless self-gravitating systems, unless already in a stable stationary regime, are expected to evolve towards a steady state after a strong mixing phase, usually designated by {\em violent relaxation} \citep{LyndenBell}.  The very existence of a convergence to some equilibrium at late time through phase-mixing is however not demonstrated in the fully general case from the mathematical point of view \citep[see, e.g., the discussion in][]{Villani}. From the physical point of view, there is no model able to predict the exact steady profile that builds up as a function of initial conditions during their evolution. The well-known statistical theory of \cite{LyndenBell} provides partial answers to this problem but its predictive power is limited. For instance, although it is partly successful \citep[see, e.g.,][]{Yamaguchi2008}, it fails to reproduce in detail the steady state of many one-dimensional systems \citep[see, e.g.,][]{Joyce2011}, due to the ``core-halo'' structure\footnote{We use quotes because the {\em core-halo} terminology is usually employed in the framework of gravo-thermal catastrophe while studying the thermodynamics of self-gravitating spherical systems~\cite[see, e.g.,][]{LyndelBellWood1968}.} that warm systems generally build during the course of the dynamics \citep[see, e.g.,][]{Yamashiro1992}.  Some promising improvements of the Lynden Bell theory have however been proposed to explain the structure of three-dimensional dark matter halos \cite[see, e.g.,][]{Hjorth2010,Pontzen2013,Carron2013}, that correspond to the case where the phase-space distribution function is initially cold. Another track relies on the derivation of solutions of the equations by conjecturing self-similarity \citep[see, e.g.,][]{Fillmore1984,Bertschinger1985,Alard2013}. Note that assuming self-similarity is one thing, proving it is a much more challenging matter. 

The only way to understand in detail how a collisionless self-gravitating system evolves according to initial conditions is therefore to resort to a numerical approach. The most widely used method by far is the $N$-body technique in its numerous possible implementations \citep[see, e.g.,][for reviews on the subject]{Bertschinger1998,Colombi2001,Dolag2008,Dehnen2011}, where the phase-space distribution function is represented by an ensemble of macro-particles interacting with each other through softened gravitational forces. However, representing the phase-space distribution function by a set of Dirac functions can have dramatic consequences on the dynamical behavior of the system \citep[see, e.g.,][]{Melott1997,Melott2007}. The irregularities introduced by this discrete representation, along with $N$-body relaxation, can eventually drive the system far from the exact solution. For instance, in the one dimensional case,  collisional relaxation is expected to drive eventually the system in thermal equilibrium \citep[see, e.g.,][]{Rybicki1971}, which is indeed obtained in $N$-body simulations after sufficient time \citep[see, e.g.,][and references therein]{Joyce2010}. Such an equilibrium is clearly not a must in the continuous limit, where there is an infinity of stable steady states to which the system can relax \citep[see, e.g.][]{Chavanis2006,Campa2009}. Such steady states, when different from thermal equilibrium, are reached at best only during a limited amount of time when using a $N$-body approach. Moreover, there is no guarantee that the steady solution given by the $N$-body simulation is the correct one. 

Fortunately, there are alternatives to the $N$-body approach, consisting in solving numerically Vlasov-Poisson equations directly in phase-space. For instance, in plasma physics, the most used solver is the so-called splitting algorithm of \cite{1976JCoPh..22..330C} --where the phase-space distribution function is sampled on a grid-- and its numerous subsequent improvements, modifications and extensions \cite[see, e.g.][but this list is far from being exhaustive]{Shoucri1978,Sonnen1999,Filbet2001,2005MNRAS.359..123A,Umeda2008,Crouseilles2009,Crouseilles2010,Campospinto2011}. In astrophysics, this method was applied successfully to one dimensional systems \citep[][]{Fujiwara1981}, to axisymmetric (3D phase-space) and non axisymmetric disks (4D phase-space) \citep[][]{Watanabe1981,Nishida1981} and to spherical systems (3D phase-space) \cite[][]{Fujiwara1983}. However, due to limitations of available computing resources, its implementation in full six-dimensional phase-space was achieved only very recently \citep[][]{Yoshikawa2013}. 
The main drawback of Eulerian methods such as those inspired from the splitting scheme of \cite{1976JCoPh..22..330C}  is to erase the fine details of the phase-space distribution at small scales as a result of coarse-graining due to finite resolution: on the long term, this coarse-graining might again lead the system far away from the exact solution. In order to fix this problem it is possible to perform adaptive mesh refinement in phase-space \cite[see, e.g.,][]{2005MNRAS.359..123A,Mehrenberger2006,Campospinto2007,Besse2008}.  

Another way to preserve all the details of the phase-space distribution function is to adopt a purely Lagrangian approach consisting in applying literally Liouville theorem, namely that the phase-space distribution function is conserved along trajectories of test particles,
\begin{equation}
f[x(t),v(t),t]={\rm constant}.
\label{eq:liouville}
\end{equation}
This property can indeed be exploited in a powerful way by decomposing the initial distribution on small patches, the {\em waterbags}, where $f$ is approximated by a constant.\footnote{Note thus that a representation of a smooth phase-space distribution function by a stepwise distribution of waterbags remains still irregular, but obviously much less than a set of Dirac functions as in the $N$-body case.}
 From equation (\ref{eq:liouville}) it follows that inside each waterbag, the value of $f$ remains unchanged during evolution, which implies that it is only needed to resolve the evolution of the boundaries of the patches. The terminology ``waterbag'' comes from the incompressible nature of the collisionless fluid in phase-space, which reflects the fact that the area of each patch is conserved. Therefore, their dynamics is analogous to that of an infinitely flexible bag full of water. In one dimension, the numerical implementation is therefore potentially very simple: one just needs to follow the boundaries of the waterbag with a polygon, which can be enriched with new vertices when the shape of the waterbag gets more involved. 

The equation of motion of the polygon vertices is the same as test particles, where the acceleration $a$ is given in one dimension by the difference between the total mass $M_{\rm right}(x)$ at the right of position $x$ and the total mass $M_{\rm left}(x)$ at the left of $x$:
\begin{eqnarray}
a(x,t) &=& -\frac{\partial \phi}{\partial x}=M_{\rm right}(x,t)-M_{\rm left}(x,t) \nonumber \\
          & = & M_{\rm tot}-2 M_{\rm left}(x,t),
\end{eqnarray}
for a total mass $M_{\rm tot}$.  We have 
\begin{equation}
M_{\rm left}(x,t)=\int_{x' \leq x} {\rm d}x' {\rm d}v' f(x',v',t).
\end{equation}
This can be rewritten, if $f$ is approximated by a constant with value $f_k$ within a patch, $P_k$, $k=1,\cdots,N_{\rm patch}$,
\begin{equation}
M_{\rm left}(x,t)=\sum_{k=1}^{N_{\rm patch}} f_k \int_{x' \leq x,\ (x',v') \in P_k} {\rm d}x' {\rm d}v'.
\end{equation}
Application of Green's theorem reads 
\begin{equation}
M_{\rm left}(x)=\sum_{k=1}^{N_{\rm patch}} f_k \oint_{x' \leq x, \partial P_k}  v(s) {\rm d}x'(s),
\label{eq:circu}
\end{equation}
where $s$ is a curvilinear coordinate. This equation represents the essence of the dynamical setting of waterbag method: if one decomposes the phase-space distribution function over a number of patches where it is assumed to be constant, resolution of Poisson equation reduces to a circulation along the contours of each individual patch. 

The waterbag model was introduced by \cite{DePackh1962} and its first numerical implementation was performed in plasma physics by \cite{Roberts1967}, followed soon in the gravitational case by \cite{1971A&A....11..188J} and \cite{1971Ap&SS..13..411C,1971Ap&SS..13..425C}.
We sketched a modern implementation of the algorithm in \cite{2008CNSNS..13...46C} that we aim to present in detail below. Although this numerical technique was one of the pioneering methods used to solve Vlasov-Poisson equations, along with the $N$-body approach \cite[see, e.g.][and references therein]{Henon1964}, it has not been used in astrophysics since the seventies, except in the cold case limit, where some developments have just started \citep{Hahn2013}. 

Although fairly easy to implement for low dimensional systems, this method indeed becomes very involved in 6 dimensional phase-space, as one has to model the evolution of 5 dimensional hypersurfaces. In the cold case, that corresponds to the initially infinitely thin waterbag limit in velocity space, the problem reduces to following the evolution of a three dimensional sheet in six-dimensional space and remains thus feasible. Another caveat of the waterbag method is that, due to mixing in phase-space induced by the relaxation of the system to a steady state, the waterbags get considerably elongated with time, which makes the cost of the scheme increasingly large with time. This is the price to pay for conserving entropy.

The purpose of this article is to describe and to test thoroughly a modern numerical implementation of the waterbag method in one dimension. One goal is to prepare upcoming extensions of this method to higher number of dimensions. As part of the tests, we study in detail the evolution of single waterbags in an attempt to perform a convergence study to the cold limit, particularly relevant to cosmology in the framework of the cold dark matter paradigm. We measure the scaling behavior of the inner part of the system. We compare it to theoretical predictions and to results obtained previously in the literature with $N$-body simulations. 

This paper is thus organized as follows. In \S~\ref{seq:algos}, we present the algorithm, of which the main ingredients were sketched briefly in \cite{2008CNSNS..13...46C}. The performances of the algorithm are tested thoroughly for systems with a carefully chosen set of initial conditions: an initially Gaussian $f(x,v)$ which is expected to evolve to a quasi-stationary state through quiescent mixing \citep{2005MNRAS.359..123A}, an initially random set of warm halos that will be seen, on the contrary, to develop chaos, and finally, an ensemble of single waterbag simulations, where the distribution function is initially supported by an ellipse of varying thickness. In \S~\ref{sec:applications}, we examine in detail the set of single waterbag simulations and study the properties of the system brought about by relaxation processes in the nearly cold regime. The cold limit was previously studied in details in one dimension with exact implementations of the $N$ body approach \citep[see, e.g.][]{Binney,Schulz2013}. It was found in particular by \cite{Schulz2013} that the projected density relaxes to a singular profile of the form $\rho(x) \propto x^{\beta-2}$ with $\beta \simeq 1.53$. We check if this property is recovered with the waterbag technique by performing a convergence study to the cold case. Our analyses are supported by analytical calculations. Finally, \S~\ref{sec:conclusions} summarizes and discusses the main results of this article. To lighten the presentation, only the most important results are presented in the core or the article: technical details are set apart in a coherent set of extensive appendices that can be found online. 
%==================================
\section{The algorithm}
\label{seq:algos}
%==================================
 Integral (\ref{eq:circu}) can be conveniently rewritten
\begin{eqnarray}
M_{\rm left}(x) &=&\oint_{x' \leq x, \partial{\cal P}} \delta f(s)  v(s) {\rm d}x'(s),
\label{eq:circu2} \\
\delta f(s) & \equiv & f^{\rm right}(s)-f^{\rm left}(s),
\end{eqnarray}
where $f^{\rm right}(s)$ and $f^{\rm left}(s)$ are the values of the phase-space distribution function when looking at the right and at the left, respectively, of the contour when facing the direction of circulation defined by the curvilinear coordinate $s$. The global contour $\partial {\cal P}$  passes through a set of orientated loops ($\partial P_k$ in equation \ref{eq:circu}),\footnote{The connecting parts between two isocontours do not contribute to the dynamics.} but without repeating twice the border common to two adjacent waterbags. In practice, it is modeled with a self-adaptive orientated polygon composed of $N$ segments joining together $N+1$ vertices following the equations of motion. 

\begin{figure}
\centerline{\hbox{
\psfig{file=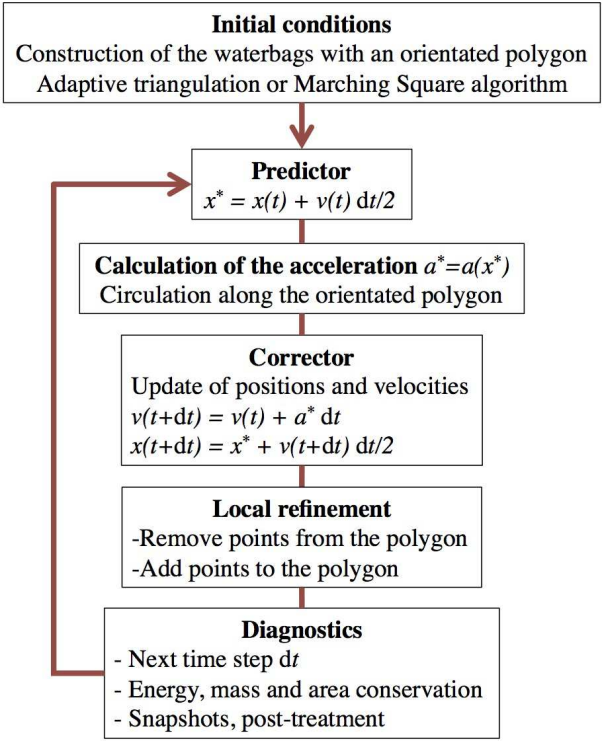,width=8cm}
}}
\caption[]{The main steps of our waterbag algorithm.}
\label{fig:algorithm}
\end{figure}
Our algorithm is summarized in Fig.~\ref{fig:algorithm}. Its important steps, already sketched briefly in \citet{2008CNSNS..13...46C}, define the structure of this section.  Section~\ref{sec:point1} explains the way the initial phase-space distribution function is sampled with the orientated polygon, which allows us to introduce the simulations performed in this paper. Section~\ref{sec:runtime} describes the dynamical component of the algorithm and is divided in five parts: \S~\ref{sec:timeintegration} and \ref{sec:point3} comment briefly on our time integration scheme and on the way we circulate along the orientated polygon to solve Poisson equation; \S~\ref{sec:myref} deals with local refinement and questions the potential virtues of unrefinement; finally, \S~\ref{sec:point5} discusses diagnostics, calculation of the value of the time step and energy conservation.  
\subsection{Initial condition generation and presentation of the simulations} 
\label{sec:point1}
A natural way to sample initial conditions consists in defining each waterbag as the area enclosed between two successive isocontours of the phase-space distribution function. The isocontours are chosen such as to bound the mean square difference between the true and the sampled (step-wise) phase-space distribution function weighted by the waterbag thickness, which means that local intercontour spacing roughly scales like $1/\sqrt{|\nabla f|}$ where $|\nabla f|$ is the magnitude of the gradient of the phase-space distribution function. To draw the isocontours, we use the so-called Marching Square algorithm, inspired from its famous three-dimensional alter-ego \citep{LoCl}. Additional technical details can be found in Appendix~\ref{sec:inicond}

Note that at the end of initial conditions generation, we recast coordinates in the center of mass frame.\footnote{Explicit expressions for the center of mass coordinates are given in Appendix~\ref{sec:kinepot}.}

Now, we introduce and comment on the three sets of simulations performed in this paper, namely an initially Gaussian $f(x,v)$ (\S~\ref{sec:gausi}) an ensemble of random halos (\S~\ref{sec:ransi}) and single waterbags of varying thickness (\S~\ref{sec:sinwai}). Additional details can be found in Appendix~\ref{app:inicond2} and its Table~\ref{tab:simuparam}, which provides the main parameters of the simulations. The large variety of these initial conditions, as shown below, should be sufficient to test thoroughly the performances of the waterbag method. 

\subsubsection{Gaussian initial conditions: Landau damping and importance of initial waterbag sampling}
\label{sec:gausi}
Our Gaussian initial conditions correspond to a phase-space distribution given by $f(x,v)=4 \exp[-(x^2+v^2)/0.08]$ smoothly truncated at $x^2+v^2 \ga 1$. The advantage of this setup is that it is not very far from the thermal equilibrium solution.\footnote{equation~(\ref{eq:apofs}).} The smoothness of the Gaussian function and the supposedly attractor nature of thermal equilibrium should, according to intuition, make this system quiescent. It was indeed previously shown numerically with a semi-Lagrangian solver that this system converges smoothly to a quasy steady state close to (but still slightly different from) thermal equilibrium \citep[]{2005MNRAS.359..123A}. Landau damping represents in plasma physics a fundamental testbed case of Vlasov codes: our Gaussian initial conditions allow us to study the analogous of it in the gravitational case. 

\begin{figure*}
\centerline{\hbox{
\psfig{file=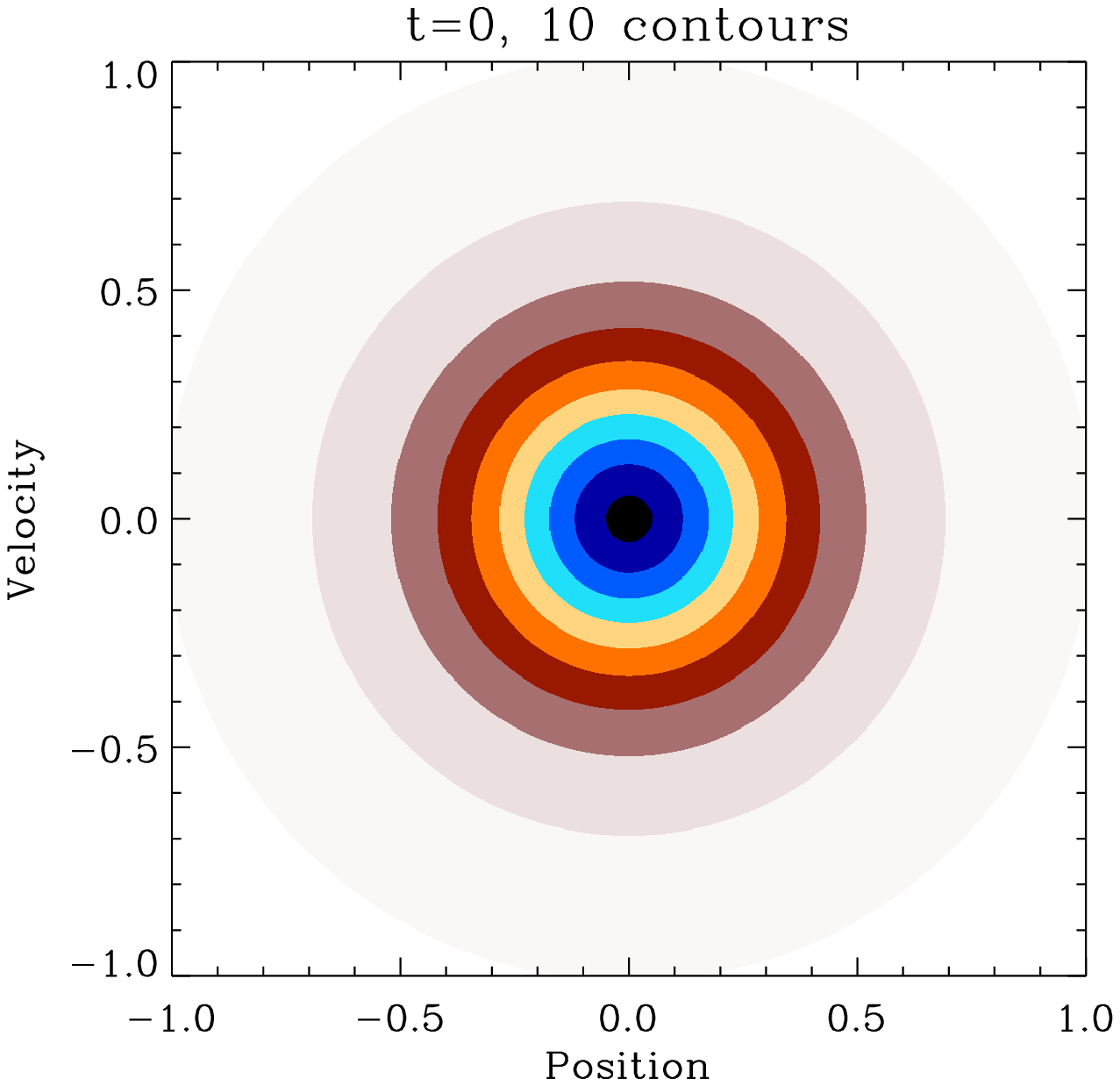,width=5.5cm,bbllx=62pt,bblly=366pt,bburx=426pt,bbury=718pt}
\psfig{file=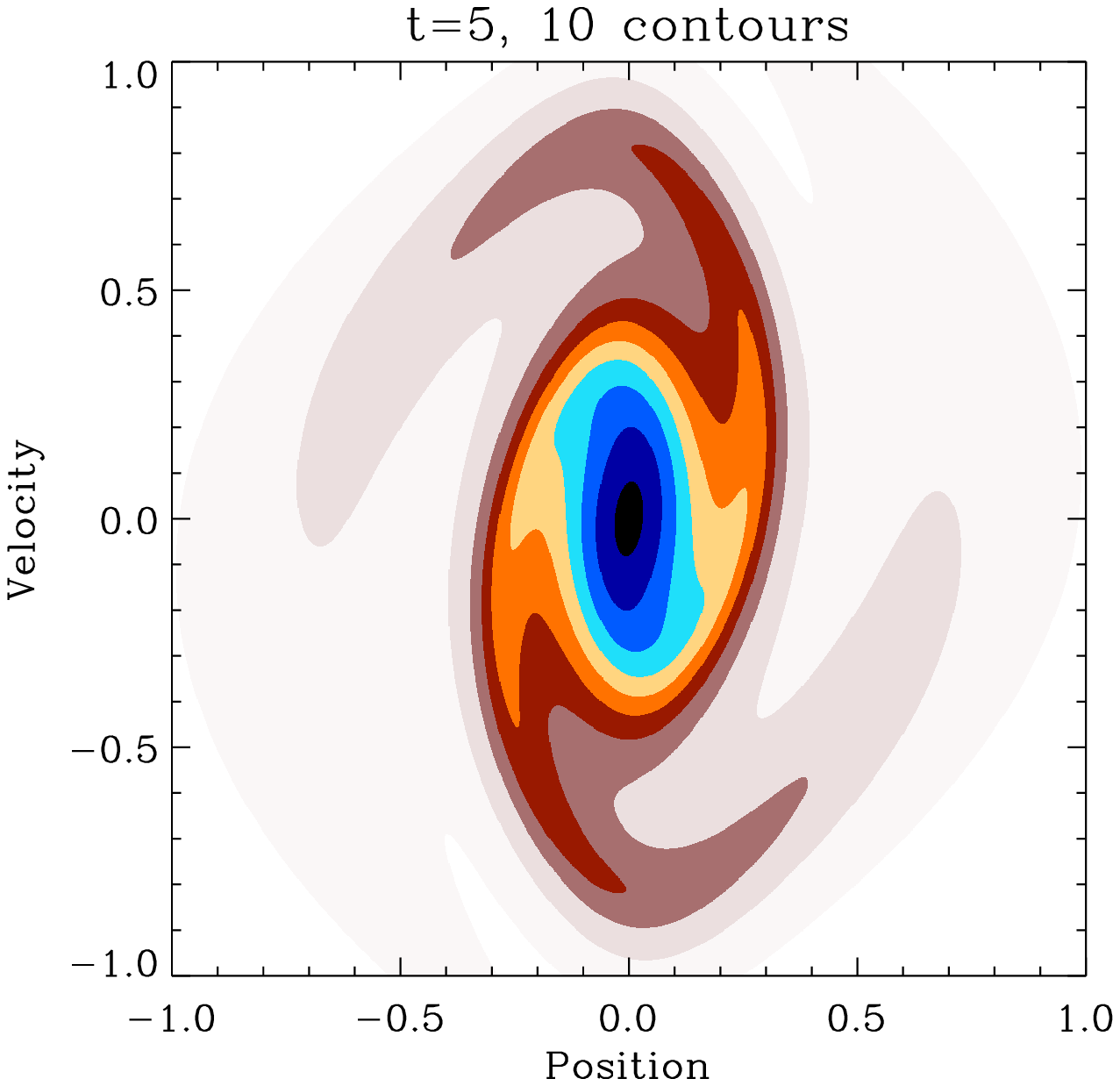,width=5.5cm,bbllx=62pt,bblly=366pt,bburx=426pt,bbury=718pt}
\psfig{file=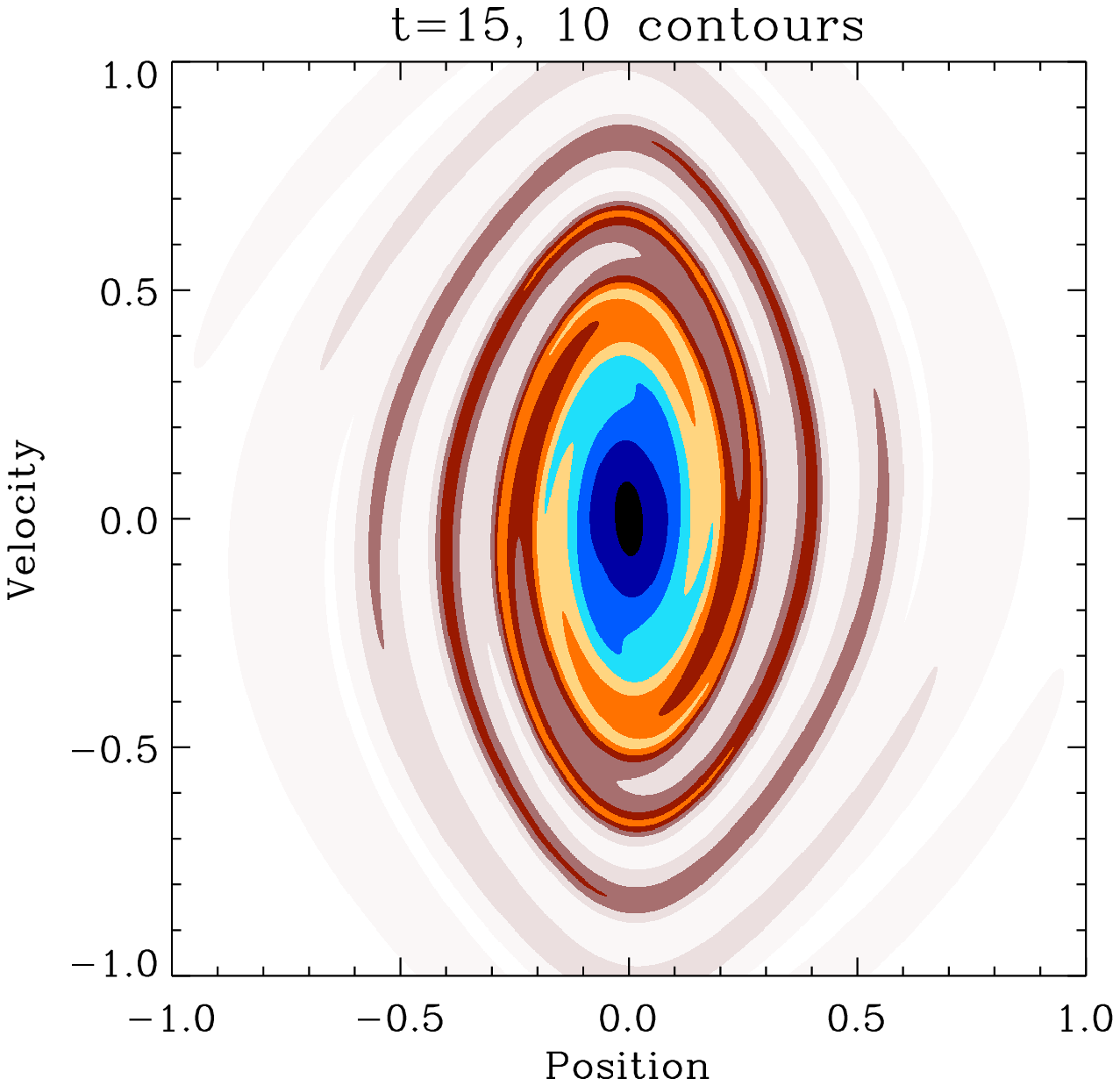,width=5.5cm,bbllx=62pt,bblly=366pt,bburx=426pt,bbury=718pt}}}
\vskip 0.2cm
\centerline{\hbox{
\psfig{file=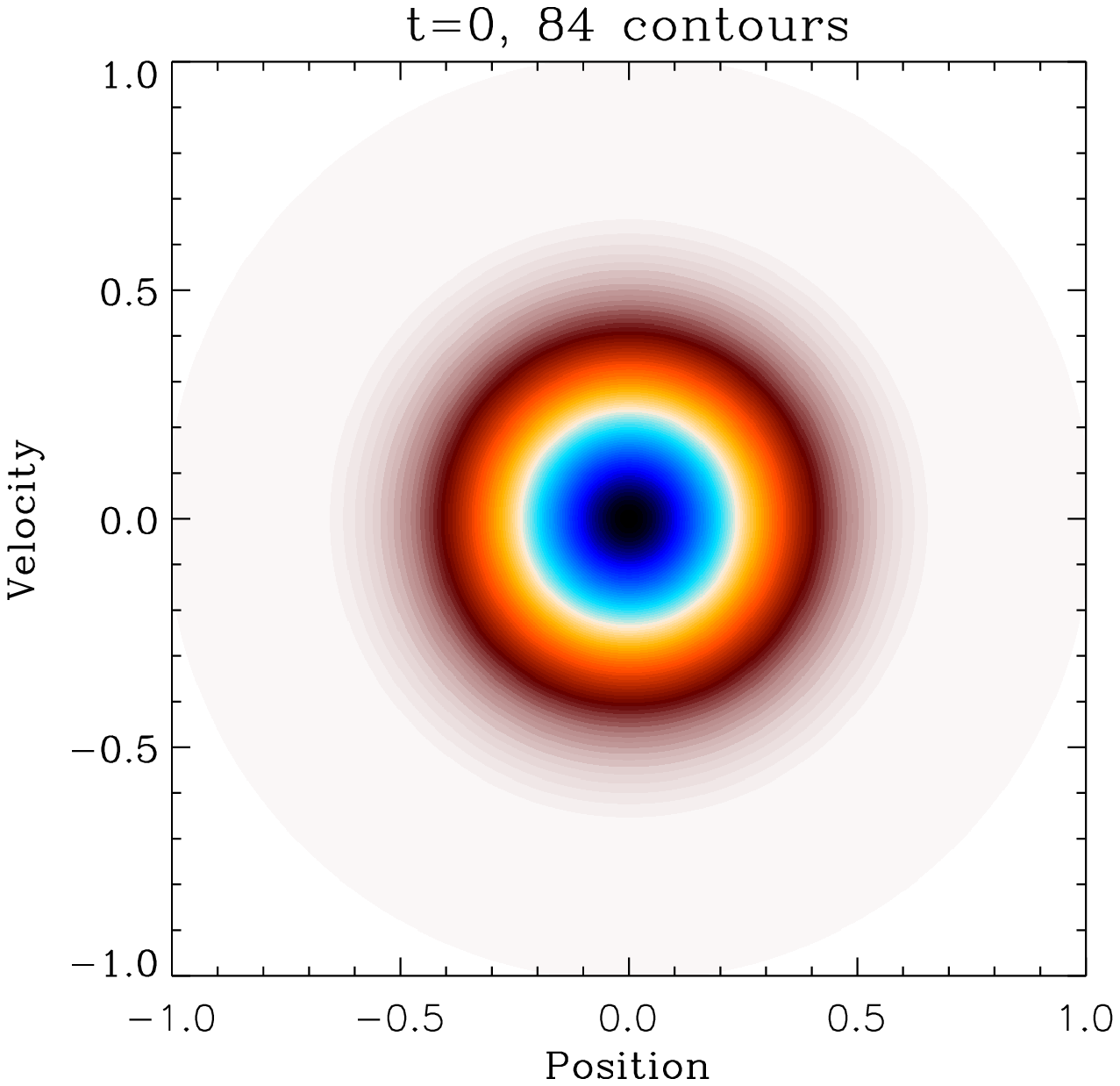,width=5.5cm,bbllx=62pt,bblly=366pt,bburx=426pt,bbury=718pt}
\psfig{file=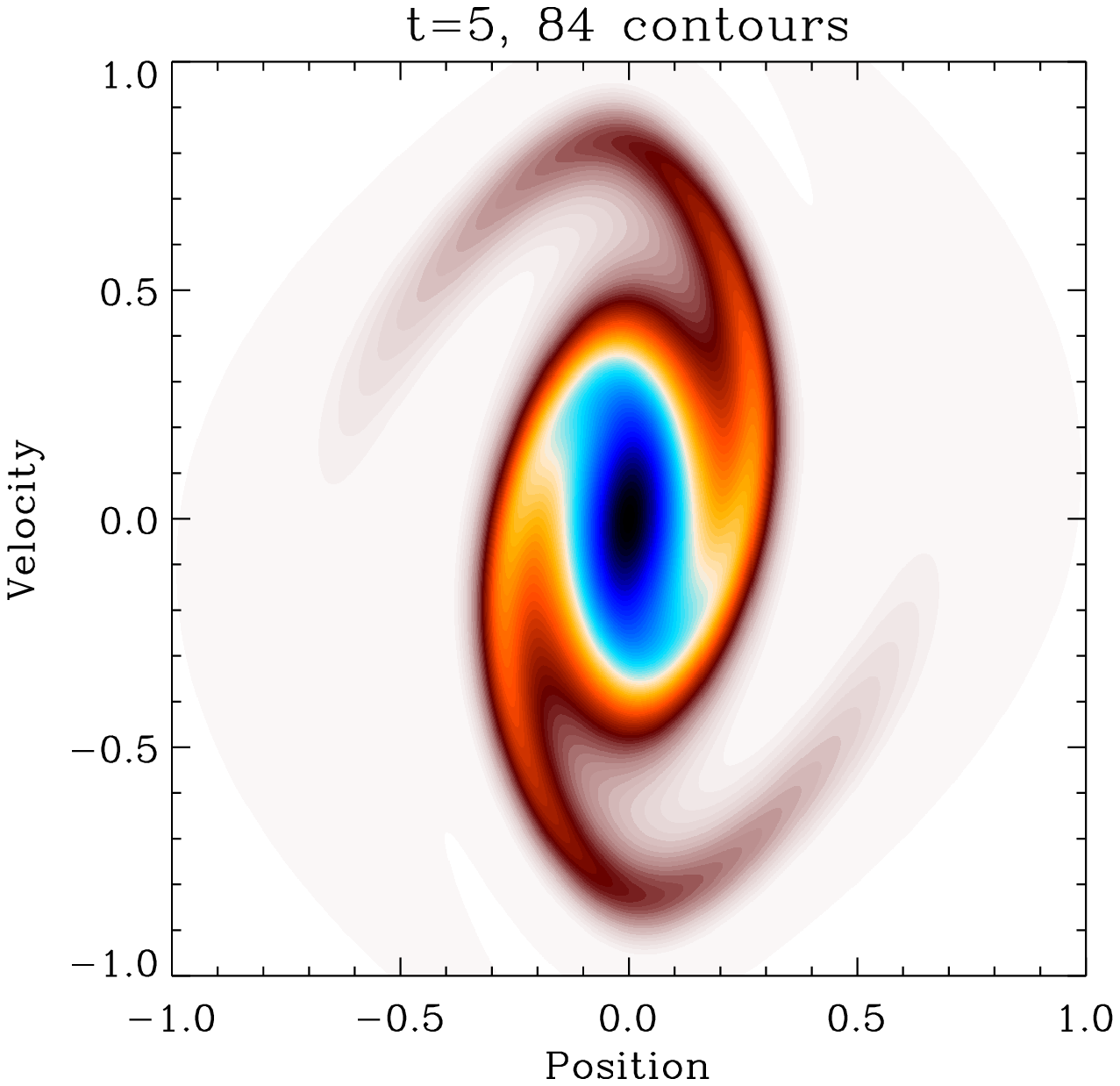,width=5.5cm,bbllx=62pt,bblly=366pt,bburx=426pt,bbury=718pt}
\psfig{file=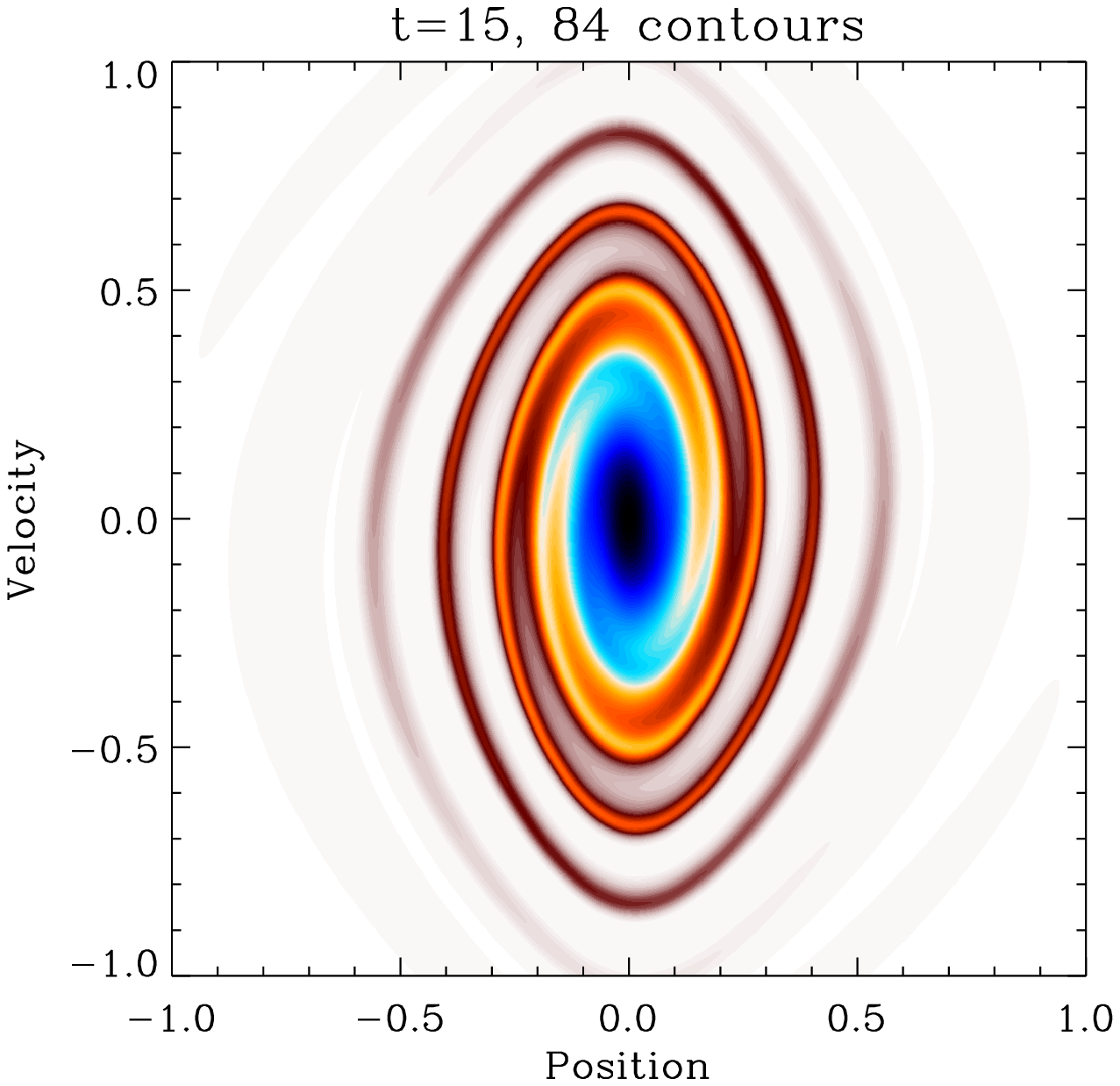,width=5.5cm,bbllx=62pt,bblly=366pt,bburx=426pt,bbury=718pt}}}
\vskip 0.2cm
\centerline{\hbox{
\psfig{file=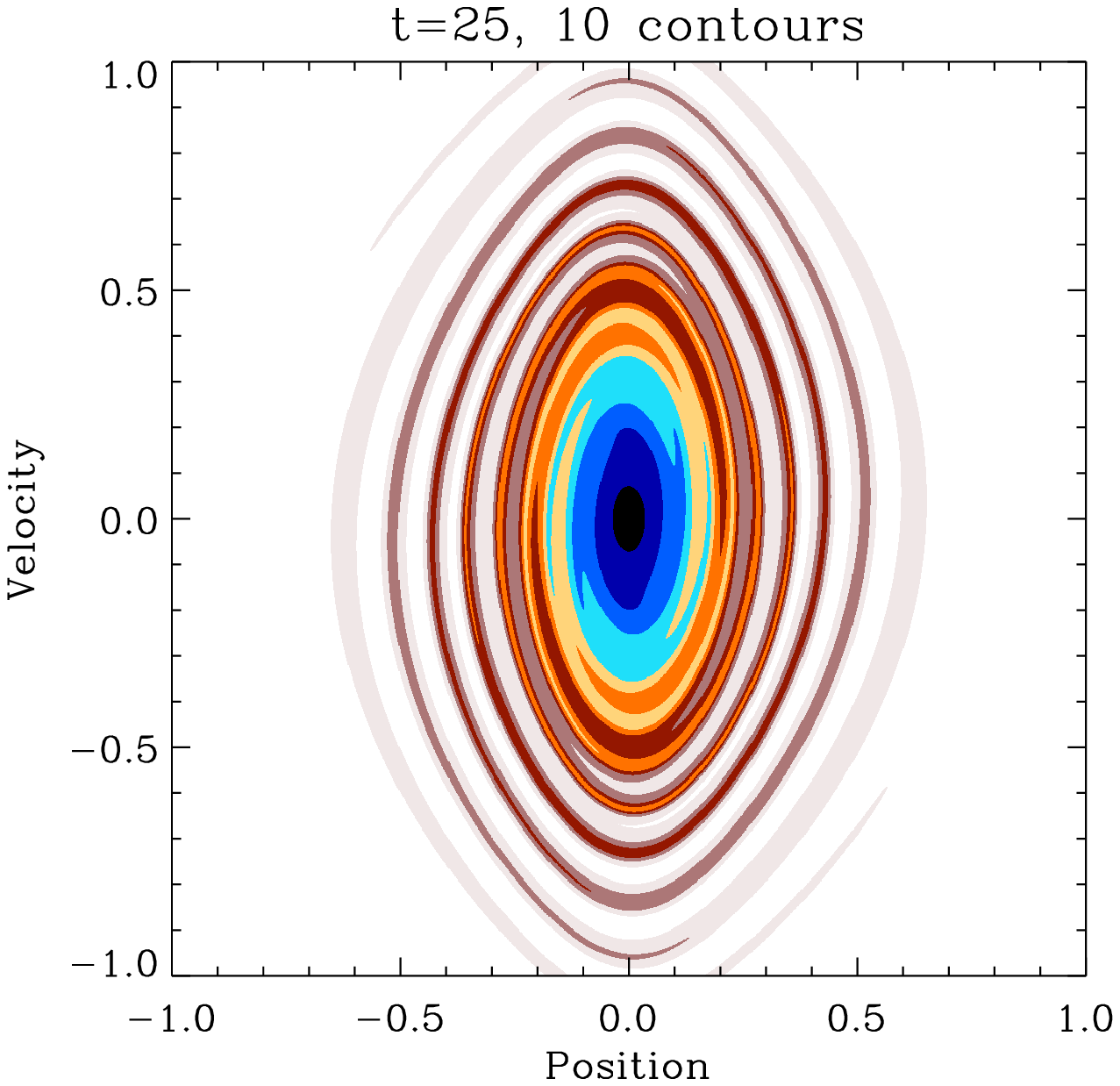,width=5.5cm,bbllx=62pt,bblly=366pt,bburx=426pt,bbury=718pt}
\psfig{file=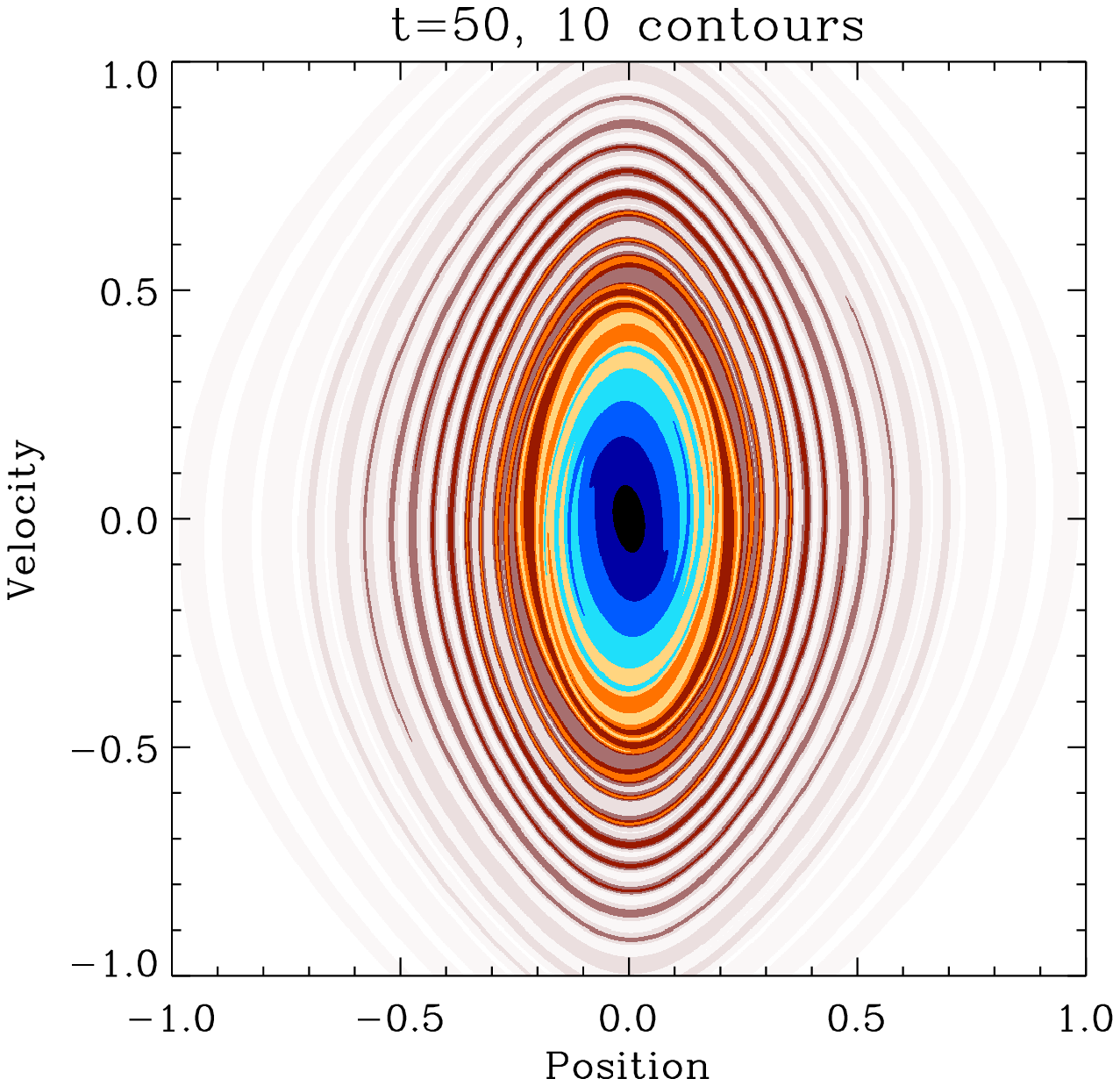,width=5.5cm,bbllx=62pt,bblly=366pt,bburx=426pt,bbury=718pt}
\psfig{file=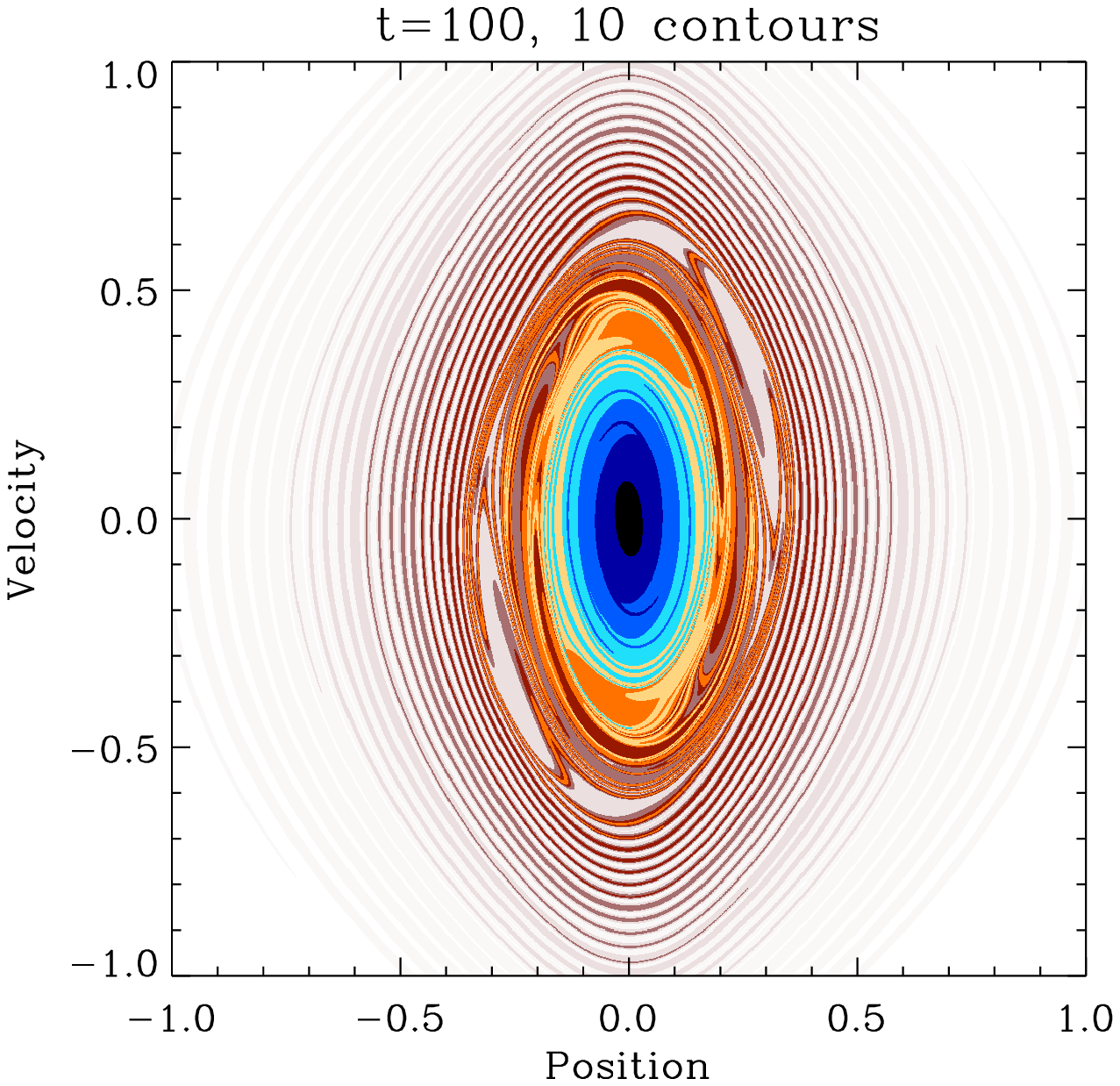,width=5.5cm,bbllx=62pt,bblly=366pt,bburx=426pt,bbury=718pt}}}
\vskip 0.2cm
\centerline{\hbox{
\psfig{file=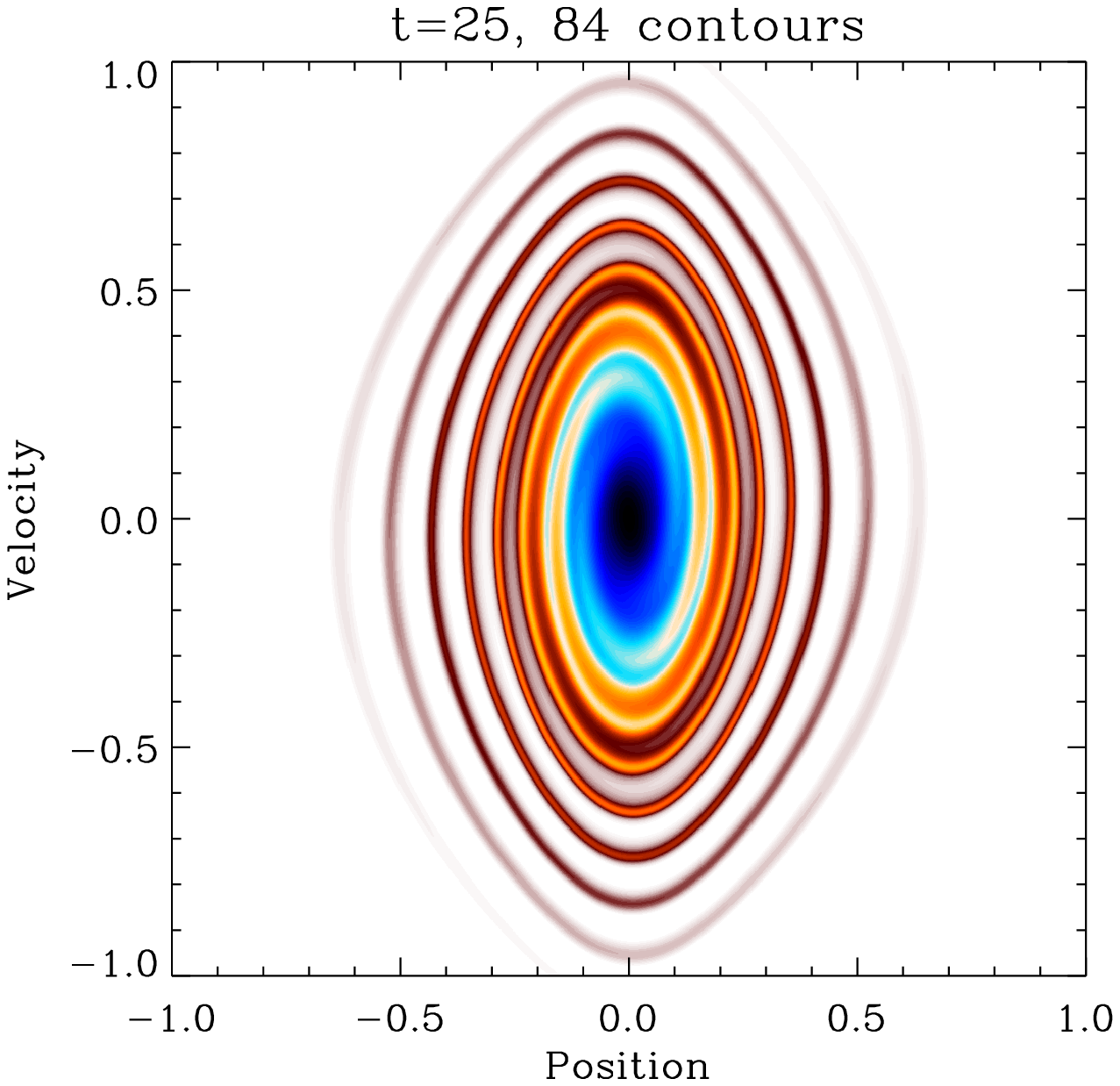,width=5.5cm,bbllx=62pt,bblly=366pt,bburx=426pt,bbury=718pt}
\psfig{file=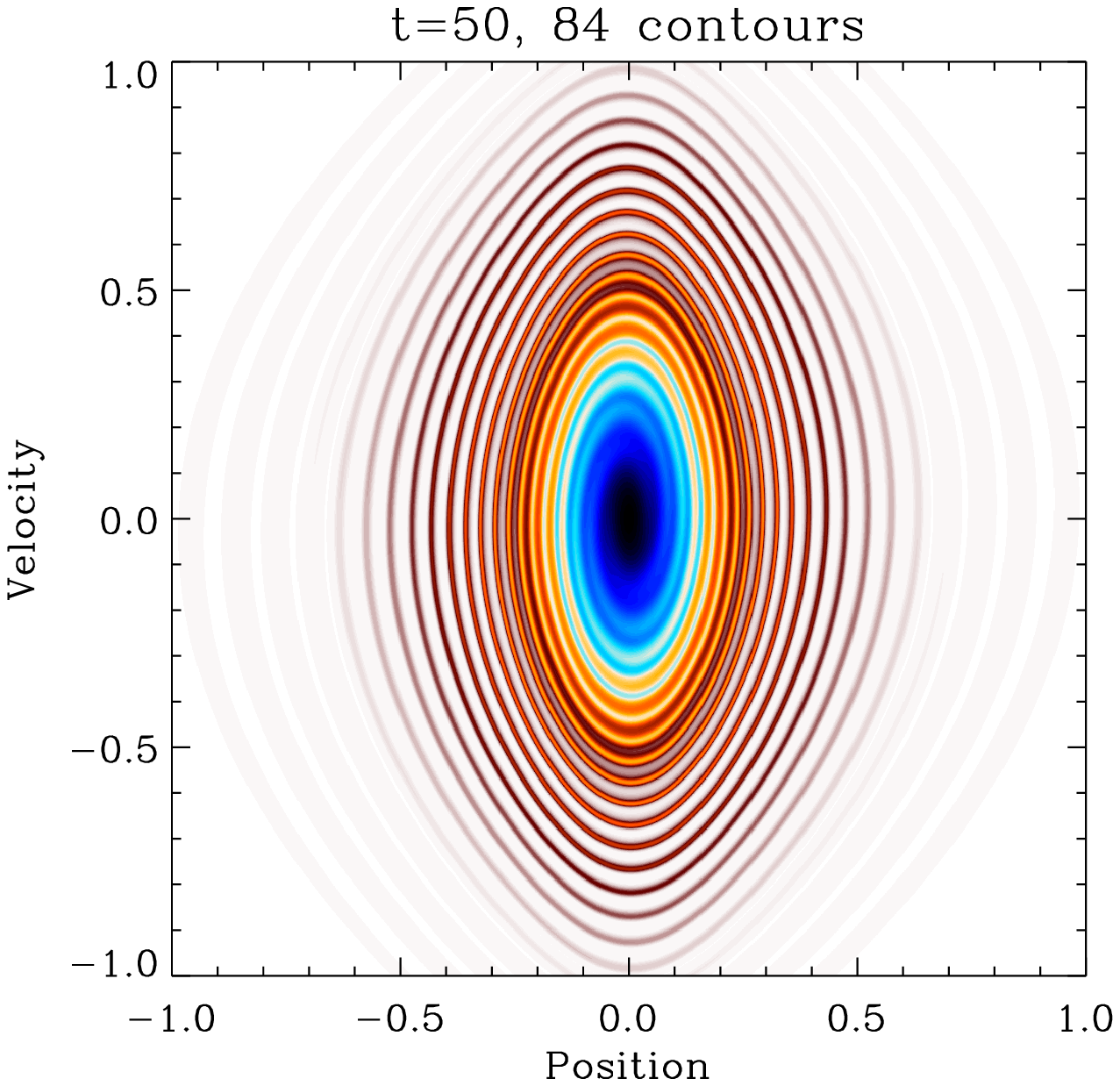,width=5.5cm,bbllx=62pt,bblly=366pt,bburx=426pt,bbury=718pt}
\psfig{file=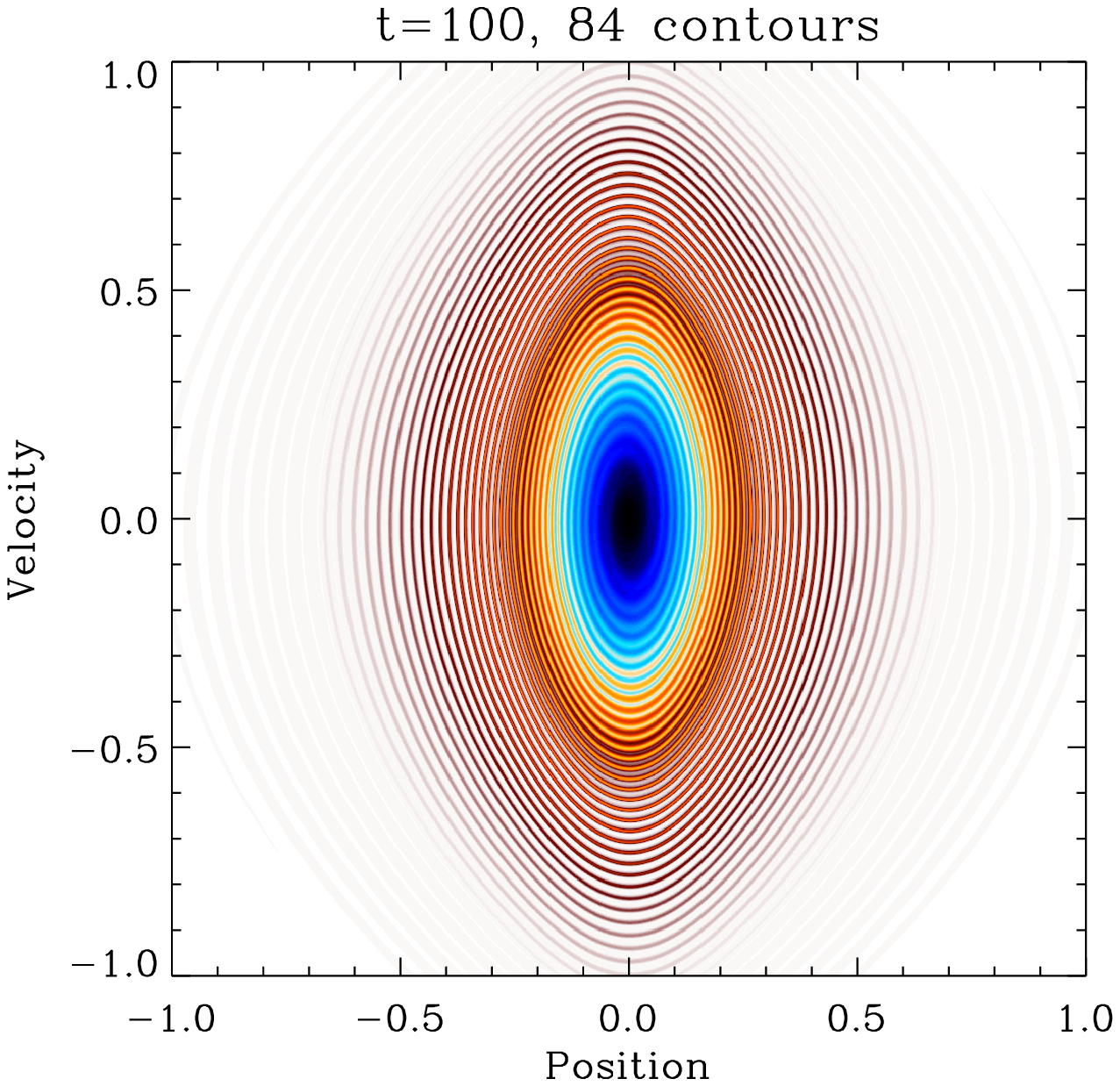,width=5.5cm,bbllx=62pt,bblly=366pt,bburx=426pt,bbury=718pt}}}
\caption[]{Simulations with Gaussian initial conditions. {\em First and third line of panels:} only 10 waterbags are used to sample initial conditions (simulation {\tt Gaussian10} in Table~\ref{tab:simuparam}). {\em Second and fourth line of panels:} 84 waterbags are used to sample initial conditions (simulation {\tt Gaussian84} in Table~\ref{tab:simuparam}). At early times, the two simulations agree very well with each other.  At late times, an instability builds up in the 10 waterbags simulation, at variance with the 84 waterbags one which still presents the expected quiescent evolution. This numerical instability appears as well when $f(x,v)$ is represented by particles as illustrated by Fig.~19 of \citet{2005MNRAS.359..123A}.
Note that these phase-space pictures are drawn using the so-called parity algorithm described in Appendix~\ref{app:drawwat}.}
\label{fig:gaussianA}
\end{figure*}

\begin{figure*}
\centerline{\hbox{
\psfig{file=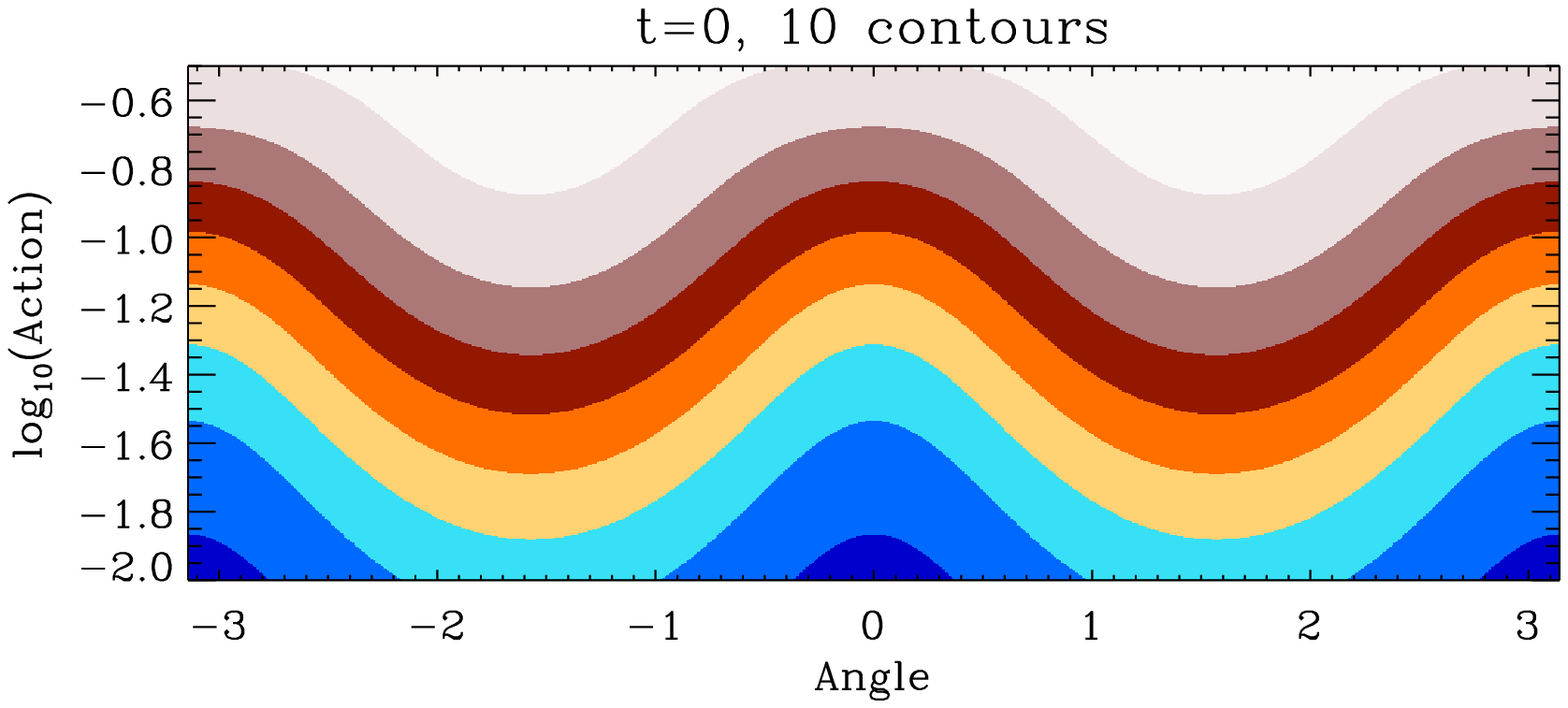,width=8.5cm,bbllx=54pt,bblly=360pt,bburx=558pt,bbury=577pt}
\hskip 0.4cm \psfig{file=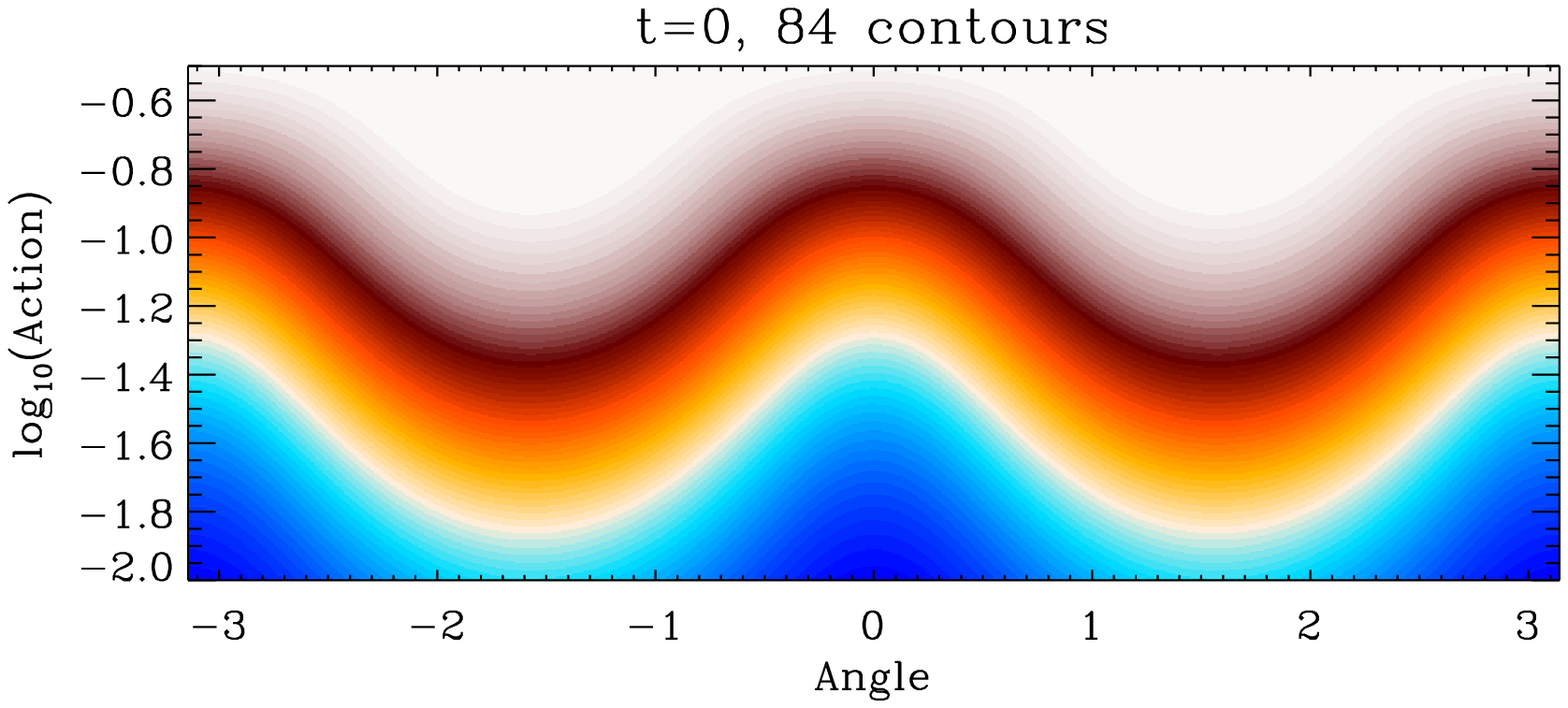,width=8.5cm,bbllx=54pt,bblly=360pt,bburx=558pt,bbury=577pt}}}
%\centerline{\hbox{
%\psfig{file=gaussian05_sparse_AE.ps,width=8.5cm,bbllx=54pt,bblly=360pt,bburx=558pt,bbury=577pt}
%\hskip 0.4cm \psfig{file=gaussian05_dense_AE.ps,width=8.5cm,bbllx=54pt,bblly=360pt,bburx=558pt,bbury=577pt}}}
\centerline{\hbox{
\psfig{file=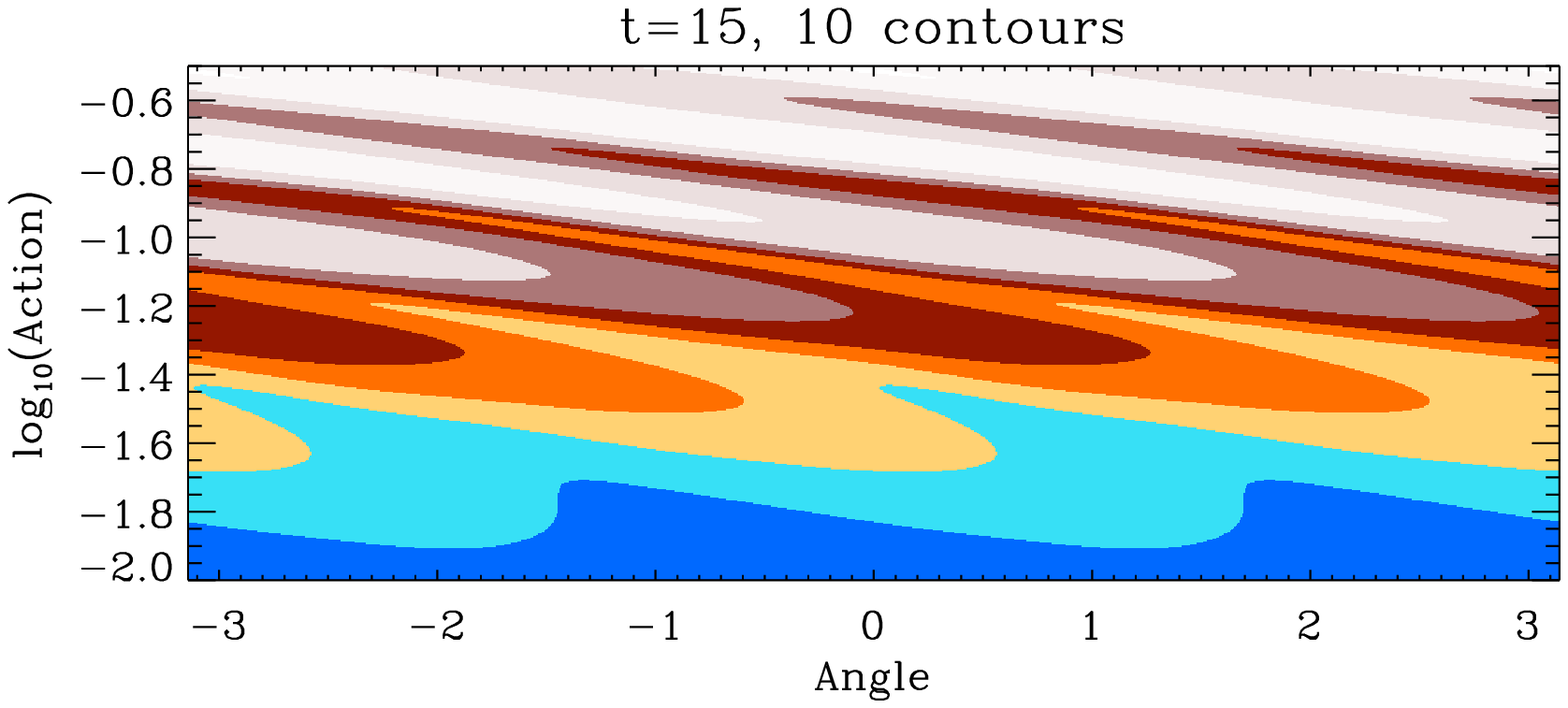,width=8.5cm,bbllx=54pt,bblly=360pt,bburx=558pt,bbury=577pt}
\hskip 0.4cm \psfig{file=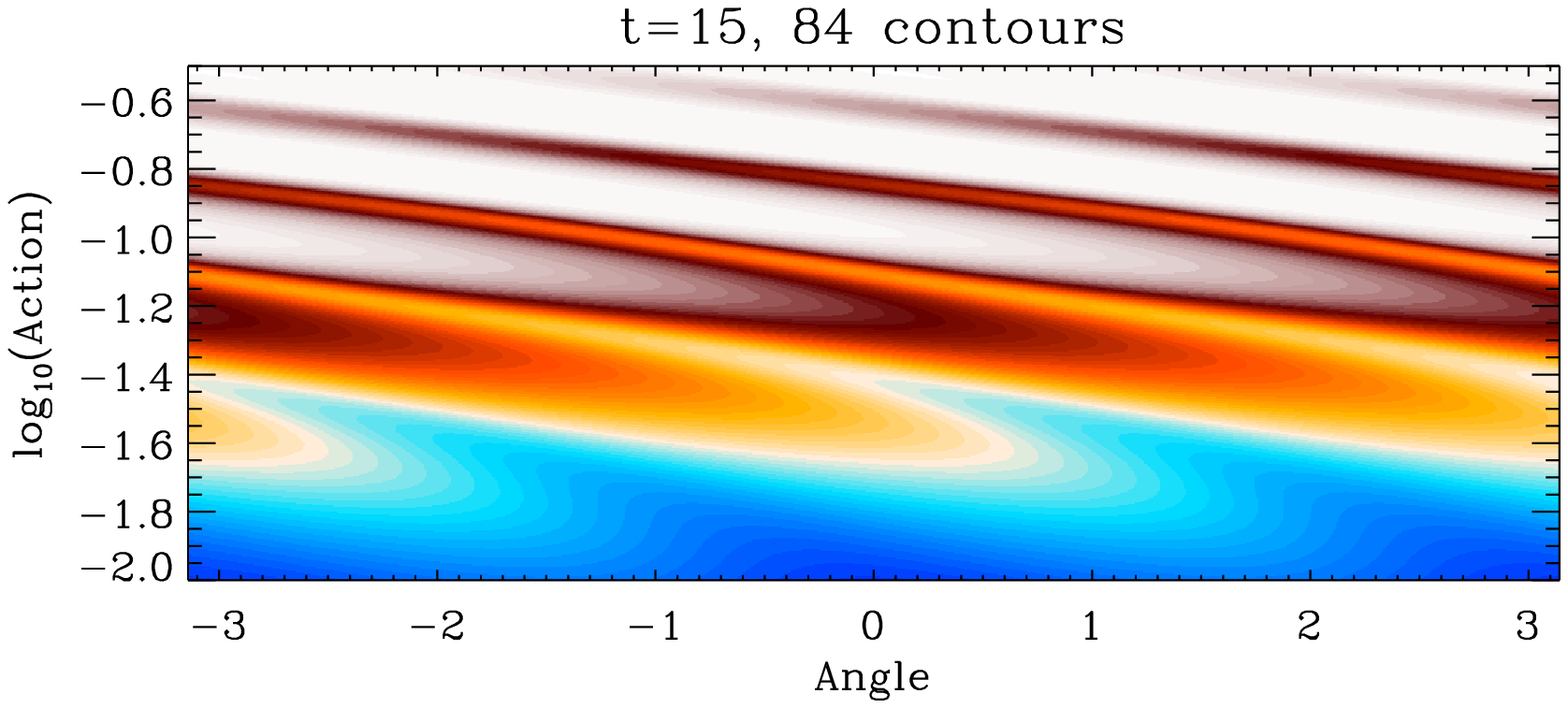,width=8.5cm,bbllx=54pt,bblly=360pt,bburx=558pt,bbury=577pt}}}
\centerline{\hbox{
\psfig{file=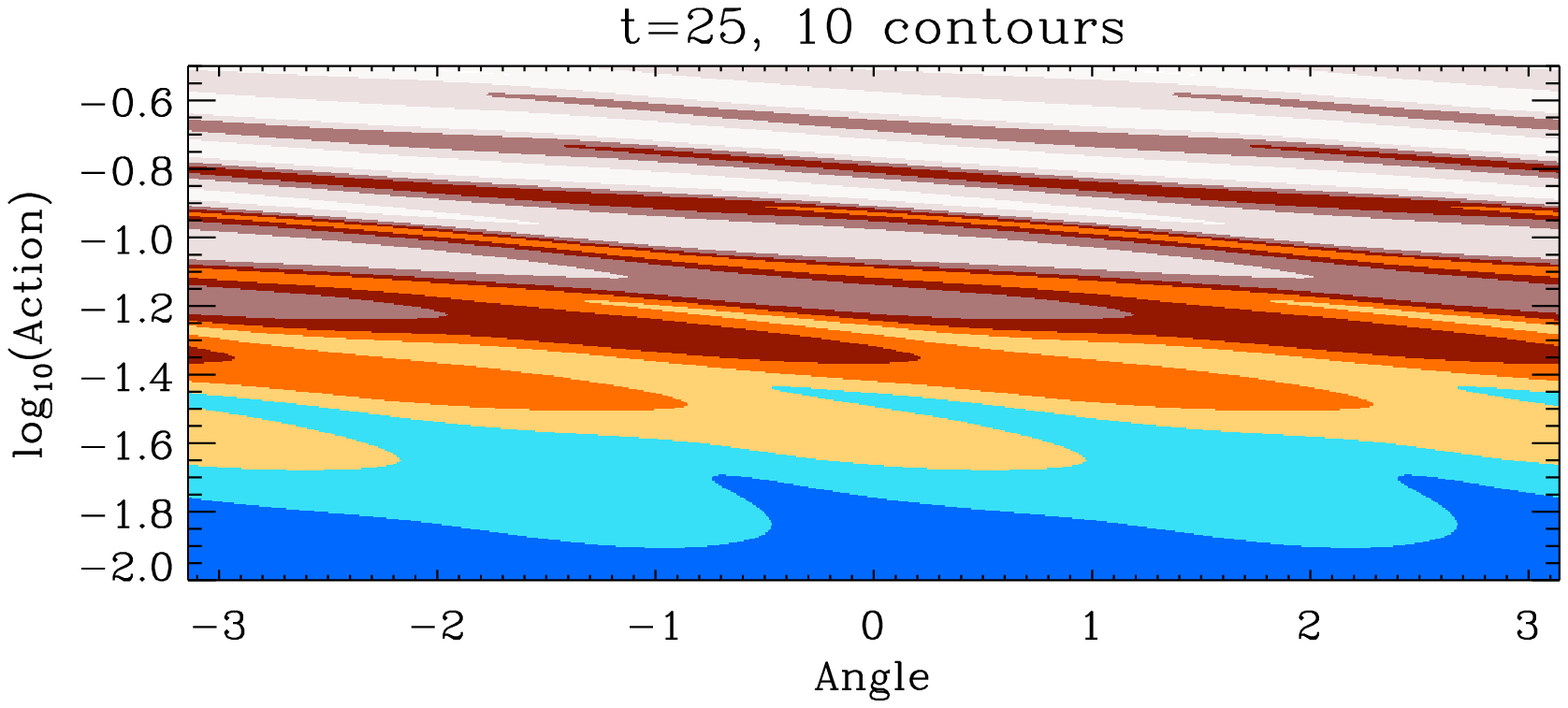,width=8.5cm,bbllx=54pt,bblly=360pt,bburx=558pt,bbury=577pt}
\hskip 0.4cm \psfig{file=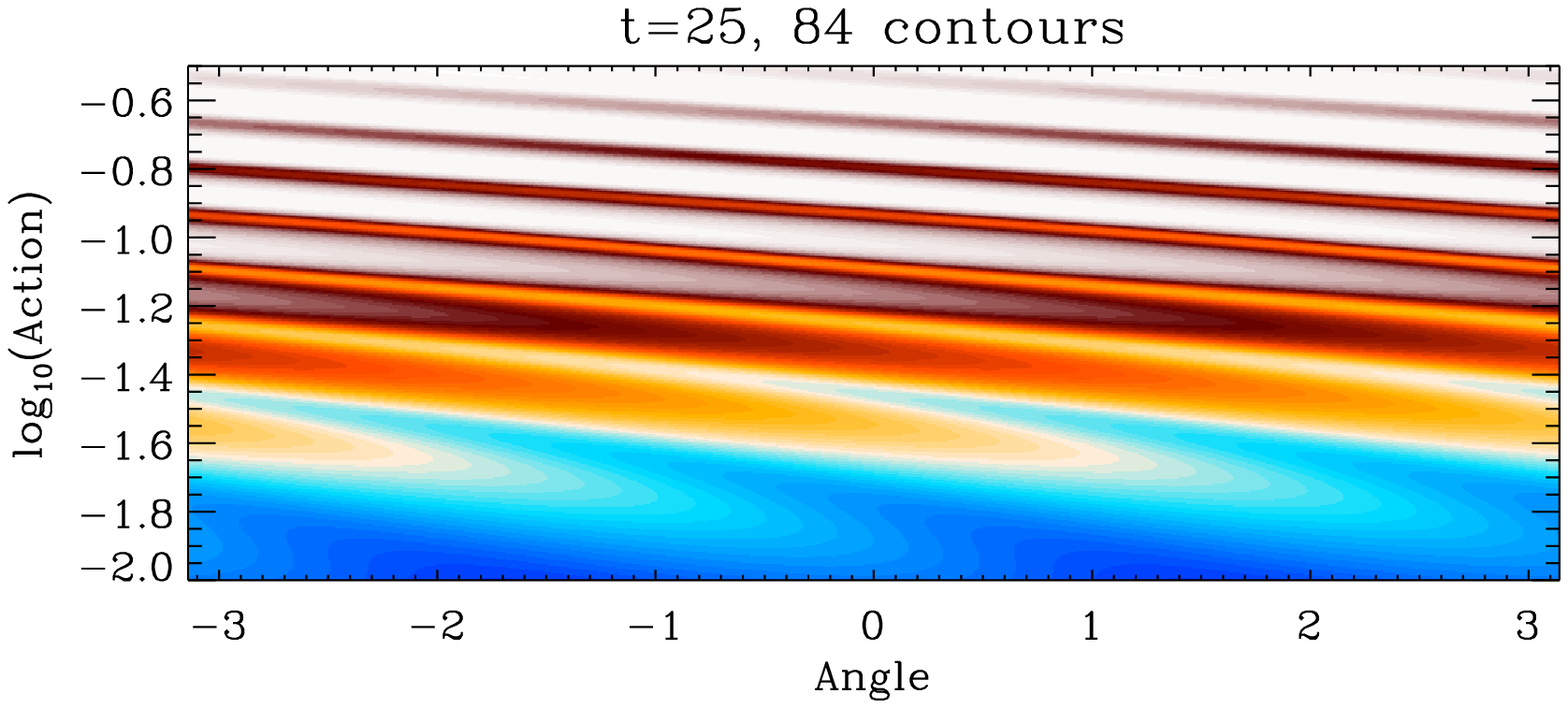,width=8.5cm,bbllx=54pt,bblly=360pt,bburx=558pt,bbury=577pt}}}
\centerline{\hbox{
\psfig{file=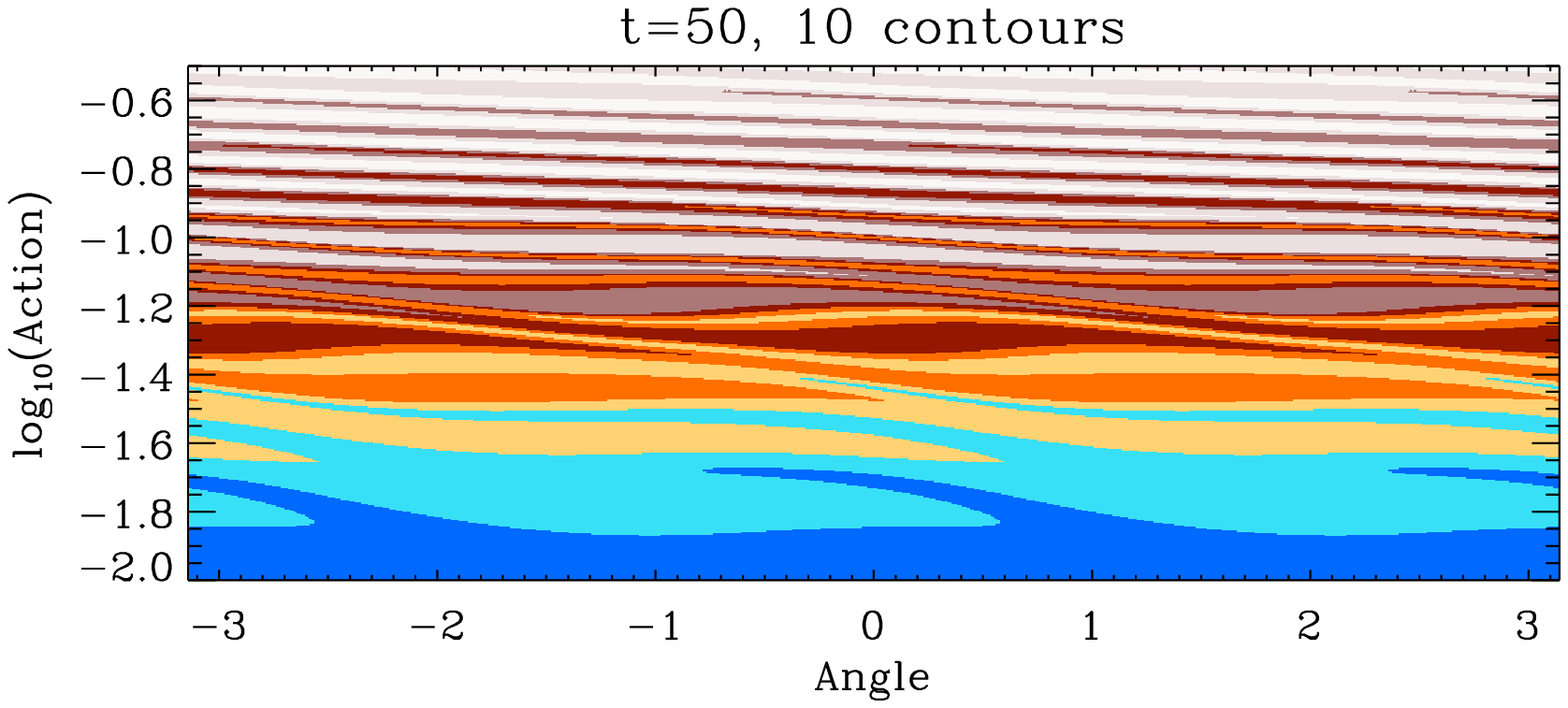,width=8.5cm,bbllx=54pt,bblly=360pt,bburx=558pt,bbury=577pt}
\hskip 0.4cm \psfig{file=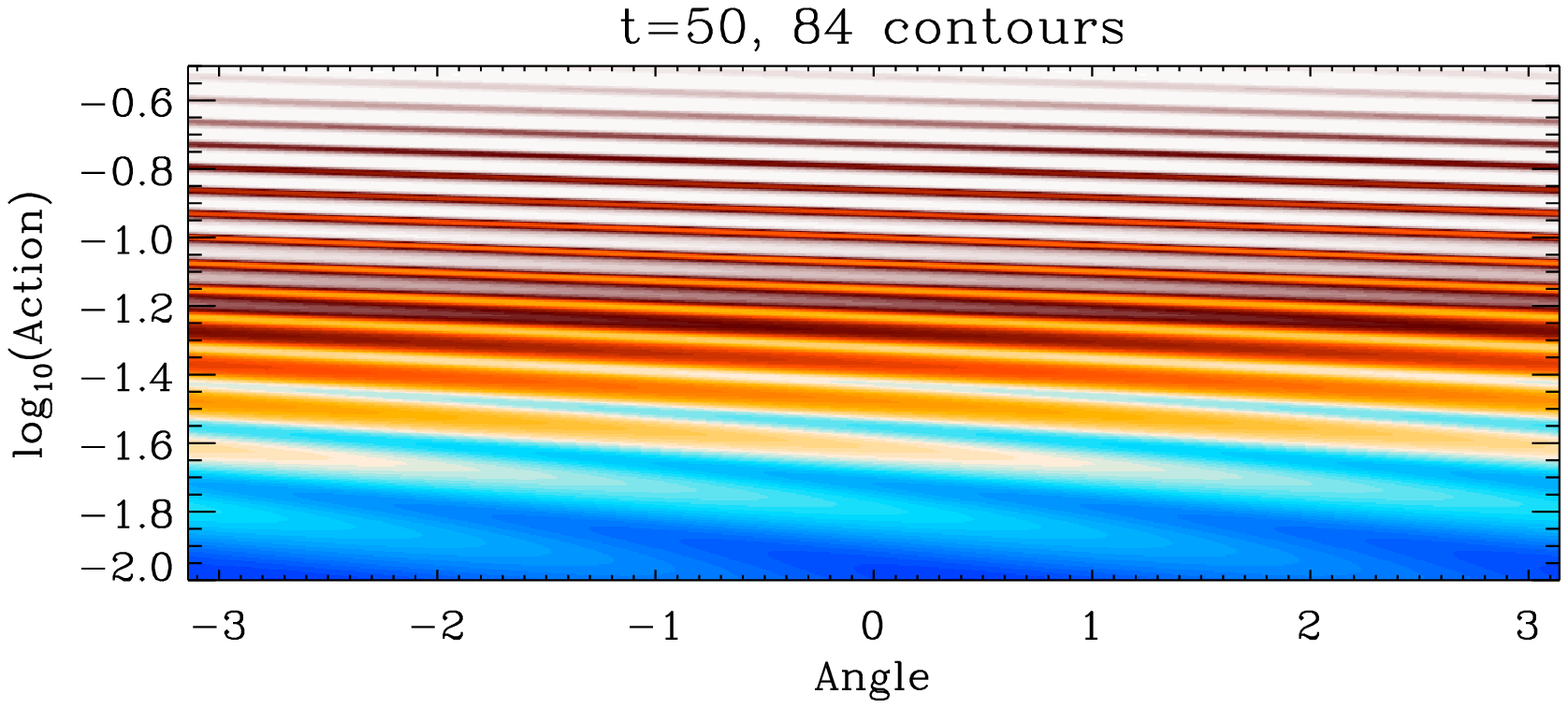,width=8.5cm,bbllx=54pt,bblly=360pt,bburx=558pt,bbury=577pt}}}
\centerline{\hbox{
\psfig{file=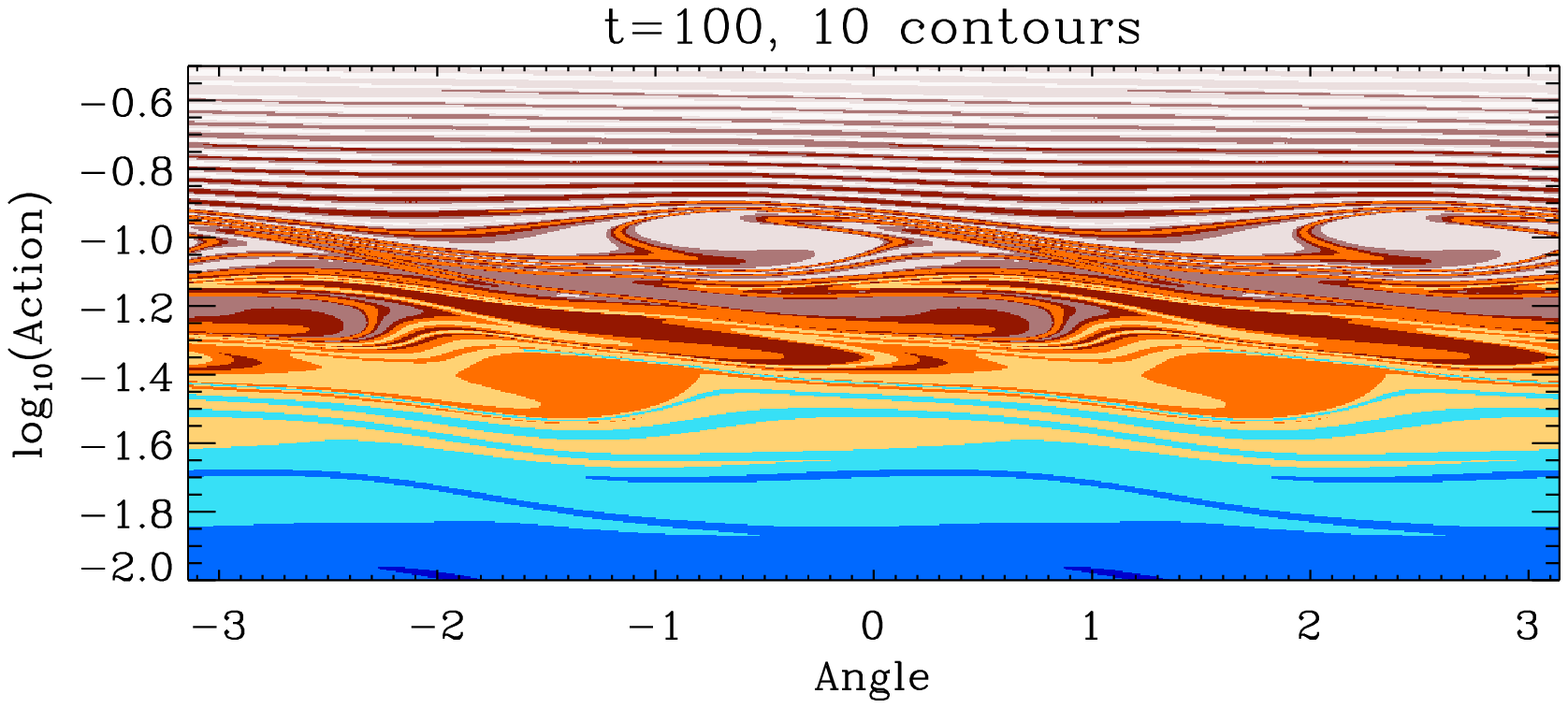,width=8.5cm,bbllx=54pt,bblly=360pt,bburx=558pt,bbury=577pt}
\hskip 0.4cm \psfig{file=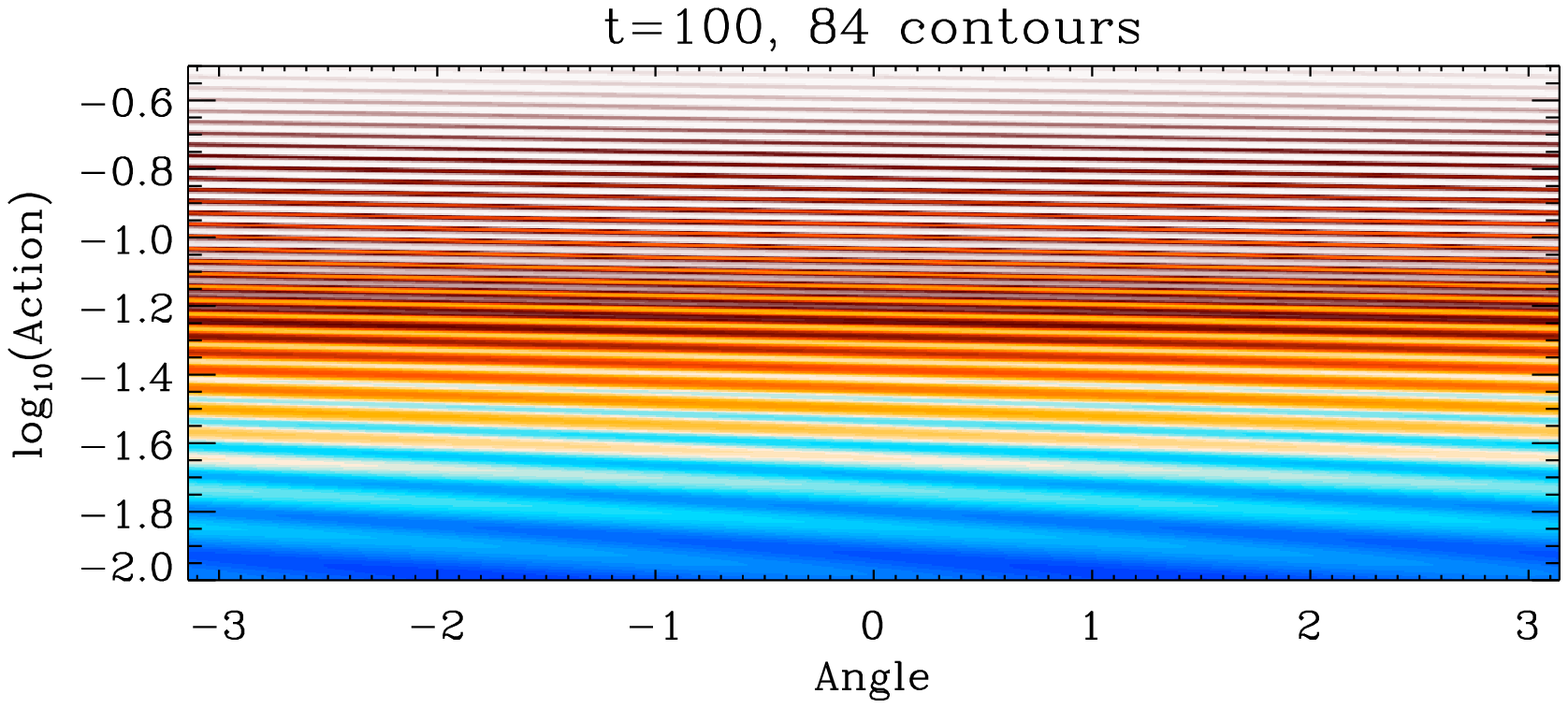,width=8.5cm,bbllx=54pt,bblly=360pt,bburx=558pt,bbury=577pt}}}
\vskip 0.2cm
\caption[]{The simulations with Gaussian initial phase-space distribution function of Fig.~\ref{fig:gaussianA} in Action-Angle space. The transformation from phase-space to Action-Angle space is described in Appendix~\ref{app:actionangles}. }
\label{fig:action_angle_gaussian}
\end{figure*}

Figure~\ref{fig:gaussianA} shows the results obtained with our waterbag code for these Gaussian initial conditions. It illustrates how important is the initial condition generation step. On the first and third line of panels, function $f(x,v)$ is sampled with only 10 waterbags, while on the second and fourth line, it is sampled with 84 waterbags. Although both simulations coincide with each other at early times, a non linear instability soon builds up in the 10 waterbags simulation, at variance with the 84 one, which remains quiescent. This is even clearer in Action-Angle coordinates, as displayed in Fig.~\ref{fig:action_angle_gaussian}: on the left column of panels, the poorness of initial waterbags sampling induces some oscillations, already visible at $t=25$, which amplify and create non-linear resonant instabilities. On the other hand, on the right column of panels, the 84 waterbags simulation presents the typical signature of Landau damping. The quiescent nature of the system is also confirmed by the fact that the total vertex number and the total length of the waterbag contours augment linearly with time (see Appendix~\ref{app:naddrem}).  Even though the instability observed in the 10 waterbags simulation might actually be present in the true system at the microscopic level, its early appearance is clearly due to the unsmooth representation of our waterbag approach. It can be delayed by augmenting the contour sampling. This effect would happen likewise in a $N$-body simulation \citep{2005MNRAS.359..123A}.

\subsubsection{Random set of warm halos: a chaotic system}
\label{sec:ransi}
\begin{figure}
\hbox{
\psfig{file=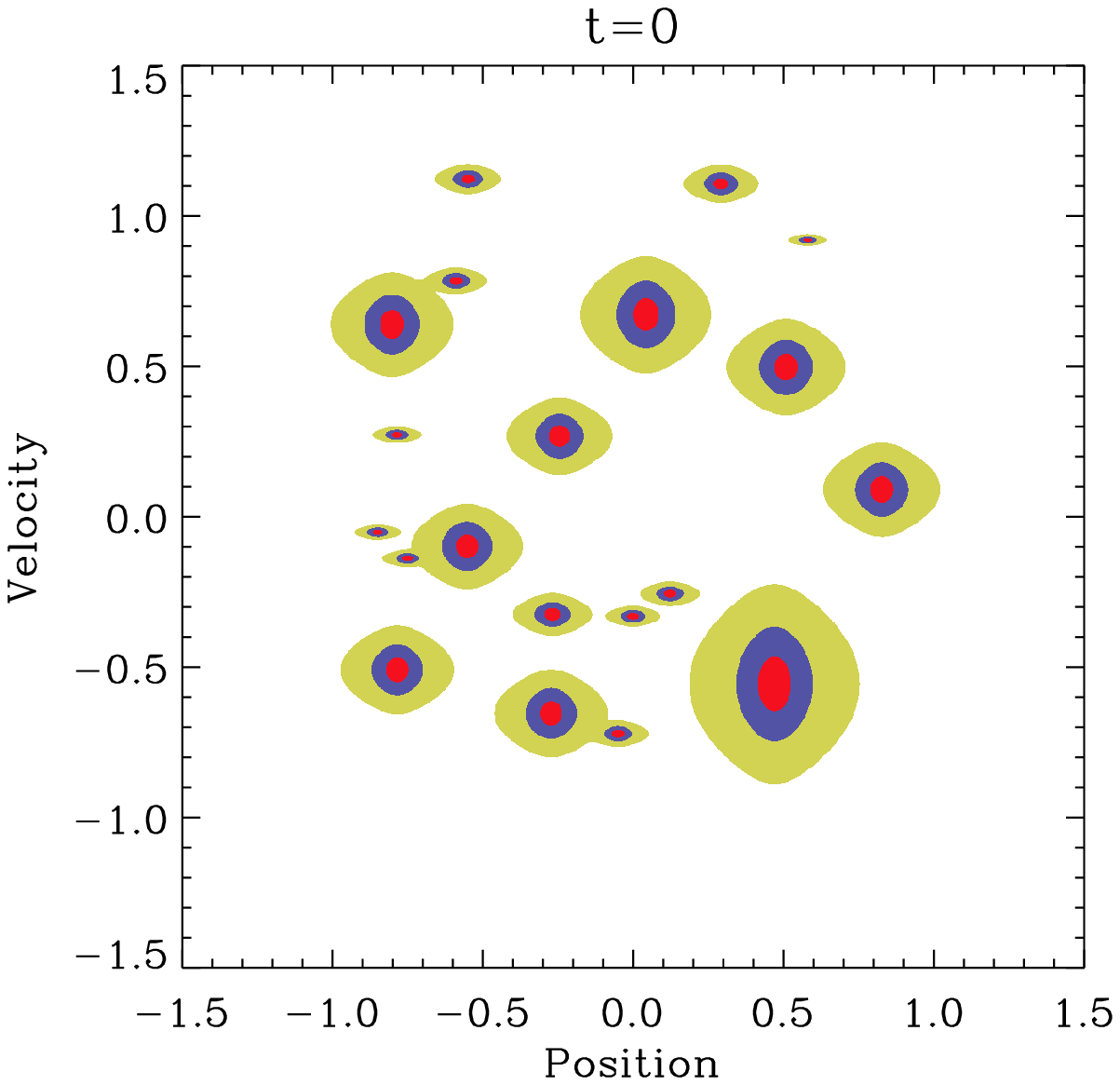,width=7.5cm,bbllx=62pt,bblly=366pt,bburx=426pt,bbury=718pt}}
\hbox{
\psfig{file=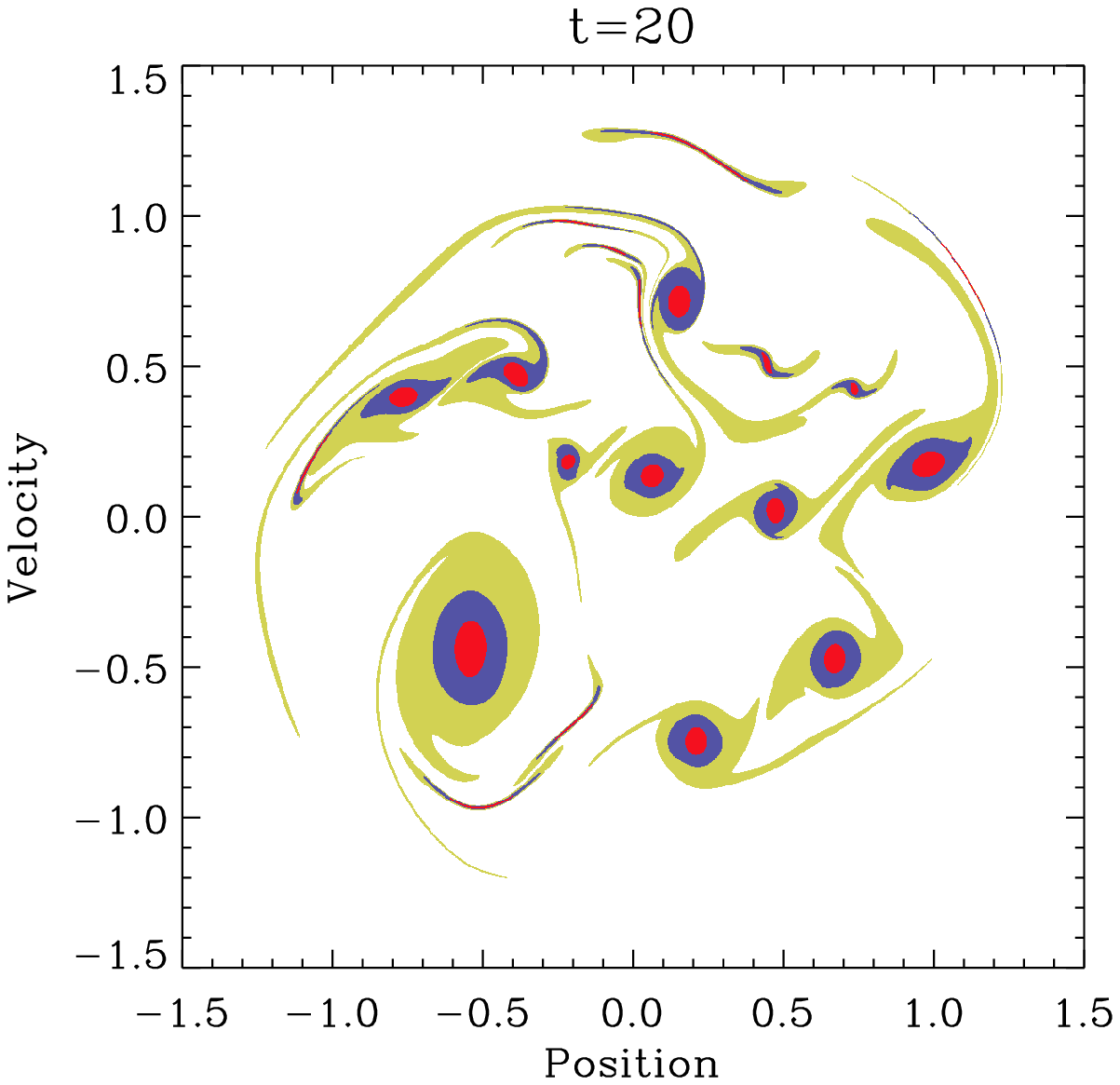,width=7.5cm,bbllx=62pt,bblly=366pt,bburx=426pt,bbury=718pt}}
\hbox{
\psfig{file=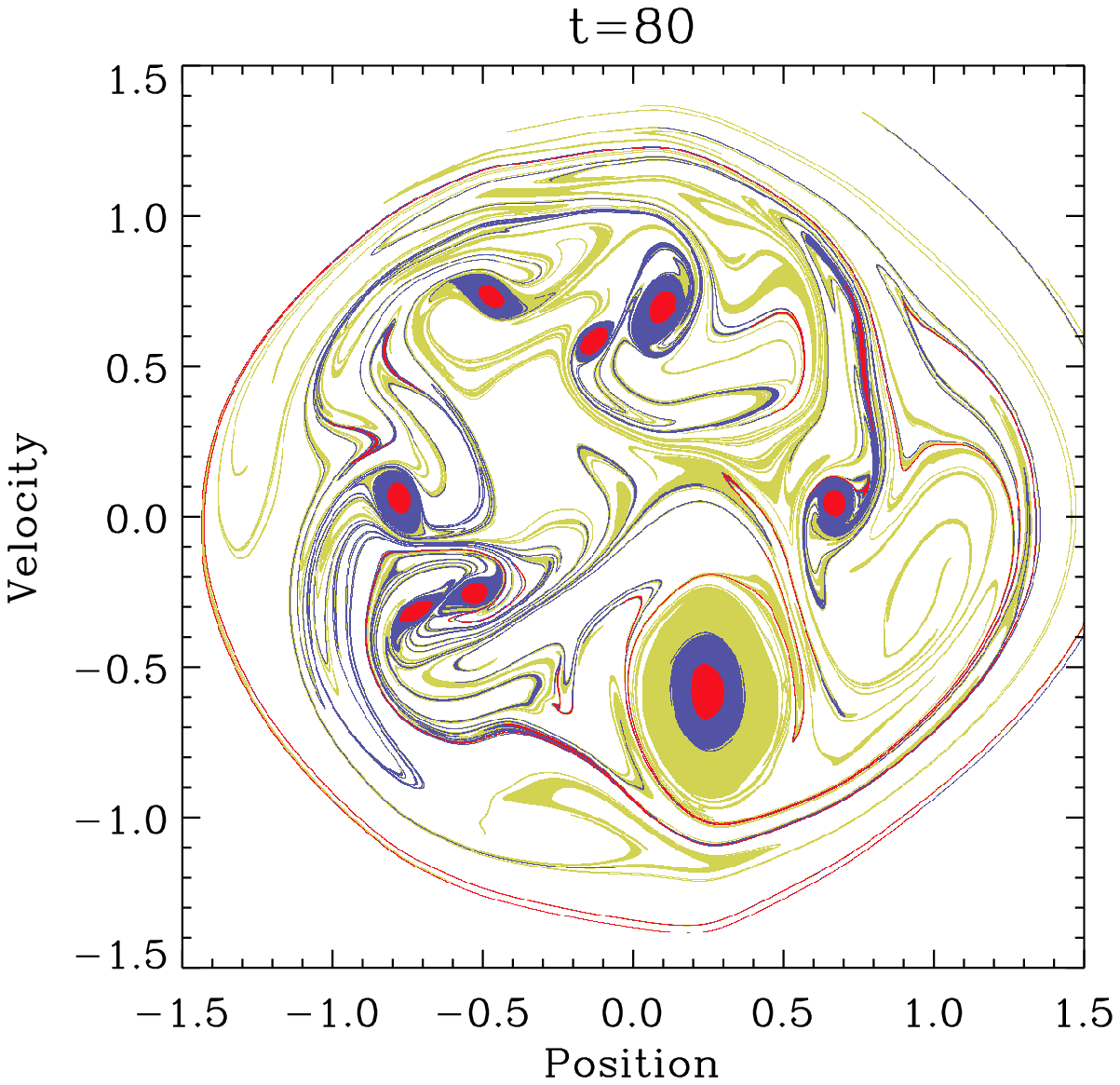,width=7.5cm,bbllx=62pt,bblly=366pt,bburx=426pt,bbury=718pt}}
\caption[]{Evolution of an initially random set of ``stationary'' halos. This system develops a chaotic behavior with a Lyapunov exponent of $0.05$ as a result of the gravitational interaction between the halos. The simulation used here corresponds to {\tt RandomU} in Table~\ref{tab:simuparam}, but other settings ({\tt RandomUT}, {\tt Random} and {\tt RandomUS}) would look exactly the same.}
\label{fig:random}
\end{figure}

Figure~\ref{fig:random} shows the case of an initially random set of halos, which represents our second test. Each halo is supposed to be at thermal equilibrium and is sampled with only three waterbags to minimize computational cost. As shown in appendix~\ref{app:naddrem}, this simulation soon builds up chaos with a Lyapunov exponent equal to $0.05$ as an effect of the gravitational interaction between the halos (this effect is dominant other instabilities that might develop due to the contour undersampling just discussed above). This numerical experiment represents thus an important test of the accuracy of the code in rather extreme conditions, somewhat opposite to the quiescent case provided by the smooth Gaussian $f(x,v)$ of previous section.

\subsubsection{Single waterbags with varying thickness: from warm to nearly cold initial conditions}
\label{sec:sinwai}
The single waterbag obviously corresponds to the simplest application of the method. It was used for instance in the seminal works of  \cite{1971A&A....11..188J} and \cite{1971Ap&SS..13..411C,1971Ap&SS..13..425C} but also subsequently in many other studies. It represents a useful way to cover a large range of initial conditions, from warm to nearly cold. The close to cold case represents by itself a challenge to simulate due to the nearly singular structures that build up in configuration space  during the course of dynamics.  

The initial configurations we consider, abusively denoted by ``top hat'', are such that the waterbag boundary is an ellipse: 
\begin{equation}
x^2+(v/\Delta p)^2=1,
\label{eq:wat}
\end{equation}
where $\Delta p$ is a parameter quantifying the initial thickness of the waterbag. Modifying $\Delta p$ is equivalent to changing the initial velocity dispersion while keeping unchanged the projected initial density profile. The total mass of the system is chosen to be unity. We performed a number of simulations with a large range of values of $\Delta p$ in the interval $[0.001,1]$. For $\Delta p=0.003$, we also performed simulations where the initial boundaries of the waterbag are perturbed randomly. The visual inspection of these simulations (Figs.~\ref{fig:convtocolda} to \ref{fig:tophatae}) will be discussed in \S~\ref{sec:tophatvisu}.  
%%%
\subsection{Runtime algorithm and tests of its performances}
\label{sec:runtime}
\subsubsection{Time integration} 
\label{sec:timeintegration}
To move the sampling points of the waterbag contours, we use the classic splitting scheme of \citet{1976JCoPh..22..330C} with a slowly varying time step: our algorithm is thus equivalent to a predictor-corrector scheme, as indicated on Fig.~\ref{fig:algorithm}. It reduces to a symplectic ``leap-frog'' when the time step is kept constant \citep[see, e.g.,][]{1988csup.book.....H}.  

Note that at the end of time integration, we recast coordinates in the center of mass frame.
%%%
\subsubsection{Poisson equation resolution} 
\label{sec:point3}
This step, of which the technical details are given in Appendix~\ref{sec:orientated}, is quite simple from the conceptual point of view, since it consists in circulating along $\partial P$ by performing a sum over the polygon edges to compute integral (\ref{eq:circu2}), after a preliminary sort of the vertices of the polygon. However, despite its apparent simplicity, it corresponds by far to the most costly part of the code from the computational point of view, because many segments of the polygon can contribute to the force exerted on one point of space. Note that the circulation technique used to compute the force can be generalized to the calculation of other useful quantities, such as the projected density, $\rho(x)$, the mass profile, $M_{\rm left}(x)$, the gravitational potential, $\phi(x)$, the bulk velocity and the local velocity dispersion, as detailed in Appendix~\ref{app:A}.
%%%
\subsubsection{Local refinement} 
\label{sec:myref}
\begin{figure}
\centerline{\hbox{
\psfig{file=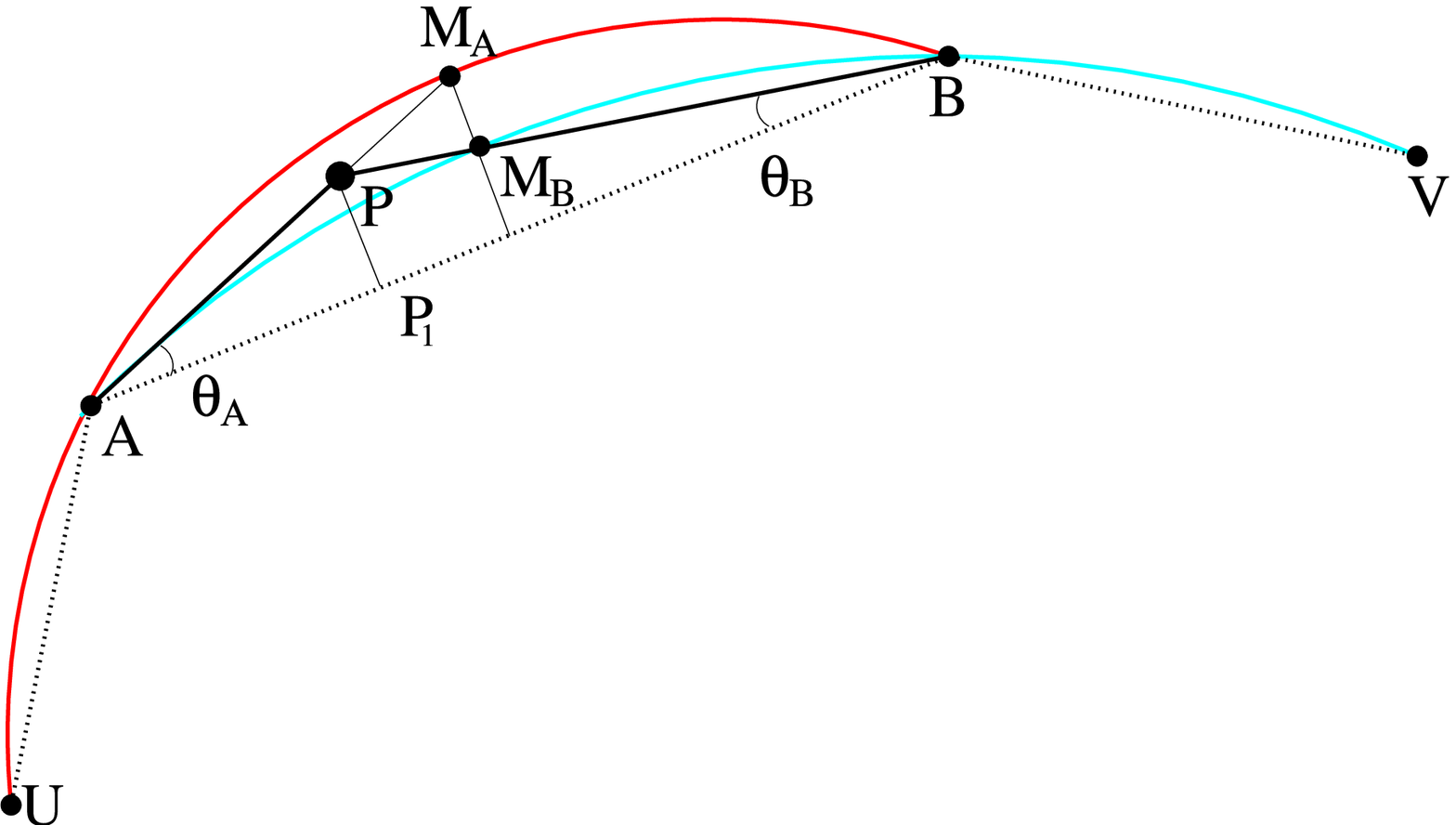,width=8cm}}}
\centerline{\hbox{
\psfig{file=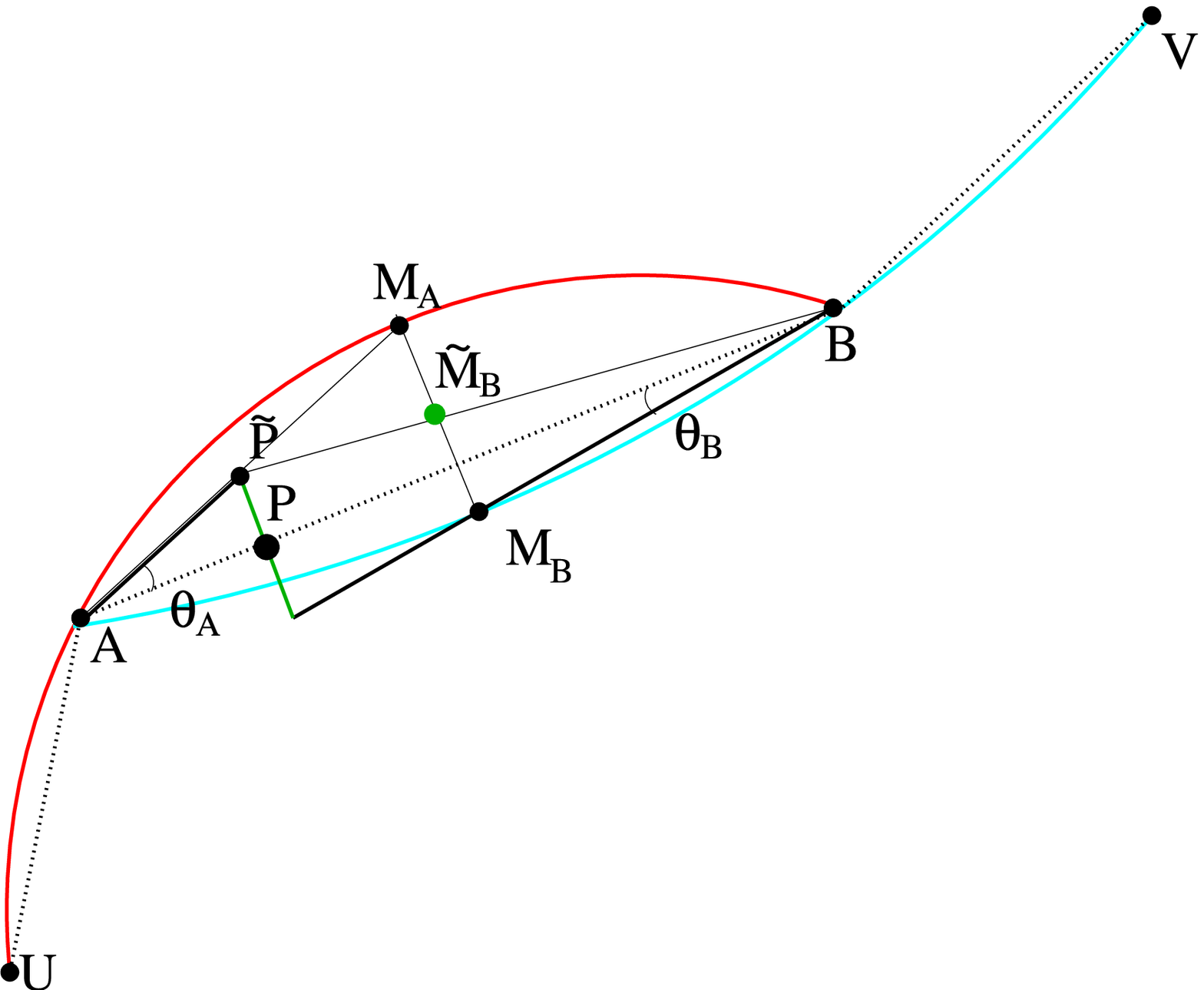,width=8cm}
}}
\caption[]{Interpolation method used for adding a new point $P$ to the orientated polygon. {\em Top panel:} using respectively the arcs of circle $C_A=\widehat{UAB}$ of radius $R_A$ and $C_B=\widehat{ABV}$ of radius $R_B$, the natural position of $P$ would be $M_A$ and $M_B$. A compromise between these two solutions is taken to be the intersection between the lines passing through segments $[A,M_A]$ and $[B,M_B]$. With this procedure, the local curvature estimated as the inverse of the radius of the arc of circle $C_P=\widehat{APB}$ is bounded by that measured at points $A$ and $B$. It converges to the usual interpolation $2/R_P=1/R_A+1/R_B$ in the small angle approximation. {\em Bottom panel:} if there is a change in the sign of local curvature, the choice of point $P$ is undefined. However, the smooth curve approximated by the four points $U$, $A$, $B$ and $V$ should intersect with segment $[A,B]$. We choose $P$ to be at the locus of this intersection: similarly as in top panel, one computes the point ${\tilde P}$ of intersection between the lines passing through $[A,M_A]$ and $[B,{\tilde M}_B]$, where ${\tilde M}_B$ is the symmetric of $M_B$ with respect to the segment $[A,B]$. Then $P$ is just the projection of ${\tilde P}$ on segment $[A,B]$. This procedure does not interpolate anymore local curvature in the small angle limit, but this is necessary to preserve the stability of refinement in terms of small rotations between successive segments of the waterbag borders. The panels of this figure are reprinted from \citet{2008CNSNS..13...46C} with permission from Elsevier.}
\label{fig:interpol}
\end{figure}
When the shape of  the waterbags contours becomes complex, it is necessary to add points to the orientated polygon to preserve all its details. Our refinement procedure is described in Fig.~\ref{fig:interpol} (see also Appendix~\ref{app:refdetails}). It consists of a geometric construct using arcs of circle passing through sets of three successive points of the polygon.  It is equivalent, in the small angle approximation, to linearly interpolating local curvature given as the inverse of the radius of these arc of circles. This refinement procedure is stable in the sense that it is ``Total Variation Preserving'' in terms of the small rotations between successive segments of waterbags borders and that it makes these borders less angular (Appendix~\ref{app:refstability}).  

Refinement is performed  when the variation of phase-space area $S$ induced by adding a refinement point exceeds some threshold $S_{\rm add}$ or when the distance between  two successive points of a contour exceeds som threshold $d_{\rm add}$, e.g., 
\begin{eqnarray}
S(\widehat{APB}) & > & S_{\rm add}, \label{eq:Saddcrit} \\
d_{AB} & > & d_{\rm add}, \label{eq:daddcrit}
\end{eqnarray}
on top panel of Fig.~\ref{fig:interpol}, where $S(\widehat{APB})$ is the area of the triangle $\widehat{APB}$ and $d_{AB}$ is the distance between $A$ and $B$.\footnote{for the bottom panel, we use $S(\widehat{A{\tilde P}B})$ instead of $S(\widehat{APB})$ in equation (\ref{eq:Saddcrit}).} The way $S_{\rm add}$ and $d_{\rm add}$ should be chosen is discussed in Appendix~\ref{sec:critraf}. Table~\ref{tab:simuparam} gives their values for the simulations we did: we have $S_{\rm add} \in [10^{-10},10^{-7}]$ and $d_{\rm add}=0.01$ or $0.02$. 

To make the algorithm more optimal, we also propose an unrefinement scheme, similarly as in \cite{1971Ap&SS..13..411C}: on Fig.~\ref{fig:interpol} points $P$ with
\begin{eqnarray}
S(\widehat{APB}) & \leq & S_{\rm rem}, \label{eq:remcrit1}\\
\min(d_{AP},d_{PB}) & \leq & d_{\rm rem}, \label{eq:remcrit2} 
\end{eqnarray}
are removed, if not violating condition (\ref{eq:Saddcrit}) and (\ref{eq:daddcrit}), of course, and if there is no local curvature sign change. In practice, $S_{\rm rem}=S_{\rm add}/2$ and $d_{\rm rem}=d_{\rm add}/2$. More technical details are given in Appendix~\ref{sec:critraf}. 

Despite its potential virtues, same accuracy for smaller computational cost, allowing unrefinement is not optimal in our 1D case if one aims to follow a system during many dynamical times. It is indeed possible to show that vertex number dynamics changes dramatically when unrefinement is activated (Appendix~\ref{app:naddrem}). In particular, unrefinement is susceptible to introduce long term noise after multiple orbital times, due to the fact that pieces of waterbag contours are alternatively refined and unrefined many times. The effects of this long term noise can evidenced by measurements of total energy conservation violation, as discussed below.  
%%%
\subsubsection{Diagnostics} 
\label{sec:point5}
Diagnostics include, of course, calculation of the value of the next time step used in the time integrator described in \S~\ref{sec:timeintegration}.  To follow accurately the evolution of the system during many orbital times, we use a classic dynamical constraint on the time step modulated by two important conditions to limit excessive refinement of the polygon due to curvature generation and contour stretching (Appendix~\ref{app:dtdetails}). Our main constraint for the time step is:
\begin{equation}
{\rm d}t \leq {\rm d}t_{\rm dyn} \equiv \frac{C}{\sqrt{\rho_{\rm max}}}, \quad C \sqrt{N_{\rm orbits}}\ll 1,
\label{eq:courant}
\end{equation}
where $\rho_{\rm max}$ is the maximum value of the projected density calculated over all the vertices and $N_{\rm orbits}$ is the number of orbital times. This dynamical criterion can be derived in a simple fashion by studying the particular case of the harmonic oscillator \citep[Appendix~\ref{sec:harmodt}; see also][]{2005MNRAS.359..123A}. Since $C$ is inversely proportional to the square root of the number of dynamical times at play, it depends strongly on the type of system studied. Table~\ref{tab:simuparam} shows that $C$ ranges from $5 \times 10^{-4}$ to $0.025$ for all the simulations we did. Because of our rather conservative choices for the values of $C$, the two other constraints on the time step related to polygon refinement, which are derived in Appendix~\ref{sec:refidt}, were found in practice to be subdominant compared to equation (\ref{eq:courant}), but it is definitely possible to construct setups where it is not the case.
\begin{figure*}
\centerline{\hbox{
\psfig{file=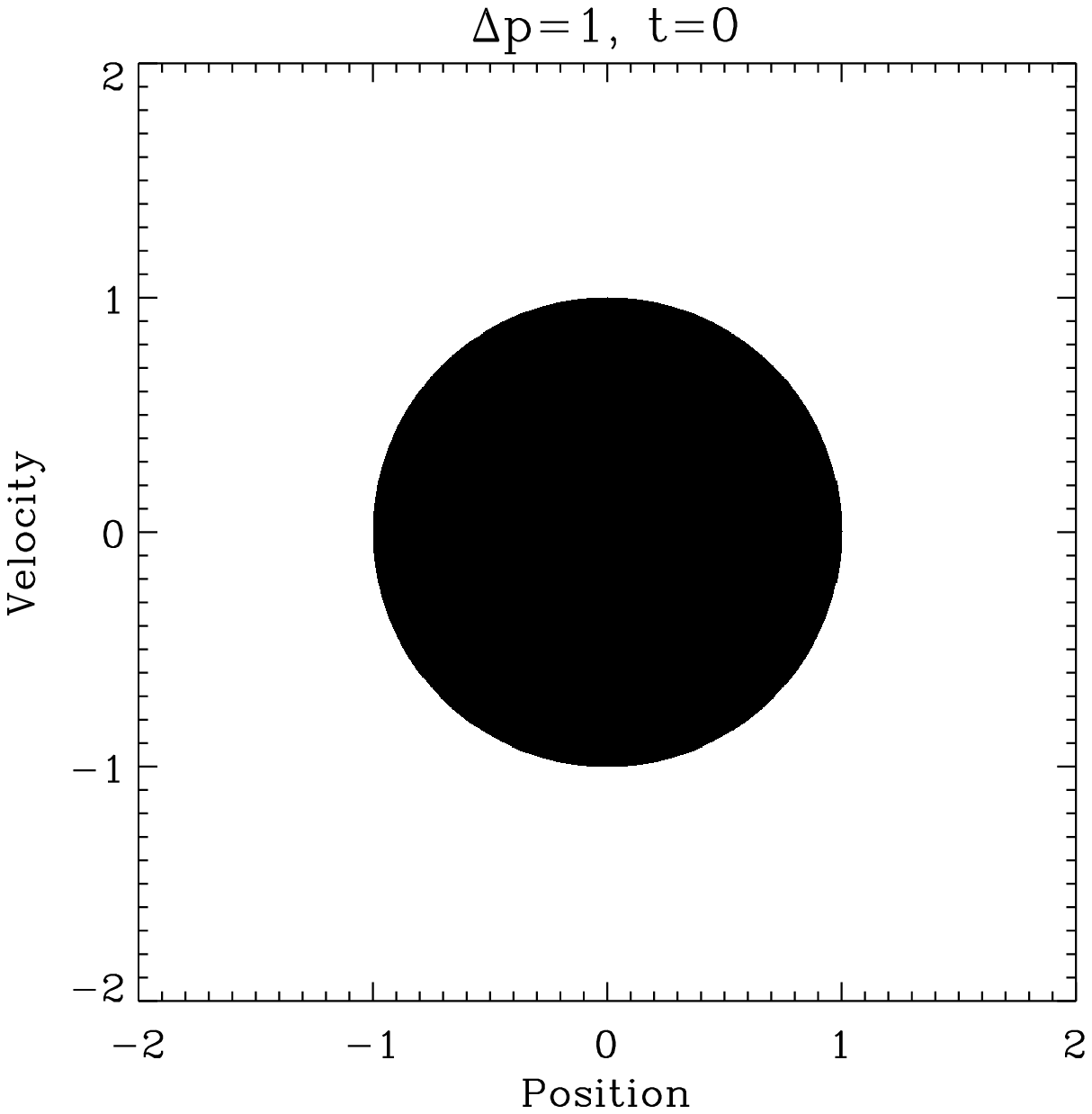,width=5.5cm,bbllx=62pt,bblly=366pt,bburx=426pt,bbury=718pt}
\hskip 0.4cm \psfig{file=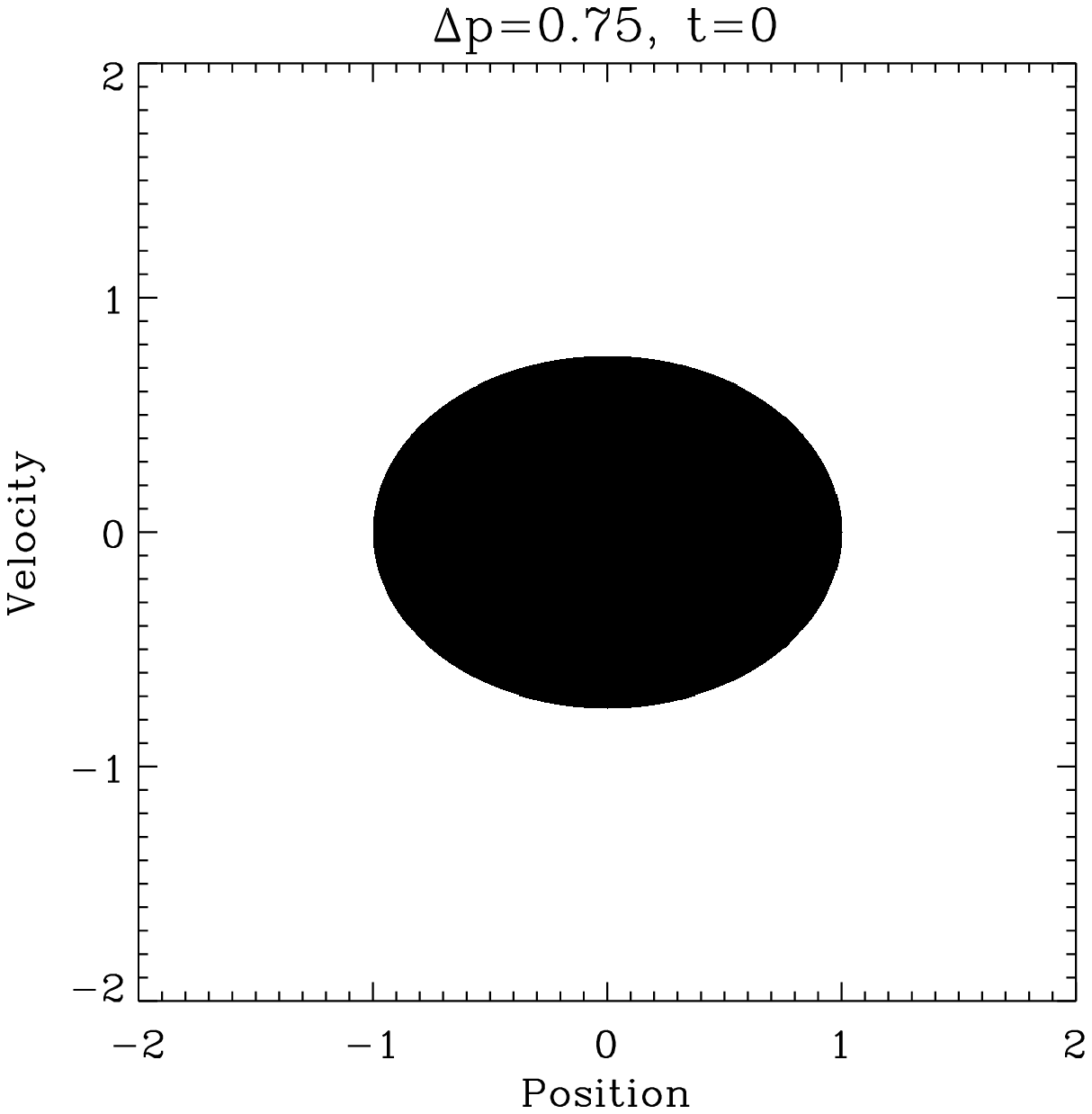,width=5.5cm,bbllx=62pt,bblly=366pt,bburx=426pt,bbury=718pt}
\hskip 0.4cm \psfig{file=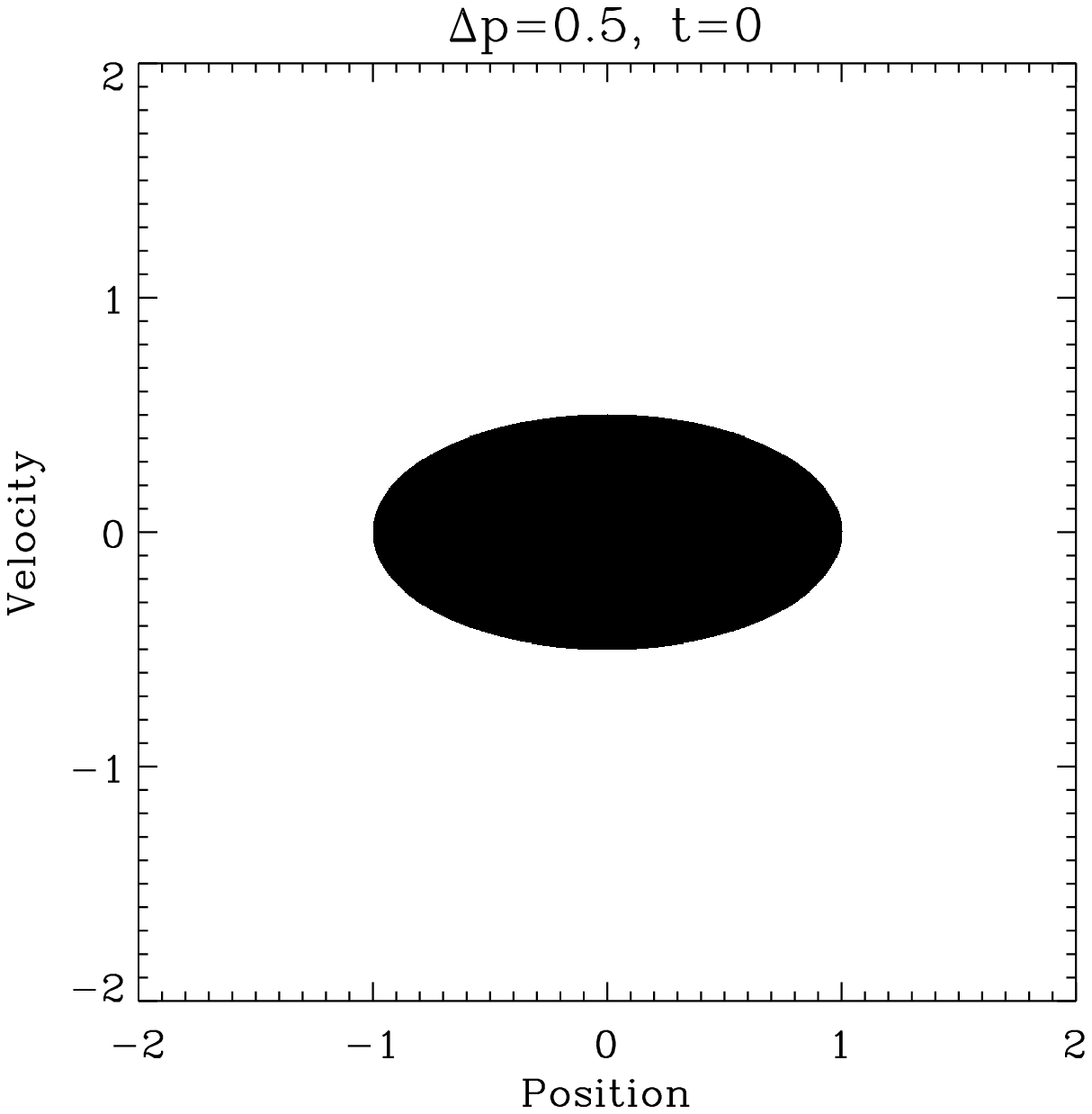,width=5.5cm,bbllx=62pt,bblly=366pt,bburx=426pt,bbury=718pt}
}}
\vskip 0.2cm
\centerline{\hbox{
\psfig{file=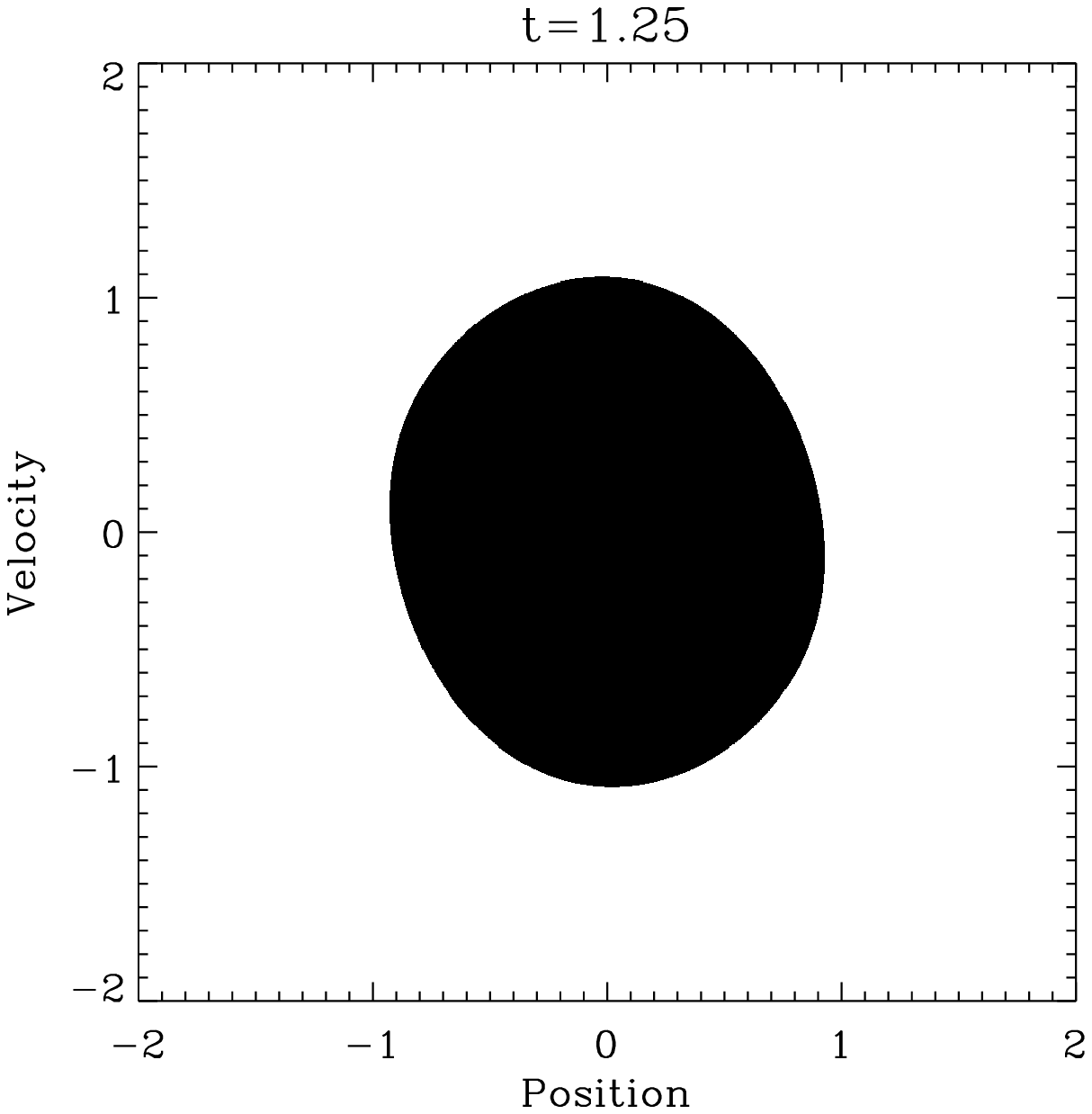,width=5.5cm,bbllx=62pt,bblly=366pt,bburx=426pt,bbury=718pt}
\hskip 0.4cm \psfig{file=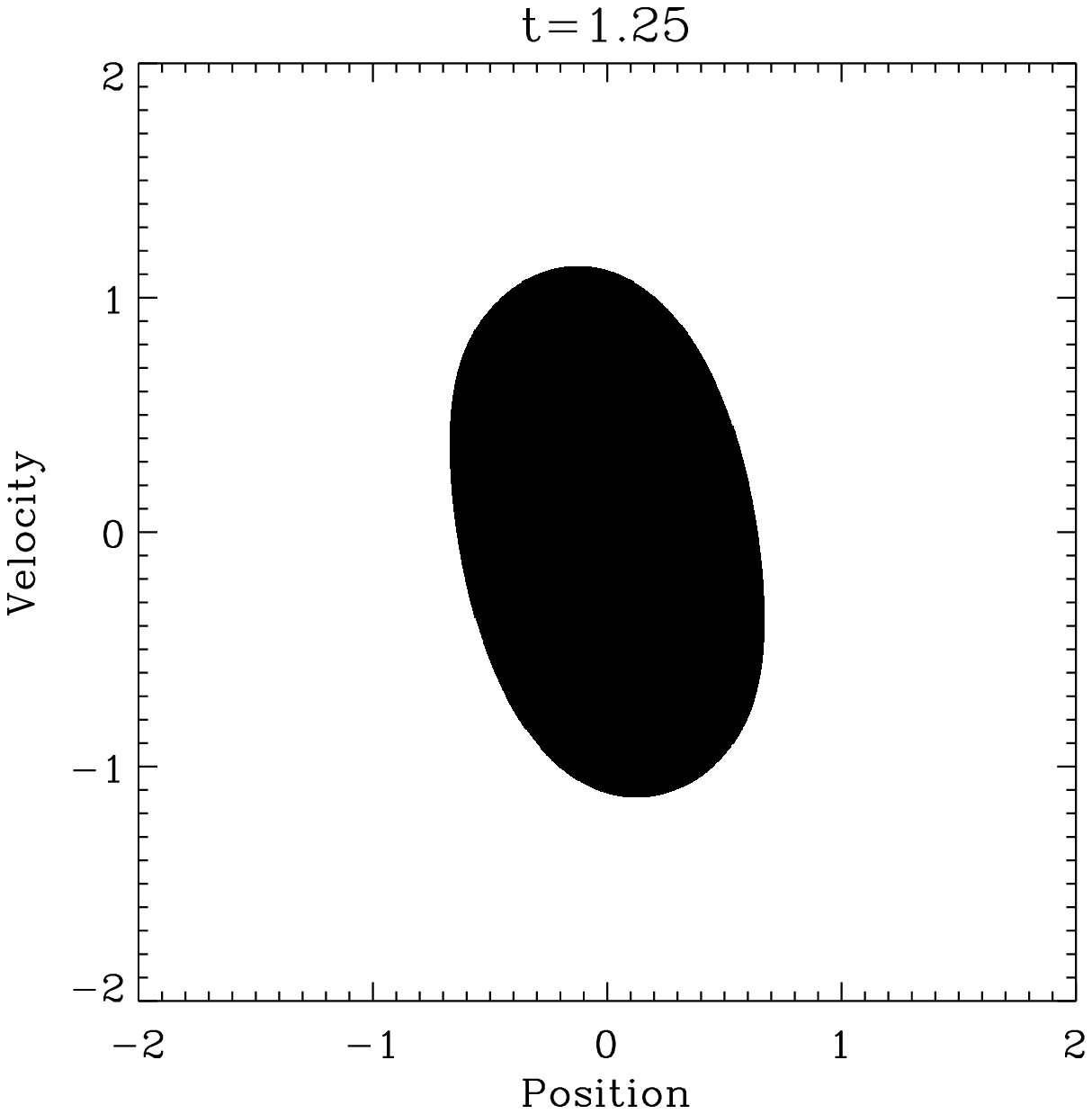,width=5.5cm,bbllx=62pt,bblly=366pt,bburx=426pt,bbury=718pt}
\hskip 0.4cm \psfig{file=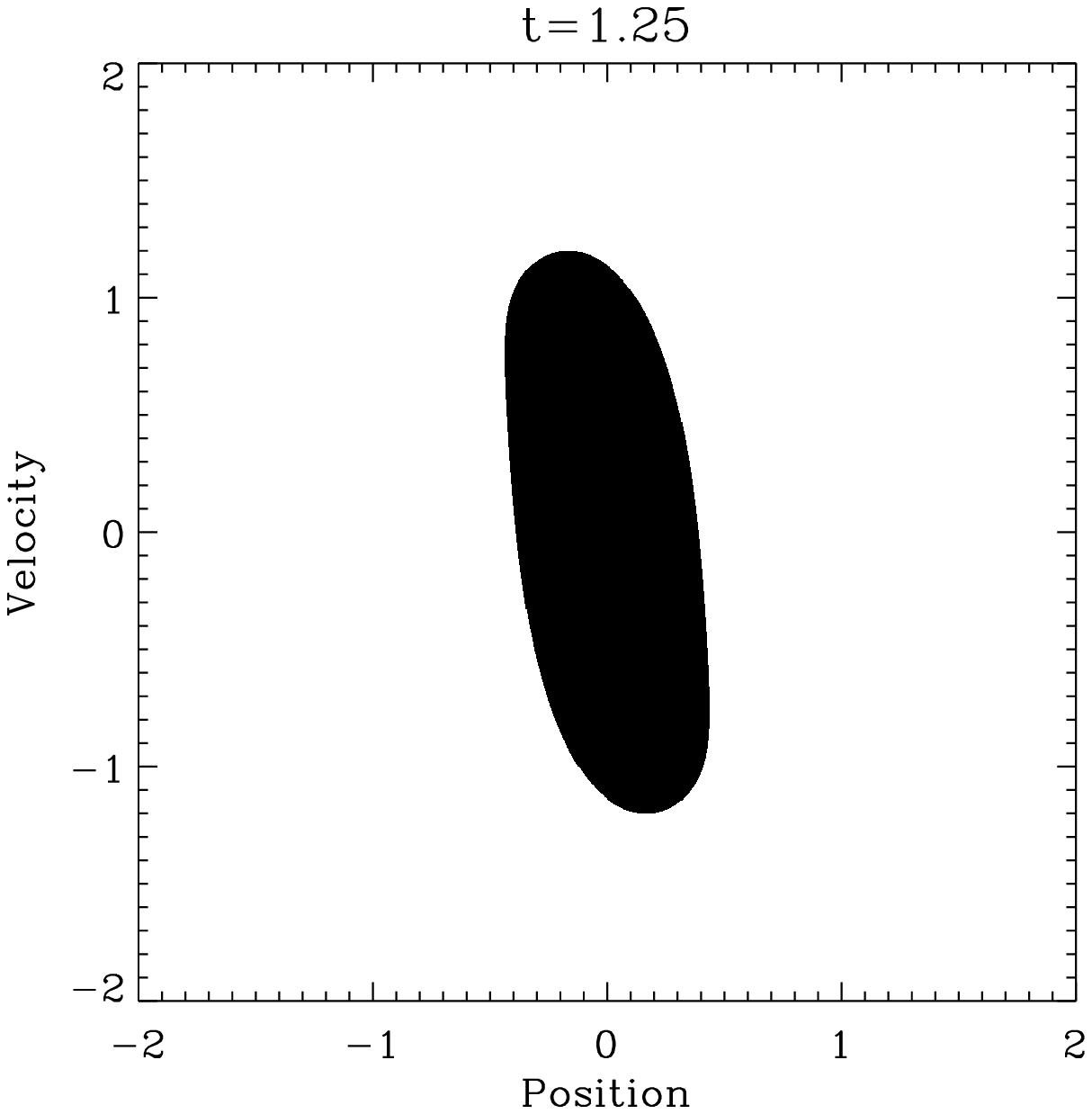,width=5.5cm,bbllx=62pt,bblly=366pt,bburx=426pt,bbury=718pt}
}}
\vskip 0.2cm
\centerline{\hbox{
\psfig{file=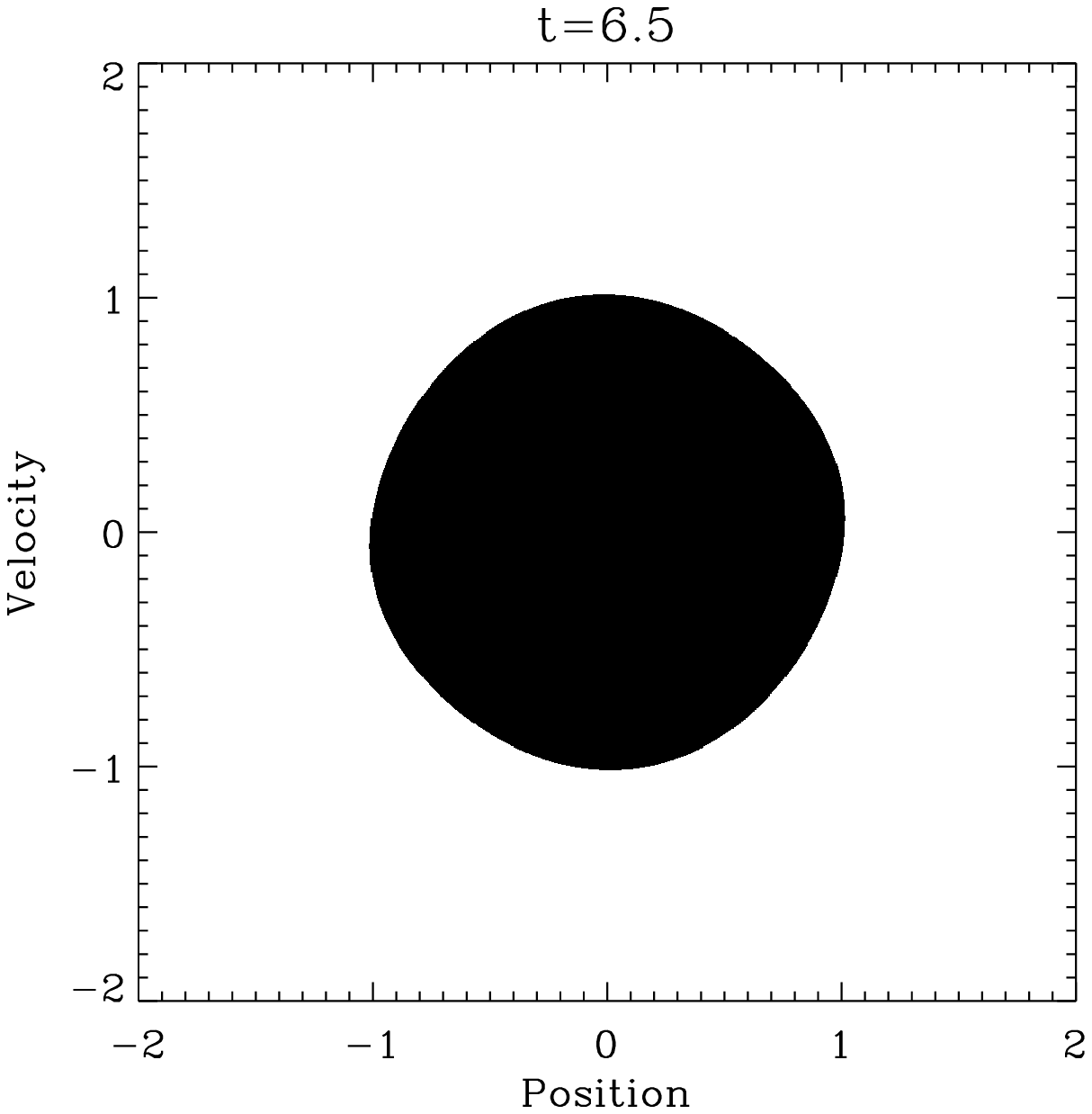,width=5.5cm,bbllx=62pt,bblly=366pt,bburx=426pt,bbury=718pt}
\hskip 0.4cm \psfig{file=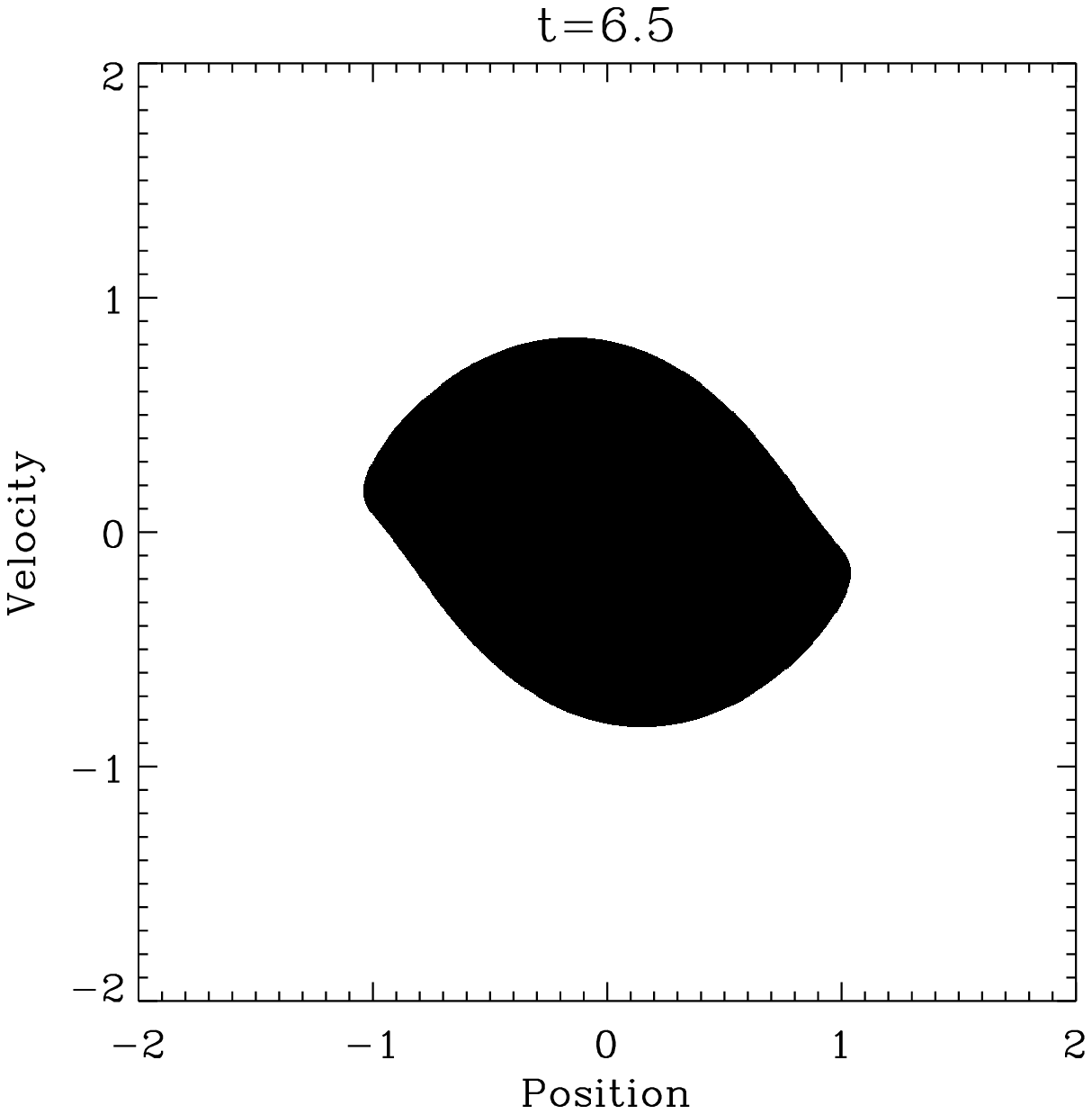,width=5.5cm,bbllx=62pt,bblly=366pt,bburx=426pt,bbury=718pt}
\hskip 0.4cm \psfig{file=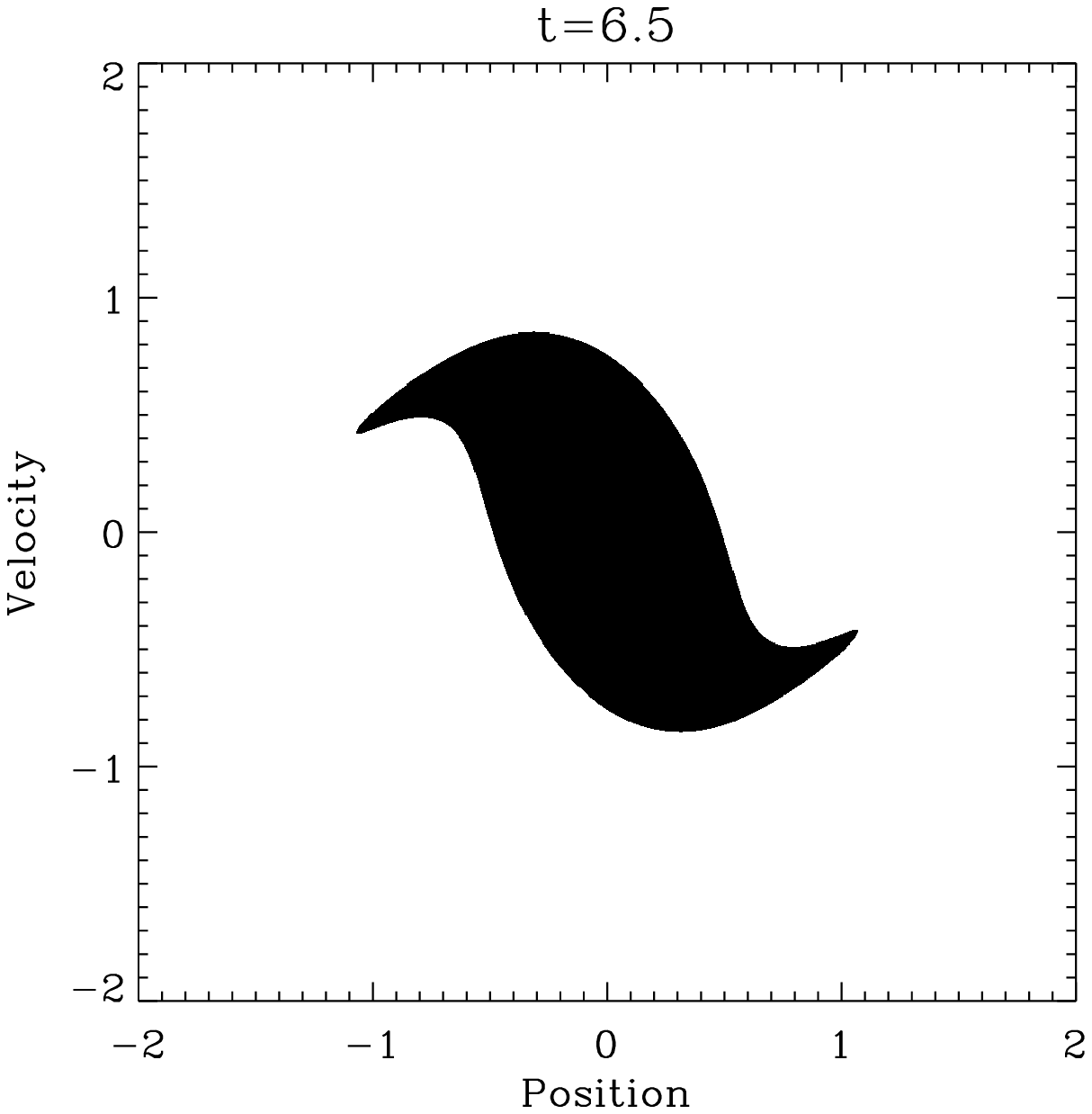,width=5.5cm,bbllx=62pt,bblly=366pt,bburx=426pt,bbury=718pt}
}}
\vskip 0.2cm
\centerline{\hbox{
\psfig{file=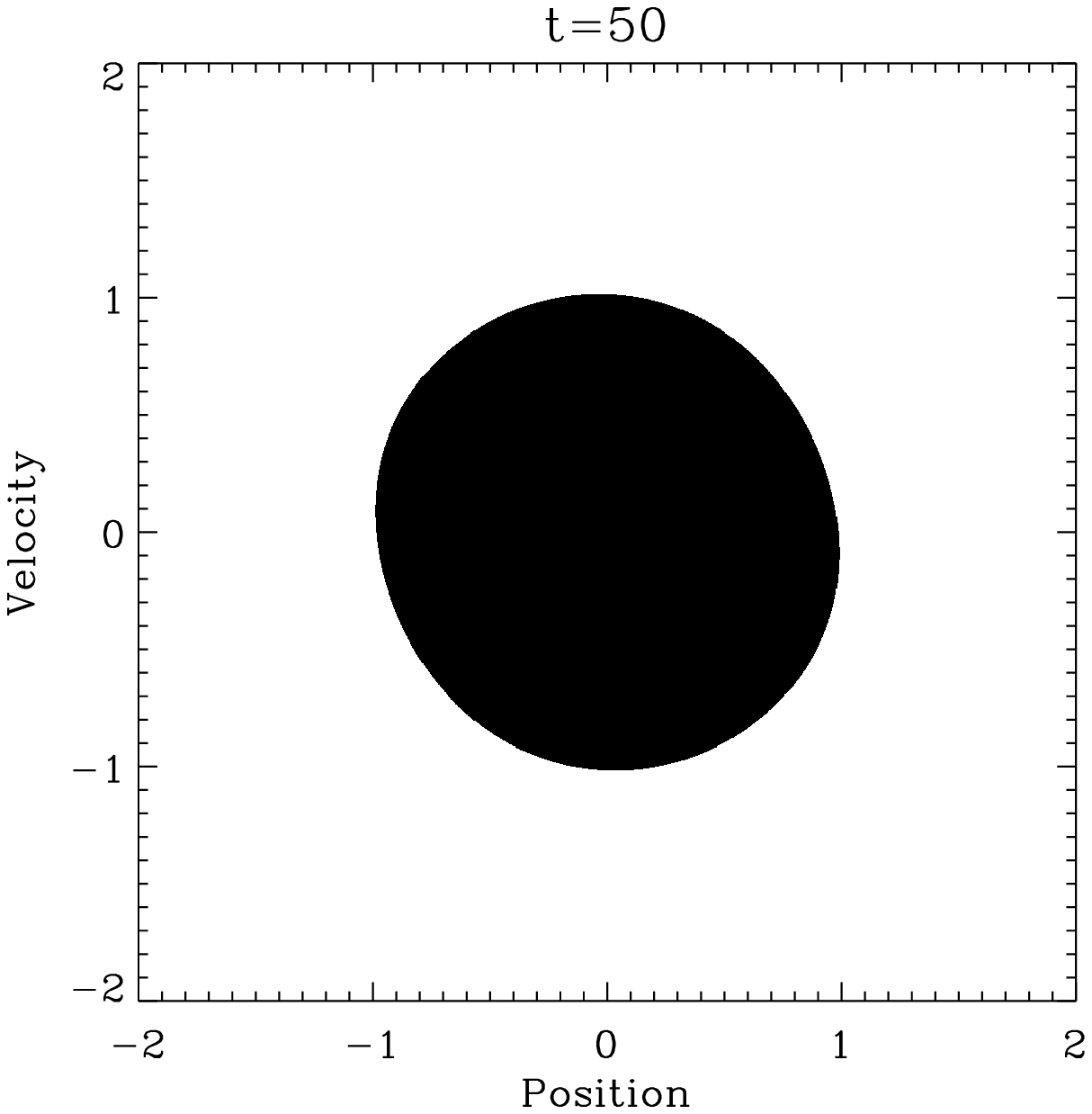,width=5.5cm,bbllx=62pt,bblly=366pt,bburx=426pt,bbury=718pt}
\hskip 0.4cm \psfig{file=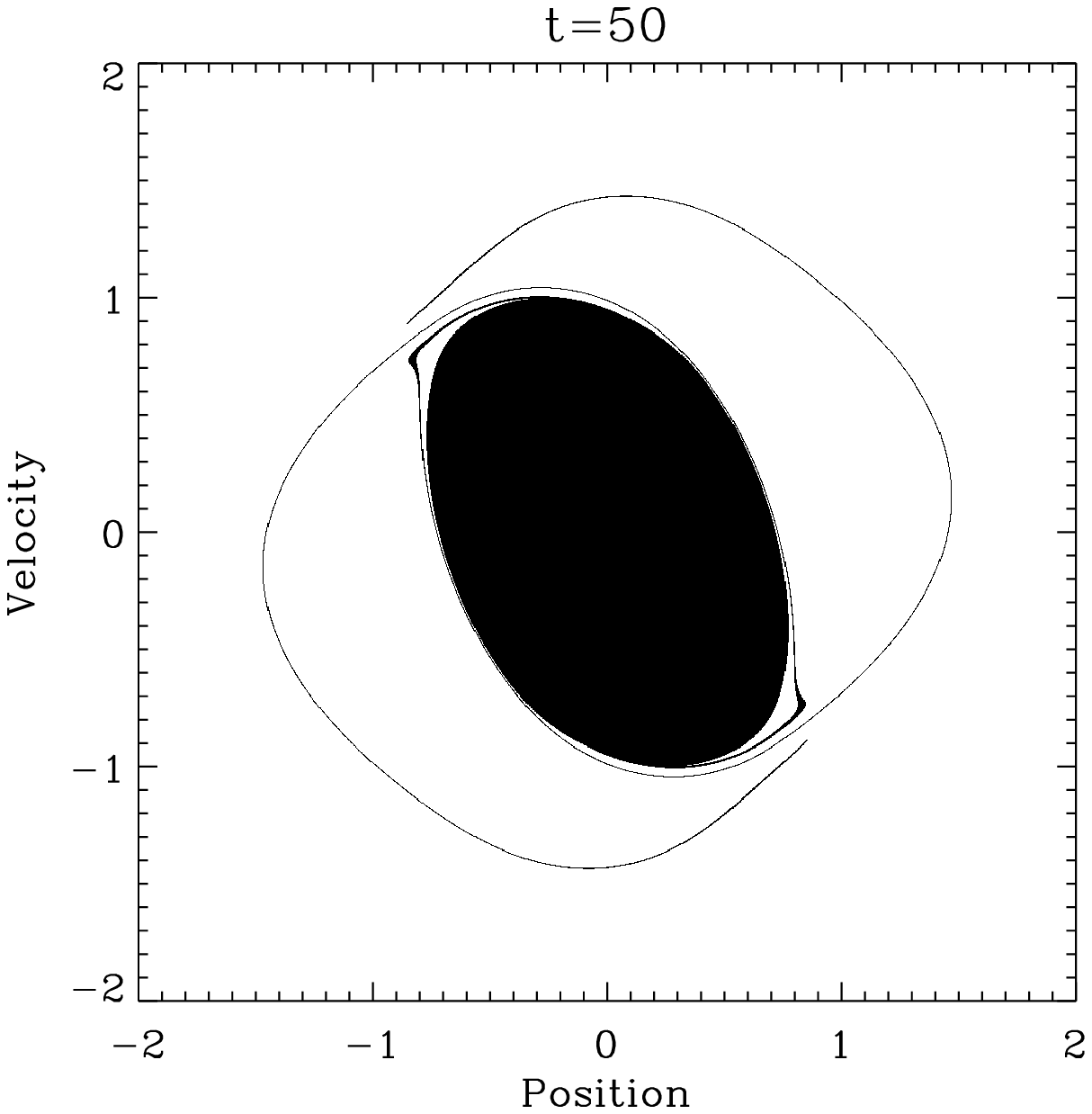,width=5.5cm,bbllx=62pt,bblly=366pt,bburx=426pt,bbury=718pt}
\hskip 0.4cm \psfig{file=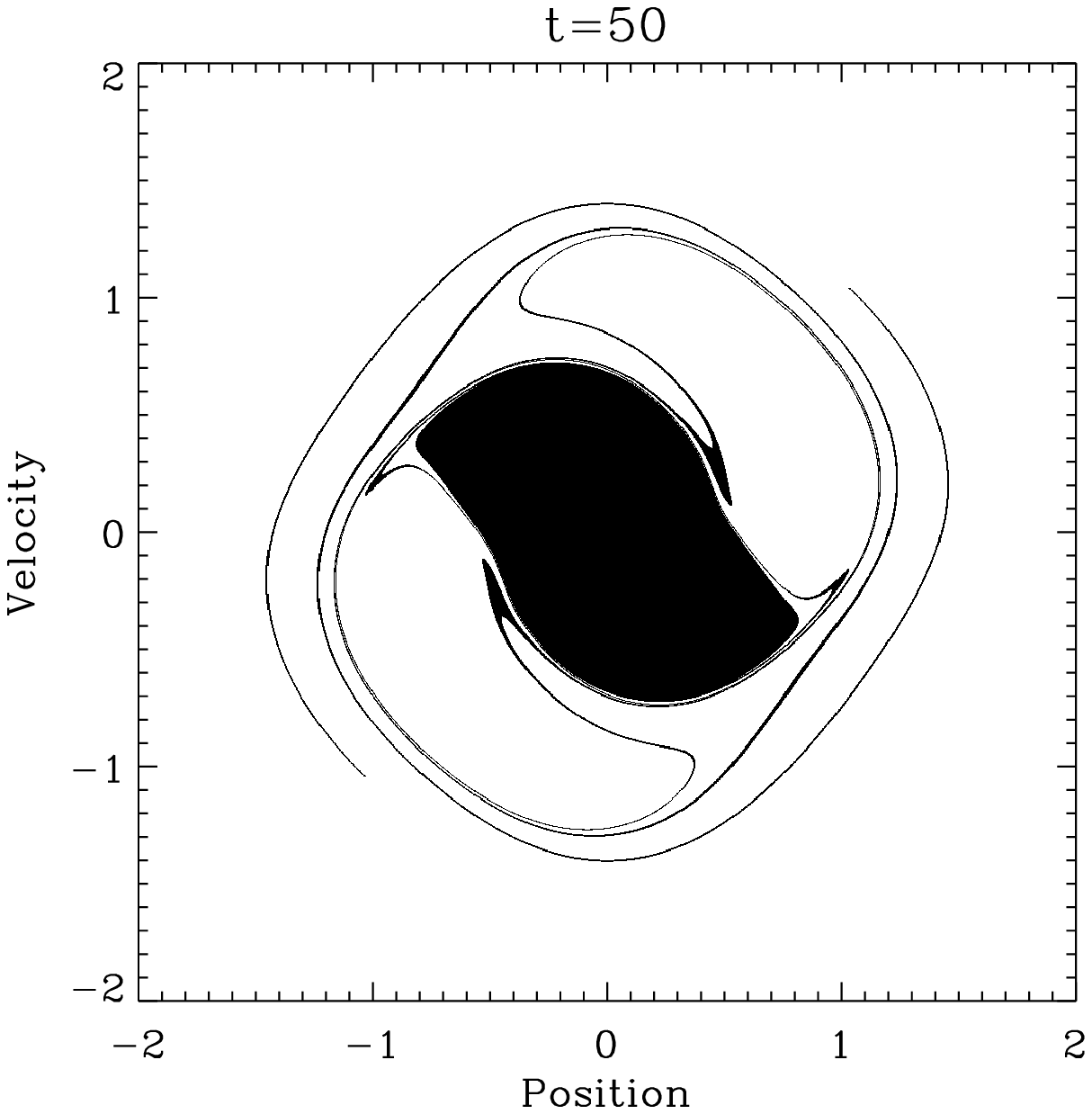,width=5.5cm,bbllx=62pt,bblly=366pt,bburx=426pt,bbury=718pt}
}}
\caption[]{The phase-space distribution function of single waterbag simulations at various times. Times increases from top to bottom, while the initial velocity dispersion, traced by the parameter $\Delta p$, decreases from left to right. The values $t=1.25$ and $t=6.5$ correspond approximately to collapse time and fourth crossing time, respectively, in the cold case ($\Delta p=0$).}
\label{fig:convtocolda}
\end{figure*}
\begin{figure*}
\centerline{\hbox{
\psfig{file=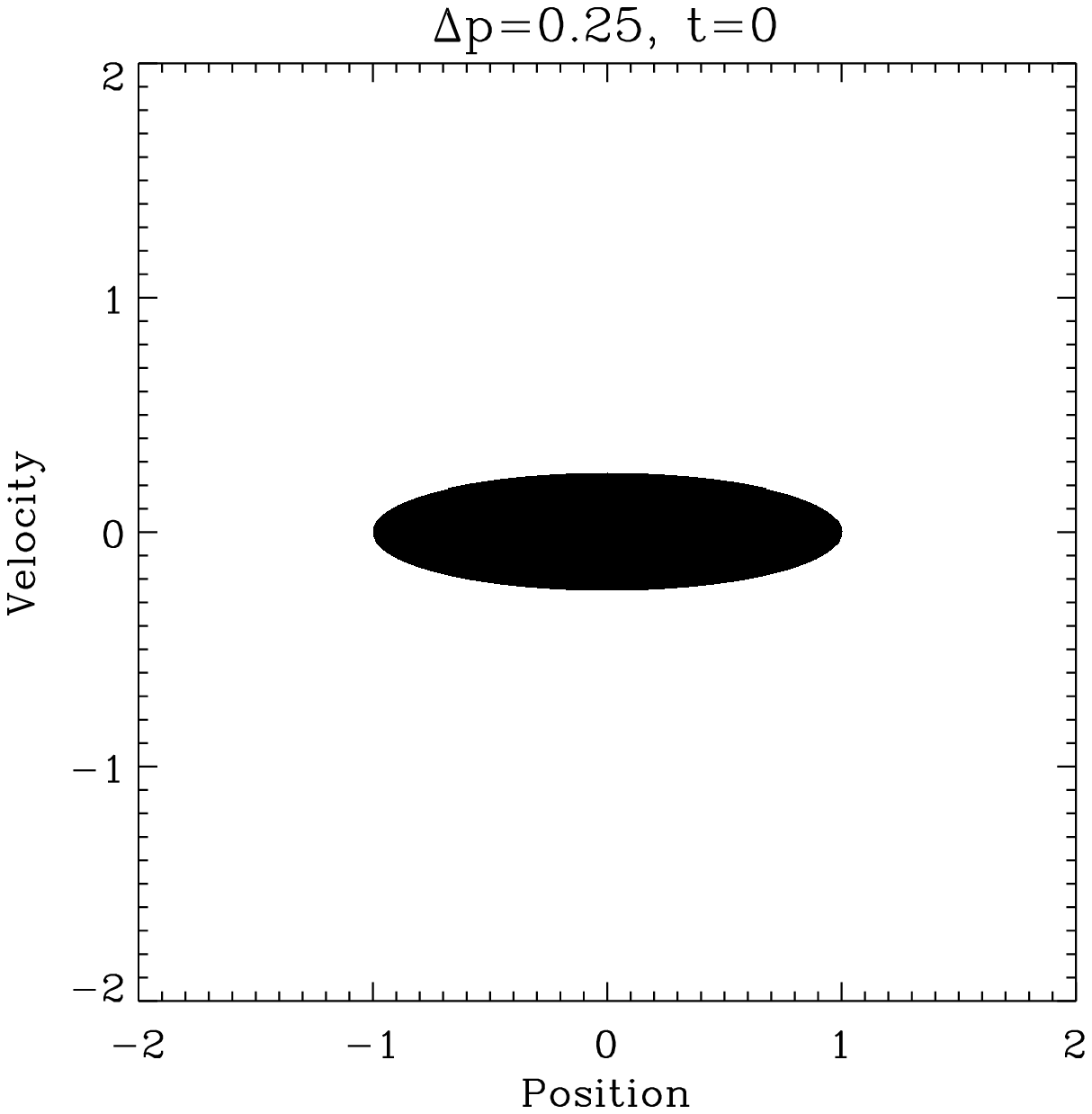,width=5.5cm,bbllx=62pt,bblly=366pt,bburx=426pt,bbury=718pt}
\hskip 0.4cm \psfig{file=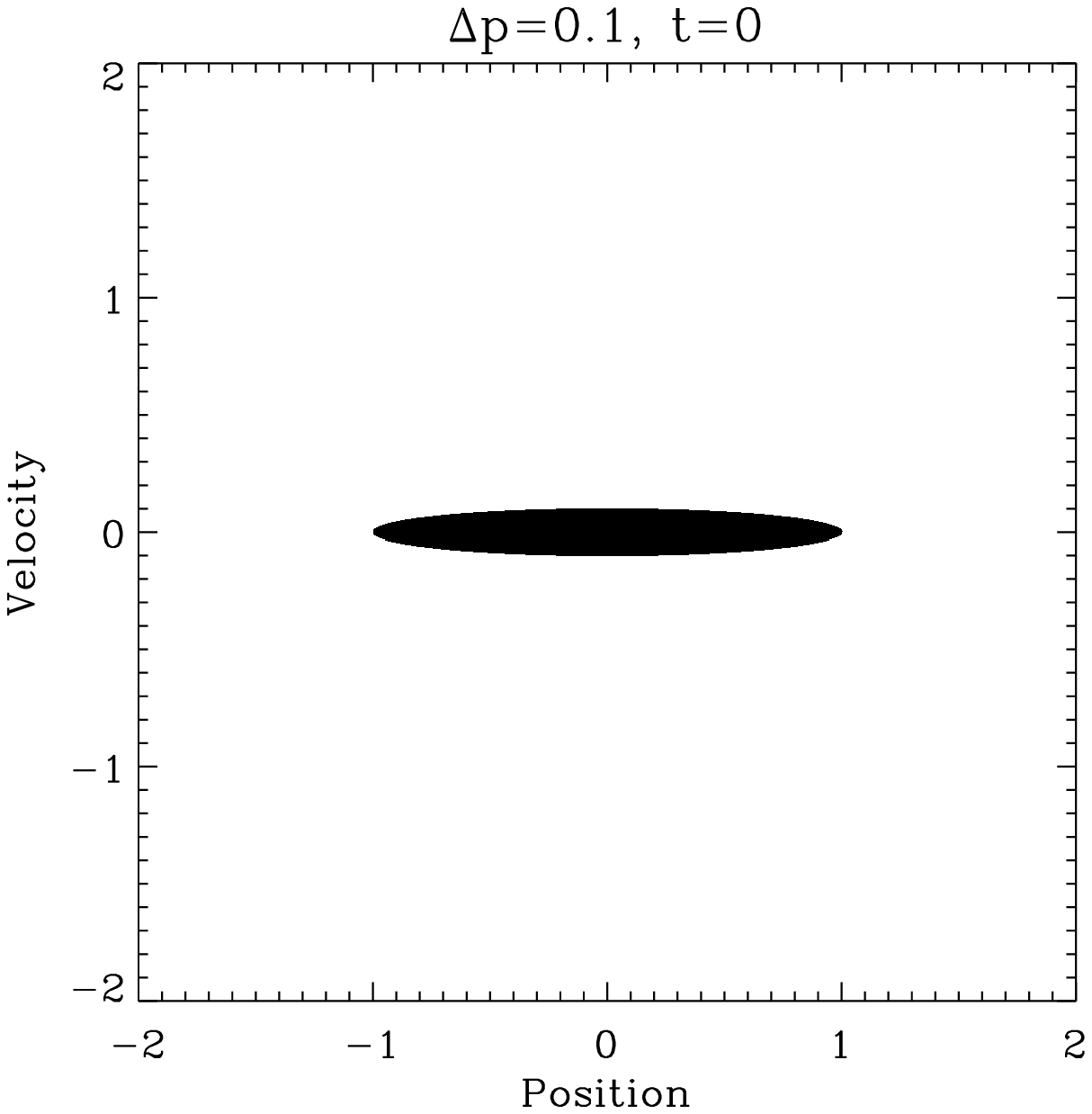,width=5.5cm,bbllx=62pt,bblly=366pt,bburx=426pt,bbury=718pt}
\hskip 0.4cm \psfig{file=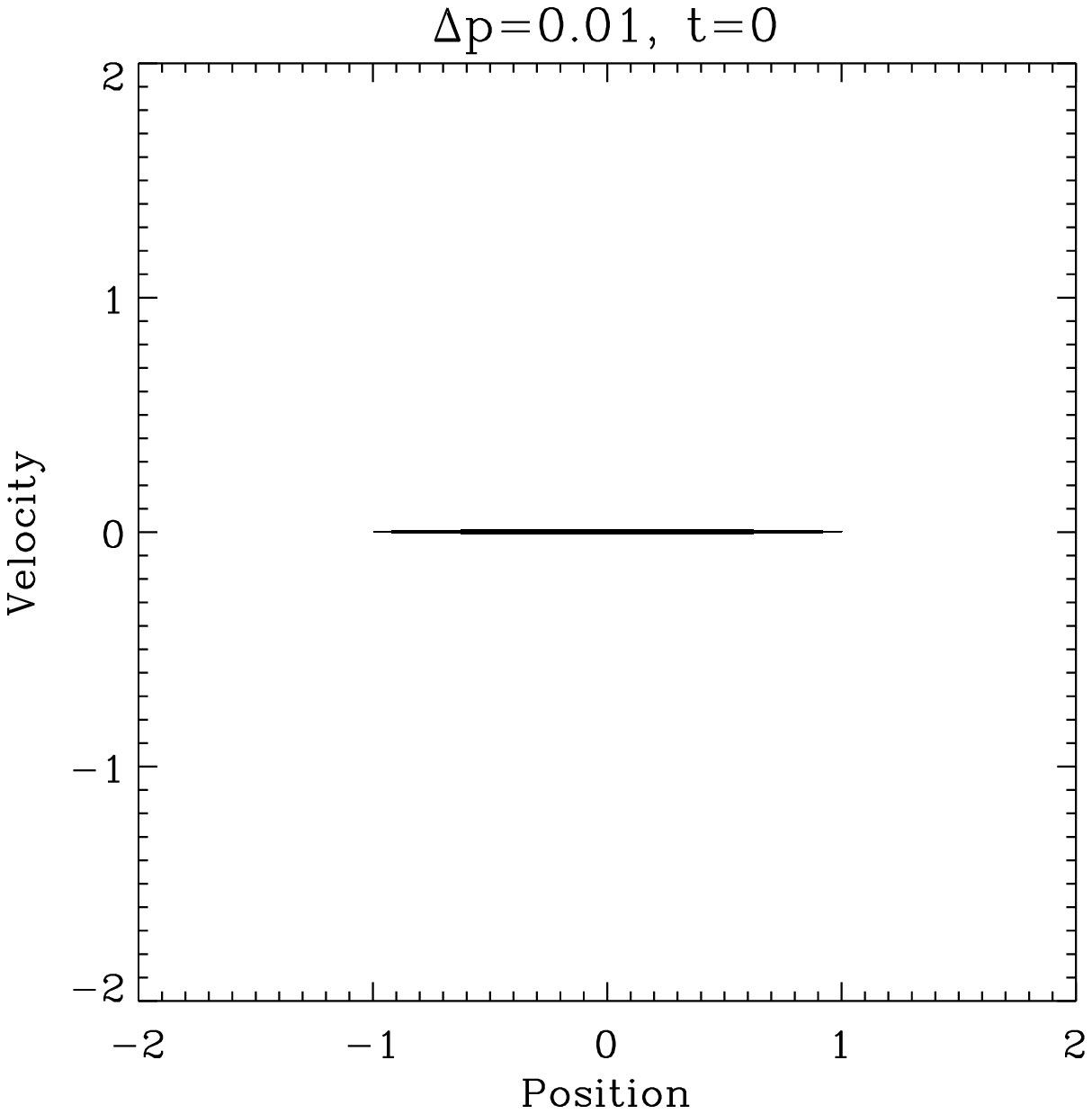,width=5.5cm,bbllx=62pt,bblly=366pt,bburx=426pt,bbury=718pt}
}}
\vskip 0.2cm
\centerline{\hbox{
\psfig{file=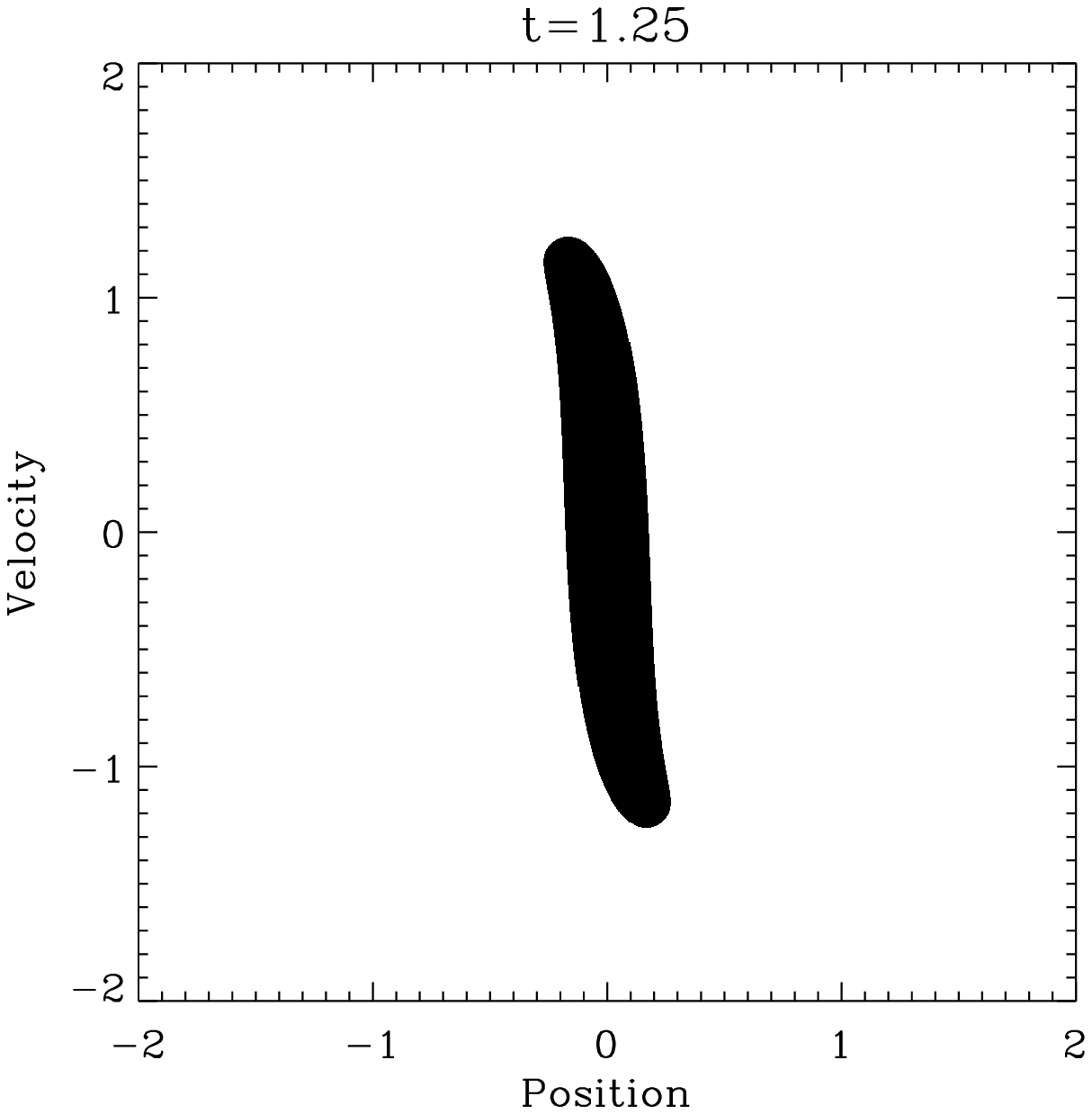,width=5.5cm,bbllx=62pt,bblly=366pt,bburx=426pt,bbury=718pt}
\hskip 0.4cm \psfig{file=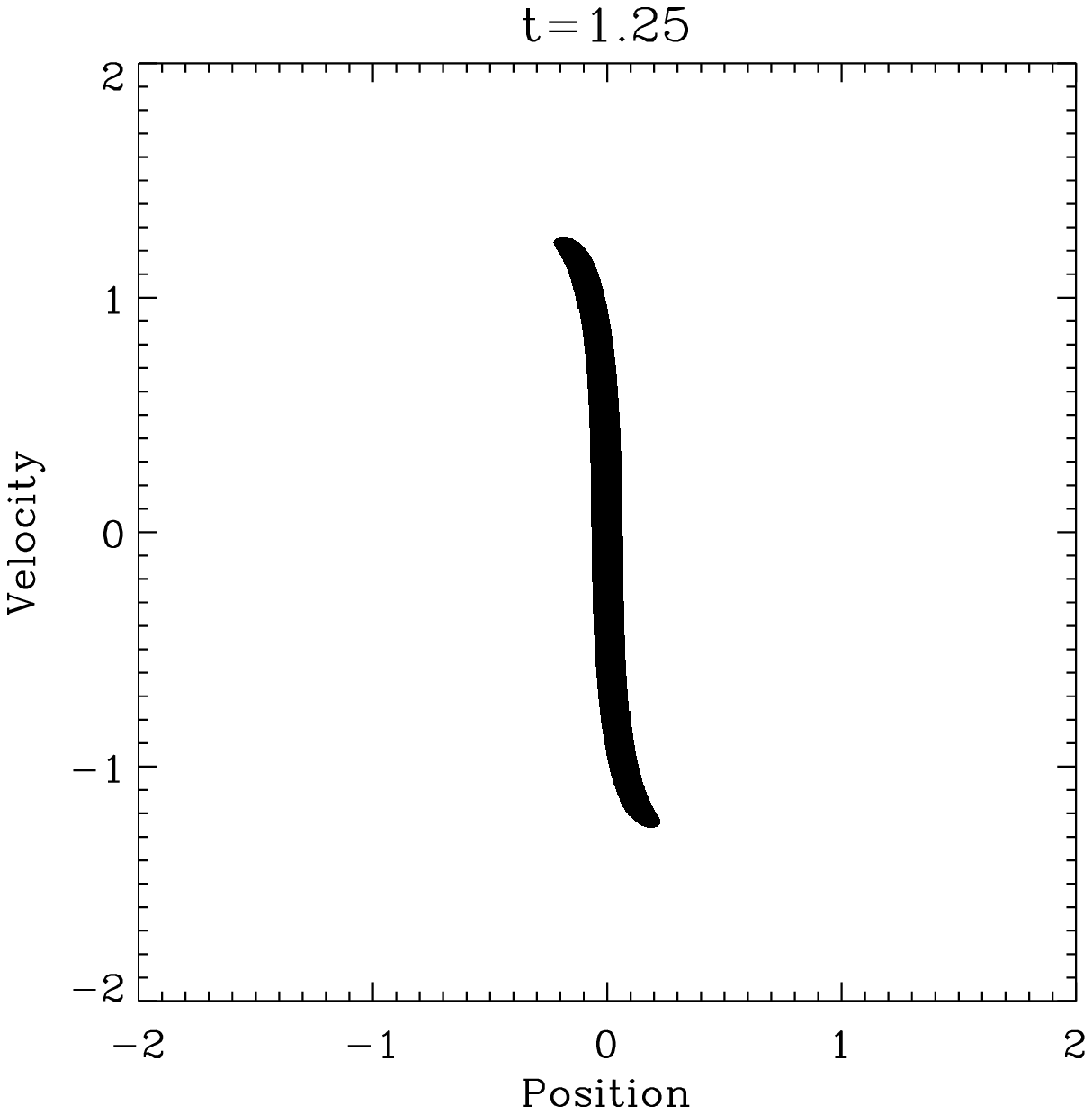,width=5.5cm,bbllx=62pt,bblly=366pt,bburx=426pt,bbury=718pt}
\hskip 0.4cm \psfig{file=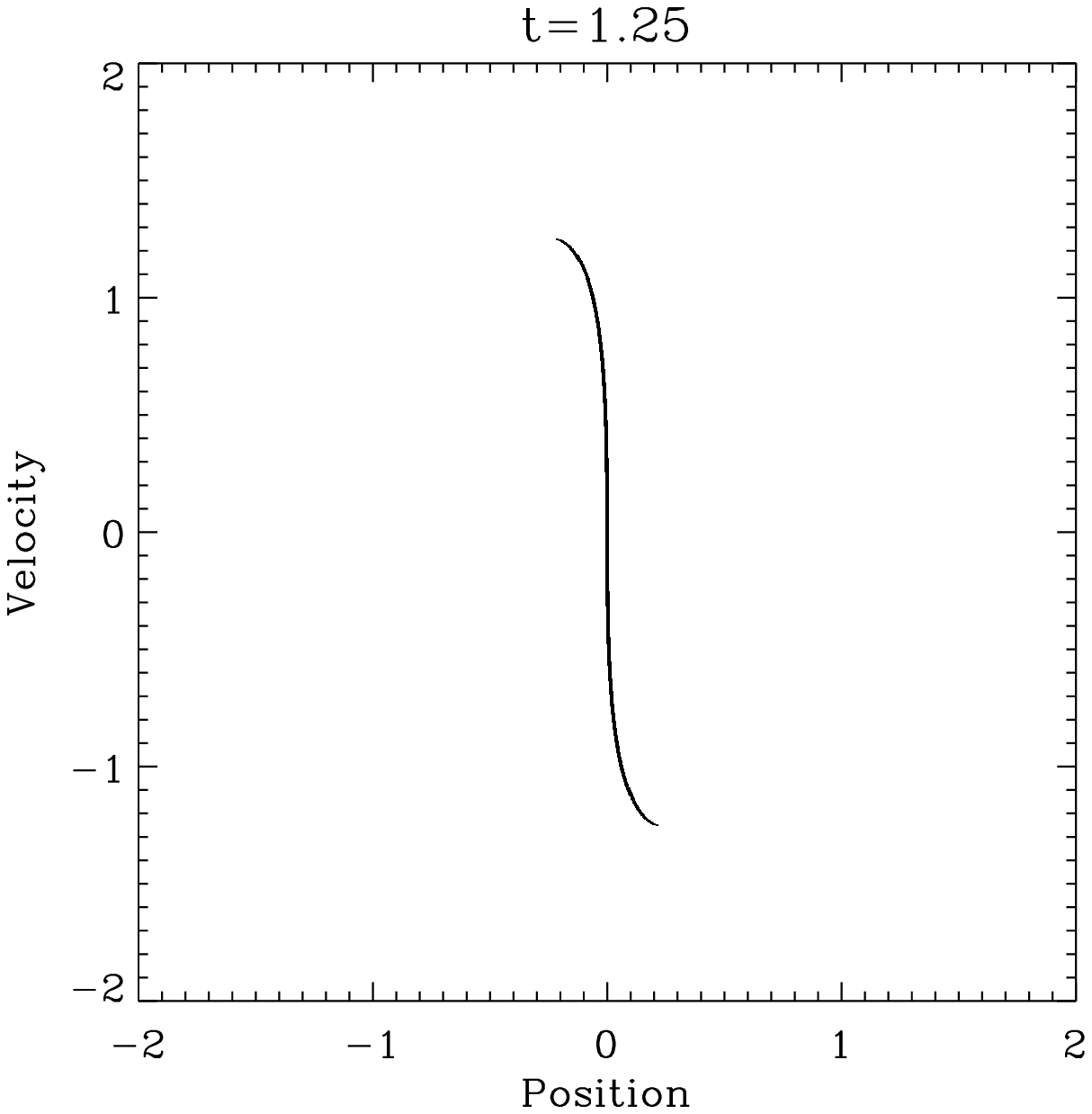,width=5.5cm,bbllx=62pt,bblly=366pt,bburx=426pt,bbury=718pt}
}}
\vskip 0.2cm
\centerline{\hbox{
\psfig{file=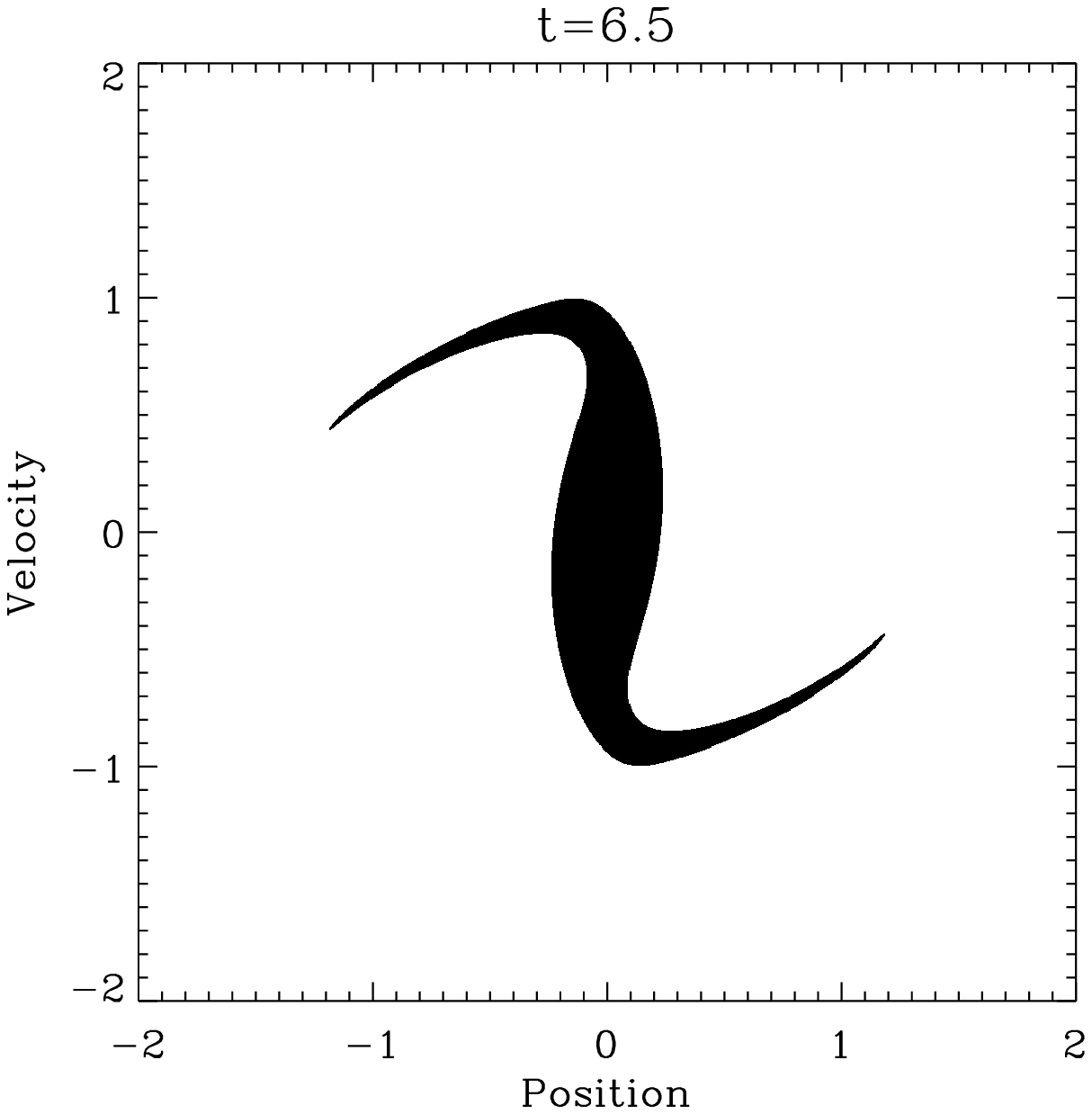,width=5.5cm,bbllx=62pt,bblly=366pt,bburx=426pt,bbury=718pt}
\hskip 0.4cm \psfig{file=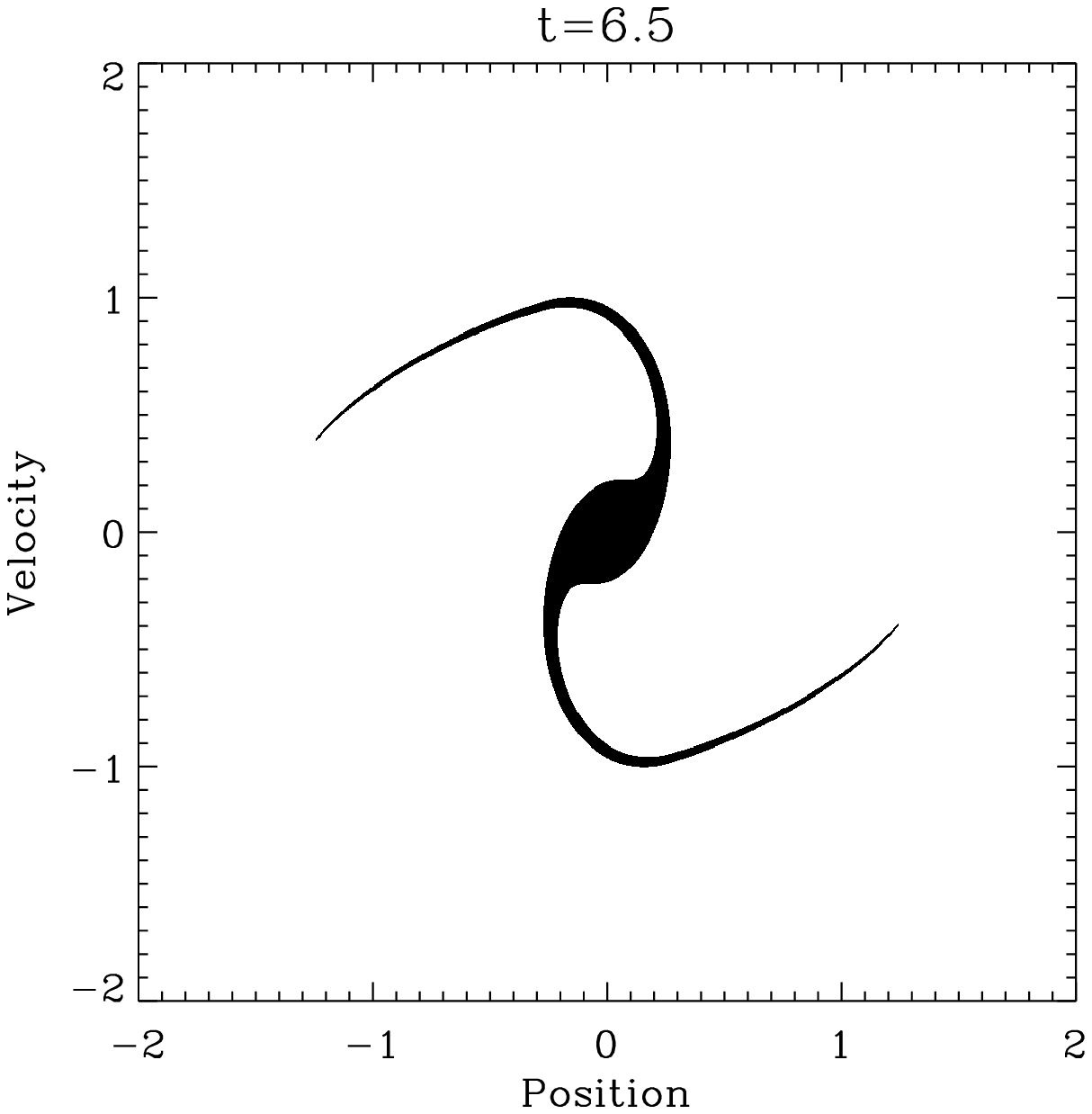,width=5.5cm,bbllx=62pt,bblly=366pt,bburx=426pt,bbury=718pt}
\hskip 0.4cm \psfig{file=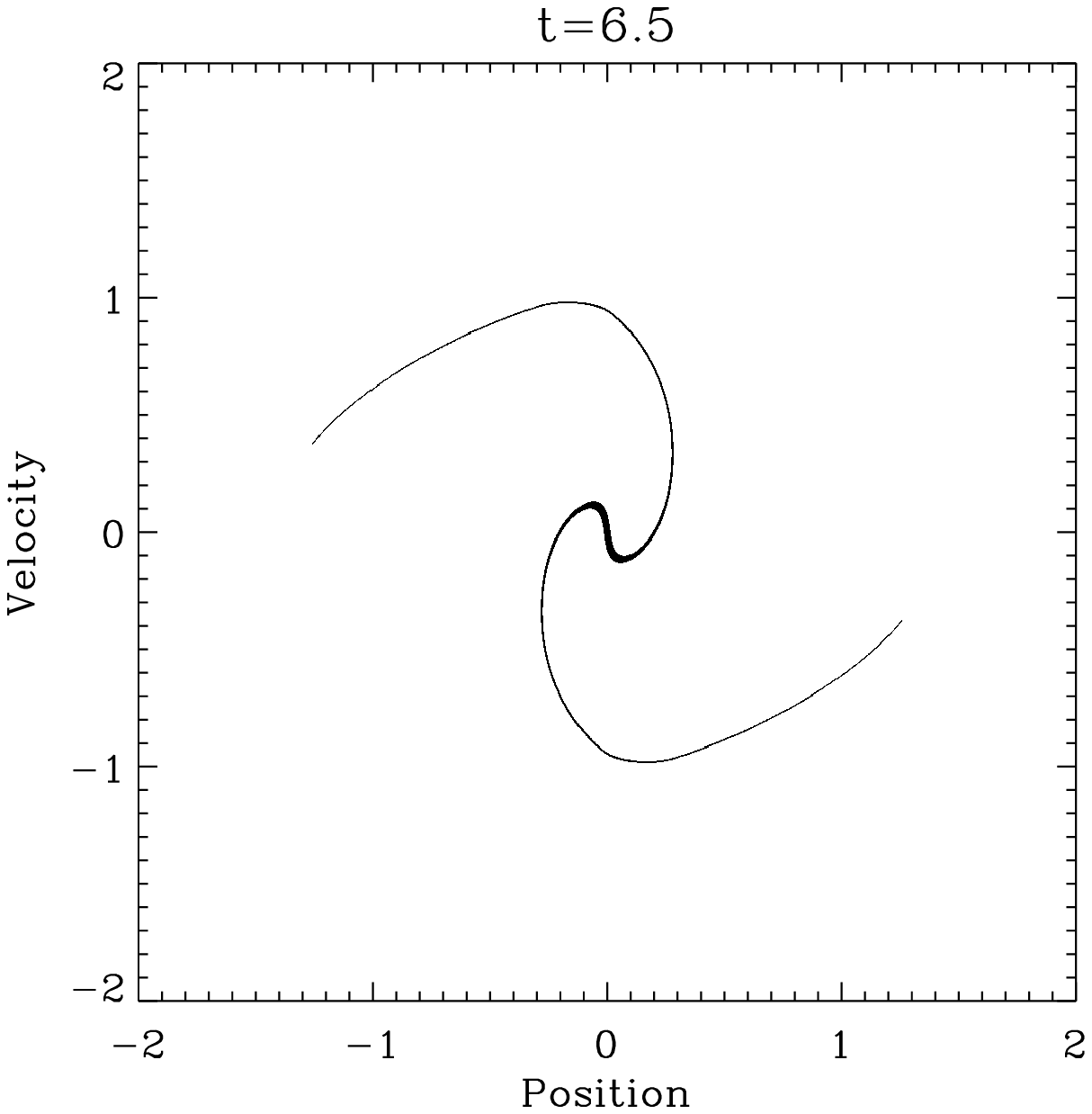,width=5.5cm,bbllx=62pt,bblly=366pt,bburx=426pt,bbury=718pt}
}}
\vskip 0.2cm
\centerline{\hbox{
\psfig{file=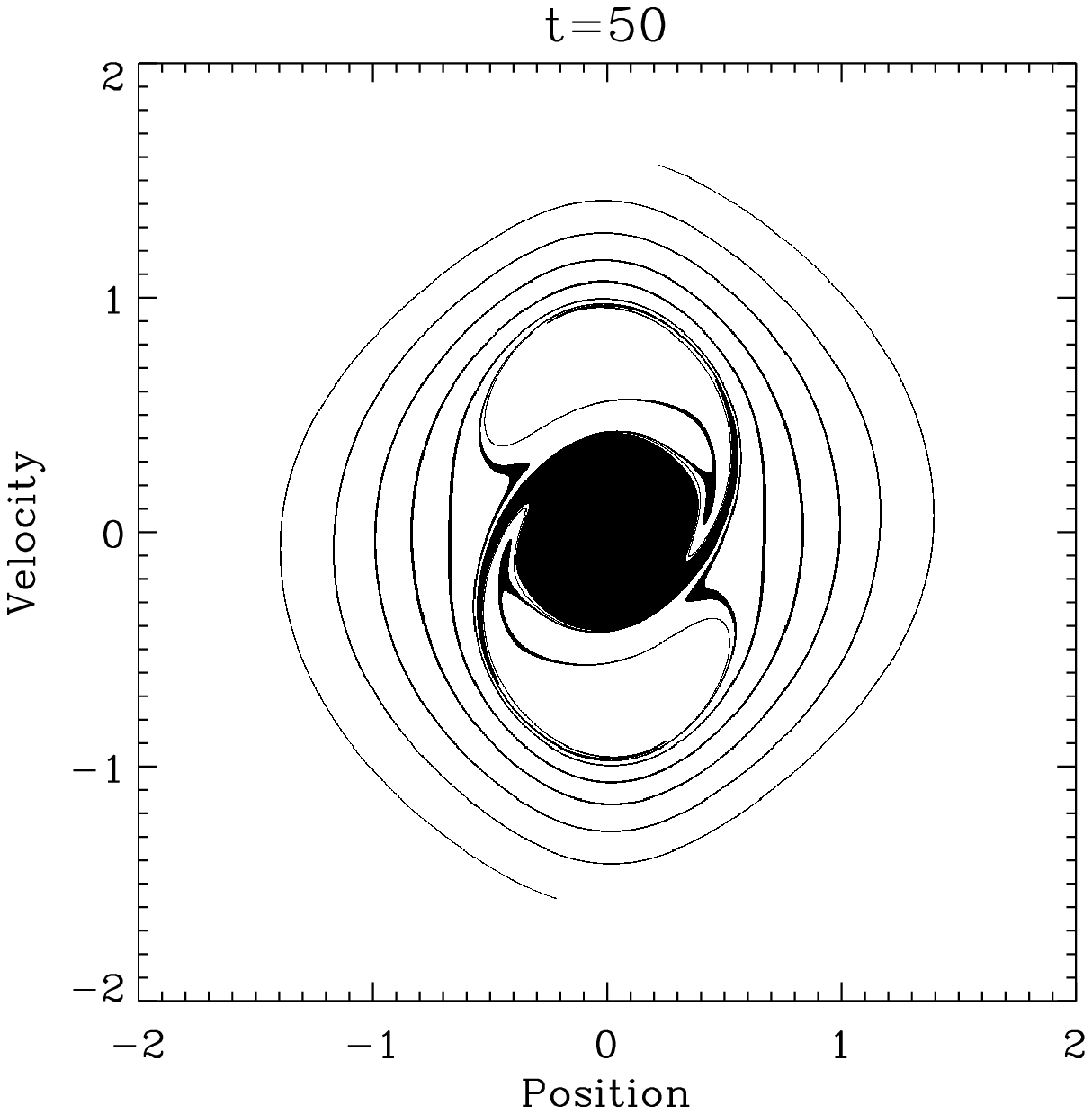,width=5.5cm,bbllx=62pt,bblly=366pt,bburx=426pt,bbury=718pt}
\hskip 0.4cm \psfig{file=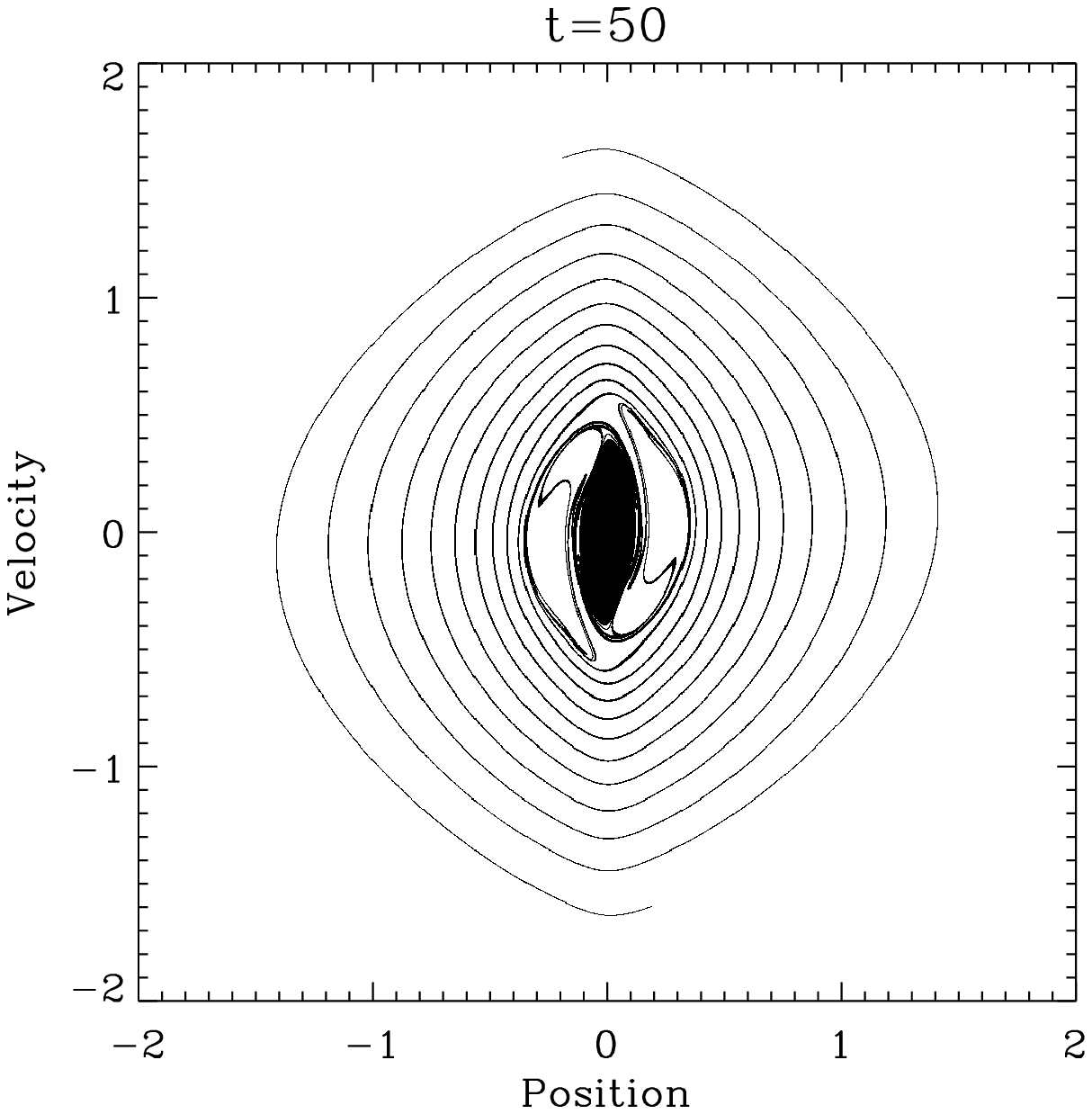,width=5.5cm,bbllx=62pt,bblly=366pt,bburx=426pt,bbury=718pt}
\hskip 0.4cm \psfig{file=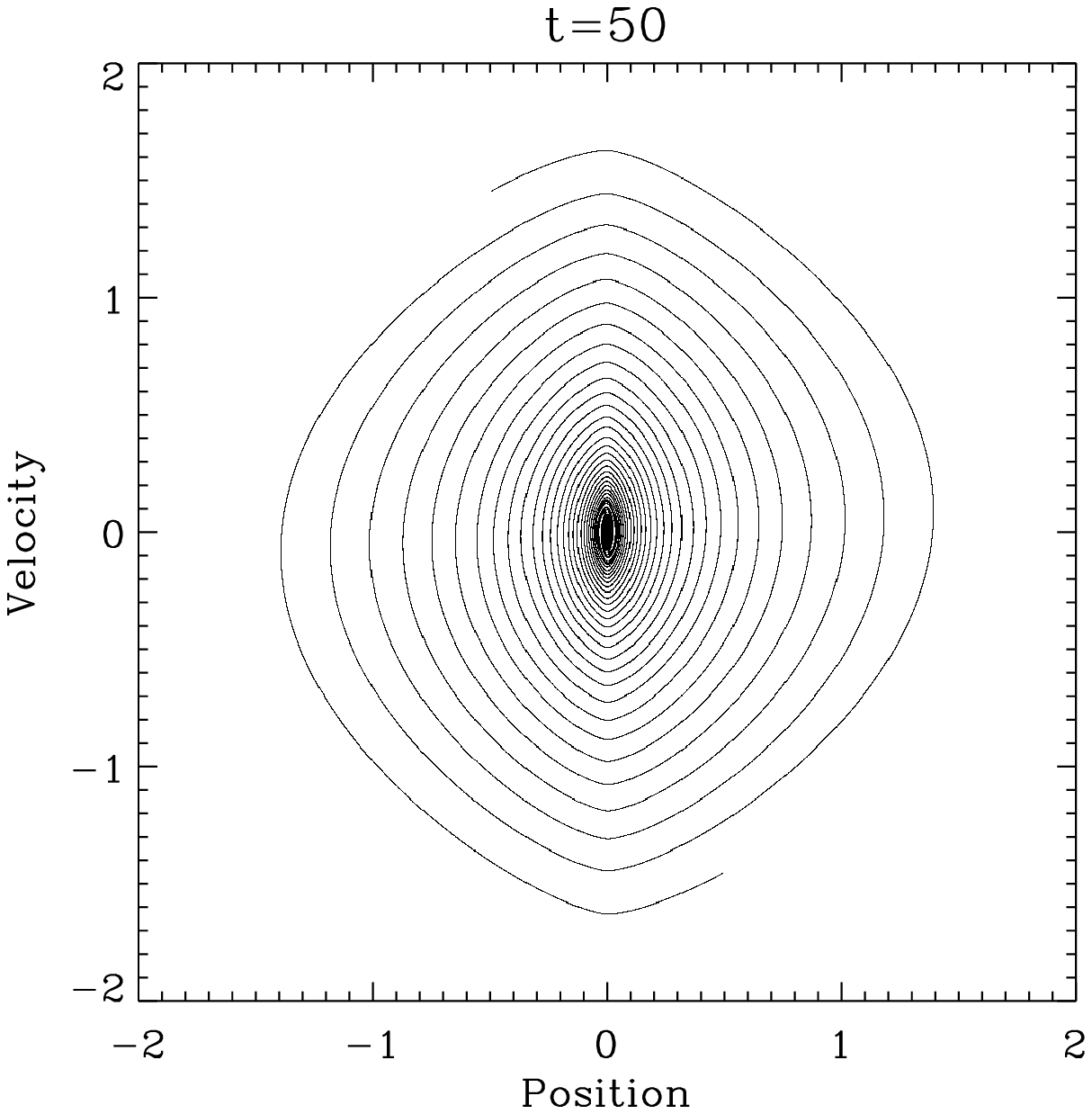,width=5.5cm,bbllx=62pt,bblly=366pt,bburx=426pt,bbury=718pt}
}}
\caption[]{Figure~\ref{fig:convtocolda}, continued, for smaller values of $\Delta p$. For $\Delta p=0.01$, we show the simulation {\tt Tophat0.010} in the nomenclature of Table~\ref{tab:simuparam}, but the other simulation ({\tt Tophat0.010U}) would not differ from this one at the level of zoom we are looking at.}
\label{fig:convtocoldb}
\end{figure*}
\begin{figure*}
\centerline{\hbox{
\psfig{file=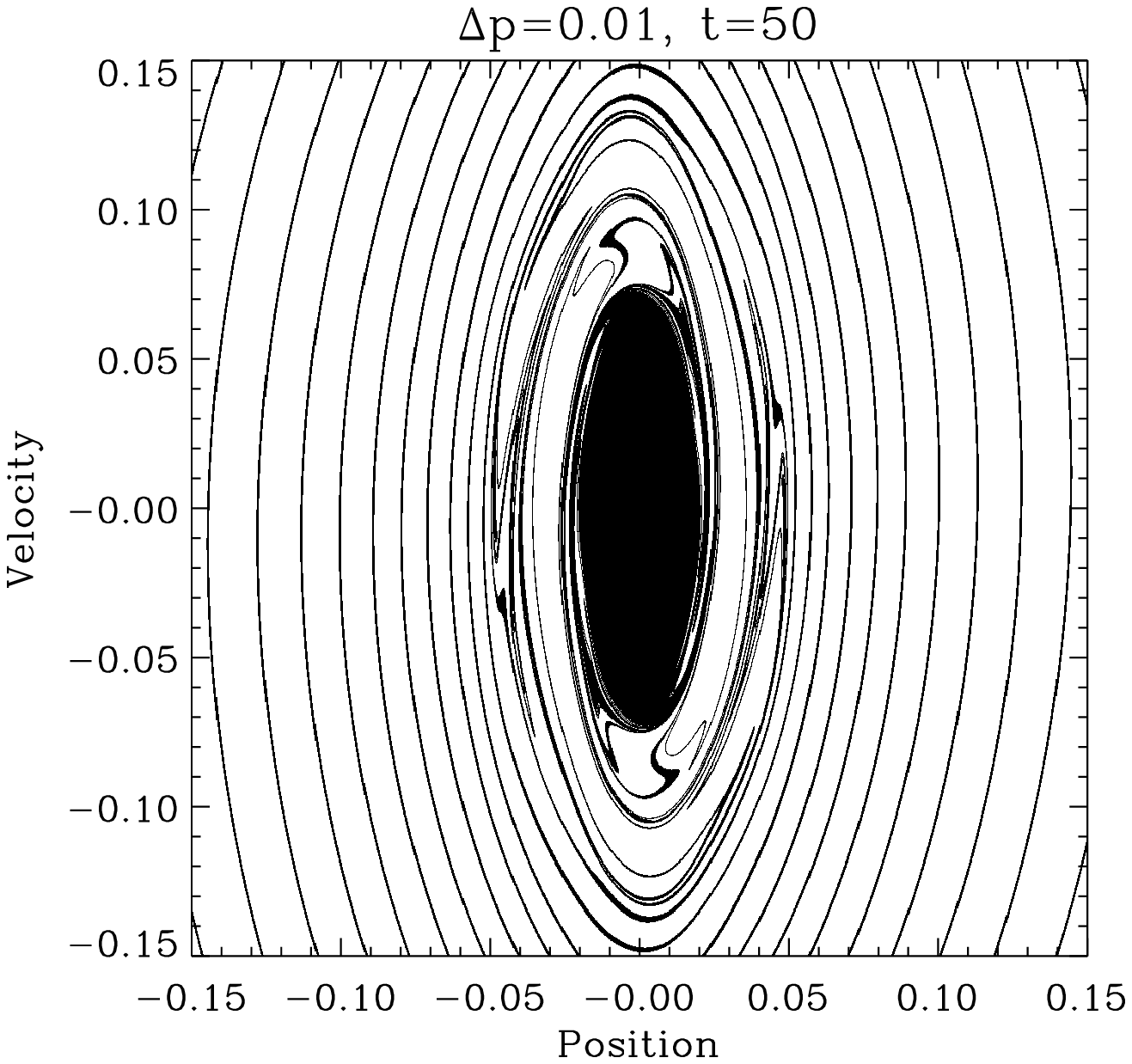,width=5.5cm,bbllx=62pt,bblly=366pt,bburx=426pt,bbury=718pt}
\hskip 0.2cm
\psfig{file=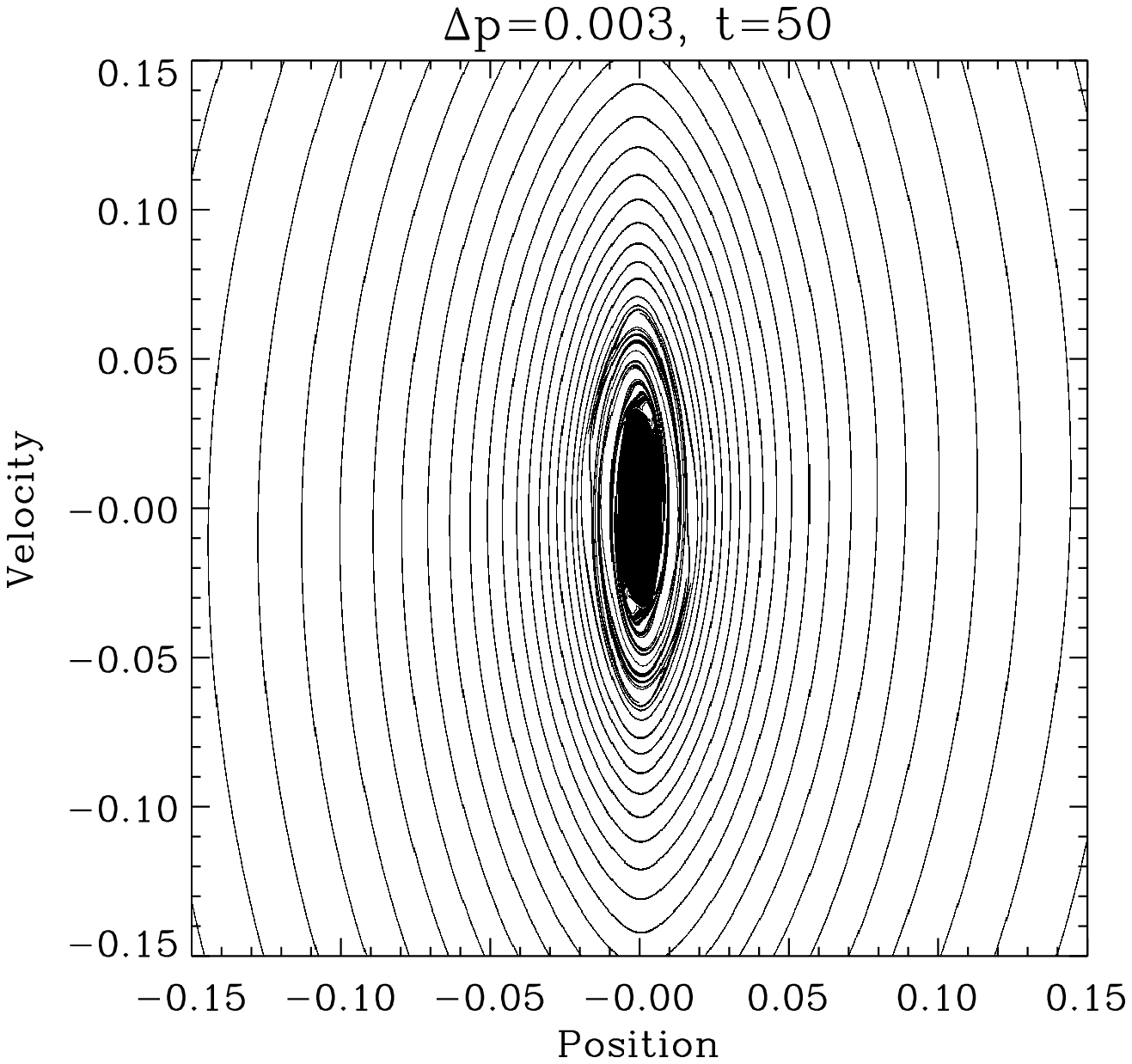,width=5.5cm,bbllx=62pt,bblly=366pt,bburx=426pt,bbury=718pt}
\hskip 0.2cm
\psfig{file=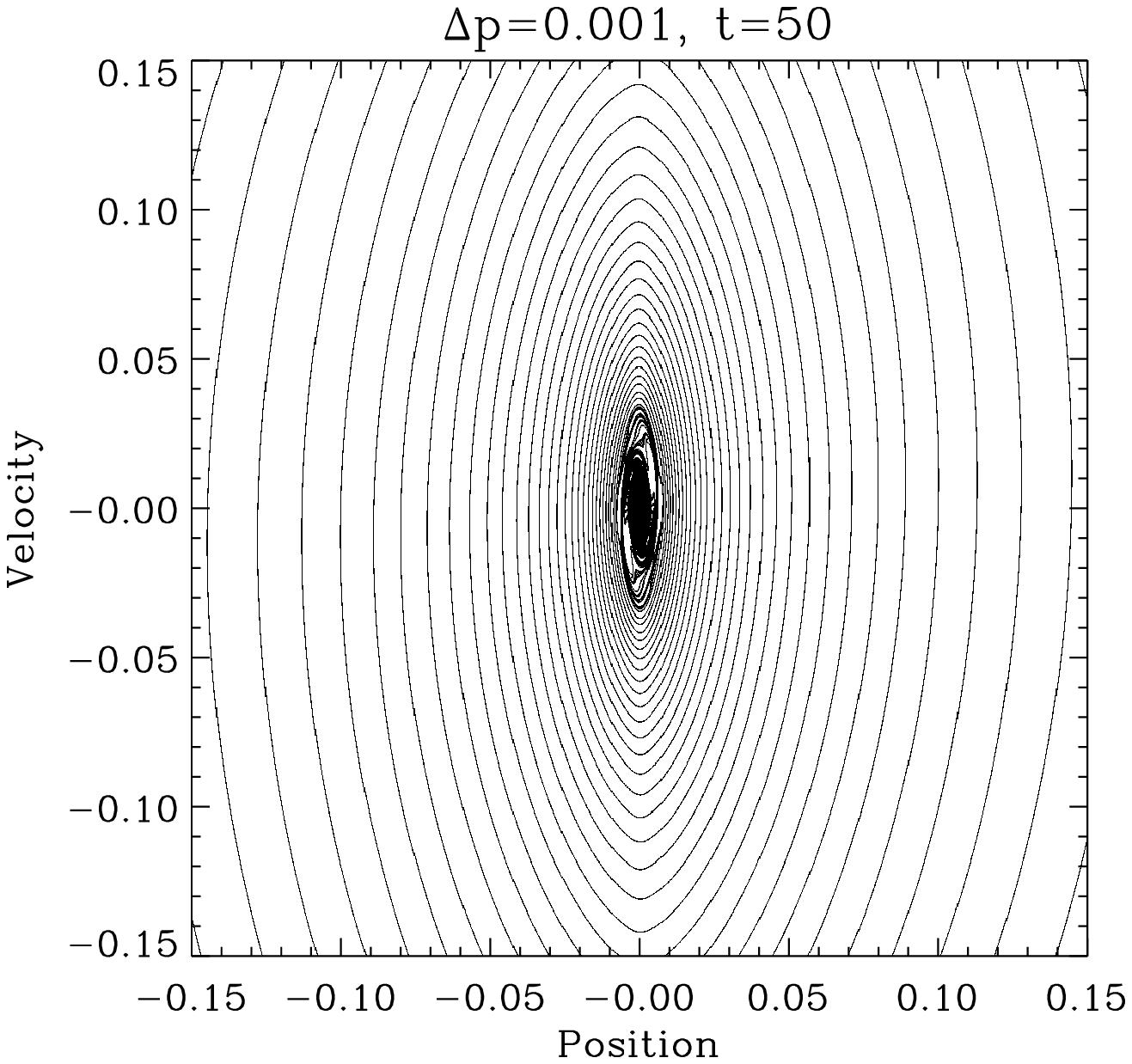,width=5.5cm,bbllx=62pt,bblly=366pt,bburx=426pt,bbury=718pt}
}}
\caption[]{A zoom in phase-space on the central part of the system at final time, for $\Delta p=0.01$ (left), $0.003$ (middle) and
$0.001$ (right). The simulations adopted here are {\tt Tophat0.010}, {\tt Tophat0.003} and {\tt Tophat0.001} in the nomenclature of Table~\ref{tab:simuparam}. With the same initial conditions but slightly different parameters for performing the simulations (as listed in Table~\ref{tab:simuparam}), some small differences can be sighted. They simply indicate a shift in effective dynamical time due to a slight change in the energetic state of the system from one simulation to another.}
\label{fig:coldzoomcenter}
\end{figure*}
\begin{figure*}
\centerline{\hbox{
\psfig{file=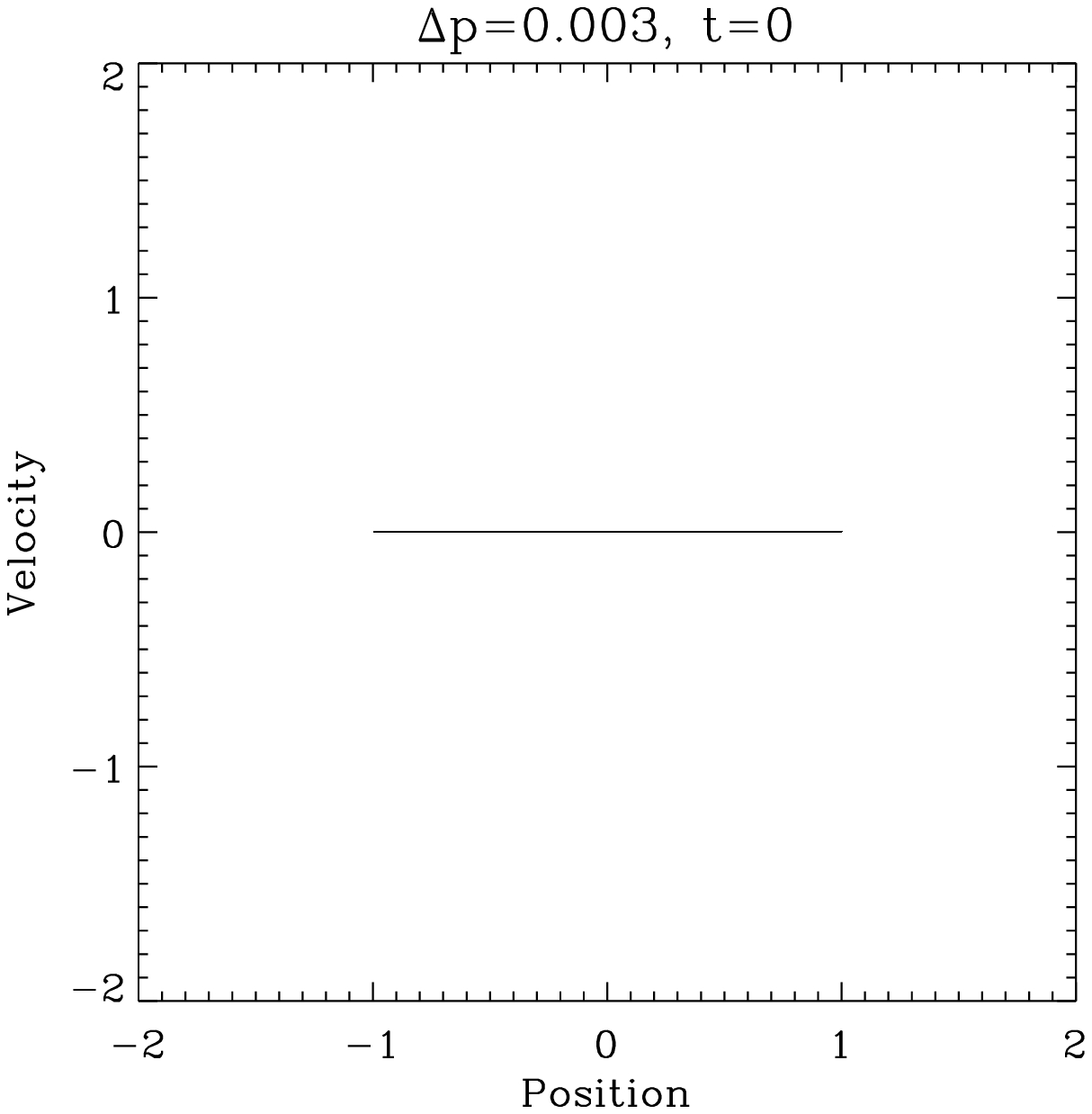,width=5.1cm,bbllx=62pt,bblly=366pt,bburx=426pt,bbury=718pt}
\hskip 0.4cm \psfig{file=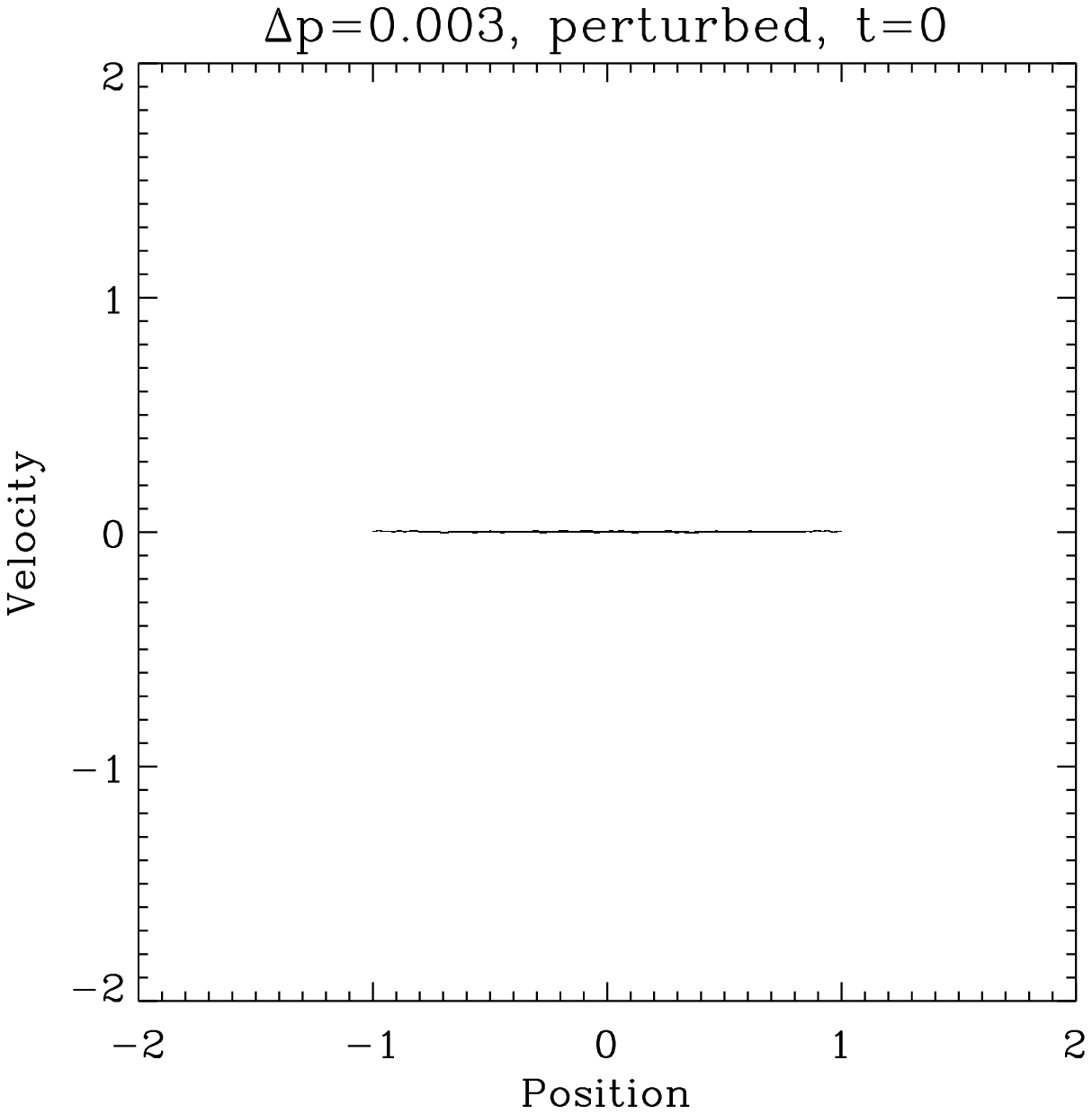,width=5.1cm,bbllx=62pt,bblly=366pt,bburx=426pt,bbury=718pt}
\hskip 0.4cm \psfig{file=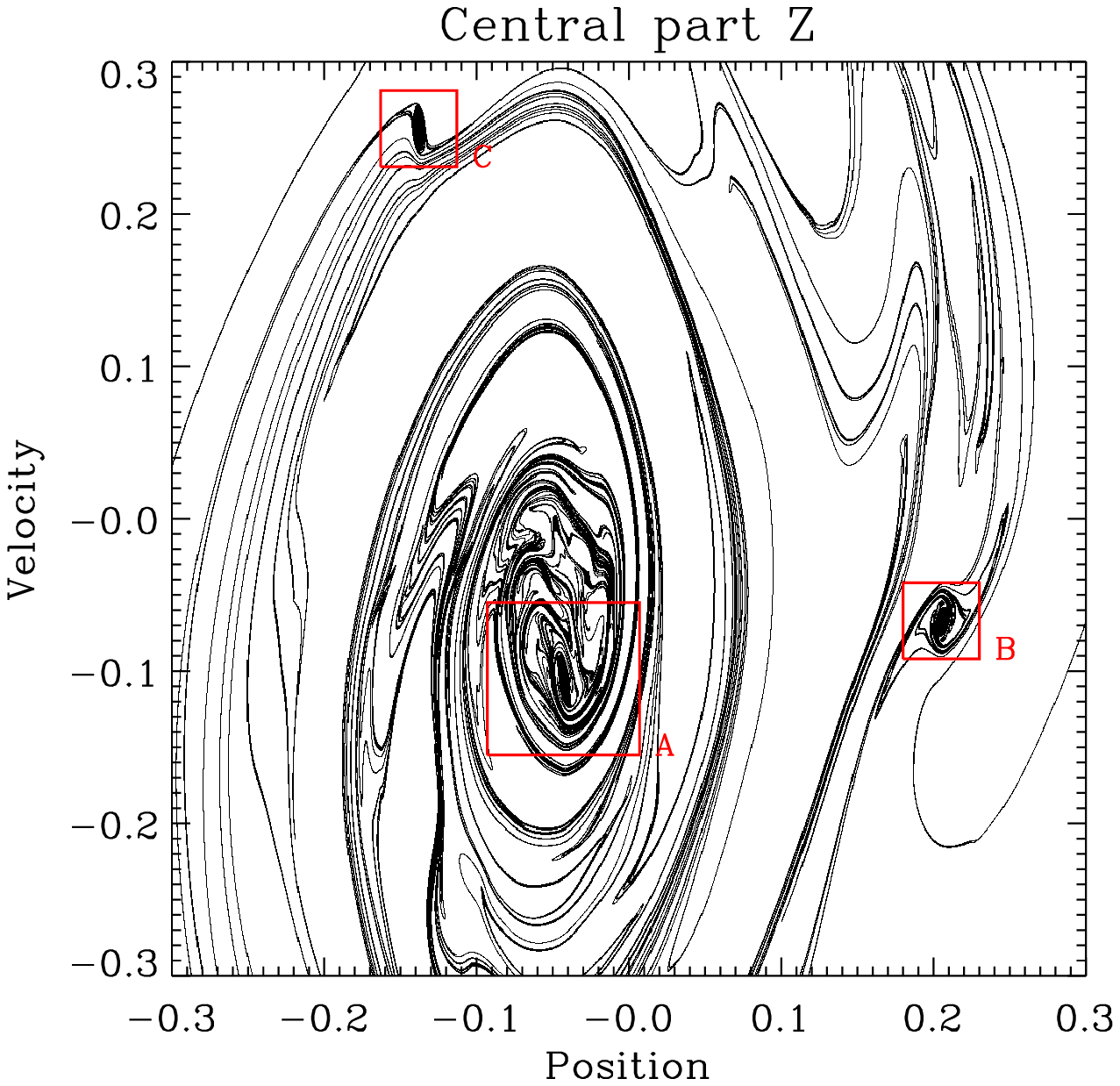,width=5.1cm,bbllx=62pt,bblly=366pt,bburx=426pt,bbury=718pt}
}}
\vskip 0.2cm
\centerline{\hbox{
\psfig{file=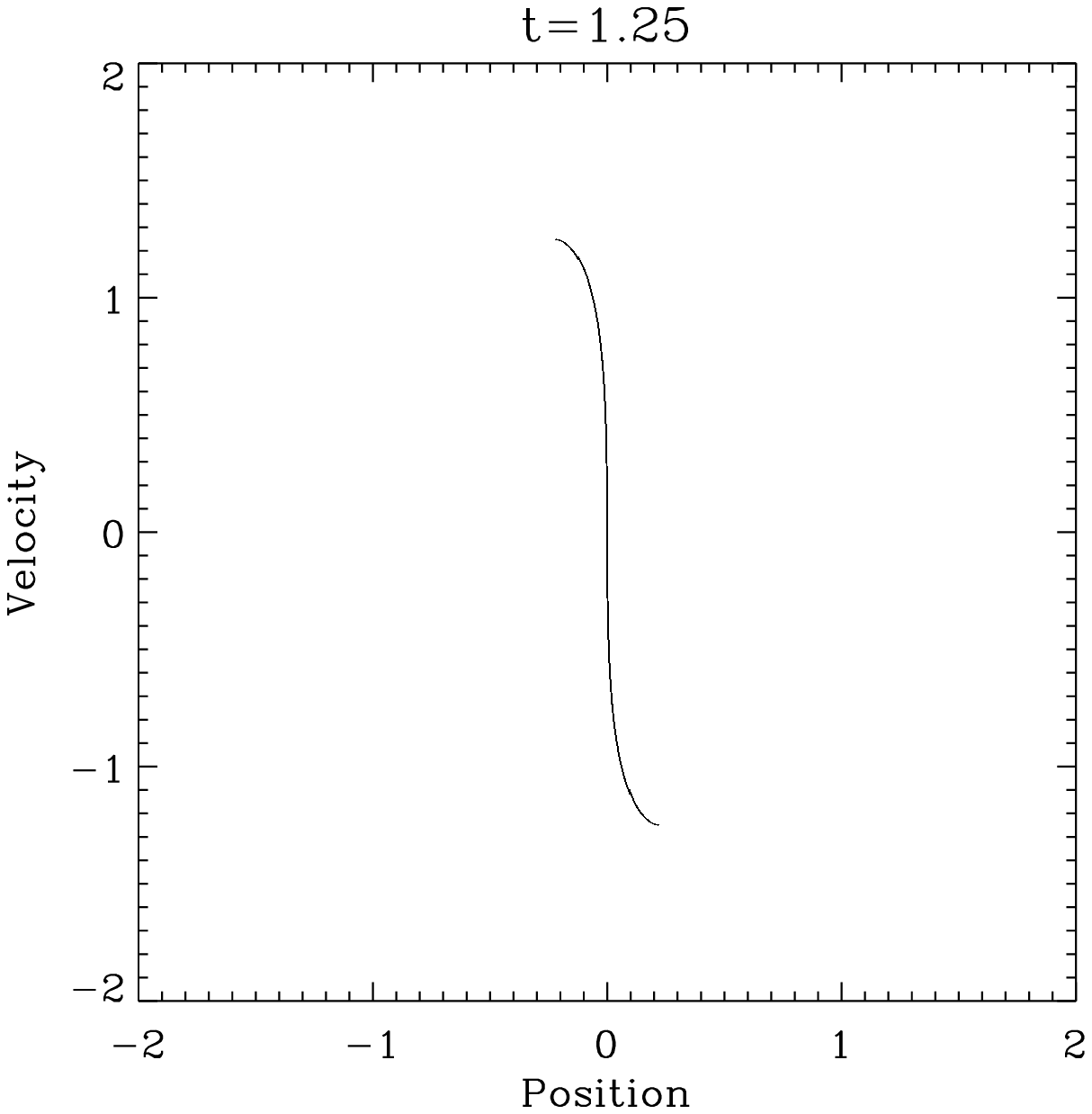,width=5.1cm,bbllx=62pt,bblly=366pt,bburx=426pt,bbury=718pt}
\hskip 0.4cm \psfig{file=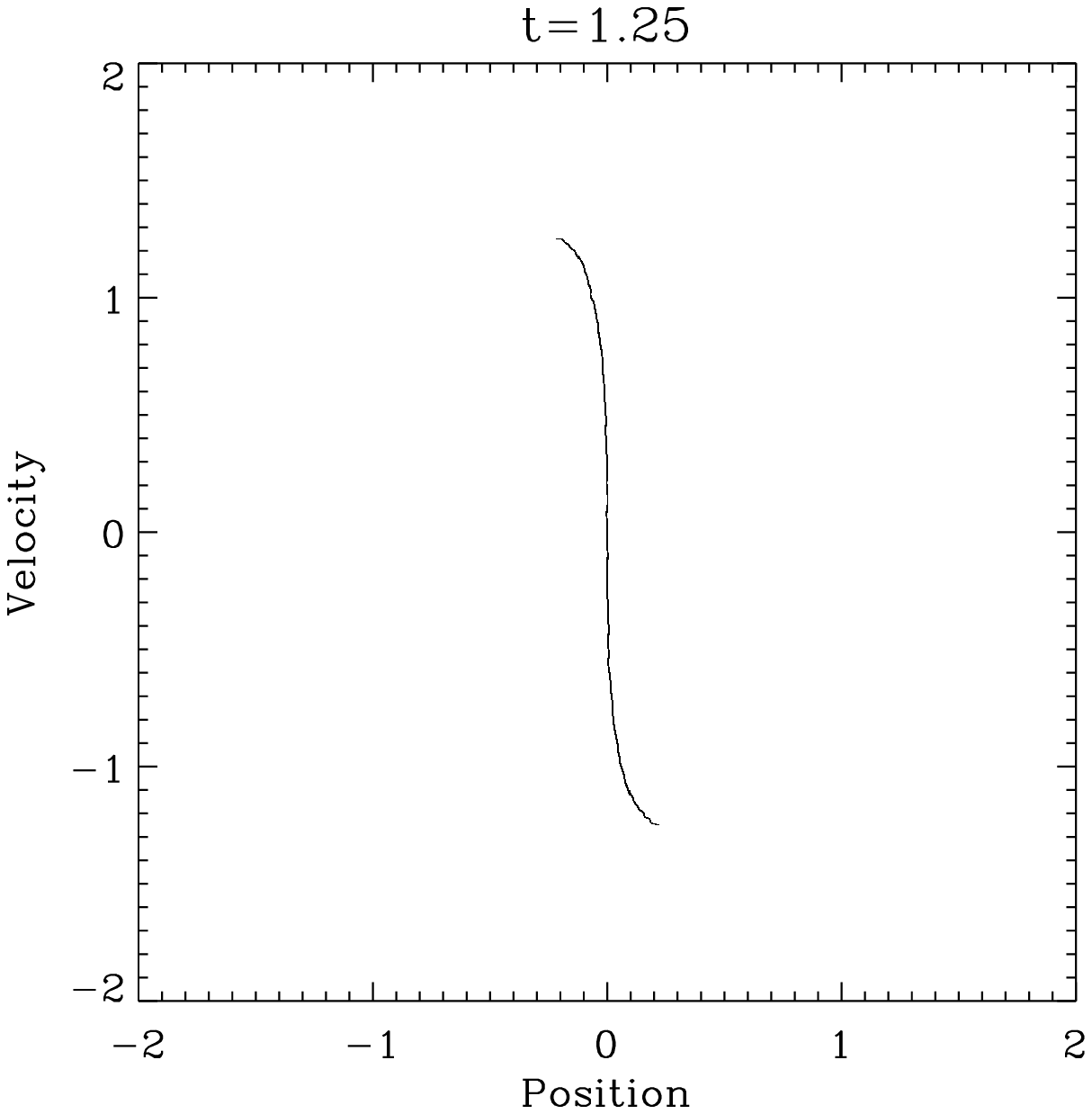,width=5.1cm,bbllx=62pt,bblly=366pt,bburx=426pt,bbury=718pt}
\hskip 0.4cm \psfig{file=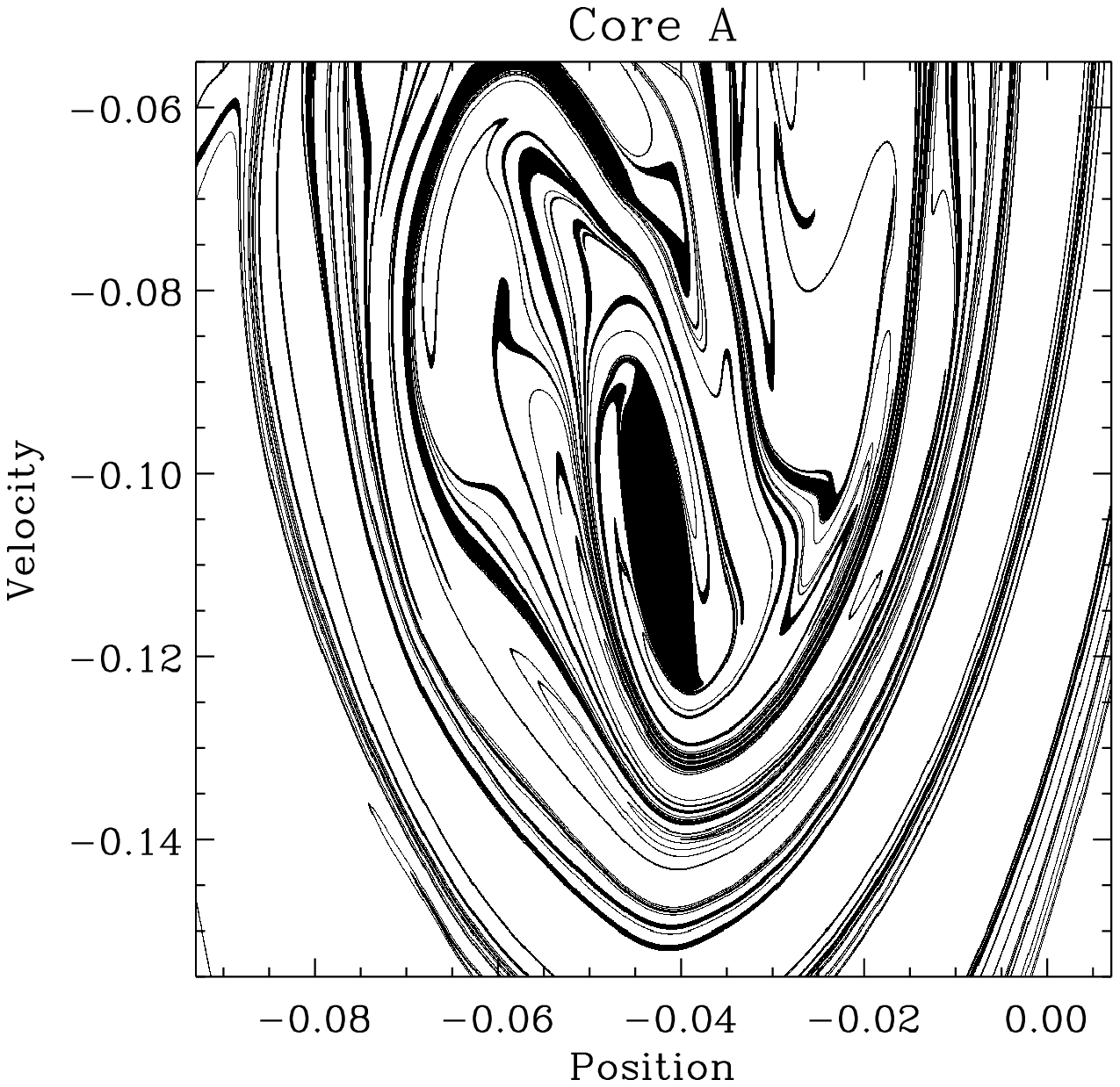,width=5.1cm,bbllx=62pt,bblly=366pt,bburx=426pt,bbury=718pt}
}}
\vskip 0.2cm
\centerline{\hbox{
\psfig{file=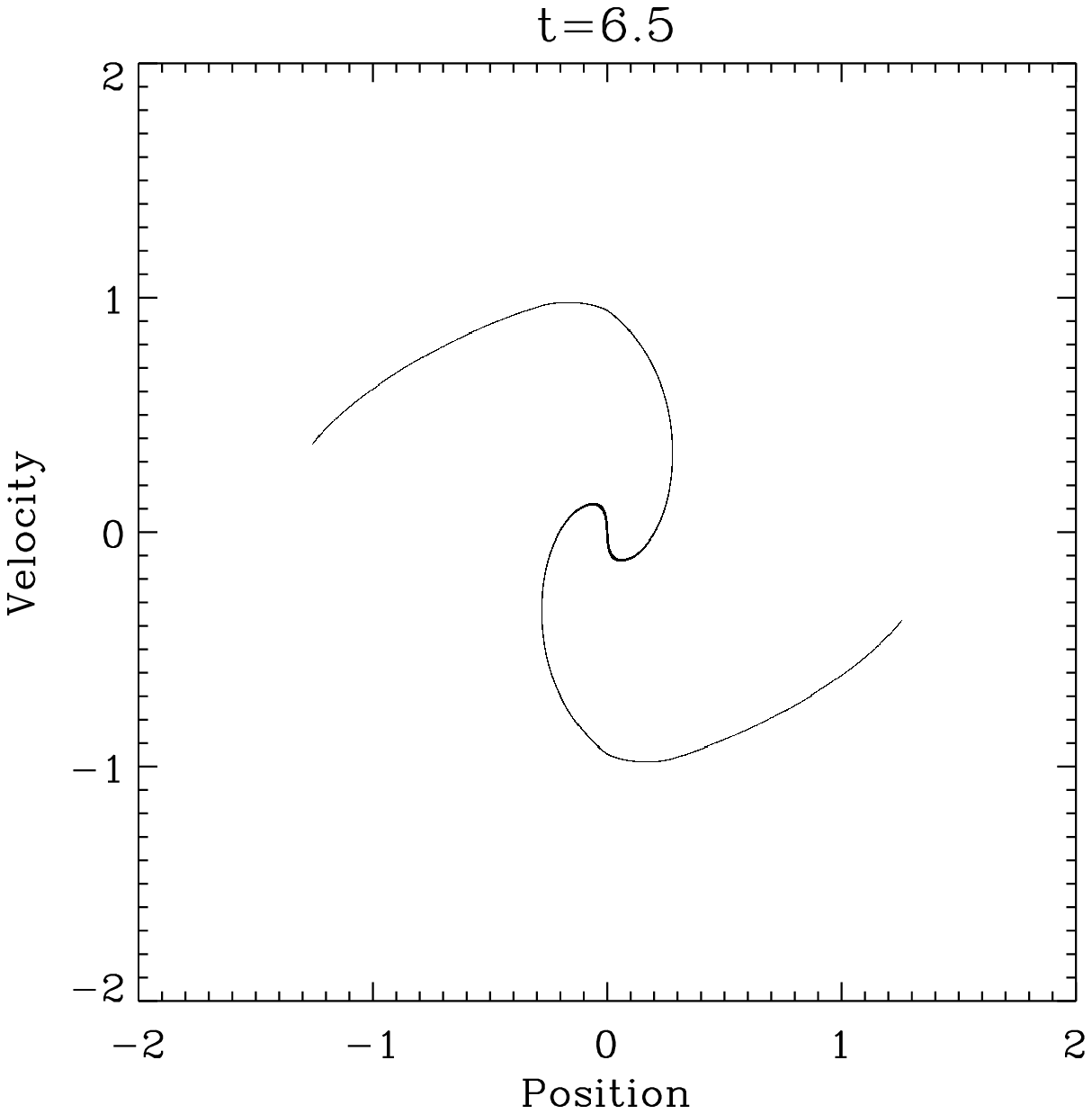,width=5.1cm,bbllx=62pt,bblly=366pt,bburx=426pt,bbury=718pt}
\hskip 0.4cm \psfig{file=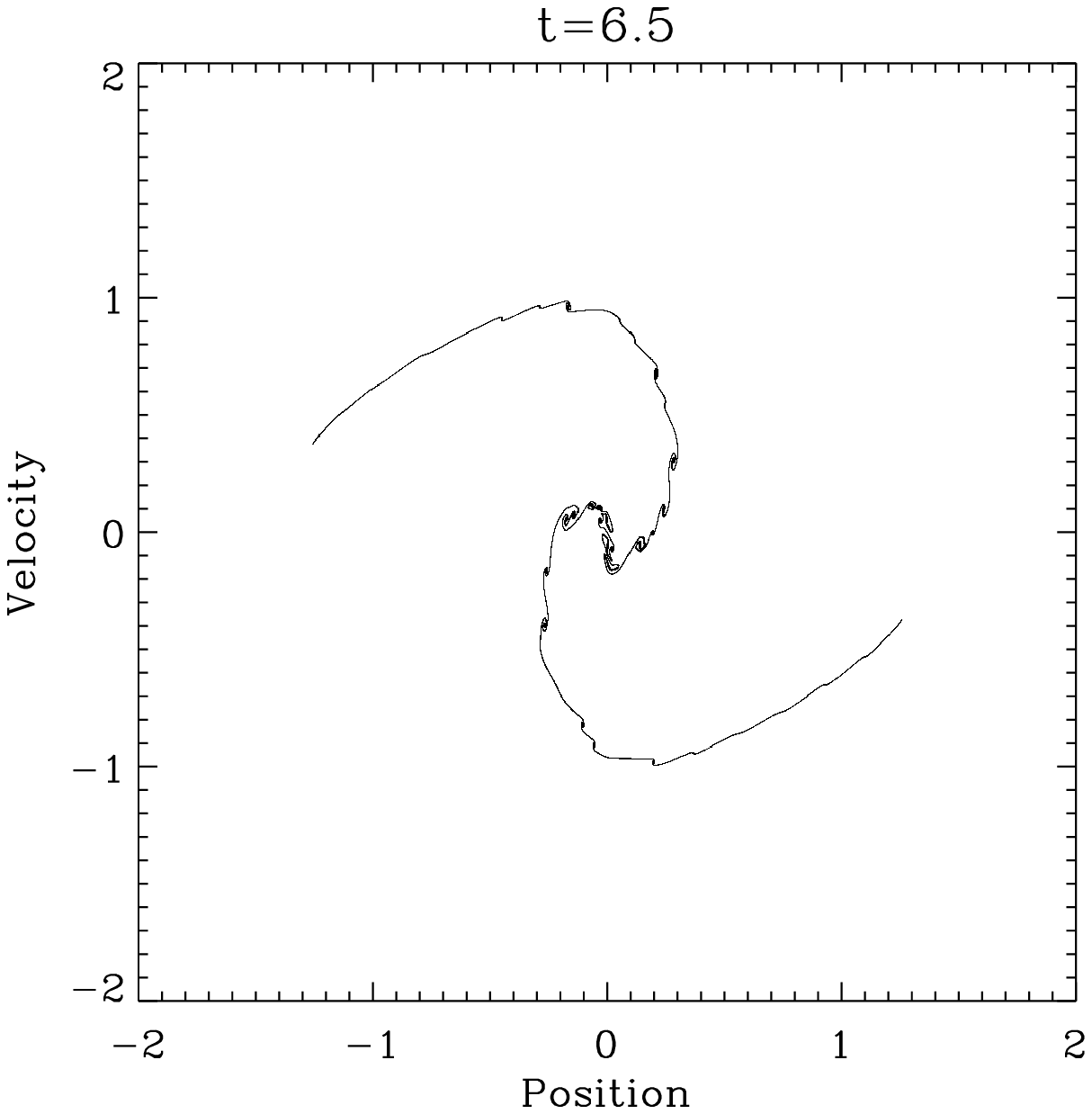,width=5.1cm,bbllx=62pt,bblly=366pt,bburx=426pt,bbury=718pt}
\hskip 0.4cm \psfig{file=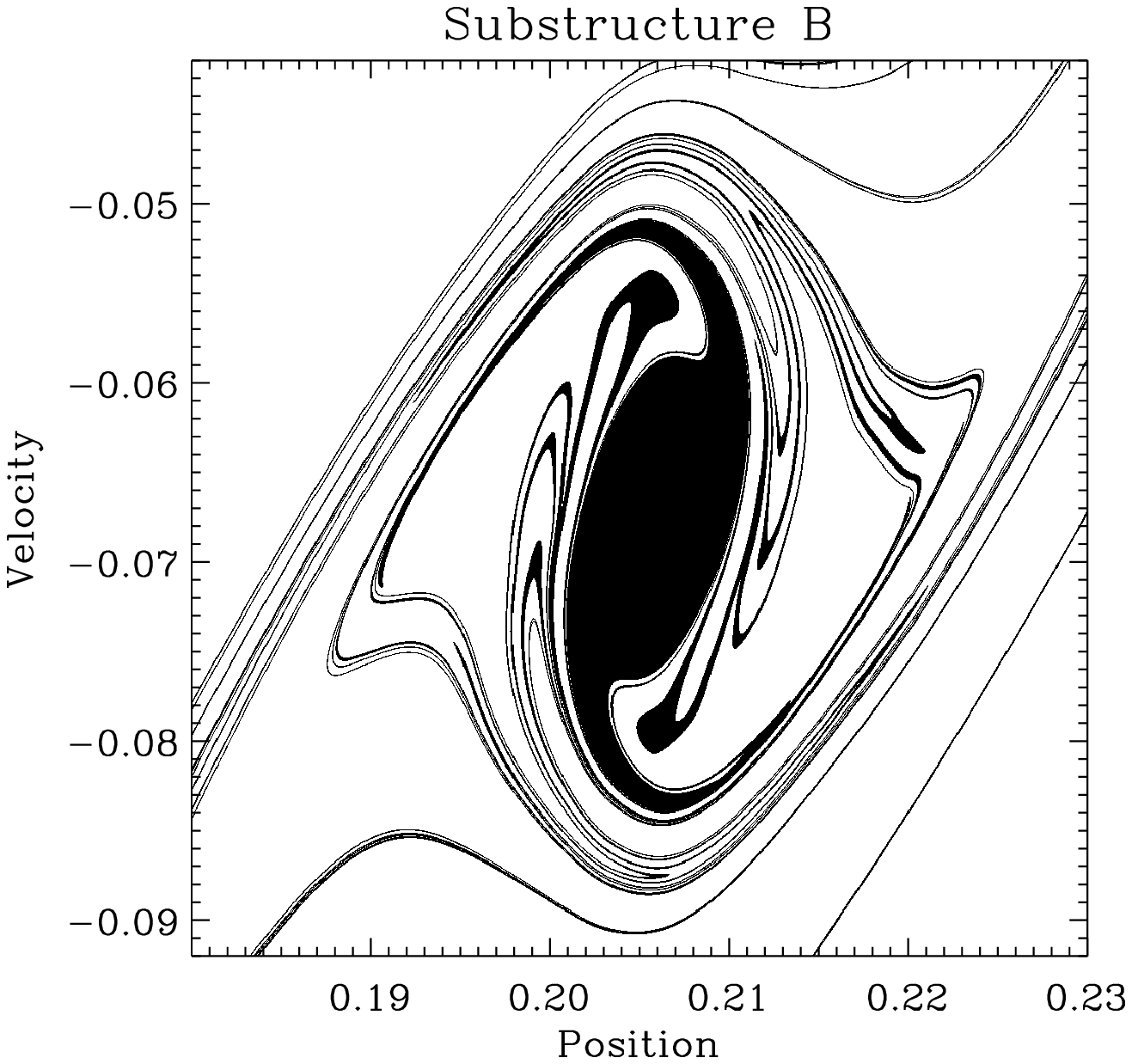,width=5.1cm,bbllx=62pt,bblly=366pt,bburx=426pt,bbury=718pt}
}}
\vskip 0.2cm
\centerline{\hbox{
\psfig{file=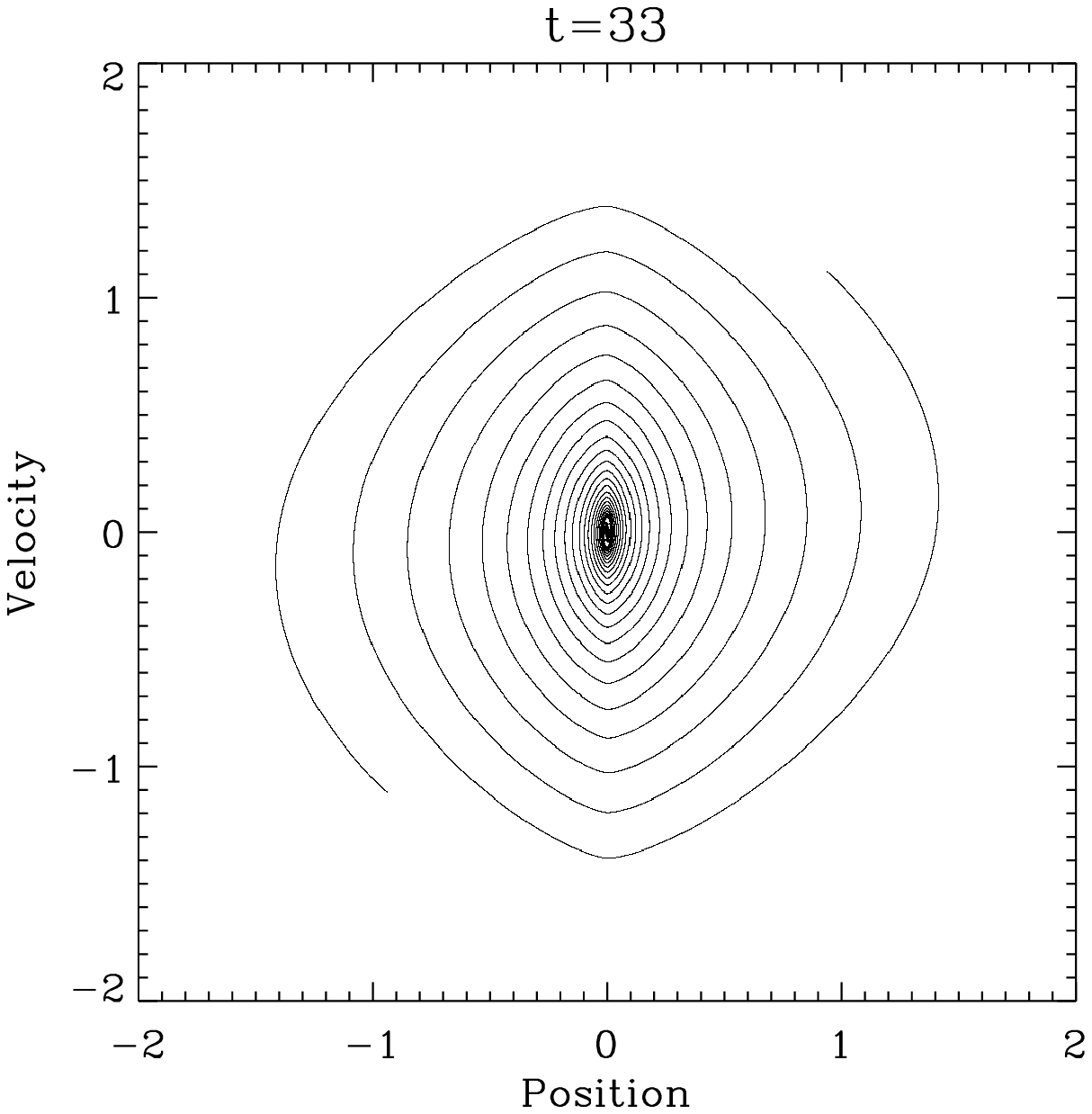,width=5.1cm,bbllx=62pt,bblly=366pt,bburx=426pt,bbury=718pt}
\hskip 0.4cm \psfig{file=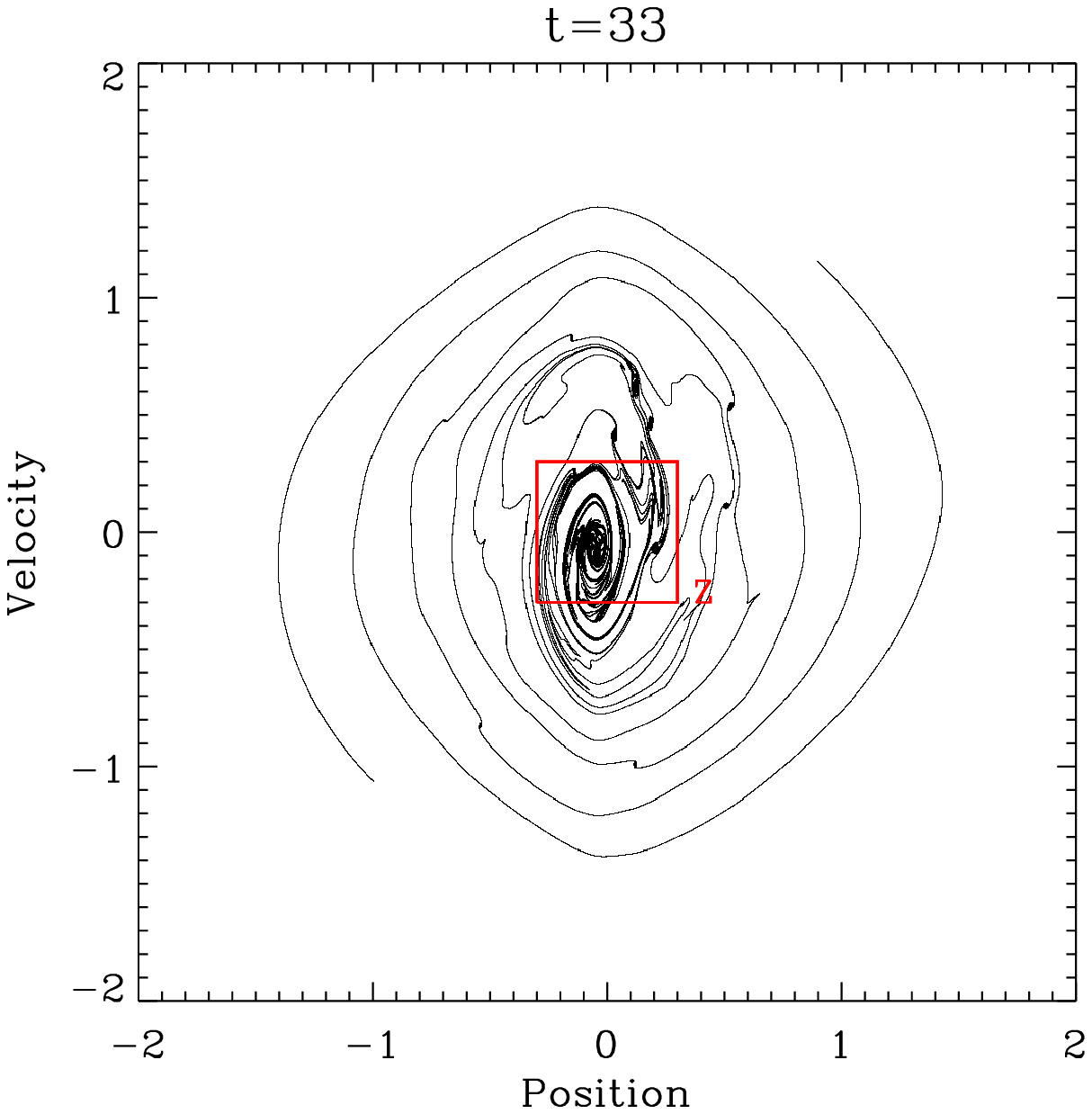,width=5.1cm,bbllx=62pt,bblly=366pt,bburx=426pt,bbury=718pt}
\hskip 0.4cm \psfig{file=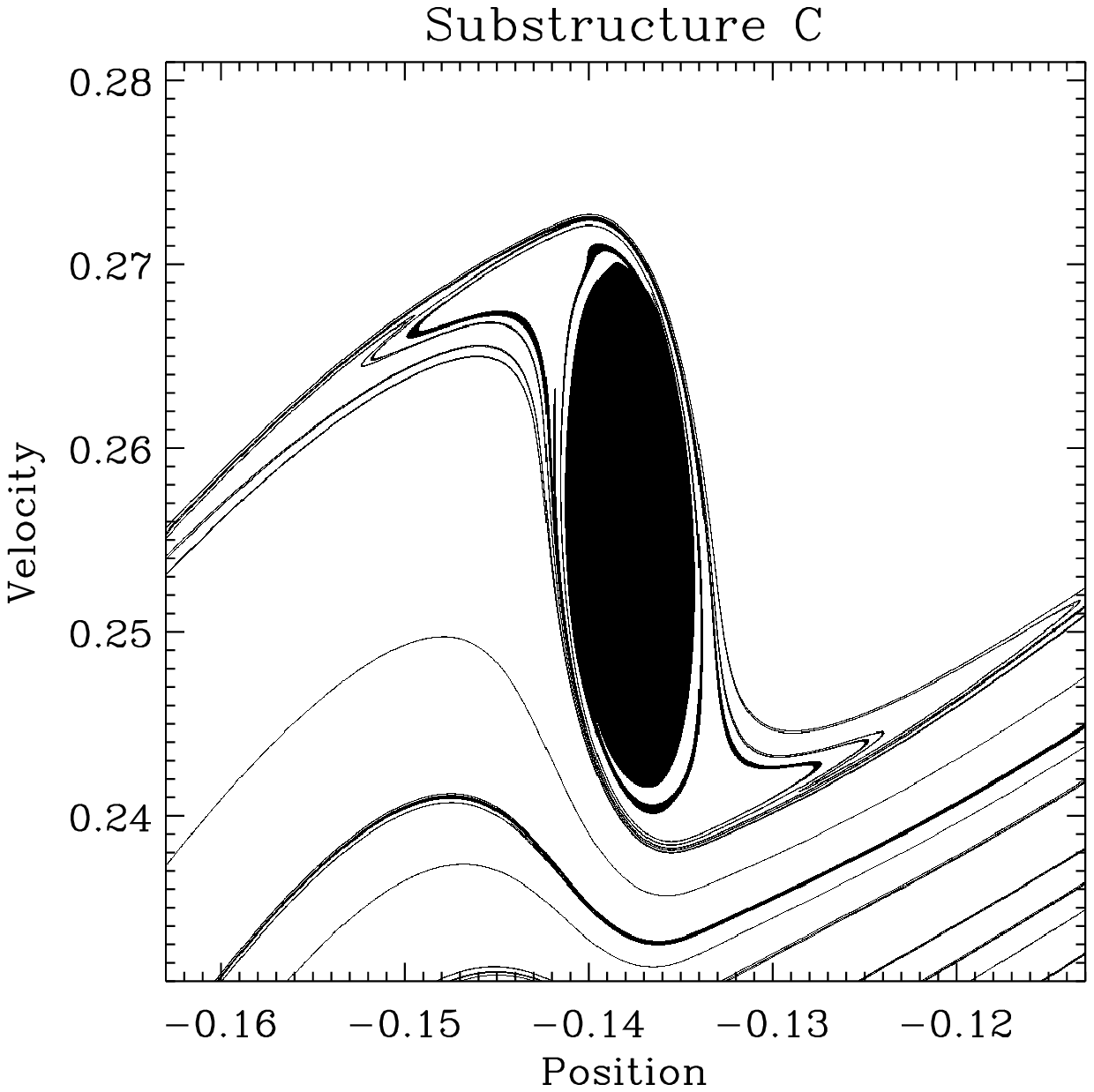,width=5.1cm,bbllx=62pt,bblly=366pt,bburx=426pt,bbury=718pt}
}}
\caption[]{The effect of random perturbations. The left column of panels shows, similarly as in Figs.~\ref{fig:convtocolda} and \ref{fig:convtocoldb}, the evolution of a waterbag with $\Delta p=0.003$. The middle column is alike, but when random perturbations have been added onto the waterbag. The four right panels correspond to successive zooms on the central part of the system (top panel), the core (second panel) and two ``subhalos'' (bottom panels). If initial conditions would be actually cold, it is reasonable to postulate that the sub-structures would present an exactly similar shape in phase-space to the unperturbed case. The simulations used here are {\tt Tophat0.003} and {\tt Perturbed} in the nomenclature of Table~\ref{tab:simuparam}. Note as discussed in the previous figure captions, other simulations would give a very similar result, except for a very slight dynamical shift.}
\label{fig:perturb}
\end{figure*}
\begin{figure*}
\centerline{\hbox{
\psfig{file=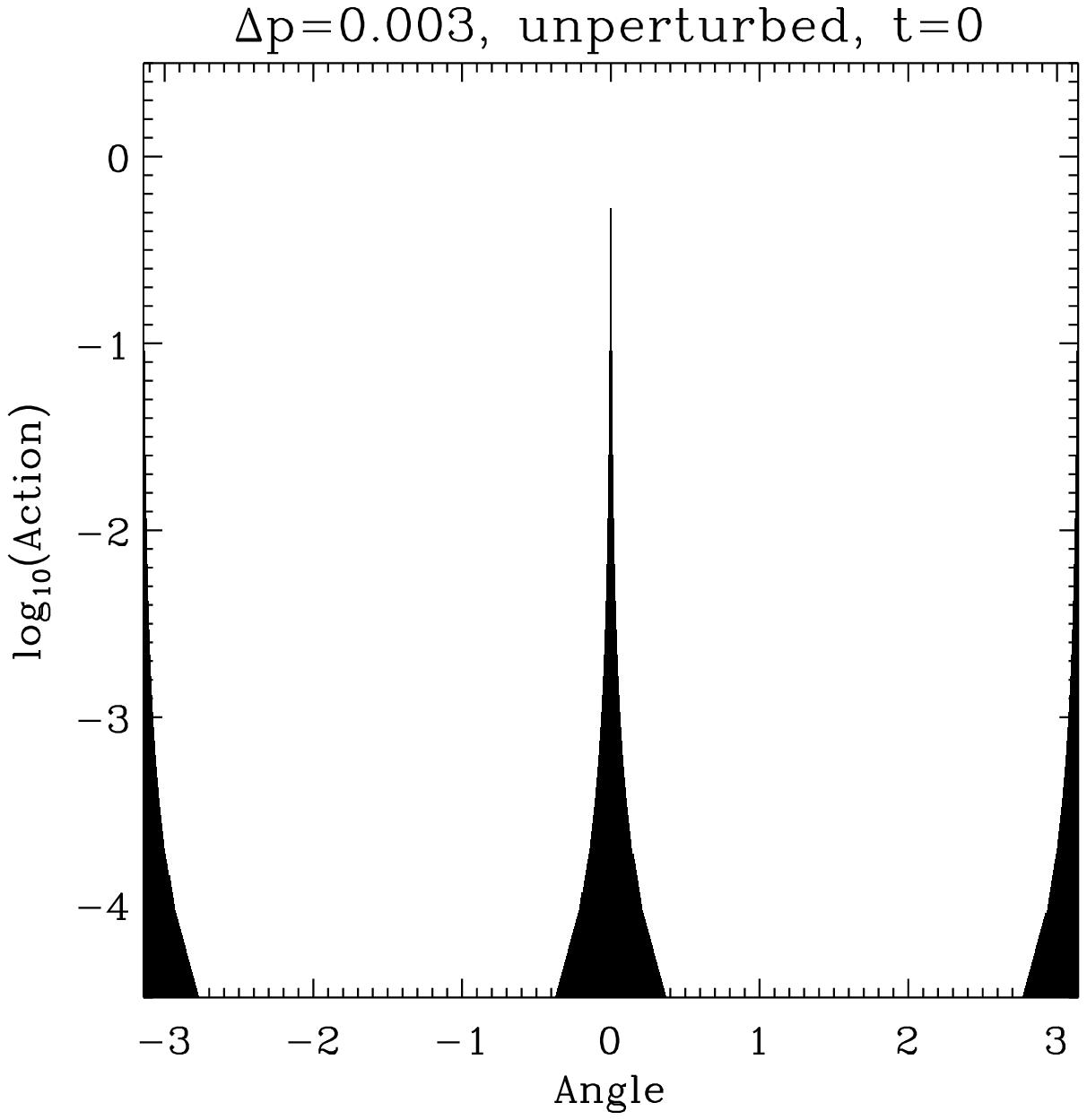,width=7.5cm,bbllx=62pt,bblly=366pt,bburx=426pt,bbury=718pt}
\hskip 0.4cm \psfig{file=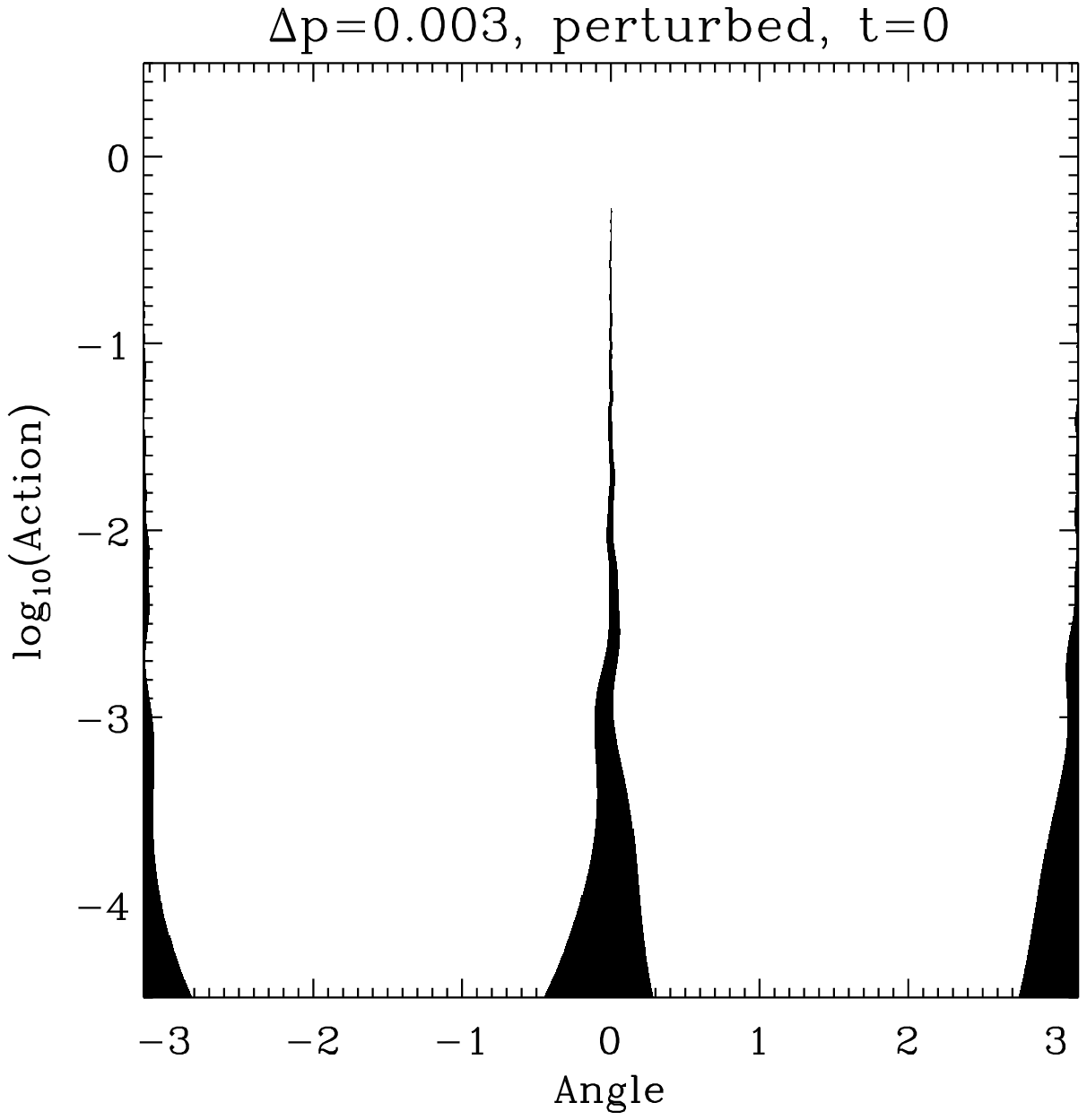,width=7.5cm,bbllx=62pt,bblly=366pt,bburx=426pt,bbury=718pt}
}}
\vskip 0.2cm
\centerline{\hbox{
\psfig{file=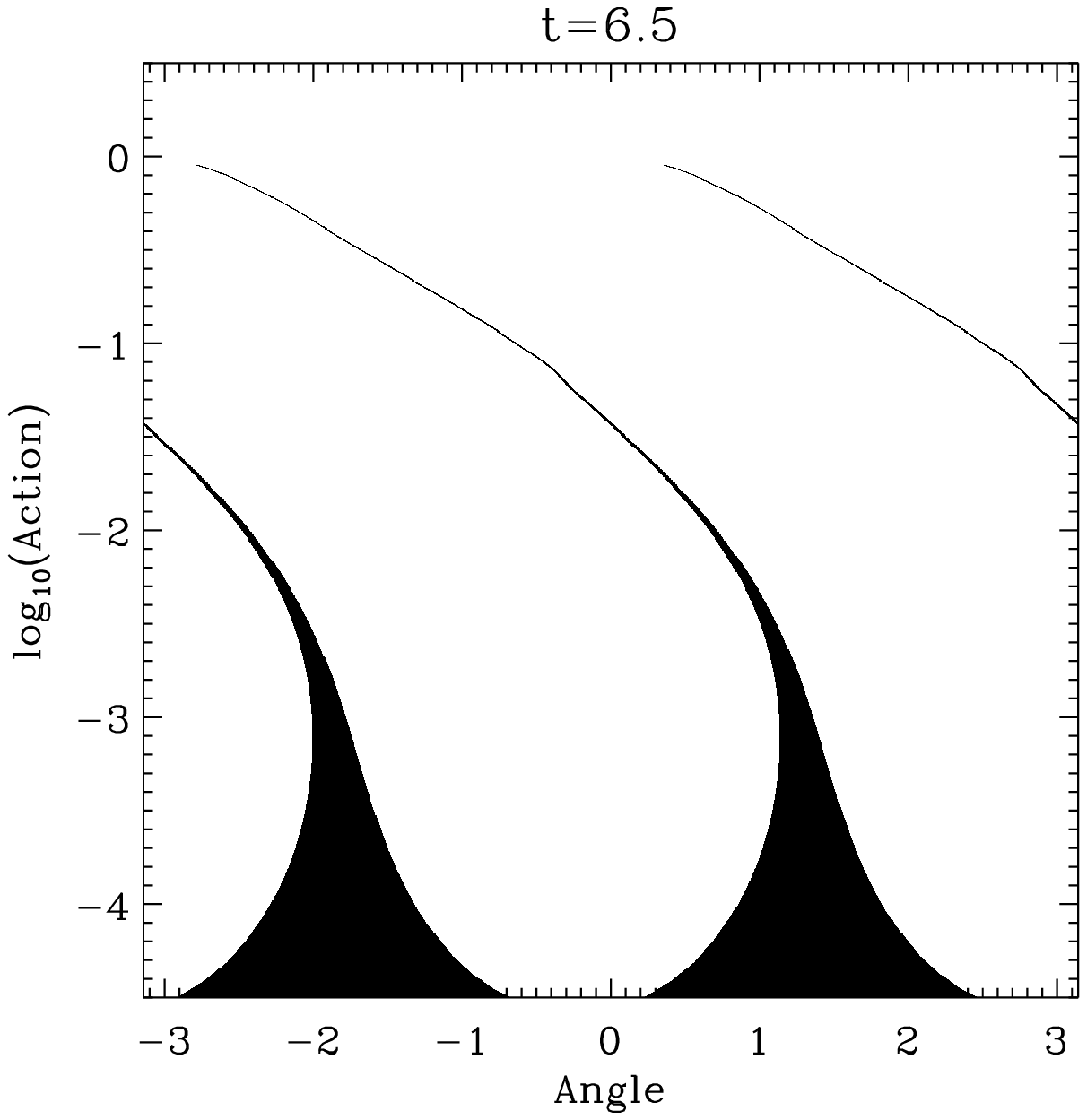,width=7.5cm,bbllx=62pt,bblly=366pt,bburx=426pt,bbury=718pt}
\hskip 0.4cm \psfig{file=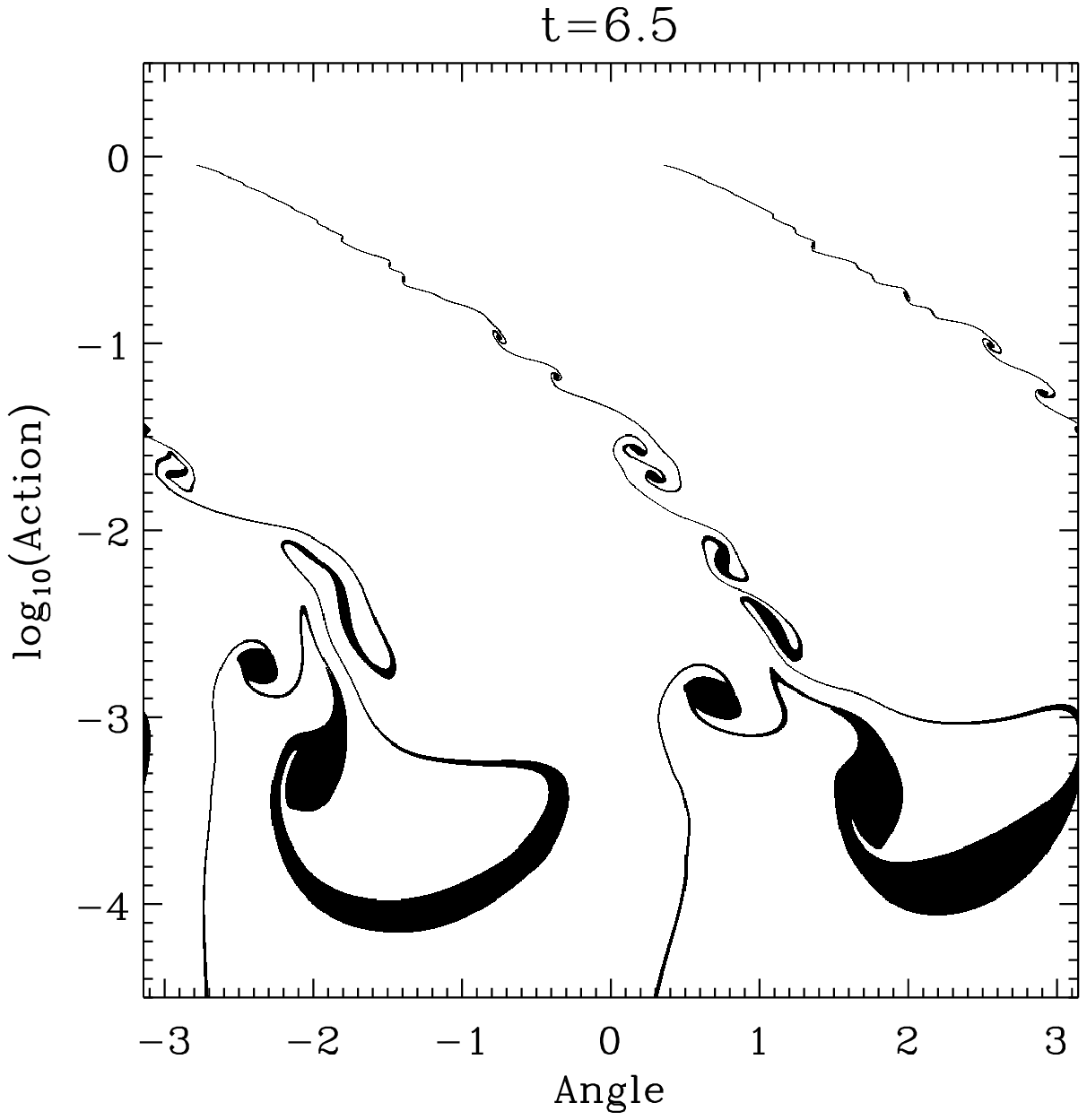,width=7.5cm,bbllx=62pt,bblly=366pt,bburx=426pt,bbury=718pt}
}}
\vskip 0.2cm
\centerline{\hbox{
\psfig{file=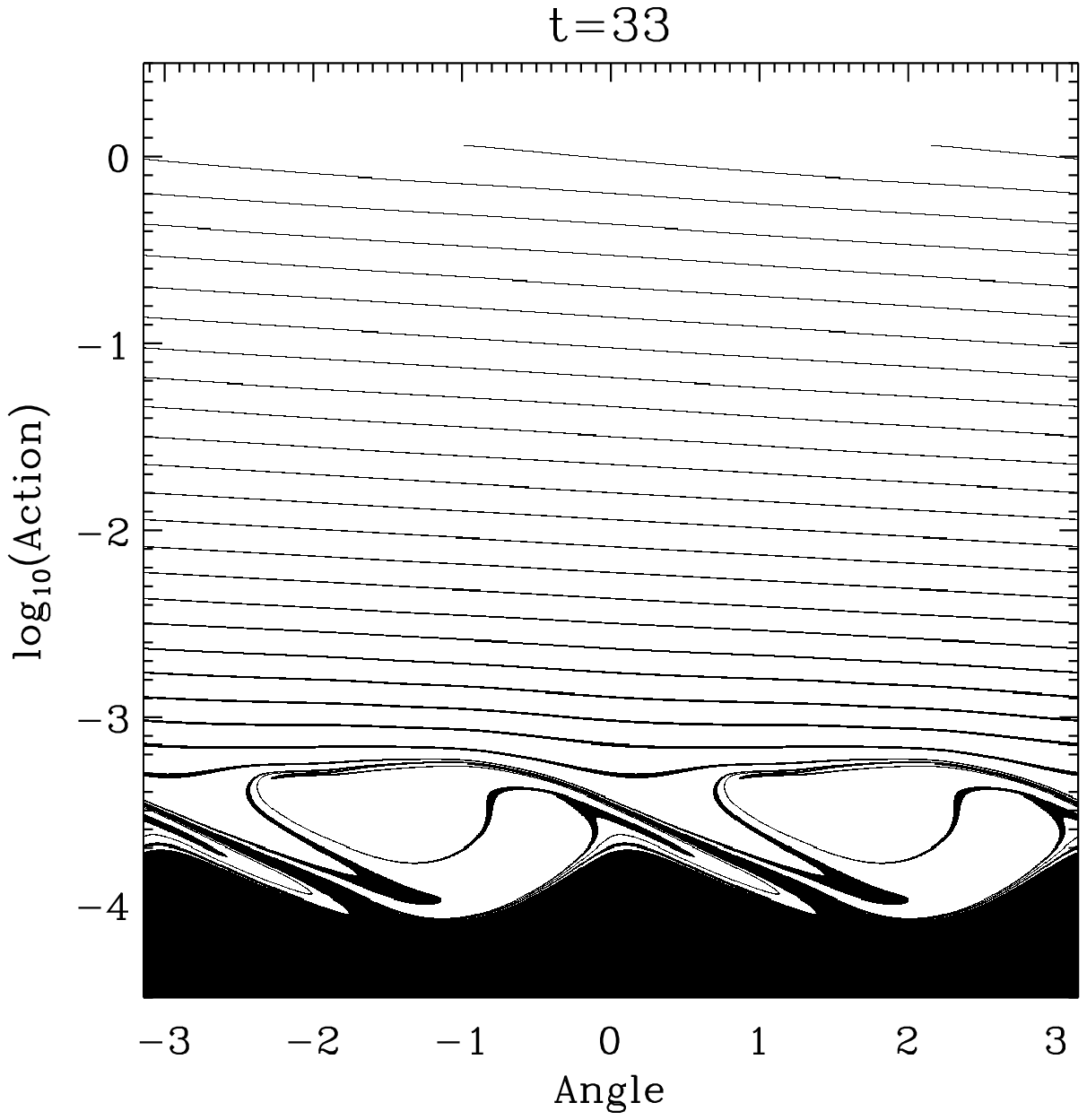,width=7.5cm,bbllx=62pt,bblly=366pt,bburx=426pt,bbury=718pt}
\hskip 0.4cm \psfig{file=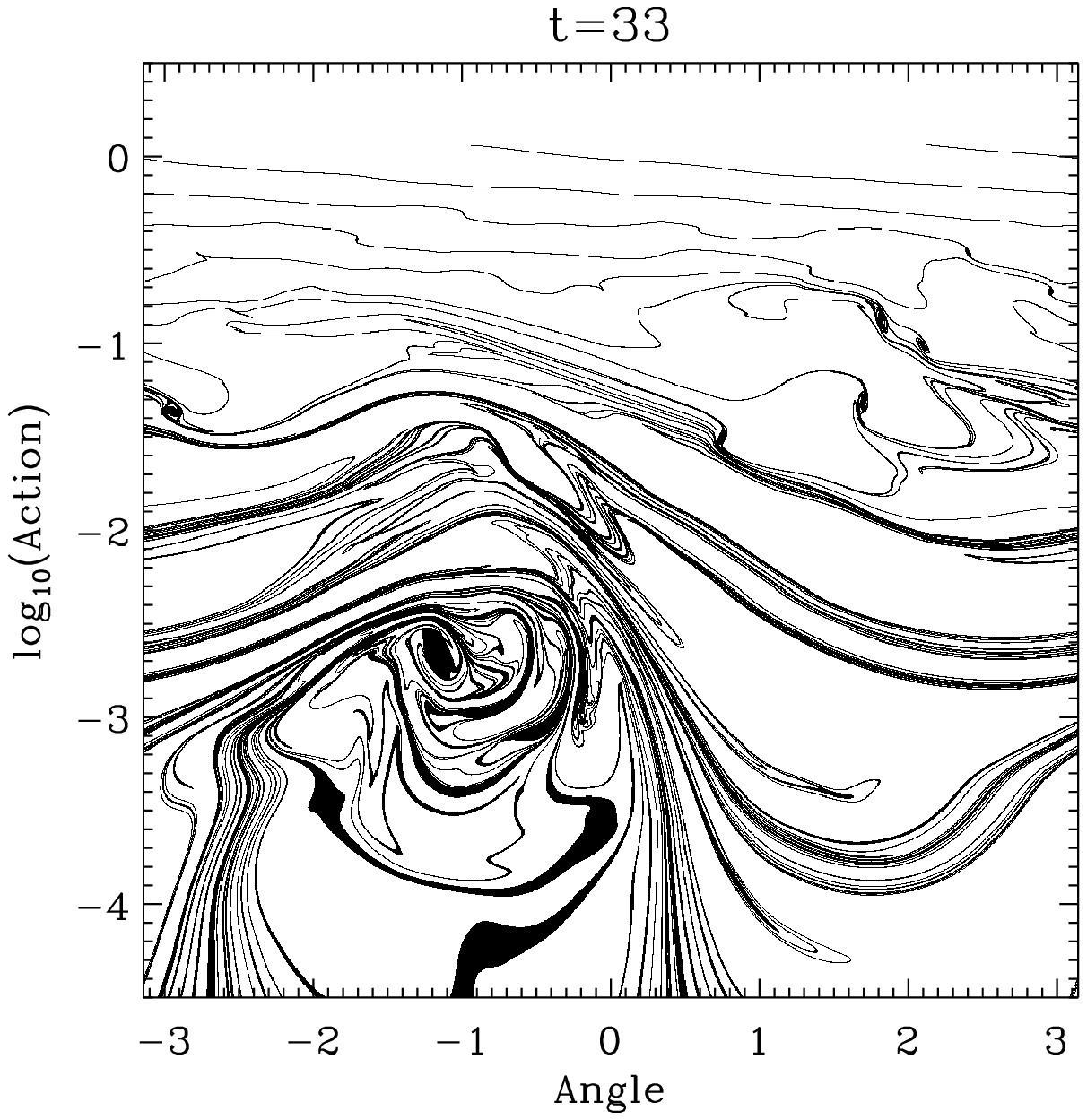,width=7.5cm,bbllx=62pt,bblly=366pt,bburx=426pt,bbury=718pt}
}}
\caption[]{Representation in Action-Angle space of the simulations of Fig.~\ref{fig:perturb}. On the left, the unperturbed single
waterbag simulation with $\Delta p=0.003$ and on the right, the randomly perturbed waterbag.}
\label{fig:tophatae}
\end{figure*}

Diagnostics also consist of performing sanity tests. Energy conservation represents a crucial test. In addition, we also tested conservation of total mass as well as the area of each individual waterbag.\footnote{The expressions for total kinetic and potential energy as well as waterbag area are given in Appendix~\ref{sec:kinepot}.} In the latter case, it is interesting to focus on the worse waterbag at a given time, because this can be used to bound violation to conservation of any casimir.\footnote{A casimir is given by
\begin{equation}
C[c] \equiv \int c[f(x,v,t)] \ {\rm d}x\ {\rm d}v =\sum_k c[f_k] V_k,
\end{equation}
where $c$ is a function assumed here to take finite values at $f_k$ and $V_k$ is the phase-space area of waterbag $k$. As a consequence of Liouville theorem, casimirs do not depend on time. With $c[f]=f$ and $c[f]=-f \ln f$, one obtains two notorious casimirs, respectively the total mass and the Gibbs entropy. The violation on conservation of $c[f]$ can be written
\begin{equation}
|\Delta C[f]| \leq  {\rm max}_k |\Delta V_k| \times \sum_l |c[f_l]|,
\end{equation}
and can thus be bounded in terms of violation to area conservation of the worse waterbag.} 
However, we found in practice that total energy conservation represents the strongest test. As studied in detail in Appendix~\ref{app:enercons}, energy conservation remains excellent for all the simulations we did, better than $\sim 2\times 10^{-4}$ in warm cases and than $\sim 10^{-3}$ in colder configurations, except for one of the randomly perturbed waterbag simulations with unrefinement allowed. As already discussed in \S~\ref{sec:myref}, unrefinement does indeed introduce long term noise that worsens energy conservation after a number of dynamical times. With unrefinement inhibited, energy can in fact be conserved at a level better than $\sim 5 \times 10^{-5}$ and $\sim 2 \times 10^{-4}$ in warm and cold cases, respectively. 
%
%
%
%----------------------------------------------------------------------------
\section{A convergence study to the cold case: single waterbags}
\label{sec:applications}
%----------------------------------------------------------------------------
%
%
In this section, we focus on the single waterbag simulations. The main purpose of this analysis is to study the relaxation of the profile to a quasi-stationary state in the limit when the waterbag becomes infinitely thin, corresponding to the cold case. After a detailed visual inspection of the simulations (\S~\ref{sec:tophatvisu}), we analyze, in the nearly cold case, the properties of the inner profile that is built during relaxation, starting first with the gravitational potential and its logarithmic slope (\S~\ref{sec:gravpot}), then proceeding with the phase-space energy distribution function (\S~\ref{sec:fofE}). In a final discussion (\S~\ref{sec:discu}), we compare our results to previous works, paying particular attention to measurements in $N$-body simulations.
%----------------------------------------------------------------------------
\subsection{Visual inspection}
\label{sec:tophatvisu}
%----------------------------------------------------------------------------
Figures~\ref{fig:convtocolda} and \ref{fig:convtocoldb} display, for each value of the thickness parameter $\Delta p$ in the range $[0.01,0.1]$, the phase-space distribution function of the single waterbag simulations at various times, showing the well known building up of a quasi-stationary profile with a core and a spiral halo \citep[e.g.,][]{1971A&A....11..188J,1971Ap&SS..13..411C,1971Ap&SS..13..425C}. The appearance of the halo arises from the filamentation of the external part of the waterbag, while a compact core survives. Figure~\ref{fig:coldzoomcenter} allows one to distinguish the core for the smallest values of $\Delta p$. 
Null for $\Delta p=1$, where the waterbag keeps a well defined oscillating balloon shape,\footnote{This is due to the fact that initial conditions are very close to a stable single waterbag stationary solution \citep[see, e.g.,][]{Severne1975}, hence the waterbag contour oscillates with a small amplitude around this solution.} the fraction of the mass feeding the halo increases with $1/\Delta p$, leaving a core of which the projected size varies roughly with ${\Delta p}^{0.8}$ for $\Delta p \la 0.1$.\footnote{Such a power-law behavior can be derived from the visual examination of top right panel of Fig.~\ref{fig:potential}.} 
In all the cases except for $\Delta p=1$, there is a region between the halo and the core where the system presents an unstable behavior. 
The extension of this region is of the same order of that of the core. Note also, from inspection of Fig.~\ref{fig:coldzoomcenter}, that the shape of the spiral remains the same whatever $\Delta p \la 0.01$ when far enough from the center: in agreement with intuition, the details of the shape of the central region in the vicinity of the core do not influence the dynamics of the outer spiral. The shape of this spiral can be computed analytically under the assumption of self-similarity \citep{Alard2013}, which, as discussed in next section, applies at least to some extent to our cold waterbags. 

Figures~\ref{fig:perturb} and \ref{fig:tophatae} focus on the perturbed waterbag, with a comparison
to its unperturbed counterpart in phase-space and in Action-Angle space, respectively. The presence of random perturbations induces the formation of sub-structures and also makes the extension of the unstable region in the center of the system much larger, as illustrated by the four right panels of Fig.~\ref{fig:perturb}. Another interesting property, is that filaments tend to pack together in phase-space, leaving larger empty regions than in the unperturbed case: this is particularly visible when comparing the two bottom panels of Fig.~\ref{fig:tophatae}. 

\begin{figure}
%\centerline{\hbox{
%\psfig{file=figures_tmp2/tophat0v001_naddremPPT.ps,width=8cm}
%}}
\centerline{\hbox{
\psfig{file=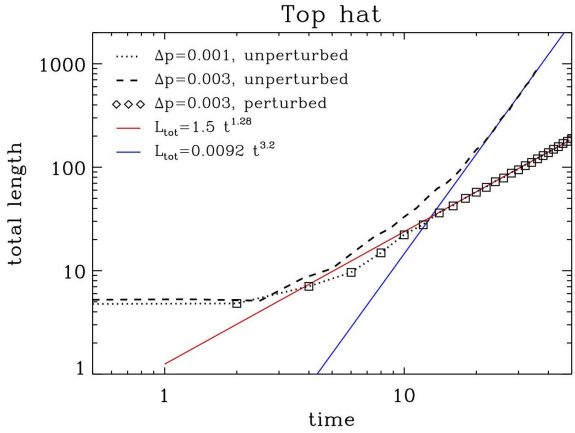,width=8.5cm}
}}
\caption[]{The total length of the waterbag contour as a function of time. We consider here our coldest set ups,
with $\Delta p=0.001$ and $0.003$, as well as the randomly perturbed $\Delta p=0.003$ waterbag. In the unperturbed case, the total length soon behaves roughly like a power-law as indicated by the red line. Random perturbations induce the appearance of numerous sub-structures and increased filamentation: the length soon augments much faster with time than for the unperturbed case. The blue line shows a late
time power-law fit, but  this is obviously only indicative, as it seems clear that the length increases with time faster than a power-law. The measurements have been made for the simulations {\tt Tophat0.001}, {\tt Tophat0.003} and {\tt Perturbed} in the nomenclature of Table~\ref{tab:simuparam}, but would not change for other runs we performed with the same initial conditions.}
\label{fig:tophatlength}
\end{figure}

Figure~\ref{fig:tophatlength} displays the total length of the waterbag as a function of time for small values of $\Delta p$.\footnote{As a complement, bottom panel of Fig.~\ref{fig:npts} gives the total number of vertices as a function of time for the all single waterbag simulations we did.} Without perturbation, the length behaves soon as a power-law of time of index $1.28$ for $t \ga 10$, a result which might again be interpreted in terms of a self-similar spiral \citep{Alard2013}. In the perturbed case, the length seems, not surprisingly, to increase faster than a power-law although we could perform an indicative fit at late time with a logarithmic slope of $3.2$. 
%----------------------------------------------------------------------------
\subsection{The gravitational potential}
\label{sec:gravpot}
\begin{figure*}
\centerline{\hbox{
\hskip -0.1cm\psfig{file=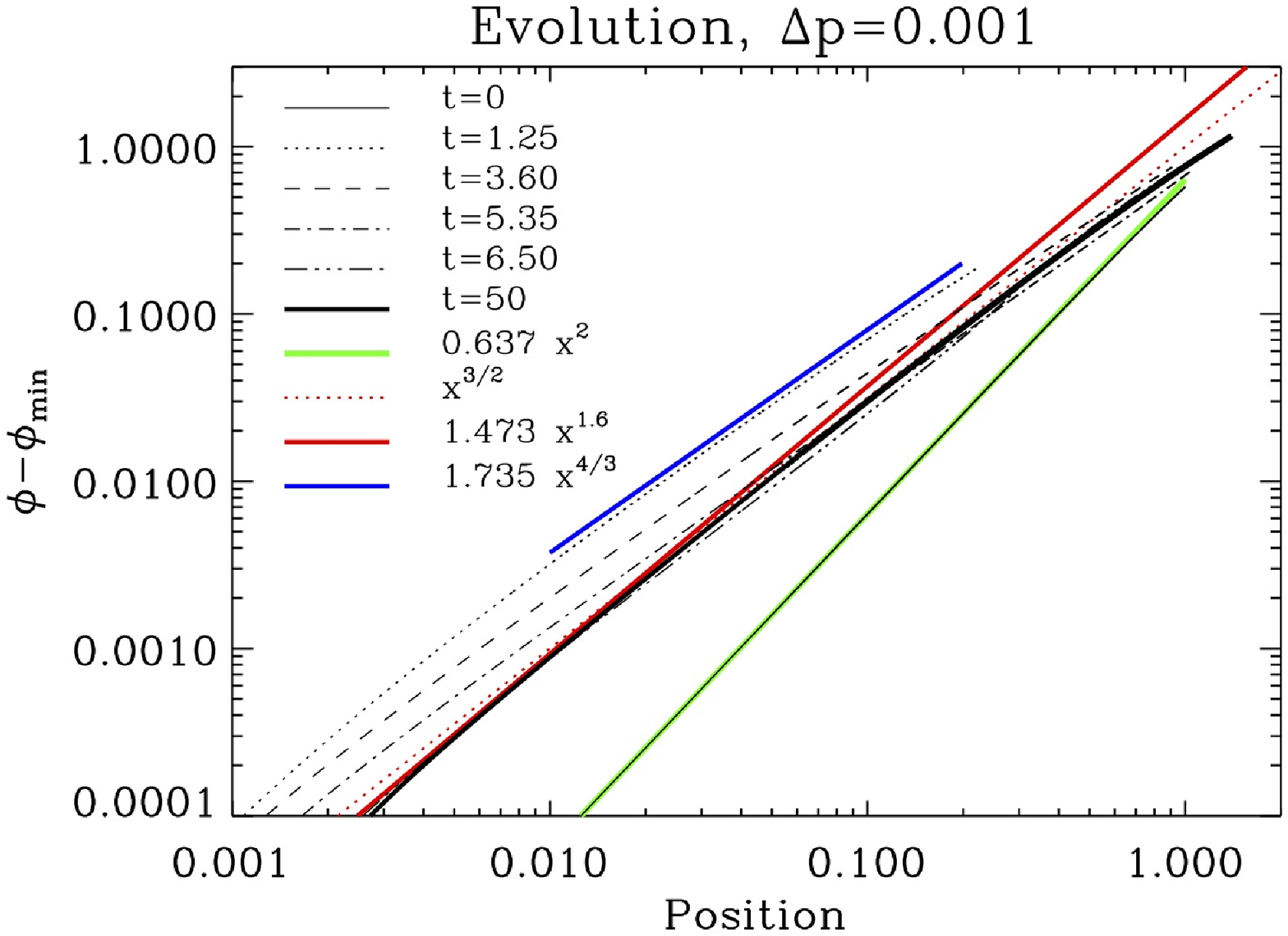,width=8.8cm}
\psfig{file=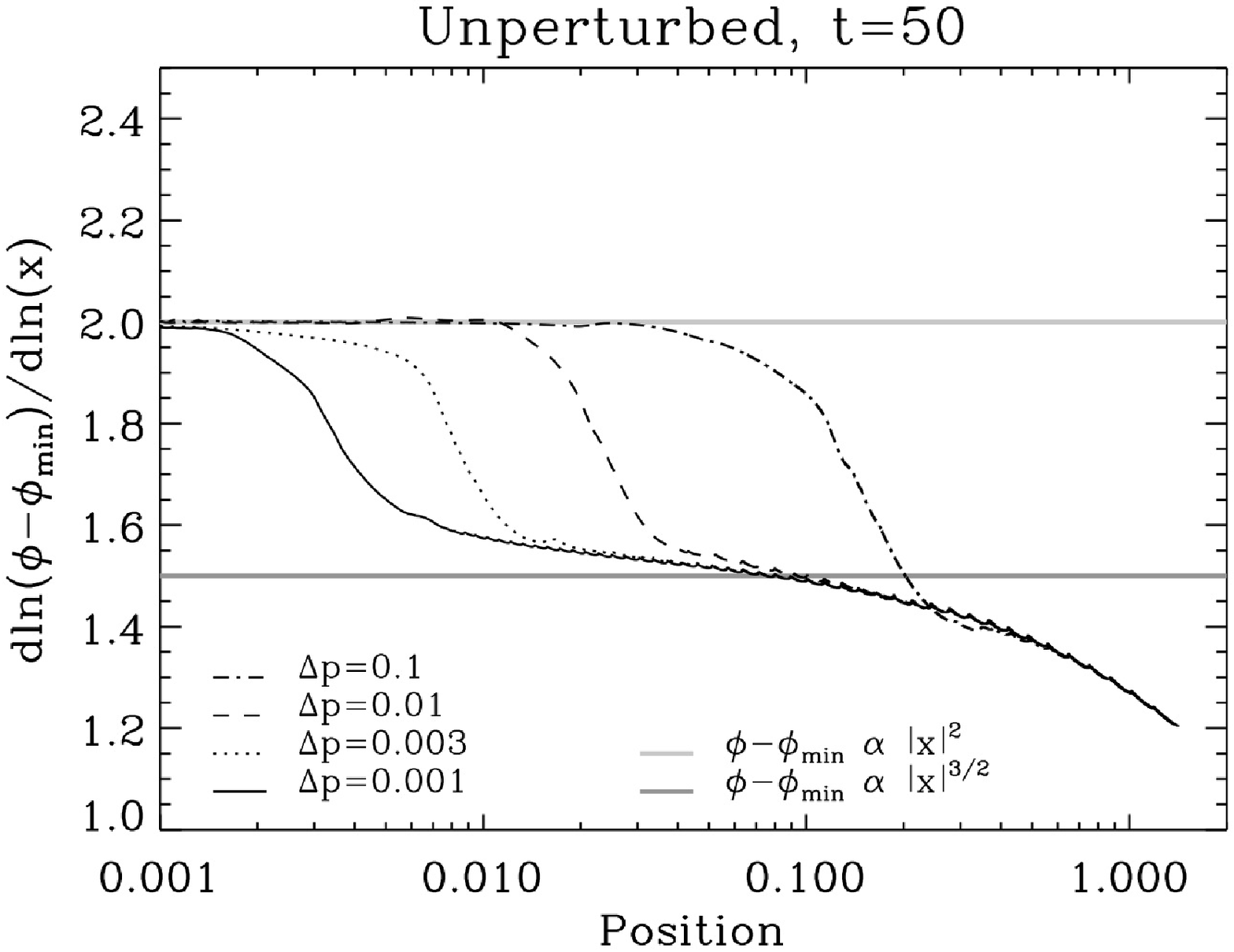,width=8.8cm}}}
\centerline{\hbox{
\psfig{file=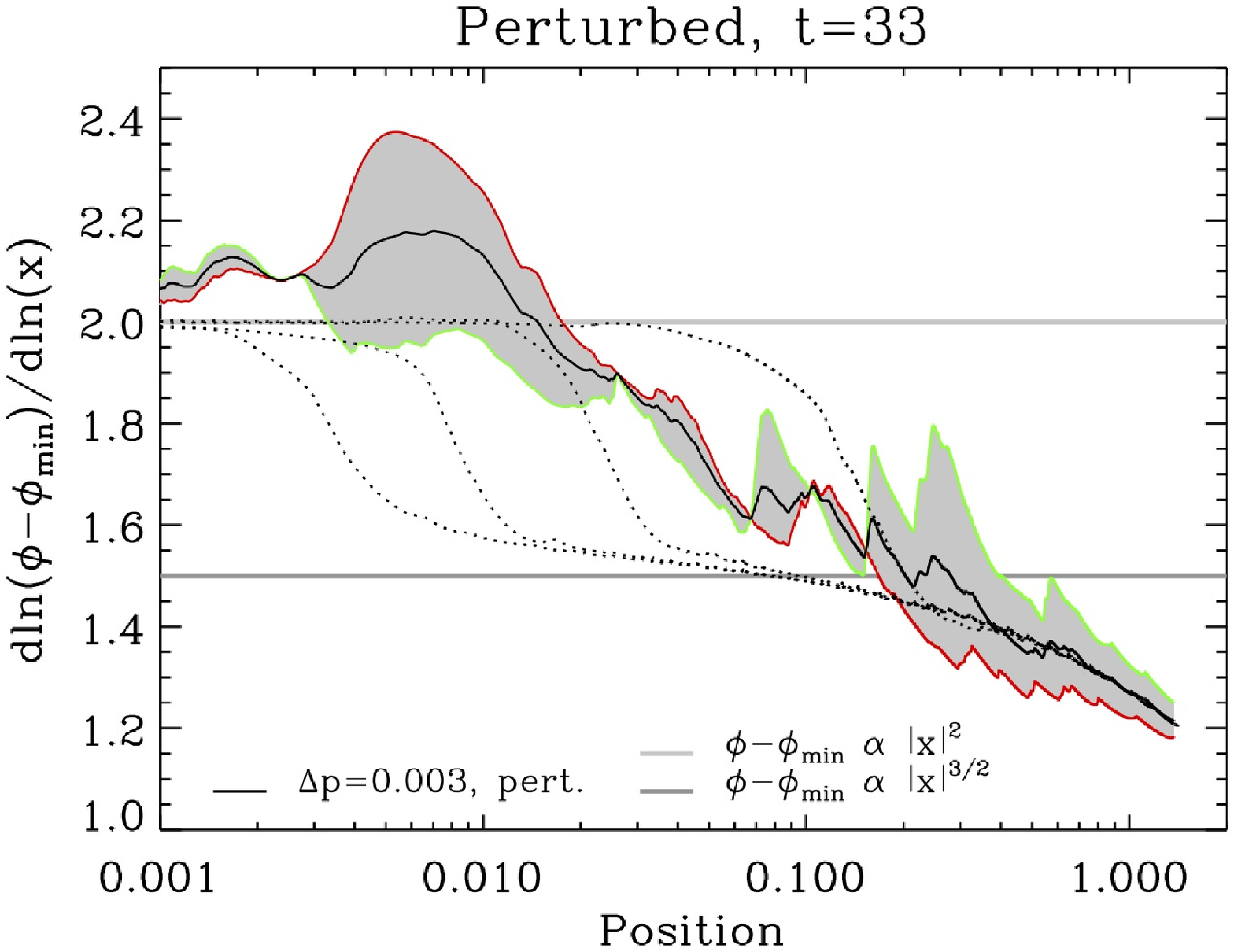,width=8.8cm}
\psfig{file=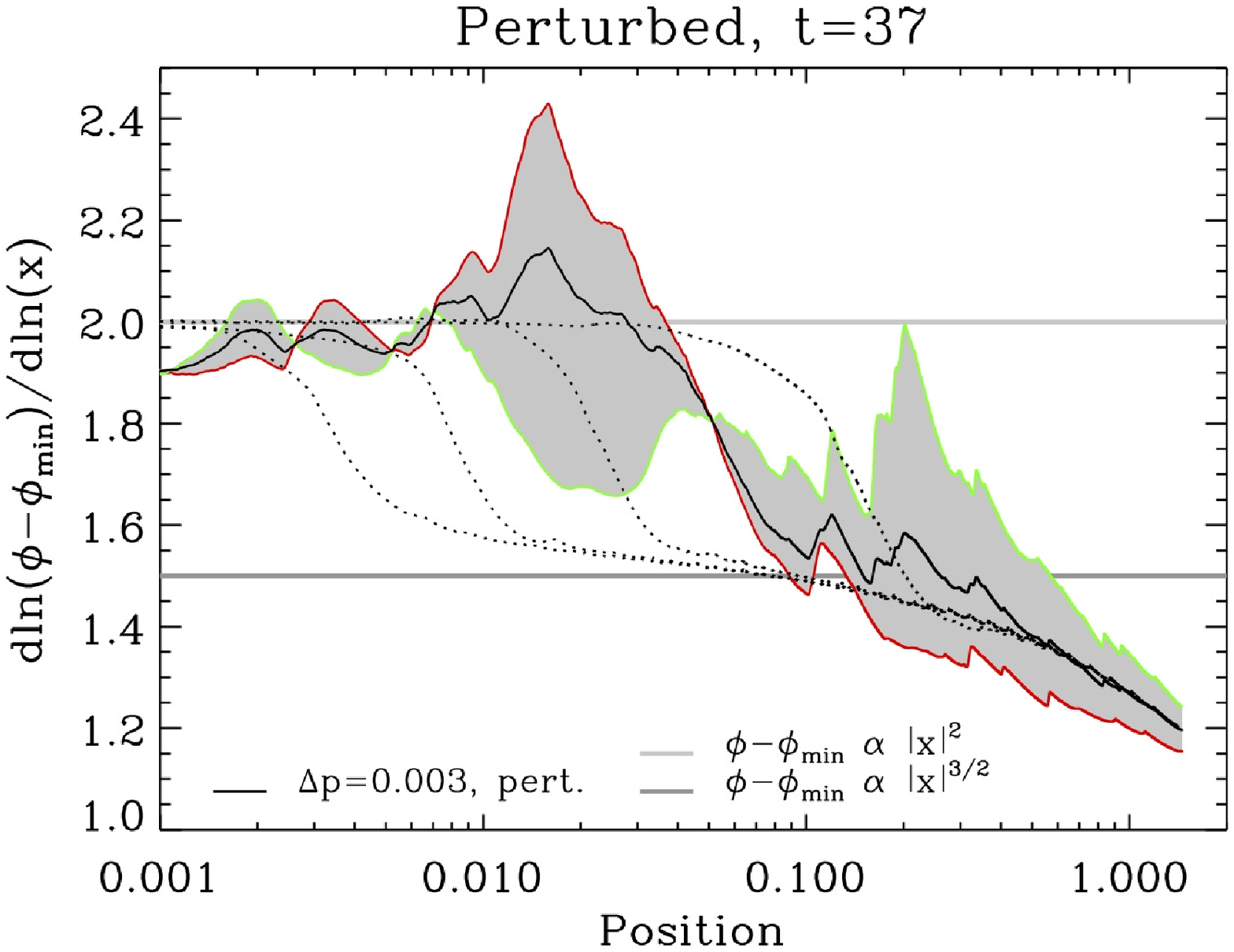,width=8.8cm}
}}
\caption[]{The gravitational potential properties in the cold case. On the {\em top-left panel}, the potential is plotted for the $\Delta p=0.001$ case as a function of scale at various times, starting from initial conditions. Except for the last snapshot of the simulation, $t=50$, the curves with $t> 0$ correspond to the first four crossing times. The green and the blue line stand for analytic predictions of \S~\ref{sec:fofE}, respectively for initial conditions and collapse time. The red power-law is the result of assuming an average phase-space density per energy level proportional to that obtained at collapse time, as discussed in \S~\ref{sec:fofE},  while the dotted one corresponds to a conjecture of \cite{Binney} based on measurement on $N$-body simulations. The {\em top-right panel} displays the logarithmic slope of the gravitational potential, for various values of $\Delta p$ in order to be able to perform a convergence study. The bottom and top gray lines correspond respectively to the index predicted by Binney and the one expected when a core dominates at the center.  The {\em two bottom panels} show the logarithmic slope measured in the perturbed case, at two different times. There is a gray shaded area bordered by a green and a red contour. These two contours correspond to the measurement of the potential on each side of its minimum, while the black curve is the average between them. In addition, the measurements displayed on top right panel are shown as dotted curves. This figure uses simulations {\tt Tophat0.001}, {\tt Tophat0.003}, {\tt Tophat0.010}, {\tt Tophat0.100U} and {\tt Perturbed} in the nomenclature of Table~\ref{tab:simuparam}, but it would not change significantly for other runs we performed with the same initial conditions. }
\label{fig:potential}
\end{figure*}

%----------------------------------------------------------------------------
The gravitational potential is shown at various times in the $\Delta p=0.001$ case on the top-left panel of Fig.~\ref{fig:potential}. The initial conditions correspond to an approximately harmonic potential with $\phi(x)-\phi_{\rm min} \propto x^2$ (green line). As discussed further in \S~\ref{sec:fofE}, in the pure cold case, the projected density presents a singularity in the center such that $\phi(x)-\phi_{\rm min} \propto x^{4/3}$ at collapse time and subsequent crossing times. This is indeed the case for our measurements if one stays sufficiently far away from the center (blue line, which superposes well to the dotted curve). However, the system relaxes very rapidly to a quasi-stationary state. The overall profile of this latter follows rather well a power-law of the form $\phi(x)-\phi_{\rm min} \propto x^{3/2}$ \citep[][red dots]{Binney}. There are some noticeable deviations from such a power-law, that we discuss now. 

To examine more in detail the scaling behavior of the potential, one can study its logarithmic slope, which can be defined as 
\begin{equation}
\beta(x)=\frac{|a(x)|}{\phi-\phi_{\rm min}},
\label{eq:betaest}
\end{equation}
where $\phi_{\rm min}$ is the minimum of the potential. Because it depends on the acceleration and on the potential, the quantity $\beta(x)$ is a well behaved estimator. It is expected be a smooth function of $x$ as shown on top right panel of Fig.~\ref{fig:potential} for $\Delta p \leq 0.1$. In our waterbag case, it shoud tend to 2 in the limit $\phi \rightarrow \phi_{\rm min}$ as a test of robustness, which is indeed the case. Finally it is rather insensitive to the presence of the core in the region where this latter should not contribute, as the superposition of the curves on top right panel of Fig.~\ref{fig:potential} demonstrate.

Using several simulations with different values of $\Delta p$ allows us to perform a convergence study to the cold case and in particular to figure accurately where the measurements are influenced by the core. For instance, for $\Delta p=0.001$, it is reasonable to state that the presence of a core does not affect the measured slope when $x \ga 0.01$, $\phi \ga 10^{-3}$, for which we find $\beta \simeq 1.57$. With the available dynamic range at our disposal, there is no clear convergence of function $\beta(x)$  to a constant at small $x$. The parameter $\beta$ seems indeed to continue slowly increasing in magnitude while reaching the smallest scales. The lack of a well defined power-law for the gravitational potential reminds us of the results obtained in the three-dimensional case, where the density profiles of dark matter halos are found in the most accurate $N$-body simulations to follow an Einasto profile \citep[see, e.g.,][]{Merritt2006,Navarro2010}. We can only set a firm lower bound for $\beta$ for small values of $x$:
\begin{equation}
\beta(x) > 1.54, \quad x \ll 1,
\end{equation}
by using the lowest possible value of $x\simeq 0.02$ for which the solid and the dotted curves still coincide on upper-right panel of Fig.~\ref{fig:potential}. 

In the randomly perturbed simulations, the results, shown on the two bottom panels of Fig.~\ref{fig:potential} for two different times, are analogous to the unperturbed case, except that they are much more noisy and that the system builds a much larger ``core'' than in the unperturbed simulations. We use quotes, because this region of approximate constant projected density is in fact quite intricate in phase-space and rather ``chaotic''. Its projected size seems to range between those of the $\Delta p=0.01$ and $\Delta p=0.1$ unperturbed simulations. Our measurements in the perturbed case are however inconclusive, because we were unable to follow the system during sufficiently many dynamical times to have reached an actually quasi-steady state and we tested only one specific kind of perturbations. So from now on, unless specified otherwise, we discuss the unperturbed case corresponding to the top panels of Fig.~\ref{fig:potential}.

%----------------------------------------------------------------------------
\subsection{The phase-space energy distribution function}
\label{sec:fofE}
%----------------------------------------------------------------------------
\begin{figure*}
\centerline{\hbox{
\psfig{file=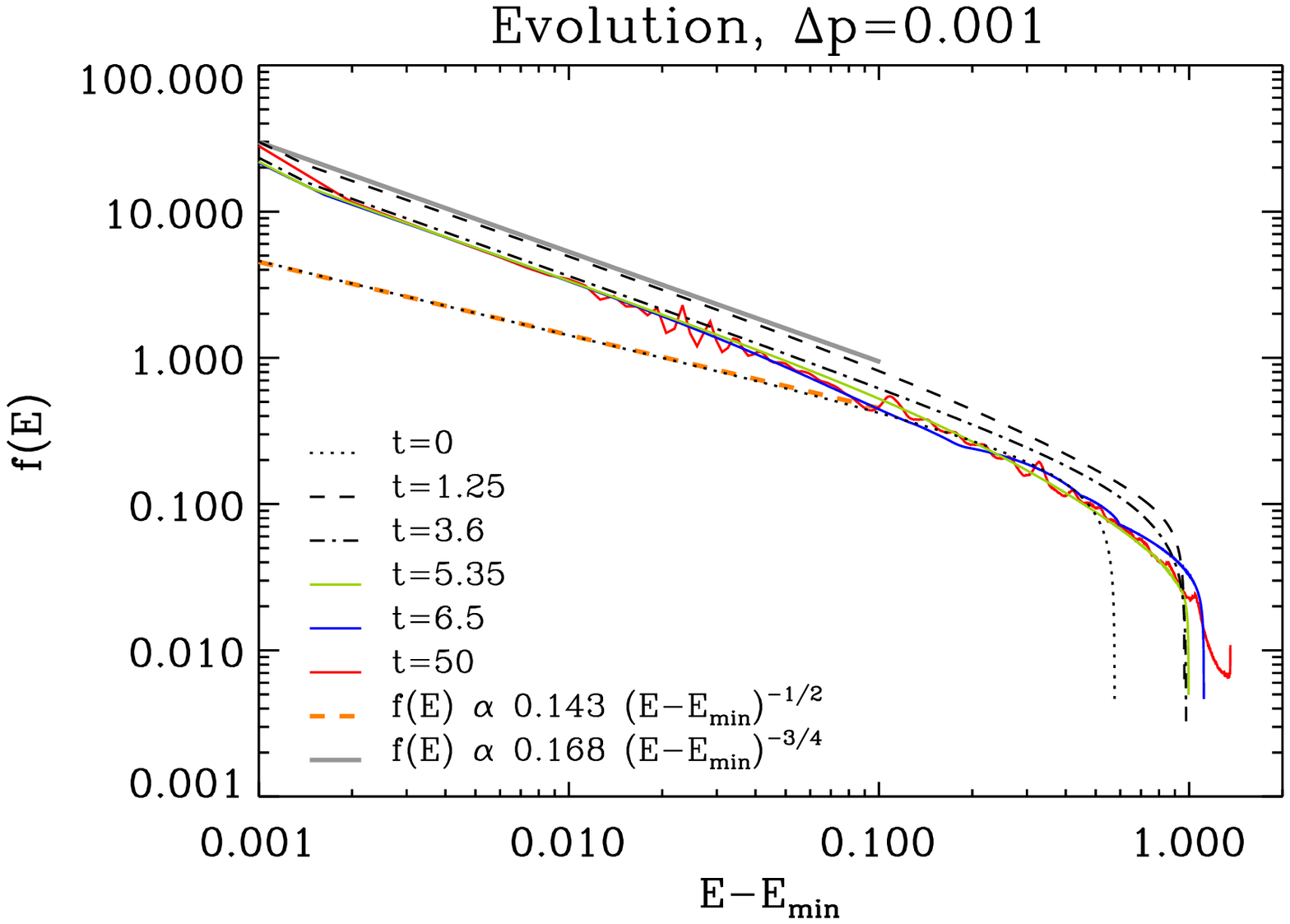,width=8.5cm}
\hskip 0.4cm
\psfig{file=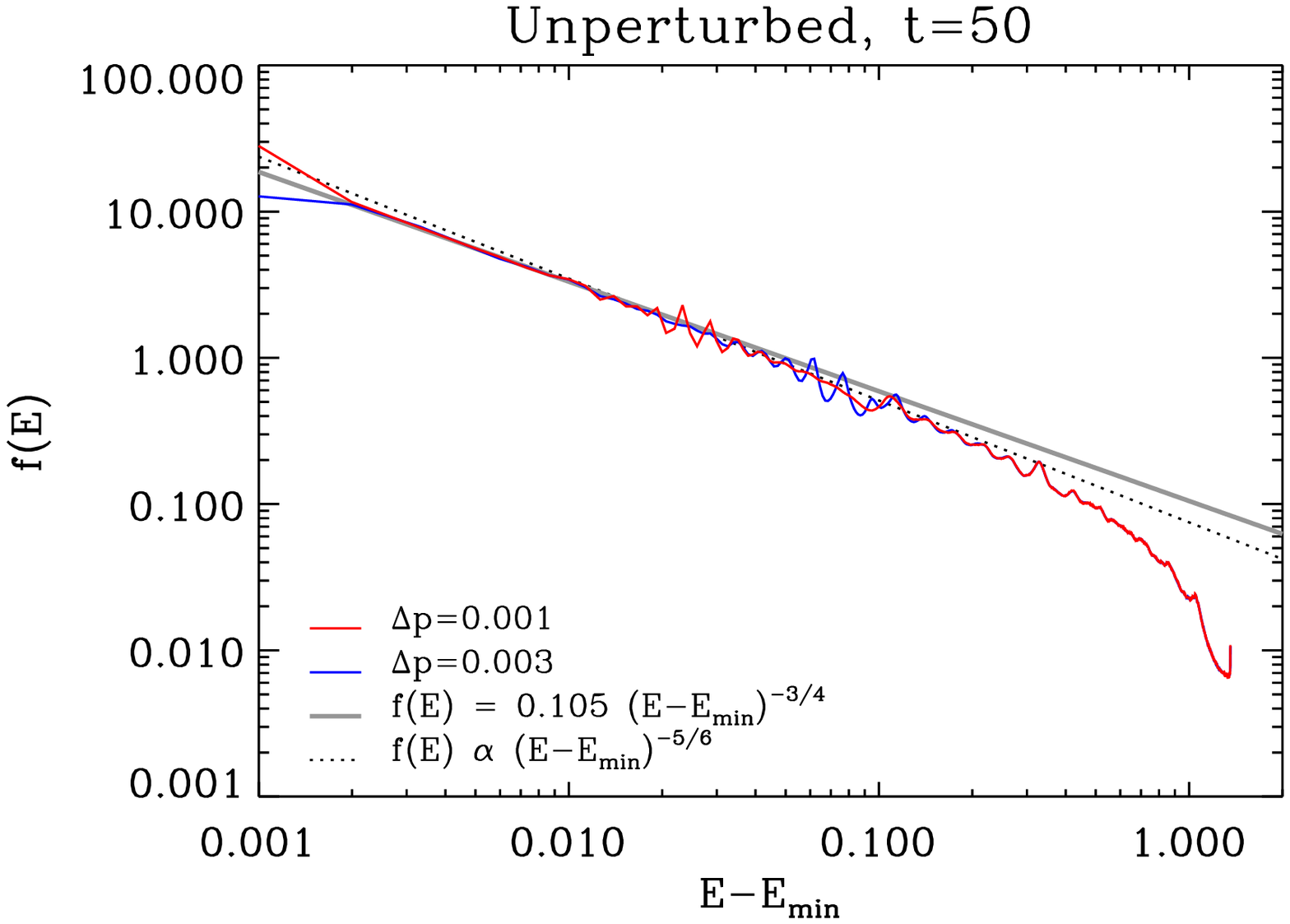,width=8.5cm}
}}
\vskip 0.2cm
\centerline{\hbox{
\psfig{file=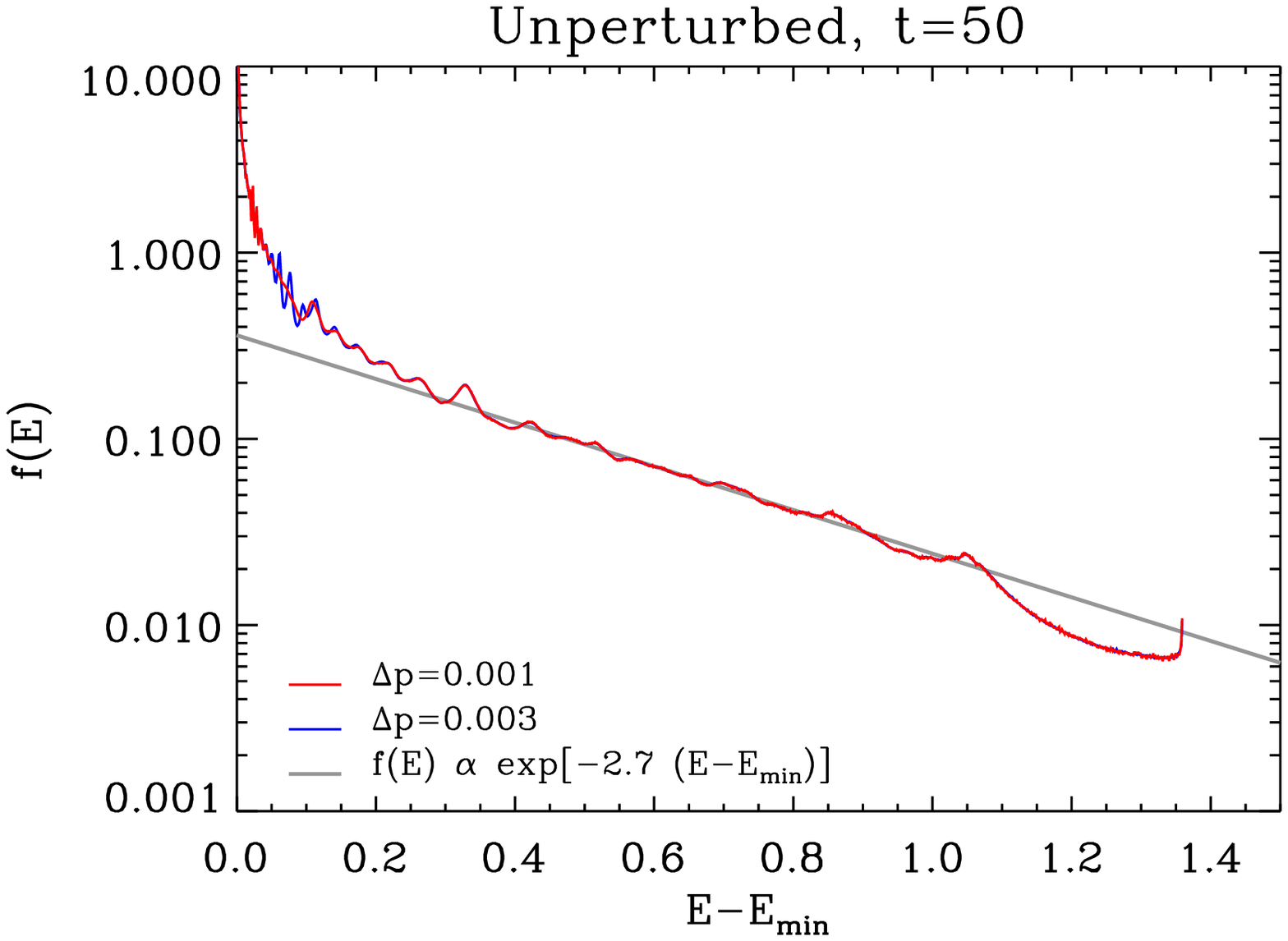,width=8.5cm}
\hskip 0.4cm
\psfig{file=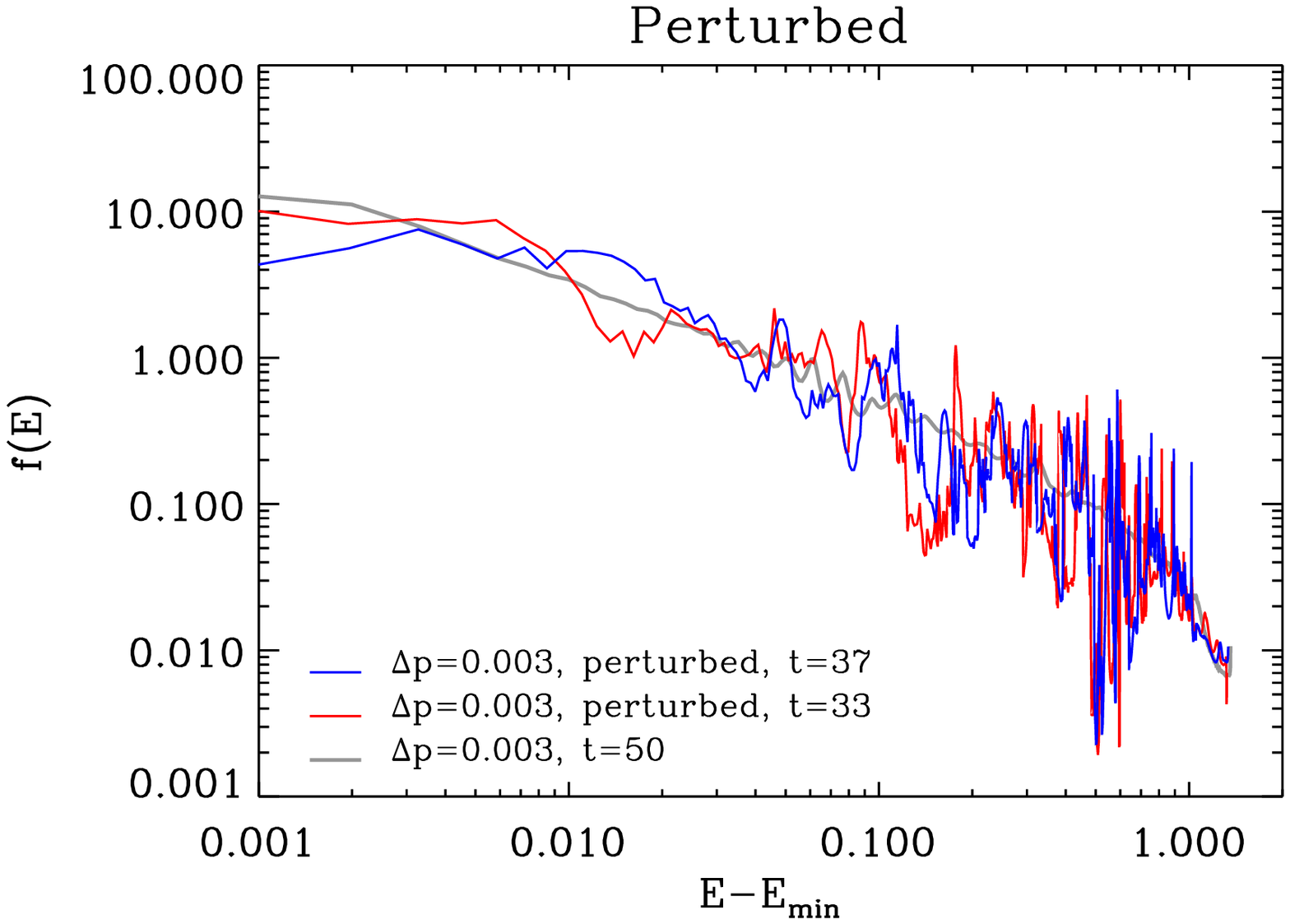,width=8.5cm}
}}
\caption[]{The phase-space energy distribution function in our close to cold waterbag simulations with $\Delta p=0.001$ and $\Delta p=0.003$. On the {\em upper-left panel}, the function $f_E(E)$ is shown for $\Delta p=0.001$ at various times, corresponding to initial conditions, first to fourth crossing times and final time. The dashed orange and solid grey lines correspond to analytic predictions (\ref{eq:powe0}) and (\ref{eq:Epred2}). On the {\em upper-right panel}, the function $f_E(E)$ is shown at last time, $t=50$, for $\Delta p=0.001$ and $\Delta p=0.003$ and fitted with power-laws of index $-3/4$ and $-5/6$, the latter value corresponding to the conjecture of \cite{Binney}. The {\em bottom-left panel} is analogous to the upper-right one, but a linear scale has been chosen for $E-E_{\rm min}$ to emphasize the exponential behavior of function $f_E(E)$ at largest energies. The {\em bottom-right} panel compares measurements of $f_E(E)$ in the randomly perturbed $\Delta p=0.003$ at two different times to the unperturbed case. 
The measurements are shown for the simulations {\tt Tophat0.001}, {\tt Tophat0.003} and {\tt Perturbed} in the nomenclature of Table~\ref{tab:simuparam}, but would not change significantly for other runs we performed with the same initial conditions. Note also, as explained in Appendix~\ref{app:fofemet}, that the measurements are performed in 1023 bins spaced linearly between the minimum and maximum of the energy. This means that function $f_E(E)$ plotted on each panel represents a smoothed version of the actual energy spectrum which has much more structure. Note also that the first bin, corresponding to $E-E_{\rm min} \sim 10^{-3}$ is expected to be spurious, because it corresponds to the smallest energy bin, which does not have a ring shape in phase space, but is homeomorphic to a disk. The measurements should thus be examined for $E \ga 2\times 10^{-3}$, which correspond to energy shells in phase-space that are not affected by the central core in the unperturbed cases.}
\label{fig:fofE}
\end{figure*}
To understand more deeply the establishment of a steady-state after relaxation, it is useful to study the {\em phase-space energy distribution function}, $f_E(E)$:
\begin{equation}
f_E(E) \equiv \lim_{\delta E\rightarrow 0} \frac{\int_{E(x,v) \in [E,E+\delta E]} f(x,v)\ {\rm d}x\ {\rm d}v}{\int_{E(x,v) \in [E,E+\delta E]}\ {\rm d}x\ {\rm d}v},
\label{eq:fofedef}
\end{equation}
which provides the average of the phase-space density per energy level. For systems where the phase-space density depends only on energy, the equality $f(x,v)=f_E[E(x,v)]$ stands. The way we compute function $f_E(E)$ is detailed in Appendix~\ref{app:fofemet}. 

Fig.~\ref{fig:fofE} displays the phase-space distribution function measured in our thinnest waterbags. The upper-left panel shows function $f_E(E)$ at various times. Except for $t=0$ and $t=50$, which correspond respectively to initial conditions and final time, the other snapshots considered have been chosen carefully to coincide with crossing times, that is to moments when the central part of the curve supporting $f(x,v,t)$ is vertical in phase-space, such as on the two middle panels of the left column of Fig.~\ref{fig:perturb}. 
The first striking result is that function $f(E)$ presents a remarkable power-law behavior at small energies, which is already present at collapse time ($t=1.25$)! Furthermore, convergence to a steady state is very fast: at the second crossing time ($t=3.6$) the energy distribution at small $E$ is already converged. The third crossing is enough to get nearly the correct shape for the full final energy spectrum. 

At this point, since collapse time seems to provide an interesting power-law slope for the energy, we might try to compute it analytically. Given the properties of the initial projected density profile, 
\begin{eqnarray}
\rho_0(x) & \equiv & \rho(x,t=0)  = \frac{2}{\pi} \sqrt{1-x^2},  \\
 &\simeq & {\bar \rho}_0 \left(1- 3 a x^2 \right), \quad x \ll 1, \label{eq:rhoapp}
\end{eqnarray}
with ${\bar \rho}_0={2}/{\pi}$ and $a = {1}/{6}$, we can easily calculate the phase-space energy distribution function in the small energy limit to understand both the power-law behaviors observed on upper-left panel of Fig.~\ref{fig:fofE} at $t=0$ and at collapse time, $t \equiv t_{\rm c}$. Details of this calculation are provided in Appendix~\ref{app:analcalc}.

Initial conditions correspond to an approximately harmonic potential
\begin{equation}
\phi-\phi_{\rm min} \simeq \frac{1}{2} {\bar \rho}_0  x^2, \quad x \ll 1
\label{eq:phini2}
\end{equation}
(green line on upper-left panel of Fig.~\ref{fig:potential}), and 
\begin{eqnarray}
f_E(E,t=0) &= &\frac{\sqrt{{\bar \rho}_0 a}}{ \sqrt{2} \pi }  \left[ \frac{a (E-E_{\rm min})}{{\bar \rho}_0}\right]^{-1/2}, \label{eq:powe0} \\
            & \simeq & 0.143 (E-E_{\rm min})^{-1/2}, 
\end{eqnarray}
for $E-E_{\rm min} \ll 1$, where $E_{\rm min}=\phi_{\rm min}$ is the minimum of energy. This result agrees perfectly with our measurements, as shown by the orange dashed line on upper-left panel of Fig.~\ref{fig:fofE}. 

At collapse time, the projected density becomes singular, $\rho(x) \propto x^{-2/3}$, corresponding to a potential of the form
\begin{equation}
\phi-\phi_{\rm min} \simeq \frac{3}{2} \frac{{\bar \rho}_0}{a} (\sqrt{a} x)^{4/3}, \quad x \ll 1
\label{eq:phicol2}
\end{equation}
(blue line on upper-left panel of Fig.~\ref{fig:potential}), and 
\begin{eqnarray}
f_E(E,t_{\rm c}) &=&\frac{ (3/2)^{3/4}\ \Gamma(5/4)\ \sqrt{{\bar \rho}_0 a}}{4 \sqrt{\pi} \ \Gamma(7/4)} 
\left[ \frac{a (E-E_{\rm min})}{{\bar \rho}_0} \right]^{-3/4}, \nonumber \\
 & & \label{eq:Epred2} \\
 & \simeq & 0.168 (E-E_{\rm min})^{-3/4}, 
\end{eqnarray}
in the limit $E-E_{\rm min} \ll 1$, again in very good agreement with our measurements as shown by the grey line on upper-left panel of Fig.~\ref{fig:fofE}. Note that the power-law index of $-3/4$ in equation (\ref{eq:Epred2}) should be obtained for small values of $E-E_{\rm min}$ at each crossing time. 

Now, suppose that mixing happens in such a way that the system relaxes to a stationary state preserving the phase-space energy distribution function obtained at crossing time: 
\begin{equation}
f(x,v) = f_E[E(x,v)]= A\ [E(x,v)-E_{\rm min}]^{-\gamma}. \label{eq:fofElaw}
\end{equation}
This implies, by solving Poisson equation,
\begin{equation}
\phi=E_{\rm min}+\phi_0\ x^\beta
\end{equation}
with
\begin{eqnarray}
\beta &=&\frac{4}{1+2\gamma},\\
\phi_0&=& \left( \frac{\pi}{2} \right)^{-\frac{1}{1+2\gamma}}\left[\frac{A (1+2 \gamma)^2 \Gamma(-1/2+\gamma)}{(3-2\gamma) \Gamma(\gamma)} \right]^{\frac{2}{1+2\gamma}}. \label{eq:com1}
\end{eqnarray}
Fitting the form (\ref{eq:fofElaw}) with the power-low index $\gamma=3/4$ on the low energy part of the final stage of our thin waterbag simulations (top right panel of Fig.~\ref{fig:fofE}) gives $A=0.105$ and indeed agrees to a great accuracy with the measured function $f_E(E)$ at small energies over about a decade. This in turns implies
\begin{equation}
\beta=8/5=1.6, 
\label{eq:slopebe}
\end{equation}
and $\phi_0=1.473$, in excellent agreement with our measurements of the potential at small scales, as indicated by the red line on top-left panel of Fig.~\ref{fig:potential} and consistent with the direct measurements of the logarithmic slope of the potential performed in \S~\ref{sec:gravpot},  which indicated $\beta(x) > 1.54$ for $x \ll 1$.  This result is clearly non trivial when examining right panel of Fig.~\ref{fig:coldzoomcenter} in regions of interest not contaminated by the core, e.g., $0.01 \la r \la 0.05$, where mixing is very strong in the form of a dense spiral structure. Note however that even though the value $\beta=8/5$ represents a good candidate for the asymptotic logarithmic slope of the gravitational potential at small scales, our measurements do not present yet the required dynamic range to provide a firm numerical proof of this.  

To complete this analysis, bottom-right panel of Fig.~\ref{fig:fofE} shows the phase space energy distribution function for the randomly perturbed waterbag with $\Delta p=0.003$. Modulo the large amount of fluctuations induced by substructures, it is interesting to notice that the energy spectrum agrees with that of the unperturbed case. However, as mentioned in \S~\ref{sec:gravpot}, we did not follow this randomly perturbed system for sufficiently long time to make any definitive conclusions. 
\subsection{Discussion}
\label{sec:discu}
Our measurements of the logarithmic slope $\beta(x)$ of the gravitational potential suggest a slowly running power-law index
with $\beta(x) > 1.54$ in the limit $x \ll 1$. They are consistent with a theoretical asymptotic value $\beta=1.6$ computed by assuming that the average phase-space density per energy level remains conserved between crossing times.  They thus disagree unarguably with the conjecture $\beta=1.5$ of \cite{Binney} as well as with the value $\beta=10/7\simeq 1.43$ obtained by \cite{Gurevich1995} by assuming adiabatic invariance from collapse time. Although we do not have sufficient dynamical range to make strong claims, this result also seems to contradict the measurements of  \cite{Schulz2013} in $N$-body simulations, who find a well defined power-law behavior of the projected density profile at small $x$ corresponding to $\beta \simeq 1.53$.  Measuring $\rho(x)$ is a difficult task for us, because of the near caustic structures that the projected density is subject to. \cite{Schulz2013} also used the interior mass profile, that is the acceleration modulus $|a(x)|$ to measure the slope, but they argue that this integral quantity is contaminated by the core up to rather large values of $x$. Note that their measurements using this estimator give slightly larger values of $\beta$, so are more consistent with ours. They also propose a Lagrangian estimator using the Action $\Omega$ as a function of enclosed mass inside the surface inside contours of constant energy. This estimator, as constructed by the authors, can be used as long as $\Omega$ remains a monotonic function of particle rank. With this estimator, they find $\beta \simeq 1.59$, in very good agreement with our theoretical predictions and consistent with our measurements! They however argue that measurements of $\beta$ based on this estimator are not determinant because they can be performed only at early times of the simulations: they prefer at the end to emphasize on the value of $\beta$ obtained from $\rho(x)$, which is measured at late times. We believe that the logarithmic slope of the gravitational potential, equation (\ref{eq:betaest}), remains a robust estimator, even if applied to a $N$-body simulation. It would be interesting to use such an estimator in the $N$-body simulations of  \cite{Schulz2013} to see if it leads to the same conclusions as their density based estimator or if it would agree better, in fact, with their Action based estimator. 

Besides the fact that we are using a different estimator for measuring the inner slope of the profile, another plausible explanation of our disagreement with \cite{Schulz2013} is that the noise introduced by their particle based approach might lead, after sufficient time, to the wrong numerical attractor. A clue to this is that they found some gaps in phase space in their simulations, which might be the signature of a resonant instability induced by the discreteness of the representation, similarly as what we found in the Gaussian simulation of Fig.~\ref{fig:gaussianA} when only a few waterbags were used to represent the phase space distribution function. Our single waterbag simulations present such features, but only in the very vicinity of the core and with negligible consequence on the measurement of the inner slope if a proper estimate of the trustable scaling range is performed. 

%----------------------------------------------------------------------------
\section{Conclusion}
\label{sec:conclusions}
%----------------------------------------------------------------------------
In this paper, we have revisited with a modern perspective the so-called waterbag method to solve numerically Vlasov-Poisson equations in one dimensional gravity, recasting in detail and testing thoroughly the method we introduced briefly in \cite{2008CNSNS..13...46C}. We have shown how to represent the phase-space distribution function with a set of waterbags sampled with an orientated polygon, to compute in a self-consistent way its dynamical evolution and to analyze its properties with the appropriate treatment of the polygonal structure. 

The method is entropy conserving so it allows one to follow  extremely accurately the evolution of a system, even in the presence of highly nonlinear instabilities.  But because it aims at preserving all the details that appear in phase-space during the course of the dynamics, the method is very costly: when there is mixing, the computational cost increases at least linearly with the number of dynamical times and becomes exponential when the system is chaotic. Our calculations were however limited by the fact our code is serial. Parallelization of the code and running it on supercomputers might alleviate partly these limitations. 

To preserve the increasing complexity of the waterbag contours, we proposed a sophisticated and robust refinement scheme to add vertices to the orientated polygon using a geometric construct interpolating local curvature, while our main refinement criterion was based on phase-space area conservation. In two dimensional phase-space, this is exactly equivalent to enforcing conservation of the following Poincar\'e invariant, which can be defined in $2N$ dimensional phase-space as
\begin{equation}
I \equiv \oint {\vec v}.{\rm d}{\vec x}(s),
\end{equation}
where the contour integral is performed on a closed curve in phase space composed of points following the equations of motion. 
This Poincar\'e invariant thus provides a natural tool to extend our refinement criterion to higher number of dimensions. 

Unrefinement, which consists of removing vertices from the polygon when they are not needed anymore, is potentially powerful, because it can decrease the computational cost of the simulation while preserving the same level of accuracy. However we showed that successive refinement/unrefinements of a waterbag contour element are unavoidable and introduce a long term noise contribution that can worsen significantly energy conservation when following a system during many dynamical times. However, all our simulations with unrefinement were still very accurate, except for one. Unrefinement might become a must in higher number of dimensions, due to the considerably larger contrasts in the various dynamical states a contour element can go through. This will be examined in a separate work on systems with spherical symmetry, which present one more dimension of angular momentum in phase-space but can also be approached with the waterbag method \citep[][]{2008CNSNS..13...46C}.

In six-dimensional phase-space, the waterbag method is very challenging to implement in the warm case due to its extreme cost in memory and computational time: indeed the waterbag contours correspond to 5-dimensional hypersurfaces. Cold initial conditions, which are relevant in cosmology, seem on the other hand approachable. In this case, the phase-space distribution is supported by a three-dimensional sheet evolving in six-dimensional phase-space.  An additional difficulty arises, however, from the fact that it is needed to soften the gravitational force to avoid numerical instabilities induced by the presence of singularities. A question then is how well the true gravitational dynamics is described by its softened counterpart.\footnote{This is the reason why, in the present work, we studied convergence to the cold case with very cold but not infinitely thin waterbags.}  In current proposed implementation, which does not yet include local refinement of the phase-space sheet \citep{Hahn2013}, the three-dimensional phase-space sheet is sampled with simplices \citep{Shandarin2012,Abel2012}. The method is thus analogous to the waterbag method in the sense that it preserves connectivity.  Again, in presence of very needed refinement, the computational cost of such simulations will increase very quickly with the number of dynamical times at play: it seems important to investigate optimal refinement algorithms, that might include unrefinement as discussed above and that should take into account of the anisotropic nature of the dynamics. 

Behavior of gravitational systems at large times in the continuous limit is still badly understood except in some very particular cases \cite[see, e.g.][]{Villani}. Even in the one dimensional gravitational case studied in this paper, the long term properties of systems as functions of initial conditions remain an open debate, because it is very challenging to follow them numerically. Particle based methods can rapidly introduce resonant instabilities that drive the system to attractors far from the exact solution. The cold case, where the initial projected density is locally of the form (\ref{eq:rhoapp}), represents a good example of this state of facts. In this paper, by studying a set of single waterbag simulations with decreasing thickness, we performed a convergence study to the cold case and analyzed in detail the inner structure of the steady state that builds up during relaxation. We measured the properties of the gravitational potential and the energy spectrum of the system. We found that the gravitational potential profile after relaxation is consistent with a running power-law
\begin{equation}
\phi(x) \propto x^{\beta(x)},
\end{equation}
where $\beta(x)$ is a slowly decreasing function of $x$, roughly averaging to $\beta \simeq 3/2$ in agreement with the conjecture of \cite{Binney}. Close to the center, we found 
\begin{equation}
\beta > 1.54,
\end{equation}
in disagreement with recent results of the literature based on $N$-body experiments \citep{Binney,Alard2013,Schulz2013}. 
In fact our measurement are consistent with  
\begin{equation}
\beta=8/5=1.6
\end{equation}
at the center of the system, a value which can be predicted explicitly by assuming that the average phase-space density per energy level is conserved between crossing times. 

Our simulations do not present sufficient dynamical range to demonstrate numerically that $\beta=8/5$ corresponds to the expected asymptotic singular behavior of the gravitational potential profile of cold systems in one dimension, but the disagreement of our measurements with the thorough $N$-body experiments of \cite{Schulz2013} is puzzling. These results are very worrying for the $N$-body approach. Indeed, in three dimensions, many important results on the structures of dark matter halos are based on measurements in $N$-body simulations \citep[see, e.g.][and references therein]{NFW1,NFW2,Navarro2010,Diemand2011}. This definitely justifies the need for developing alternative methods to solve Vlasov-Poisson without resorting to particles.

\section*{Acknowledgements}
We thank Tom Abel, Christophe Alard, James Binney, Walter Dehnen, Christophe Pichon and Scott Tremaine for useful discussions.
The analytic calculations of \S~\ref{sec:fofE} and Appendix~\ref{app:analcalc} have been performed with {\tt Mathematica}.  
JT acknowledges the support of an Arab Fund Fellowship for the year 2013-2014.
This work has been funded in part by ANR grant ANR-13-MONU-0003 as well as NSF grants AST-0507401 and AST-0206038. 
%
%----------------------------------------------------------------------------

\appendix
%=======================================================
\section{Initial conditions and simulation settings}
\label{app:inicond2}
%=======================================================
\begin{table*}
\begin{tabular}{l|llllll}
\hline
Designation           &  Initial conditions                                                               & $S_{\rm add}$          & $S_{\rm rem}$      & $d_{\rm add}$ & $d_{\rm rem}$ & $C$ \\ \hline
{\tt Gaussian10U}  & Gaussian, 10 contours, unrefinement allowed                       & $10^{-8}$              & $S_{\rm add}/2$  & $0.01$        & $0.005$      & $0.025$ \\
{\tt Gaussian10}    & Gaussian, 10 contours, no unrefinement, larger $S_{\rm add}$ & $2\times 10^{-8}$ & $0$                   & $0.01$         & $0$             & $0.025$ \\
{\tt Gaussian84U}  & Gaussian, 84 contours, unrefinement allowed                       & $10^{-8}$              & $S_{\rm add}/2$  & $0.01$         & $0.005$      & $0.025$ \\
{\tt Gaussian84}    & Gaussian, 84 contours, no unrefinement, larger $S_{\rm add}$ & $2\times 10^{-8}$  & $0$                 & $0.01$        & $0$             & $0.025$ \\ 
 & & & & & & \\
{\tt RandomU}       & Random set of halos, unrefinement allowed                         & $10^{-8}$              & $S_{\rm add}/2$  & $0.01$         & $0.005$       & $0.005$ \\
{\tt Random}         & Random set of halos, nounrefinement, larger $S_{\rm add}$    & $2\times 10^{-8}$  & $0$                  & $0.01$         & $0$             & $0.005$ \\
{\tt RandomUT}     & Random set of halos, unrefinement allowed, smaller time step  & $10^{-8}$            & $S_{\rm add}/2$  & $0.01$         & $0.005$       & $0.0025$ \\
{\tt RandomUS}     & Random set of halos, unrefinement allowed, smaller $S_{\rm add}$  & $10^{-9}$       & $S_{\rm add}/2$  & $0.01$         & $0.005$       & $0.005$ \\ 
 & & & & & & \\
{\tt Tophat1.000U}  & Waterbag, $\Delta p=1$, unrefinement allowed                 & $10^{-7}$               & $S_{\rm add}/2$  & $0.02$         & $0.01$        & $0.0025$ \\
{\tt Tophat0.750U}  & Waterbag, $\Delta p=0.75$, unrefinement allowed            & $0.75 \times 10^{-7}$ & $S_{\rm add}/2$ & $0.02$   & $0.01$        & $0.0025$ \\
{\tt Tophat0.500U}  & Waterbag, $\Delta p=0.5$, unrefinement allowed              & $0.5 \times 10^{-7}$ & $S_{\rm add}/2$ & $0.02$      & $0.01$         & $0.0025$ \\
{\tt Tophat0.250U}  & Waterbag, $\Delta p=0.25$, unrefinement allowed            & $0.25 \times 10^{-7}$ & $S_{\rm add}/2$ & $0.02$   & $0.01$         & $0.0025$ \\
{\tt Tophat0.100U}  & Waterbag, $\Delta p=0.1$, unrefinement allowed              & $10^{-8}$                & $S_{\rm add}/2$  & $0.02$         & $0.01$         & $0.0025$ \\
 & & & & & & \\
{\tt Tophat0.010U}  & Waterbag, $\Delta p=0.01$, unrefinement allowed            & $10^{-9}$                & $S_{\rm add}/2$  & $0.02$         & $0.01$         & $0.0025$ \\
{\tt Tophat0.010}  & Waterbag, $\Delta p=0.01$, no unrefinement, larger $S_{\rm add}$  & $10^{-9}$      & $S_{\rm add}/2$  & $0.02$     & $0.01$         & $0.0025$ \\
{\tt Tophat0.003U}  & Waterbag, $\Delta p=0.003$, unrefinement allowed         & $0.3\times 10^{-9}$  & $S_{\rm add}/2$  & $0.02$      & $0.01$         & $0.001$ \\
{\tt Tophat0.003}  & Waterbag, $\Delta p=0.003$, no unrefinement, larger $S_{\rm add}$  & $2.4\times 10^{-9}$  & $0$     & $0.02$         & $0$            & $0.001$ \\
{\tt Tophat0.001U}  & Waterbag, $\Delta p=0.001$, unrefinement allowed         & $10^{-10}$             & $S_{\rm add}/2$  & $0.02$      & $0.01$         & $0.001$ \\
{\tt Tophat0.001}  & Waterbag, $\Delta p=0.001$, no unrefinement, larger $S_{\rm add}$  & $8\times 10^{-10}$  & $0$     & $0.02$         & $0$            & $0.001$ \\
{\tt Tophat0.001S}  & Waterbag, $\Delta p=0.001$, no unrefinement                 & $10^{-10}$              & $0$                         & $0.02$         & $0$            & $0.001$ \\
 & & & & & & \\
{\tt PerturbedU}  & Waterbag, $\Delta p=0.003$, perturbed, unrefinement allowed   & $10^{-10}$             & $S_{\rm add}/2$  & $0.02$      & $0.01$         & $0.0005$ \\
{\tt Perturbed}  & Waterbag, $\Delta p=0.003$, perturbed, no unrefinement, larger $S_{\rm add}$  & $8\times 10^{-10}$  & $0$     & $0.02$         & $0$            & $0.0005$ \\
{\tt PerturbedS}  & Waterbag, $\Delta p=0.003$, perturbed, no unrefinement          & $10^{-10}$              & $0$                         & $0.02$         & $0$            & $0.0005$ \\
\hline
\end{tabular}
\caption[]{The designation of the simulations according to the important parameters used to performed them: type of initial conditions, refinement/unrefinement criteria parameters introduced in \S~\ref{sec:myref} (equations \ref{eq:Saddcrit}, \ref{eq:daddcrit}, \ref{eq:remcrit1} and \ref{eq:remcrit2}) and the time-step parameter $C$ introduced in \S~\ref{sec:point5}  (equation \ref{eq:courant}).}
\label{tab:simuparam}
\end{table*}
In this appendix, we provide a full description of the initial conditions of the simulations performed in this work, while Table~\ref{tab:simuparam} gives all the simulation settings. 
\begin{itemize}
\item {\em The Gaussian initial conditions} are created as follows: setting
\begin{equation}
G(x,v)\equiv \rho_{\rm G} \exp\left( -\frac{1}{2} \frac{x^2+v^2}{\sigma_{\rm G}^2} \right), \label{eq:gauini}
\end{equation}
we write
\begin{eqnarray}
f(x,v) & = & G(x,v), \quad x^2+v^2 \leq {\cal R}^2, \label{eq:monfg} \\
        & = &  G(x,v) \nonumber \\
         &  &\times  \max\left[ 1+2\ {\rm th}\left( \frac{{\cal R}-\sqrt{x^2+v^2}}{\eta_{\rm G}}\right), 0 \right], \nonumber \\
 & &\quad \quad \quad \quad \ \ x^2+v^2 > {\cal R}^2. \label{eq:monfg2}
\end{eqnarray}
Our initial distribution function is thus a truncated Gaussian. The practical choice of the parameters corresponds to ${\cal R}=1$, $\rho_{\rm G}=4$, $\sigma_{\rm G}=0.2$ and $\eta_{\rm G}=0.02$, which makes the total mass of the system approximately equal to unity for a Gaussian truncated at 5 sigmas.
\item {\em The ensemble of stationary clouds} initial conditions are created as follows. 
Each of these halos initially approximates the stationary solution corresponding to thermal equilibrium \citep{1942ApJ....95..329S,Camm1950,Rybicki1971}:
 \begin{equation}
f_{\rm S}(x,v)=\frac{\rho_{\rm S}}{[{\rm ch}(\sqrt{\sqrt{2\pi} \rho_{\rm S}/\sigma_{\rm S}} x)]^2} \exp\left[ -\frac{1}{2} \left( \frac{v}{\sigma_{\rm S}}\right)^2\right].
\label{eq:apofs}
\end{equation}
The individual components are generated at random positions in a phase-space disk of radius unity (prior to recasting with respect to center of mass). Their profile follows equation (\ref{eq:apofs}) with $\rho_{\rm S}=6$ and individual random values for the velocity dispersion $\sigma_{\rm S}$, ranging in the interval $[0.005,0.1]$. To make sure that the clouds do not overlap too much with each other, we impose the distance in phase-space between the center of any two clouds $i$ and $j$ to be larger than $4 [\sigma_{\rm S}(i)+\sigma_{\rm S}(j)]$. 
Then, the components are added on the top of each other in phase-space, to obtain the desired distribution function $f_{\rm r}(x,v)$.  Finally, apodization is performed as follows
\begin{eqnarray}
f(x,v) & = & f_{\rm r}(x,v), \quad \quad f_{\rm r}(x,v) \ge \eta_{\rm r}, \label{eq:fra}\\
         & = & \eta_{\rm r} \max\left\{ 1+2\ {\rm th}\left[ \frac{f_{\rm r}(x,v)-\eta_{\rm r}}{\eta_{\rm r}}\right],0 \right\}, \nonumber \\
         &    & \quad \quad \quad \quad \quad \quad  f_{\rm r}(x,v) < \eta_{\rm r}, \label{eq:fra2}
\end{eqnarray}
with $\eta_{\rm r}=0.05$. 
\item {\em Our single waterbag simulations} have the following initial vertices coordinates for the orientated polygon:
\begin{eqnarray}
x_i  & = & \cos( 2\pi i/N), \label{eq:xiw}\\
v_i  & = & \Delta p \sin( 2\pi i/N), \label{eq:viw}
\end{eqnarray}
with $i \in [0,\cdots,N]$ and a total mass unity,  which implies $f^{\rm left}  = {1}/{(\pi \Delta p)}$ and $f^{\rm right}=0$ in equation (\ref{eq:circu2}). 
As listed in Table~\ref{tab:simuparam}, we consider several values of the thickness parameter $\Delta p$ ranging in the interval $[0.001,0.1]$. In all the cases, we take $N=1000$. 

For $\Delta p=0.003$, we also performed simulations where the initial configuration is perturbed randomly as follows:
\begin{equation}
v \rightarrow v+\delta v,
\label{eq:vip}
\end{equation}
\begin{eqnarray}
\delta v &=& 0.0006 \sum_{k=-50}^{50} |k|^{-1/2}\left[ G_{2k} \cos( \pi k x) \right.\nonumber \\
 & &\left. + G_{2k+1} \sin(\pi k x) \right]
\label{eq:deltavip}
\end{eqnarray}
where $G_i$ is a Gaussian random number of average zero and variance unity. In this case, we take $N=10000$. 

The simulations were run up to $t=50$, except for the perturbed waterbag simulations which ended earlier, due to their computational cost. 
\end{itemize}

%=======================================================
\section{Initial conditions with the isocontour method} 
\label{sec:inicond}
%=======================================================
To construct the orientated polygon following isocontours $C_k$ of the phase-space distribution function, we propose to proceed in five steps: 
\begin{description}
\item[(1)] Sampling of $f(x,v)$ on a rectangular mesh of dimensions $n_x$ and $n_v$;
\item[(2)] Choice of the isocontours $C_k$, $k=\{0,\cdots,N_{\rm patch}\}$ and calculation of the value $f_k$, $k=\{0,\cdots,N_{\rm patch}\}$ associated with each waterbag delimited by two successive isocontours, which is easily given, using mass conservation by
\begin{equation}
f_k=\frac{\int_{f\in[C_{k-1},C_{k}]} f(x,v)\ {\rm d}x\ {\rm d}v}{\int_{f\in[C_{k-1},C_{k}]} {\rm d}x\ {\rm d}v}.
\label{eq:fkint}
\end{equation}
with the convention $C_1\equiv 0$ and $C_{N_{\rm patch}}\equiv \max_{x,v} f(x,v)$. We have, evidently, $C_1=0 \leq f_1 \leq C_2 \leq \cdots \leq C_{N_{\rm patch}} \leq f_{N_{\rm patch}}$, with a convenient choice $f_0\equiv 0$. 
\item[(3)] Identification of the cells of the mesh intersecting with $C_k$; 
\item[(4)] Construction of closed loops of the orientated polygon associated to $C_k$ by walking on the mesh using the identified sites as a footpath; 
\item[(5)] Connection between each individual loop with ``null'' segments (not contributing to the dynamics) to finish building the orientated polygon as a full closed curve. 
\end{description} 

Step (1) is fairly straightforward. Given the dimensions of the area covered by the mesh, one just has to take large enough values of  $n_x$ and $n_v$ to be able to catch all the variations of $f(x,v)$ in phase-space. In addition, the choice of the initial sampling must be harmonious with refinement, as discussed in \S~\ref{sec:refinement}. Indeed, using too sparse a mesh for constructing the orientated polygon will trigger refinement at the very beginning of the simulation: it is clearly better to use a thinner grid to create the orientated polygon than to trigger refinement. The point of this latter is indeed to account for creation of curvature as an effect of dynamical evolution. In practice, we used an initial grid with $n_x=n_v=1024$ covering the range $(x,v) \in [-1.2,1.2]$ for sampling the Gaussian initial conditions, while $n_x=n_v=4096$ and $(x,v) \in [-1.4,1.4]$ were used for the ensemble of stationary clouds.

Step (2) is difficult if the goal is to achieve an optimal set up, except in the trivial case when the phase-space distribution function is actually a finite set of waterbags. For a smooth $f(x,v)$, the waterbag description can only approach the true phase-space distribution function in an approximate way and the best isocontours sampling is unknown. In this paper, to chose the initial isocontours, we adopt a local scheme consisting in bounding the error measured in each waterbag as follows:
\begin{equation}
{\cal R}_k \sigma_k \simeq  E_{\rm th},\label{eq:prescrip}
\end{equation}
where $E_{\rm th}$ is a control parameter. In this equation, ${\cal R}_k$ is an estimate of the width of the waterbag
\begin{equation}
{\cal R}_k \equiv  (C_{k}-C_{k-1}) \left\langle \frac{1}{|\nabla f|} \right\rangle_k,
\end{equation}
where $\langle {1}/{|\nabla f|} \rangle_k$ is the average of the inverse of the magnitude of the gradient of the phase-space distribution function over the waterbag:
\begin{equation}
\left\langle \frac{1}{|\nabla f|} \right\rangle_k  \equiv \frac{\int_{f \in [C_{k-1},C_k]} \frac{1}{|\nabla f|} {\rm d}x\ {\rm d}v}{\int_{f \in [C_{k-1},C_{k}]} {\rm d}x\ {\rm d}v}. \label{eq:overnab}
\end{equation}
The quantity $\sigma_k$ corresponds to an estimate of the average error in the waterbag:
\begin{equation}
\sigma_k^2 \equiv  \frac{\int_{f \in [C_{k-1},C_{k}]} \left[ f_k - f(x,v) \right]^2 {\rm d}x \ {\rm d}v}{\int_{f \in [C_{k-1},C_{k}]} {\rm d}x\ {\rm d}v}.
\label{eq:quadr}
\end{equation}
For thin waterbags, on has $\sigma_k \simeq (C_{k}-C_{k-1})/(2 \sqrt{3})$. The criterion (\ref{eq:prescrip}) reads thus, approximately,
\begin{equation}
{\cal R}_k (C_{k}-C_{k-1}) \simeq {\rm constant}.
\end{equation}
This means that the typical distance between two isocontours is typically proportional to $1/\sqrt{|\nabla f|}$, instead of $1/|\nabla f|$ for the ``natural'' setting, 
\begin{equation}
C_{k+1}-C_{k}={\rm constant}.
\label{eq:natset}
\end{equation} 
Even though equation (\ref{eq:natset}) remains a possible choice in our code, we prefer in practice to use the prescription (\ref{eq:prescrip}), which provides a denser isocontour sampling in regions where $|\nabla f|$ is small. 

In practice, we used $E_{\rm th}=0.01$ and $E_{\rm th}=0.001$ in equation (\ref{eq:prescrip}) respectively for the 10 and 84 waterbags simulations with Gaussian initial conditions, while $E_{\rm th}=0.05$ was used for the random set of halos. 

As a final remark for the implementation of step (2), enforcing criterion (\ref{eq:prescrip}) is easy if (a) the number of waterbags $N_{\rm patch}$ is left free while $E_{\rm th}$ is the control parameter of choice, (b) the calculations are performed following the lexicographic order given by increasing values of $f(x,v)$ sampled on the grid, (c) integrals such as in equations (\ref{eq:fkint}), (\ref{eq:overnab}) and (\ref{eq:quadr}) are performed using simple sums over the pixels of the grid that verify $f \in ]C_{k-1},C_k]$. 

Steps (3) and (4) can be performed with the so-called ``Marching Square'' algorithm, which, to work properly, requires $f(x,v)$ to be smooth at the scale of the mesh cell size. 

Firstly (step 3), one identifies the sites of the mesh intersecting with $C_k$. To do so, we compute the values $f_{i,j}$ of $f$ at positions $(x_i,v_j)$ corresponding to the corners of each cell of the mesh. A contour $C_k$ intersects a square composed of 4 corners $(i,j)$, $(i+1,j)$, $(i,j+1)$, $(i+1,j+1)$ if either 
\begin{eqnarray}
f_{i,j} \leq & C_k & < f_{i+1,j}, \nonumber\\
f_{i,j} > & C_k & \geq f_{i+1,j}, \nonumber \\
f_{i,j} \leq & C_k & < f_{i,j+1}, \nonumber \\
f_{i,j} > & C_k & \geq f_{i,j+1}, \nonumber \\
f_{i+1,j} \leq & C_k & < f_{i+1,j+1}, \nonumber \\
f_{i+1,j} > & C_k & \geq f_{i+1,j+1}, \nonumber \\
f_{i,j+1} \leq & C_k & < f_{i+1,j+1}, \nonumber \\
f_{i,j+1} > & C_k & \geq f_{i+1,j+1}.
\end{eqnarray}
Each condition above, if fulfilled, defines an intersection along one of the edges of a cell. topologically, there can be 0, 2 or 4 intersections. Four intersections means either that two disconnected parts of the isocontour are very close to each other or that there is
a saddle point in the cell. At the level of accuracy defined by the grid resolution, these two statements are equivalent. 

\begin{figure}
\centerline{\hbox{
\psfig{file=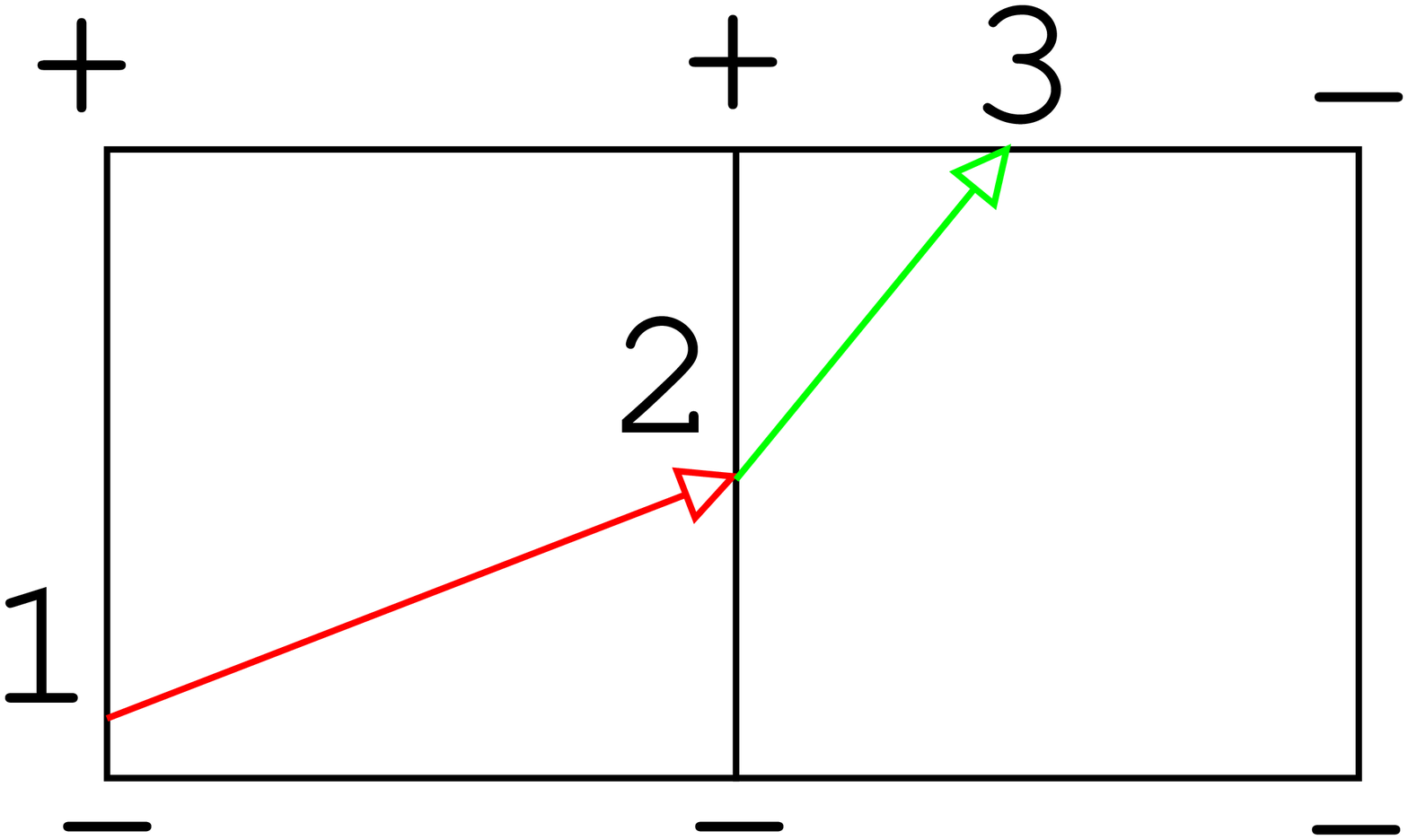,width=4cm}}}
\centerline{\hbox{
\psfig{file=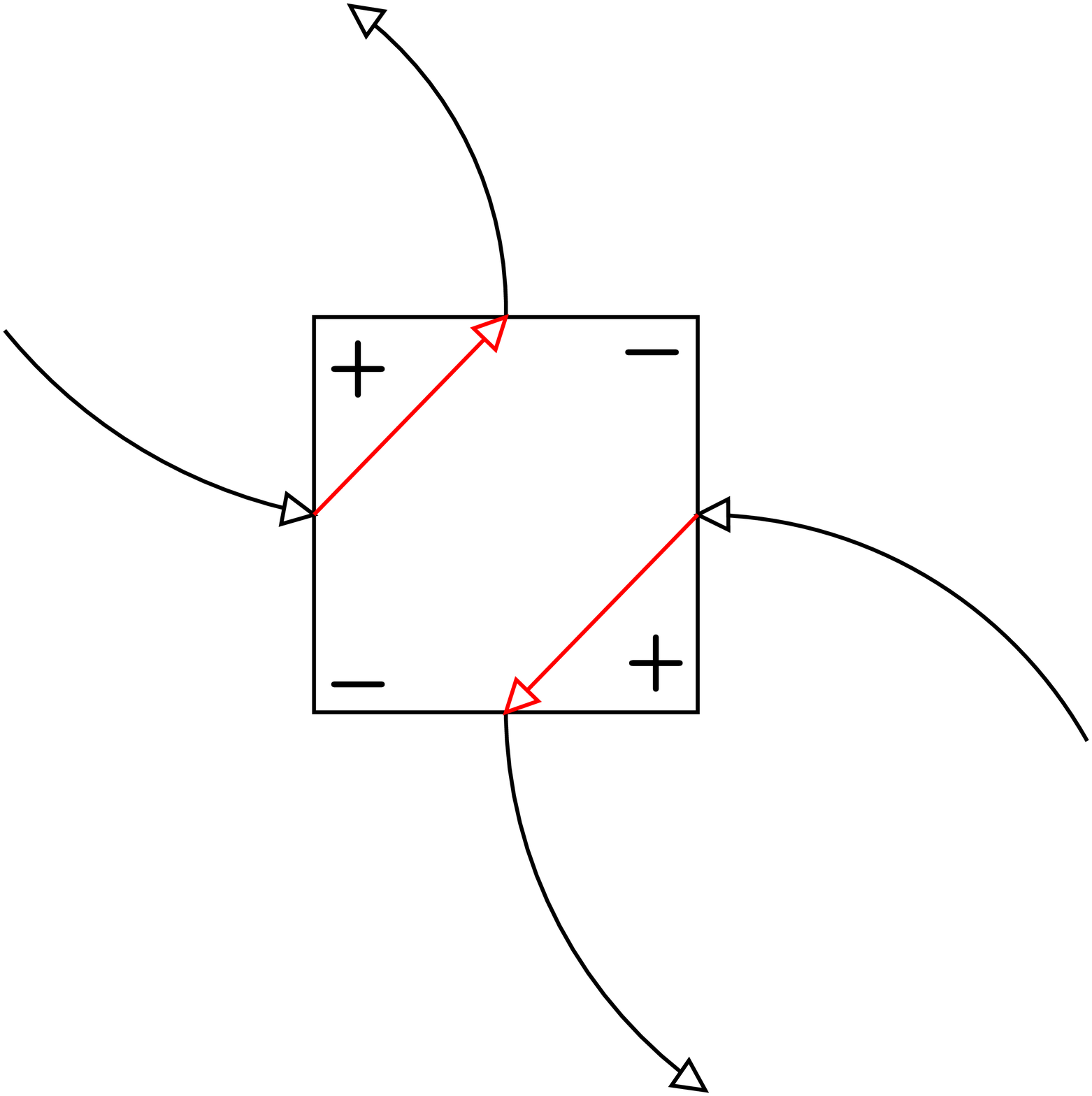,width=5.33333cm}}}
\caption[]{Sketch of the method used to draw orientated isocontours on the cells of a regular rectangular grid.  The signs ``+'' and ``-'' indicate whether the measured distribution function at the corners of the cells is larger or smaller than the isocontour value.  In general, there are only 2 points of intersection of an isocontour level with a cell (upper panel). If circulation is chosen such that $f^{\rm right} < f^{\rm left}$, the construction of the part of the polygon belonging to the left cell is straightforward (red arrow). Then from the left cell, we have to proceed to the right cell, to continue drawing the polygon in a similar way to obtain the green arrow. Some cells might contain four intersection points as illustrated by lower panel. They simply have to be flagged in a particular way to make sure that one can pass through them twice.}
\label{fig:method}
\end{figure}

Secondly (step 4), one walks on the sites identified previously to construct closed loops of the orientated polygon. Let us imagine we chose to circulate along isocontours in such a way that $f^{\rm right} < f^{\rm left}$. Once each site of the grid intersecting with isocontour $C_k$ has been identified, one starts at random with one of the flagged sites, which contains 2 or 4 intersections. 
If it contains 2 intersections, the direction of circulation within the cell is straightforward: a segment is drawn unambiguously with a starting point and an ending point by using the condition $f^{\rm right} < f^{\rm left}$ to find the direction of circulation. The positions of these points is found by bilinear interpolation, or if more accuracy is needed, iteratively (e.g. by dichotomy) to match the actual location of the intersections of the cell with $C_k$. The important property of this exact positioning is to be able to achieve a high level of smoothness of the constructed contour to avoid introducing artificial curvature variations that might trigger unnecessary refinement. From the end point of the segment, one can easily find the neighboring cell containing it and start again the process, as illustrated by upper panel of Fig.~\ref{fig:method}. 
 Each time a cell is treated, it is flagged again, to avoid passing twice a the same place. At some point, since isocontours are sets of closed curves, one comes back to the starting cell: a connected part of isocontour $C_k$ is achieved. An abstract link is created to close the loop in order to be able to circulate along it for future use, such as local refinement: in practice, we associate to each vertex $i$ of the polygon two integer numbers $(J^{\rm F}_i, J^{\rm B}_i)$ which give the index $j$ of the next point of the closed contour under examination while walking on it forward and backwards, respectively. 

The grid is then scanned again to find a new component of isocontour $C_k$ until all the cells intersecting with it have been treated appropriately. There might be cells containing four intersections (see lower panel of Fig.~\ref{fig:method}). They just have to be flagged in a particular way, since one has to pass through them twice. Note that step (3) and step (4) can be performed simultaneously: we presented them separately for clarity.

Step (5) is cosmetic and trivial enough. 
%=======================================================
\section{Details on refinement}
\label{sec:refinement}
%=======================================================
%
In this Appendix, we first provide a number of useful formulas that can be easily derived from elementary geometrical analysis of Fig.~\ref{fig:interpol}. Then, we study the properties of our refinement procedure in terms of small rotations along waterbag contours. Finally we discuss about the evolution of the number of vertices during the course of dynamics, with explicit measurements in simulations. 
%=======================================================
\subsection{Useful formulae}
\label{app:refdetails}
%=======================================================
Given the distance $d_{XY}$ between points $X$ and $Y$, the following formula can be easily derived from Fig.~\ref{fig:interpol}:
\begin{eqnarray}
d_{PP_1} & = & d_{AB} \frac{\tan \theta_A\tan\theta_B}{\tan \theta_A+\tan\theta_B}, \\
d_{AP_1} & = & d_{AB} \frac{\tan\theta_B}{\tan \theta_A+\tan\theta_B}, \\
d_{P_1B} & = & d_{AB} \frac{\tan \theta_A}{\tan \theta_A+\tan\theta_B},
\end{eqnarray}
where $P_1$ is the projection of $P$ on segment $[A,B]$. This gives us the relative position of refined point $P$ with respect to segment $[A,B]$. To decide whether it has to be located on the right side or on the left side of $[A,B]$ is determined by e.g. the sign of the vector product ${\overrightarrow{UA}} \wedge {\overrightarrow{AB}}$.\footnote{In the very unlikely case when either $U$, $A$ and $B$ or $A$, $B$ and $V$ are aligned, point $P$ is set at the mid point of $[A,B]$.}  The quantities $\tan \theta_A$ and $\tan\theta_B$ are given by
\begin{eqnarray}
\displaystyle
\tan \theta_A & = & \frac{2 D_A}{d_{AB}}\left[ \sqrt{1+\left( \frac{d_{AB}}{2D_A} \right)^2}-1\right], \label{eq:e1}\\
\displaystyle
\tan\theta_B & = & \frac{2 D_B}{d_{AB}}\left[ \sqrt{1+\left( \frac{d_{AB}}{2D_B} \right)^2}-1\right], \label{eq:e2}
\end{eqnarray}
with
\begin{eqnarray}
D_A^2 & = & R_A^2 - (d_{AB}/2)^2, \label{eq:e3} \\
D_B^2 & = & R_B^2 - (d_{AB}/2)^2. \label{eq:e4}
\end{eqnarray}
In these equations, $R_A$ and $R_B$ are the radii of the arc of circles $\widehat{UAB}$ and $\widehat{ABV}$, given by the usual formula:
\begin{eqnarray}
R_A & = & \frac{d_{UB}}{2 \sin( \alpha_A )}, \label{eq:RA} \\
R_B & = & \frac{d_{AV}}{2 \sin( \alpha_B )}, \label{eq:RB}
\end{eqnarray}
where $\alpha_A$ ($\alpha_B$) is the angle between vectors $\overrightarrow{UA}$ and $\overrightarrow{AB}$ ($\overrightarrow{AB}$ and $\overrightarrow{BV}$). 

If the local curvature does not change sign, the expression for the interpolated curvature radius, i.e. the radius of the arc of circle $\widehat{APB}$, reads:
\begin{eqnarray}
R_P&= & \frac{1}{2} \frac{d_{AB}}{\sin(\theta_A +\theta_B)}.
\label{eq:curvsimple}
\end{eqnarray}
In the small angle regime, from equations (\ref{eq:e1}), (\ref{eq:e2}), (\ref{eq:e3}) and (\ref{eq:e4}), $\theta_A \simeq  d_{\rm AB}/(4R_A)$ and $\theta_B \simeq d_{\rm AB}/(4 R_B)$, it follows that
\begin{equation}
\frac{1}{R_P} \simeq \frac{1}{2R_A}+\frac{1}{2R_B}, \quad \frac{d_{AB}}{R_A} \ll 1,
\quad \frac{d_{AB}}{R_B} \ll 1,
\label{eq:smallangle}
\end{equation}
the usual interpolation formula for local curvature. 

When the curvature changes sign (bottom panel of Fig.~\ref{fig:interpol}), we have $R_P = \infty$, in disagreement with equation (\ref{eq:smallangle}). Still, the following property remains true
\begin{equation}
\min\left( \frac{s_A}{R_A}, \frac{s_B}{R_B} \right) \leq \frac{s_P}{R_P} \leq \max\left( \frac{s_A}{R_A}, \frac{s_B}{R_B} \right),
\label{eq:RPbounded}
\end{equation}
where  
\begin{equation}
s_A \equiv \frac{{\overrightarrow{UA}}}{d_{UA}} \wedge \frac{{\overrightarrow{AB}}}{d_{AB}},
\end{equation}
denotes the sign of the rotation between vectors $\overrightarrow{UA}$ and $\overrightarrow{AB}$, and analogously
for $s_B$ and $s_P$. This means that the curvature of the new point $P$ is bounded by that of its neighbors, which is crucial for preserving the stability of the algorithm, as studied more in details in \S~\ref{app:refstability}. 

Note finally that implementation of refinement is facilitated from the algorithmic point of view by using the connectivity information arrays $J^{\rm F}$ and $J^{\rm B}$ introduced in \S~\ref{sec:inicond}. 
%==================================================
\subsection{Stability of refinement}
\label{app:refstability}
%==================================================
The stability of our refinement procedure can be demonstrated in terms of the (signed) angle $\alpha_i$ measured at vertex $i$ between segments $[i-1,i]$ and $[i,i+1]$ of a closed contour. We have
\begin{equation}
s_i \sin \alpha_i =\frac{d_{i-1,i+1}}{2 R_i},
\end{equation}
where $d_{i-1,i+1}=\sqrt{(x_{i+1}-x_{i-1})^2+(v_{i+1}-v_{i-1})^2}$ is the distance between point $i-1$ and $i+1$ and $R_i$ is the radius of the circle passing through points $i-1$, $i$ and $i+1$. Therefore, note that the variations of angle $\alpha_i$ are directly related to those of local curvature. 
We can define $\alpha_i^{\rm bef}$ and $\alpha_i^{\rm aft}$ as corresponding to the states of the orientated polygon before and after refinement. With the scheme described in Fig.~\ref{fig:interpol} we have, when looking at top panel of this figure, 
\begin{eqnarray}
|\alpha_A^{\rm bef}| & \geq  & 2 \theta_A,\\
|\alpha_B^{\rm bef}| & \geq  & 2 \theta_B,\\
|\alpha_P^{\rm aft}| & =  & \theta_A+\theta_B.
\end{eqnarray}
From this we can deduce
\begin{eqnarray}
|\alpha_A^{\rm aft}|  & \leq  & |\alpha_A^{\rm bef}|, \quad {\rm sgn}(\alpha_A^{\rm aft})={\rm sgn}(\alpha_A^{\rm bef}) \\
|\alpha_B^{\rm aft}|  & \leq  & |\alpha_B^{\rm bef}|, \quad {\rm sgn}(\alpha_B^{\rm aft})={\rm sgn}(\alpha_B^{\rm bef}) \\
|\alpha_P^{\rm aft}|  & \leq &  \frac{1}{2} (|\alpha_A^{\rm bef}|+|\alpha_B^{\rm bef}|), \nonumber \\
                             &  & \quad {\rm sgn}(\alpha_P^{\rm aft})=\frac{1}{2}[ {\rm sgn}(\alpha_A^{\rm bef})+{\rm sgn}(\alpha_B^{\rm bef})],
\end{eqnarray}
even when curvature locally changes sign (bottom panel). In other words, our refinement scheme makes the border of the waterbags less angular. Furthermore, we have
\begin{equation}
|\alpha_A^{\rm aft}|+|\alpha_P^{\rm aft}|+|\alpha_B^{\rm aft}|=|\alpha_A^{\rm bef}|+|\alpha_B^{\rm bef}|,
\end{equation}
hence, 
\begin{equation}
\sum_j |\alpha_j^{\rm aft}|=\sum_i |\alpha_i^{\rm bef}|,
\end{equation}
a property that demonstrates that our refinement algorithm is ``Total Variation Preserving'' in term of the small rotations between successive segments of waterbags borders.
%
%=====================================
\subsection{Refinement/unrefinement criteria}
\label{sec:critraf}
%=====================================
%
%
%
As already discussed in the main text (\S~\ref{sec:myref}), our refinement criterion is the following. On Fig.~\ref{fig:interpol}, the orientated polygon is augmented with candidate point $P$ if
\begin{eqnarray}
S(\widehat{APB}) & > & S_{\rm add}, \label{eq:Saddcrita} \\
d_{AB} & > & d_{\rm add}, \label{eq:daddcrita}
\end{eqnarray}
where $S(\widehat{APB})$ is the surface of the triangle composed of the points $A$, $B$ and $P$ (${\tilde P}$ when the local curvature sign changes) and $d_{AB}$ the distance between $A$ and $B$. 

In equation (\ref{eq:Saddcrita}), the choice of $S_{\rm add}$ controls, along with time stepping implementation, the overall accuracy of phase-space area conservation. It has to be taken as a very small fraction of the area $S_{\rm tot}$ occupied in phase-space by the system during the various stages of its evolution. Our choice is to have $S_{\rm add}$ ranging from about $10^{-10} S_{\rm tot}$ to $10^{-7} S_{\rm tot}$ and depends in practice on how mixing becomes dramatic: in particular, if the borders of the waterbags become very close to each other, it is necessary to use a smaller value of $S_{\rm add}$.\footnote{Hence, it seems fair to think that optimally, $S_{\rm add}$ should be chosen according to environment, but this would be a rather complex, non local procedure, far beyond the scope of this paper.} The additional criterion (\ref{eq:daddcrita}) is optional, although justified by the fact that a series of successive points along the border of a waterbag can become, at least temporarily, aligned. The control parameter $d_{\rm add}$ is taken to be small fraction of the total size $L$ of the system during various stages of its evolution, typically $d_{\rm add} \in [L/100, L/50]$. 

To keep track of the amount of local refinement compared to initial conditions, a refinement level $\ell$ is associated to each point of the polygon. Initial vertices are all flagged with $\ell=1$. Refinement level of point $P$ is then given by $\ell_P=\max(\ell_A,\ell_B)+1$. This information is needed if one aims to preserve the points of the polygon up to some level $\ell_{\rm min}$ when unrefinement is performed, as discussed below.

Point removal is in fact performed before refinement. To do this, a scheme dual to that used for refinement is adopted. For a triangle $\widehat{APB}$ composed of three successive points along the border of a waterbag, point $P$ is removed if all the following conditions are fulfilled:
\begin{eqnarray}
P & \neq & {\rm inflection\ point}, \label{eq:noinflexion} \\
S(\widehat{APB}) & \leq & S_{\rm rem}, \label{eq:aremcrit1}\\
\min(d_{AP},d_{PB}) & \leq & d_{\rm rem}, \label{eq:aremcrit2} \\
d_{AB} & < & d_{\rm add}, \label{eq:addicon} \\
\ell_P & > & \ell_{\rm min}, 
\end{eqnarray}
with $S_{\rm rem} < S_{\rm add}$ and $d_{\rm rem} < d_{\rm add}$.
Our practical choice is 
\begin{eqnarray}
S_{\rm rem} &=& S_{\rm add}/2,\\
d_{\rm rem} &=&d_{\rm add}/2.
\end{eqnarray}
Forbidding inflexion point removal is just an approximation of the test dual to $S(\widehat{A{\tilde P}B}) > S_{\rm add}$ when there is a change of sign of local curvature. 

Vertex removal might be performed in a certain order to improve accuracy: for instance, our choice is to first examine the points $P$ with the smallest values of $S(\widehat{APB})$.\footnote{Note that the points which are removed have to be flagged to take into account the corresponding change of the polygon structure. Once a point $P$ is removed, the potential decision to also remove its direct neighbors has to be reexamined and the tests (\ref{eq:noinflexion}), (\ref{eq:aremcrit1}), (\ref{eq:aremcrit2}) and (\ref{eq:addicon}) have to be performed again. Rigorously speaking, sorting of the arrays of values of $S(\widehat{APB})$ should be performed again each time a point is removed: for simplicity we skip this operation, but this should have little consequence on the results.} 

To have access to part of the Lagrangian information on the system, the points of the polygon with refinement level $\ell \leq \ell_{\rm min}$ are always preserved. In practice, we set $\ell_{\rm min}=1$ which corresponds to keeping the vertices generated during initial set up.\footnote{Note, following this reasoning, that preserving the Lagrangian information suggests that unrefinement should be performed first following decreasing values of $\ell$ and then, for a given $\ell$, increasing values of $S(\widehat{APB})$.} 

Note that one might perform several passes when refining or when unrefining. Although this is an option in our code, we do not adopt it in practice. The necessity to perform several passes can indeed hide another defect, such as undersampling of the initial waterbag contours, or a time step too large resulting in a large amount of curvature generated/reduced between two successive states of the system.

As a final trivial but important algorithmic remark, after removal/insertion of vertices on the orientated polygon, we reorder the data structure so that it does not have any hole due to point removal and so that the newly added points are located in memory nearby their actual neighbors. 

%=======================================================
\subsection{Refinement/unrefinement: number of vertices ``dynamics''}
\label{app:naddrem}
%=======================================================
\begin{figure*}
\centerline{\hbox{\psfig{file=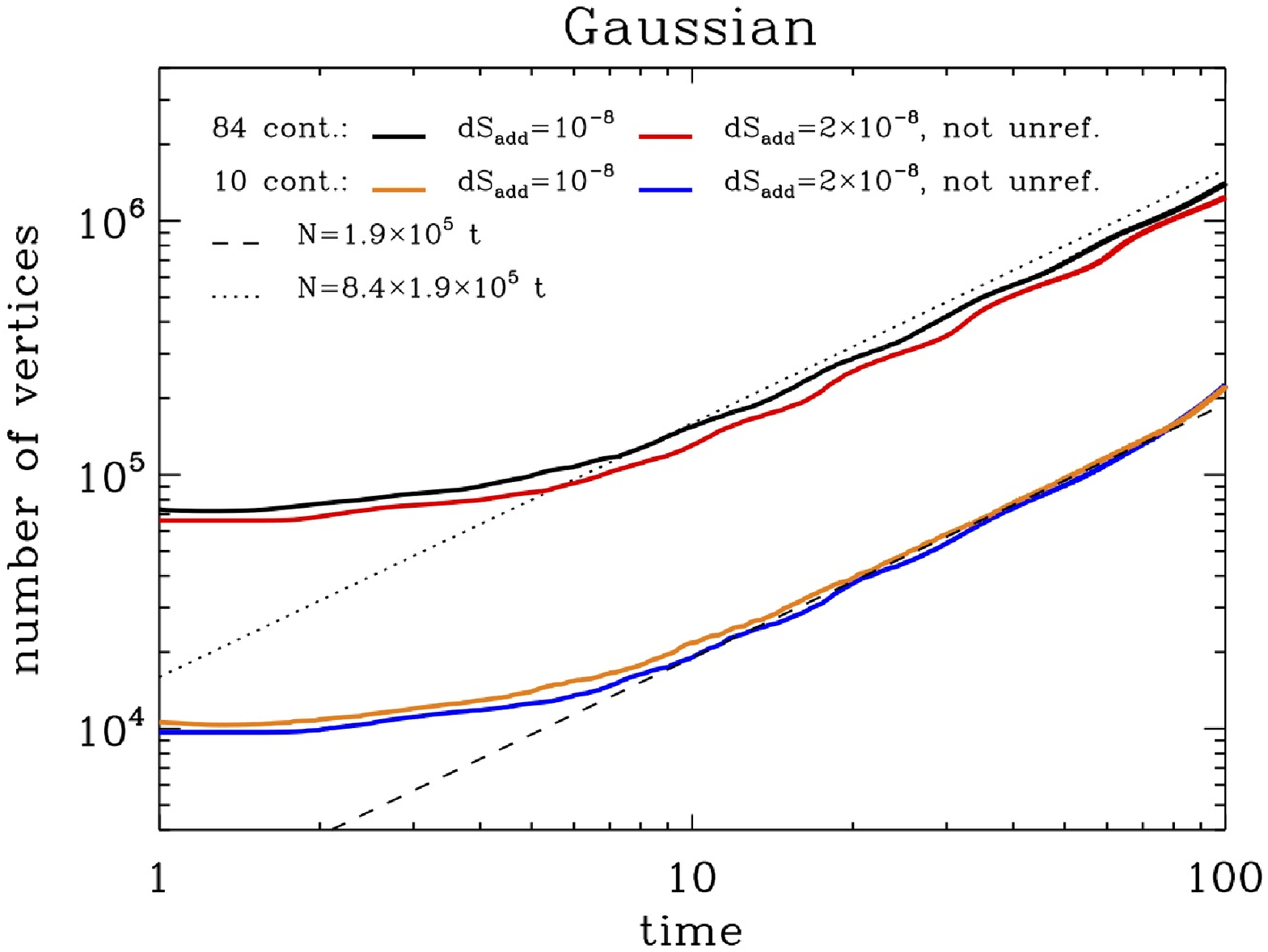,width=9cm}\psfig{file=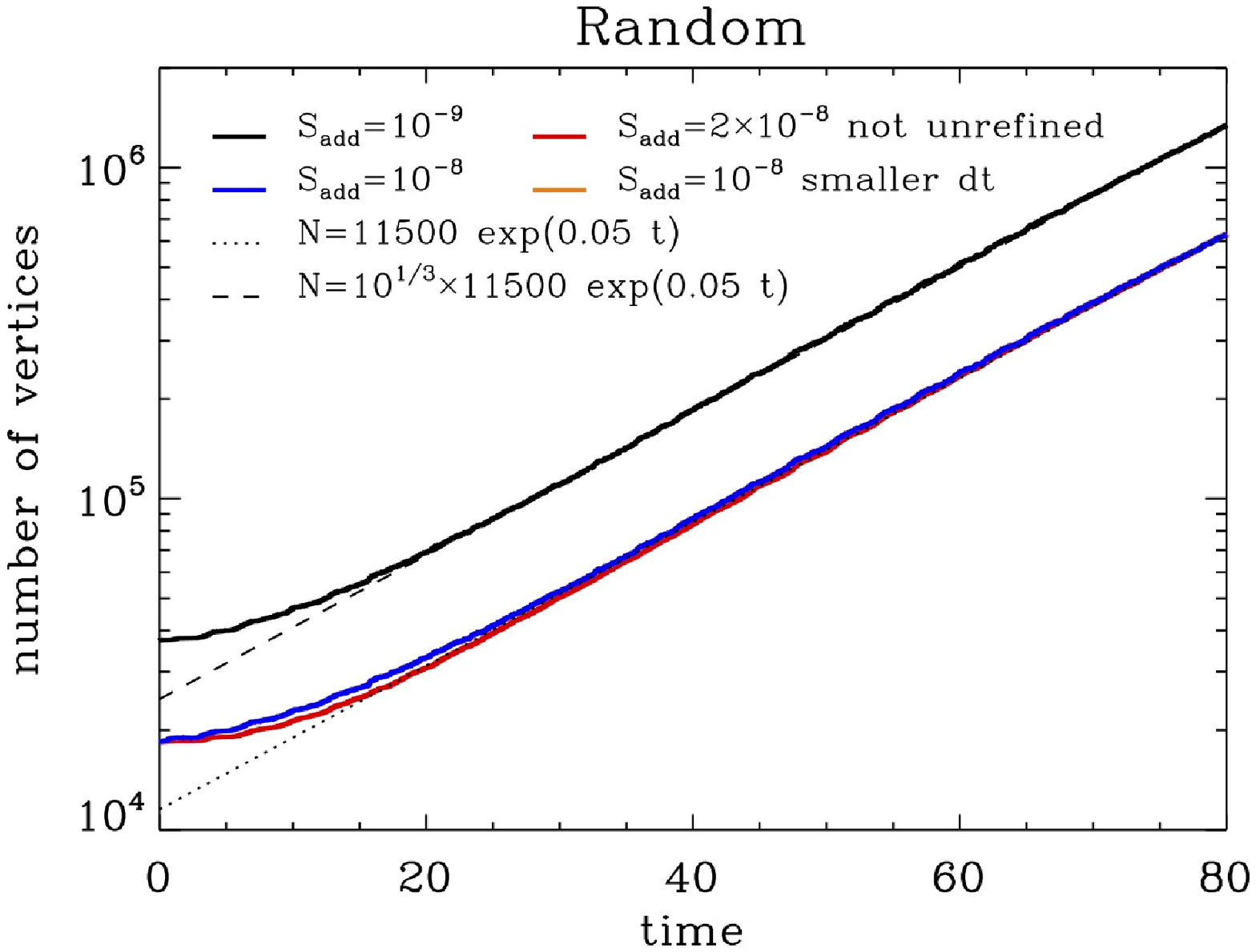,width=9cm}}}
\centerline{\hbox{\psfig{file=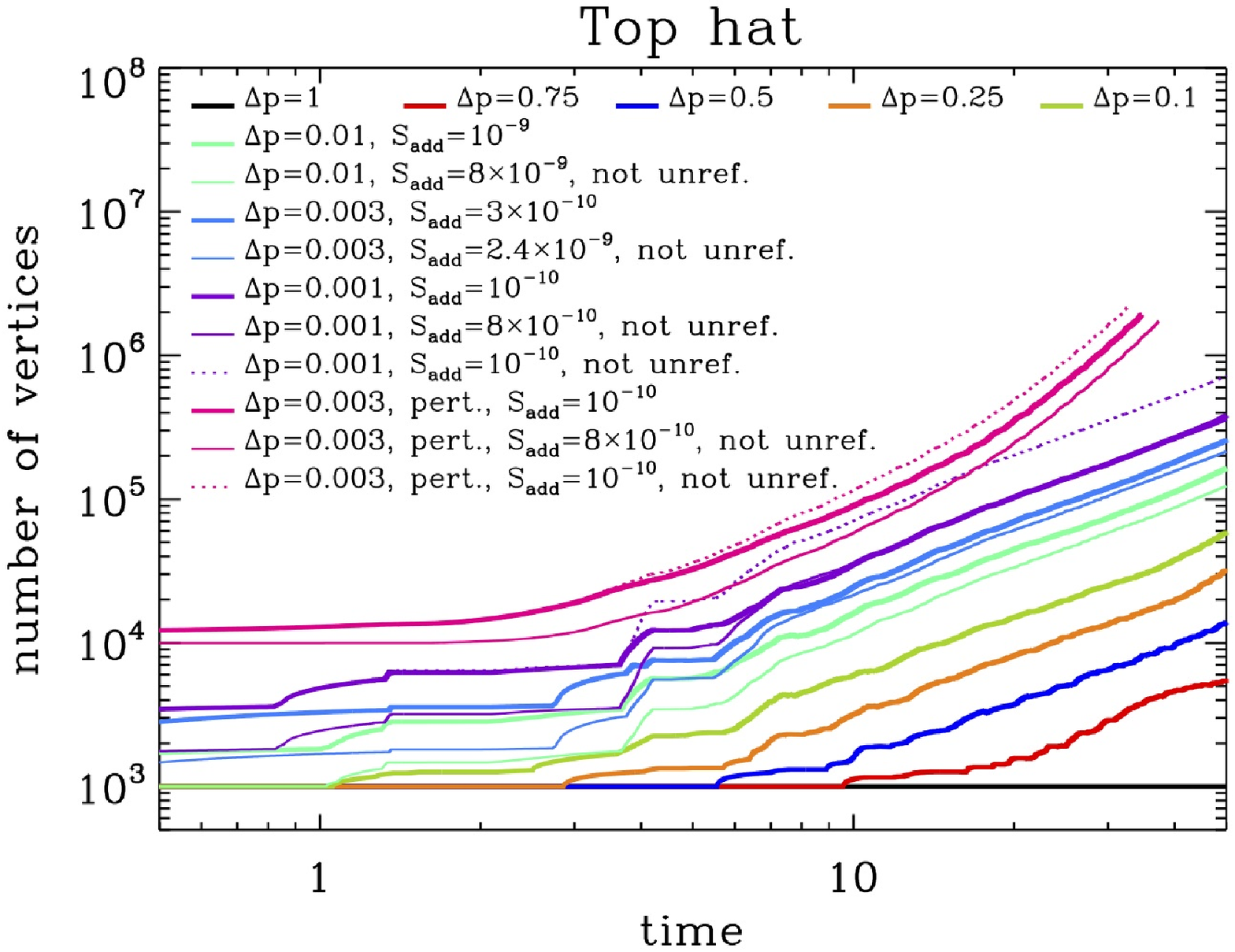,width=12cm}}}
\caption[]{Number of vertices as functions of time, measured in the simulations with Gaussian initial conditions (upper-left panel), those with the initially random distribution of halos (upper-right panel) and the single waterbag simulations (bottom panel). In addition to the measurements, a linear fit is performed after relaxation in the Gaussian case and an exponential one for the random case, as indicated on each panel and discussed in the main text.  In the Gaussian case, the 84 contours curves lie roughly about a factor 8 above the 10 contours curves, as expected from a simple rule of three. In the random case, the black curve is a factor $10^{1/3}$ above the blue one. This steams from the fact that in the small angle approximation, the surface of a triangle composed of 3 equidistant successive points $A$, $P$ and $B$ along a waterbag contour is $S \simeq ({1}/{16}) d^3/R$, where $R$ is the curvature radius and $d$ is the distance between points $A$ and $B$. Changing both $S_{\rm add}$ and $S_{\rm rem}$ by a factor $\alpha$ in equations (\ref{eq:Saddcrit}) and (\ref{eq:daddcrit})  thus reduces the typical distance between successive points of the contour by a factor $\alpha^{1/3}$, hence the corresponding increase of the number of vertices. In general, the points are not equidistant, but the reasoning still stands from statistical averaging. Note that in upper-right panel, the orange curve coincides exactly with the blue one, and is thus invisible.}
\label{fig:npts}
\end{figure*}
 Figure~\ref{fig:npts} shows the vertex count as a function of time for the all the simulations we performed in this work. It illustrates well the variety of the cases we have at hand: the simulations with Gaussian initial conditions present, after relaxation, a linear behavior of the total number of vertices with time which is the expected signature of quiescent mixing,\footnote{Note, however, in the case with 10 contours, that the number of vertices starts to depart from linearity at late times (between $t=70$ and $t=100$), due to the increasing contribution of the unstable region.} the simulations with random initial conditions develop chaos with a Lyapunov exponent equal to $0.05$, while most of the single waterbag simulations relax to a power-law behavior. It is important to notice as well that the number of vertices scales as expected with the waterbag density (upper-left panel) and with the values of the refinement parameters (upper-right panel). 

\begin{figure}
\centerline{\hbox{\psfig{file=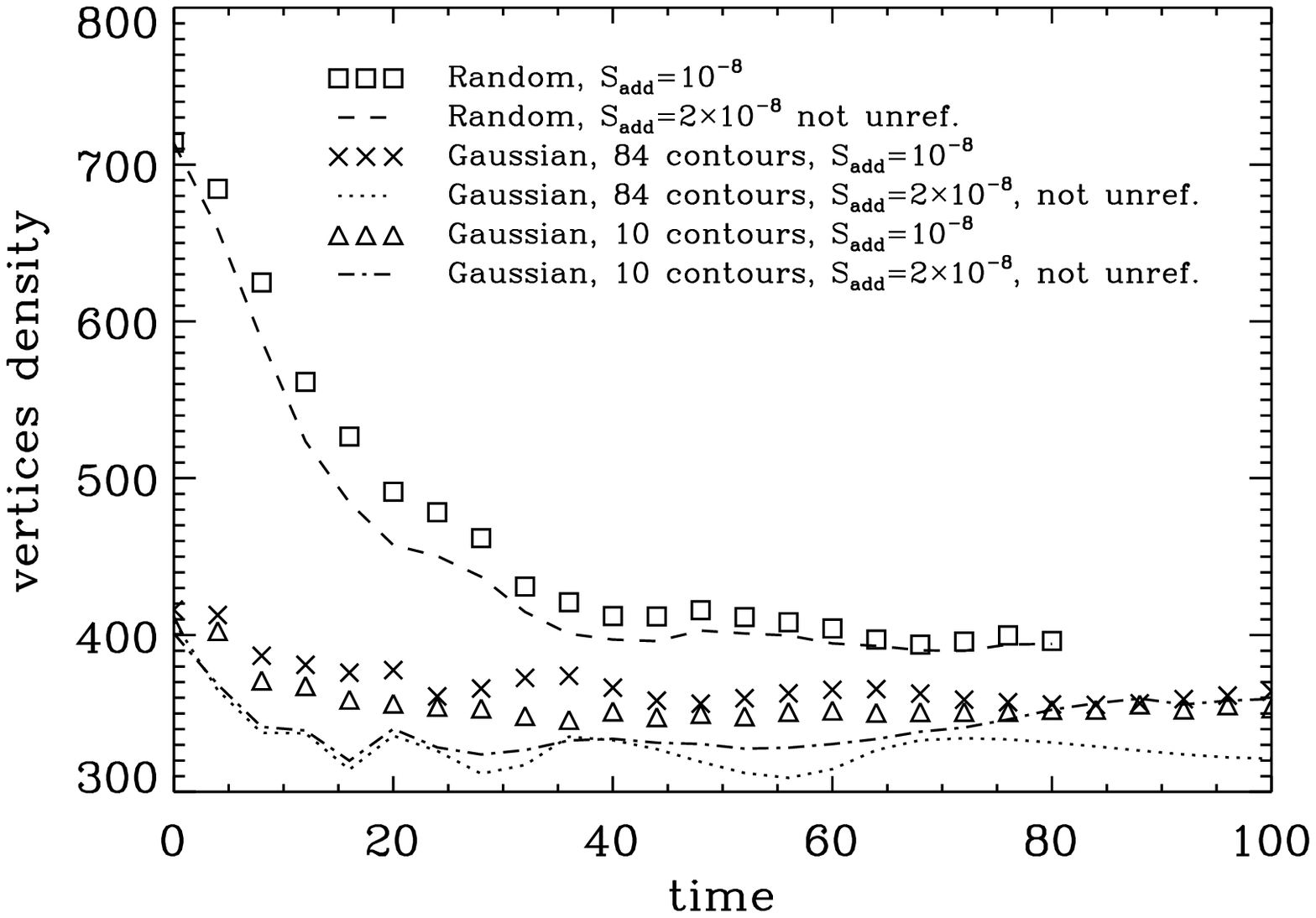,width=8.8cm}}}
\caption[]{The vertex number density as a function of time for the various simulations we realized with Gaussian and the random set of halos initial conditions, except for the random simulation with $S_{\rm add}=10^{-9}$, for clarity. }
\label{fig:numdens}
\end{figure}

The linear vertex density is shown on Fig.~\ref{fig:numdens} as a function of time for the Gaussian and random set of waterbags simulations, on which we focus from now on. It becomes rapidly steady, of the order of a few hundred points per unit length: as expected from a well behaved numerical behavior, vertex number is a good tracer of the waterbag length, whatever refinement strategy employed. 

\begin{figure}
\centerline{\hbox{\psfig{file=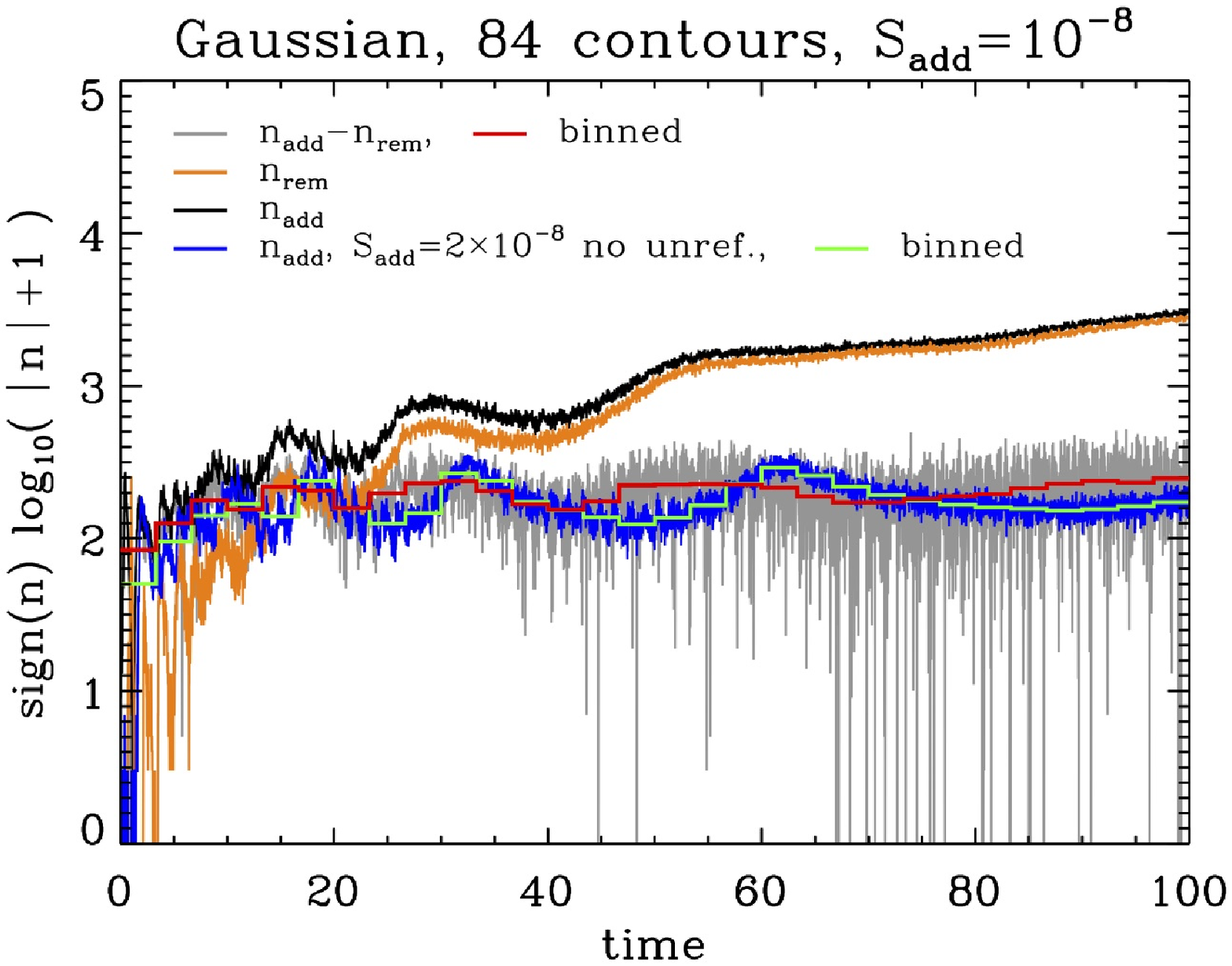,width=9cm}}}
\centerline{\hbox{\psfig{file=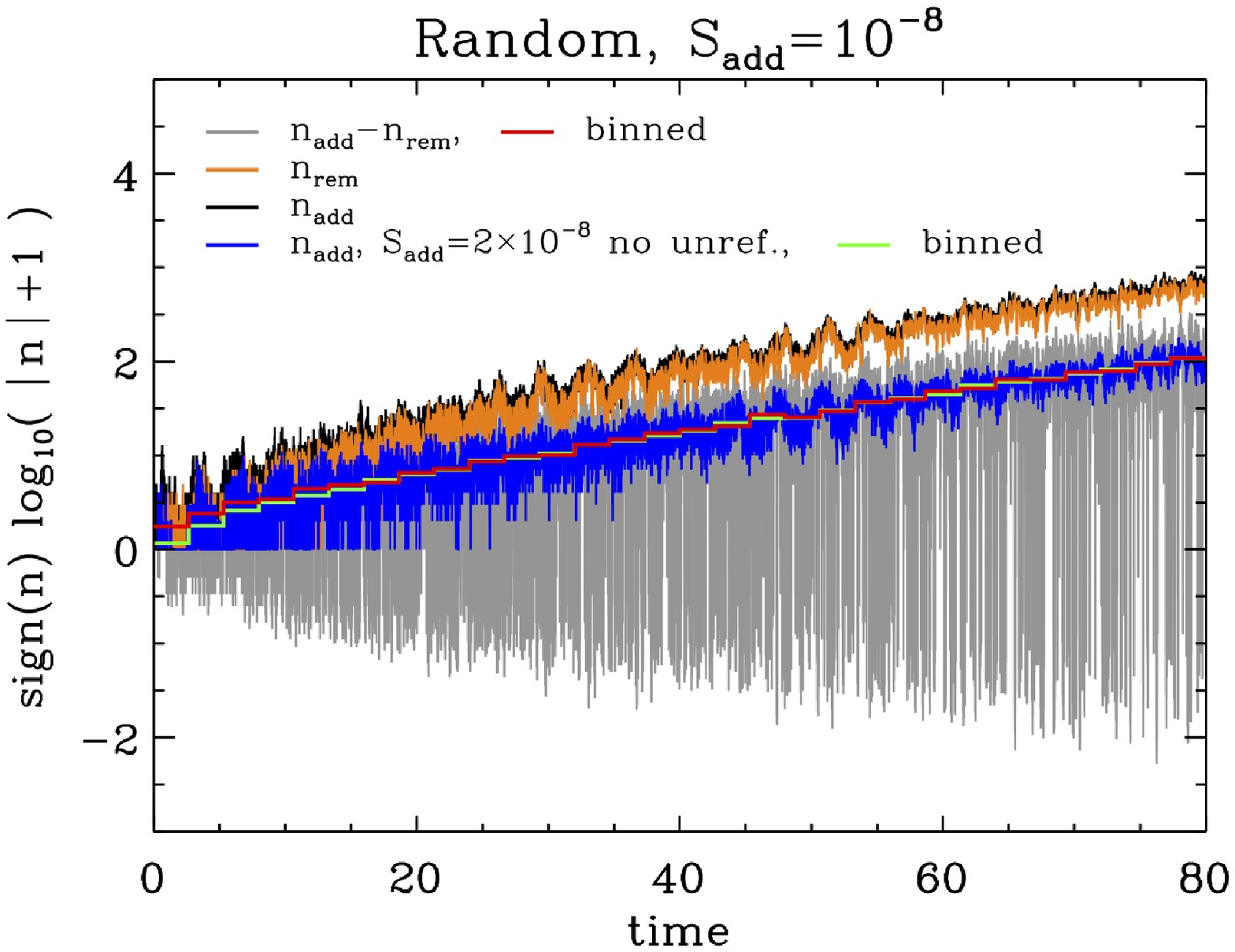,width=9cm}}}
\caption[]{Vertex creation/destruction. This figure shows $n_{\rm add}$, $n_{\rm rem}$ and $n_{\rm add}-n_{\rm rem}$ as functions of time, where $n_{\rm add}$ and $n_{\rm rem}$ are respectively the numbers of vertices added and removed to the orientated polygon at each time step. These quantities are plotted for simulations with $S_{\rm add}=10^{-8}$ and unrefinement allowed. In addition,  $n_{\rm add}$ is plotted for simulations with a twice larger value of $S_{\rm add}$ but unrefinement inhibited. Because of the important noise on the measurements, binned versions of $n_{\rm add}-n_{\rm rem}$ and $n_{\rm add}$ are plotted for these respective setups. The top and bottom panel correspond respectively to Gaussian initial conditions with 84 contours and to the random initial distribution of halos.}
\label{fig:npadr}
\end{figure}

As can be deduced from top panels of Fig.~\ref{fig:npts}, for each simulation with $S_{\rm add}=10^{-8}$ and unrefinement allowed, we performed a simulation with unrefinement inhibited and a twice larger value of $S_{\rm add}$ such that the vertex number count/number density, hence computational time, becomes approximately the same in both simulations after relaxation.\footnote{Note that this tuning was not obtained by a mathematical reasoning, but by trying several values of $S_{\rm add}$.}  It is therefore interesting to compare more in detail these two setups in terms of number of vertices dynamics, in particular to see how many points ($n_{\rm add}$) are added at each time step, how many ($n_{\rm rem}$) are removed in the case unrefinement is allowed, and what is the net result ($n_{\rm add}-n_{\rm rem}$). Figure~\ref{fig:npadr} shows these quantities as functions of time for the Gaussian case with 84 contours (top panels) and the random halos (bottom panels). 

When unrefinement is allowed, $n_{\rm rem}$ becomes quickly of the same order of $n_{\rm add}$. The net result $n_{\rm add}-n_{\rm rem}$ is of course globally positive but very noisy. This can be interpreted as follows. 
During the course of dynamics, contours are submitted to two effects: 
\begin{description}
\item[(i)] {\em Variation of the distance $\ell$ between too successive points $A$ and $B$ of a waterbag border, essentially due to the variations of the force.} Indeed, it is easy to write (see Appendix~\ref{sec:refidt})
\begin{equation}
\frac{{\rm d}\ell}{{\rm d}t} \simeq \frac{1}{\ell} (v_A-v_B)(x_A-x_B)(1-2\rho),
\end{equation}
a quantity which can be either locally positive or negative according to the time considered and the value of $\rho$. According to criteria (\ref{eq:daddcrit}) and (\ref{eq:remcrit2}), this can thus induce $n_{\rm add}=0$ and $n_{\rm rem} > 0$ or reversely. For a system which covers phase-space approximately evenly on the coarse level, one can thus expect $n_{\rm add}$ of the same order of $n_{\rm rem}$. For instance, waterbag contours following a quiescent dynamics such as in the Gaussian case (Fig.~\ref{fig:gaussianA}) have $(v_A-v_B)(x_A-x_B) < 0$ in the lower left and upper-right quadrants of phase-space, and $(v_A-v_B)(x_A-x_B) >0$ for the two other quadrants. Of course, this symmetry is not exactly verified: one expects a net positive effect from mixing, due to the fact that two distinct points of a contour generally have different average orbital speeds. This can be easily understood for instance by assuming that the two points $A$ and $B$ correspond to two harmonic oscillators with slightly different frequencies. 
\item[(ii)] {\it Variation of the surface $S$ of the triangle composed by three successive points $A$, $P$ and $B$ of a waterbag border, essentially due to the variations of the derivative of the force, i.e. the gradient of the projected density.} Again, as shown in appendix~\ref{sec:refidt}, we indeed have
\begin{eqnarray}
\frac{{\rm d}S}{{\rm d}t}  & \simeq & \frac{1}{4} 
(x_P-x_A)(x_B-x_P)(x_A-x_B)\frac{\partial \rho}{\partial x} \nonumber \\
 & & \times {\rm sgn}\left(\overrightarrow{AP} \wedge \overrightarrow{PB} \right).
\end{eqnarray}
The same argument of symmetry made in point (i) applies and once again, the regions of the contours where $n_{\rm add} \geq 0$ and $n_{\rm rem}=0$ should be compensated by other regions of the contour where $n_{\rm add} =0$ and $n_{\rm rem} >0$, according to criteria (\ref{eq:Saddcrit}) and (\ref{eq:remcrit1}). For instance, in the quiescent case represented by our Gaussian simulation, a simple geometric analysis shows that, in general, ${\rm d}S/{\rm d}t > 0$ in the upper-left and the lower-right quadrants of phase-space, and ${\rm d}S/{\rm d}t < 0$ in the two other quadrants, in agreement with intuition. 
\end{description}
If the criterion on $S_{\rm add}$ is aggressive, effect (ii) is dominant over effect (i), which is the case in our simulations with unrefinement. 

When unrefinement is inhibited, both effects (i) and (ii) induce $n_{\rm add} > 0$. Hence we set a less stringent constraint on $S_{\rm add}$ to have approximately the same net effect $n_{\rm add}-n_{\rm rem}$ than in the case when unrefinement is allowed. However one has to be aware of the fact that local sampling of the contours is not the same in both configurations. It would go beyond the scope of this paper to perform detailed geometric comparisons of local sampling in both methods, but it is important to notice the following. In a steady state regime, an element of contour will pass regularly through regions where effects (i) and (ii) are alternatively positive and negative, implying in the case unrefinement is triggered, that this element of contour will be alternatively refined and unrefined: from a Lagrangian point of view, where one would locally set the waterbag border under consideration at rest, the refined areas can be assimilated to waves propagating along the contour. In regions when orbital speed is large, this can induce a large source of noise. While being potentially a powerful option, performing unrefinement thus does not seem to be the best choice in our one dimensional case if one aims to follow the evolution of a system during many dynamical times. This is illustrated quantitatively by the energy conservation diagnostics performed in \S~\ref{app:enercons}. 

%=======================================================
\section{Details on the calculation of the time step}
\label{app:dtdetails}
%=======================================================
Here, we give details on the way the time step is calculated, restricting our analyses to the predictor corrector scheme displayed on Fig.~\ref{fig:algorithm}. In \S~\ref{sec:harmodt}, a dynamical prescription is derived in the framework of a harmonic potential. In \S~\ref{sec:refidt}, two additional criteria on the time step are computed to make it harmonious with refinement. Finally \S~\ref{sec:dtplots} discusses the practical implementation and the evolution of the time step. 
%================================
\subsection{Time step and long term evolution: a case study of the harmonic oscillator} 
\label{sec:harmodt}
%================================
In the harmonic case, the potential is given by
\begin{equation}
\phi(x)=\rho x^2,
\end{equation}
where $\rho$ is a constant projected density. The solution of such a system is given by the harmonic oscillator, 
\begin{equation}
x(t)={\bar x}(t) \equiv x_0 \cos( \omega t + \psi), 
\end{equation}
with the frequency $\omega=\sqrt{2\rho}$. 

At step $n$ of the simulation, the predictor-corrector algorithm reads, given the acceleration $a_n(t+{\rm d}t/2)=-\omega^2 (x_n+v_n {\rm d}t/2)$,
\begin{equation}
\left( \begin{array}{c}
x_{n} \\
v_{n}  
\end{array} 
\right) = M^n  \left( \begin{array}{c}
x_{0} \\
v_{0}  
\end{array}\right)
\end{equation}
with
\begin{equation}
M = \left( \begin{array}{cc}
\cos\theta & \frac{1}{\omega^2 {\rm d}t} \sin^2\theta \\
-\omega^2 {\rm d}t & \cos\theta
\end{array}\right)
\end{equation}
\begin{equation}
%  \theta=\arccos\left[ 1-2 (\omega {\rm d}t)^2
  \theta=\arccos\left[ 1-\frac{(\omega {\rm d}t)^2}{2}
  \right], 
\label{eq:thetaarccos}
\end{equation}
and $(x_0,v_0)$ correspond to initial conditions. Using a standard diagonalization procedure, one can write 
\begin{equation}
M^n  = \left( \begin{array}{cc}
\cos(n\theta) & \frac{\sin\theta}{\omega^2 {\rm d}t} \sin(n\theta) \\
-\frac{\omega^2 {\rm d}t}{\sin \theta} \sin(n\theta)& \cos(n\theta)
\end{array}\right),
\end{equation}
to be compared to the exact solution 
\begin{equation}
E_n = \left( \begin{array}{cc}
\cos(n \omega {\rm d}t) & \frac{1}{\omega} \sin(n\omega {\rm d}t) \\
-\omega \sin(n\omega {\rm d}t)& \cos(n\omega {\rm d}t)
\end{array}\right).
\end{equation}
Enforcing a relative error on $x_n$ and $v_n$ of the order of at most $\epsilon$ is roughly equivalent to
\begin{equation}
|n \theta - n \omega {\rm d}t| \leq \epsilon, \quad \epsilon \ll 1.
\end{equation}
After Taylor expanding equation~(\ref{eq:thetaarccos}) at second order and assuming that the system is followed during $N_{\rm orbits}$ orbital times, corresponding to a total time of $T=(2 \pi/\omega) N_{\rm orbits}=n\ {\rm dt}$, one obtains the following constraint on the time step
\begin{eqnarray}
{\rm d}t & \leq & {\rm d}t_{\rm dyn} \\
{\rm d}t_{\rm dyn} & \equiv & \frac{C}{\sqrt{\rho}}, \quad C = \sqrt{\frac{ 6 \epsilon}{\pi N_{\rm orbits}}}.
\label{eq:courantosci}
\end{eqnarray}
This expression shows, importantly, that ${\rm d}t_{\rm dyn}$ is inversely proportional to $1/\sqrt{N_{\rm orbits}}$: the larger the number of orbital times, the smaller the time step should be. Equation (\ref{eq:courantosci}) is generalized to the non harmonic case by just taking the maximum of the projected density (equation~\ref{eq:courant}), with $C$ ranging from typically $10^{-2}$ (corresponding roughly to, e.g., $\epsilon=10^{-3}$ and $N_{\rm orbits}=20$) to $10^{-4}$ (corresponding roughly to, e.g., $\epsilon=10^{-6}$ and $N_{\rm orbits}=200$). 
%================================
\subsection{Time step and refinement}
\label{sec:refidt}
%================================
Since our refinement is based on the measurement of the areas $S$ of the triangles formed by 3 successive points along the orientated polygon (equation~\ref{eq:Saddcrit}) and the distance $d$ between successive points of the polygon (equation~\ref{eq:daddcrit}), it is sensible to set constrains on the time step that bound the variations of $S$ and $d$. 
\begin{itemize}
\item {\em Constraint on the time step from triangle area variations:}
the area of the triangle composed by 3 successive points $A$, $P$ and $B$ of the polygon is
\begin{equation}
S_{(0)}\equiv S(\widehat{APB})=\frac{1}{2} \left| \overrightarrow{AP} \wedge \overrightarrow{PB} \right|.
\end{equation}
The time derivative of $S_{(0)}$ is thus given by
\begin{eqnarray}
 \frac{{\rm d}S_{(0)}}{{\rm d}t}  &=& \frac{1}{2} \left[ (x_P-x_A) a_B + (x_B-x_P) a_A \right. \nonumber \\
 & & \quad \quad \left. + (x_A-x_B) a_P \right]\ {\rm sgn} \left( \overrightarrow{AP} \wedge \overrightarrow{PB} \right),
\end{eqnarray}
where $a_X$, $X=A,P,B$, denotes the acceleration. It is useful to perform a Taylor expansion of the acceleration around point $P$:
\begin{eqnarray}
 a_A &= &a_P+\frac{\partial a}{\partial x} (x_A-x_P)+ \frac{1}{2} \frac{\partial^2 a}{\partial x^2} 
(x_A-x_P)^2 \nonumber \\
& & \quad \quad \quad \quad \quad \quad \quad \quad \quad \quad \quad \quad \quad \quad + {\cal O}(\delta x^3), \\
 a_B &= &a_P+\frac{\partial a}{\partial x} (x_B-x_P)+ \frac{1}{2} \frac{\partial^2 a}{\partial x^2} 
(x_B-x_P)^2 \nonumber \\
& & \quad \quad \quad \quad \quad \quad \quad \quad \quad \quad \quad \quad \quad \quad + {\cal O}(\delta x^3),
\end{eqnarray}
where $\delta x=\max(|x_P-x_A|,|x_B-x_P|,|x_A-x_B|)$, to see that ${\rm d}S_{(0)}/{\rm d}t$ cancels at leading order in $\delta x$, as expected from the symplectic nature of the system. The second order term reads
\begin{eqnarray}
\frac{{\rm d}S_{(0)}}{{\rm d}t}  & \simeq & \frac{1}{4} 
(x_P-x_A)(x_B-x_P)(x_B-x_A)\frac{\partial^2 a}{\partial x^2}   \nonumber \\
  & & \quad \quad \quad \quad \quad \times {\rm sgn} \left( \overrightarrow{AP} \wedge \overrightarrow{PB} \right)+ {\cal O}(\delta x^4),
\label{eq:dsdt}
\end{eqnarray}
showing, as expected, that the second derivative of the force, hence the derivative of the projected density, controls the variations of $S$. 

\begin{figure}
\centerline{\hbox{
\psfig{file=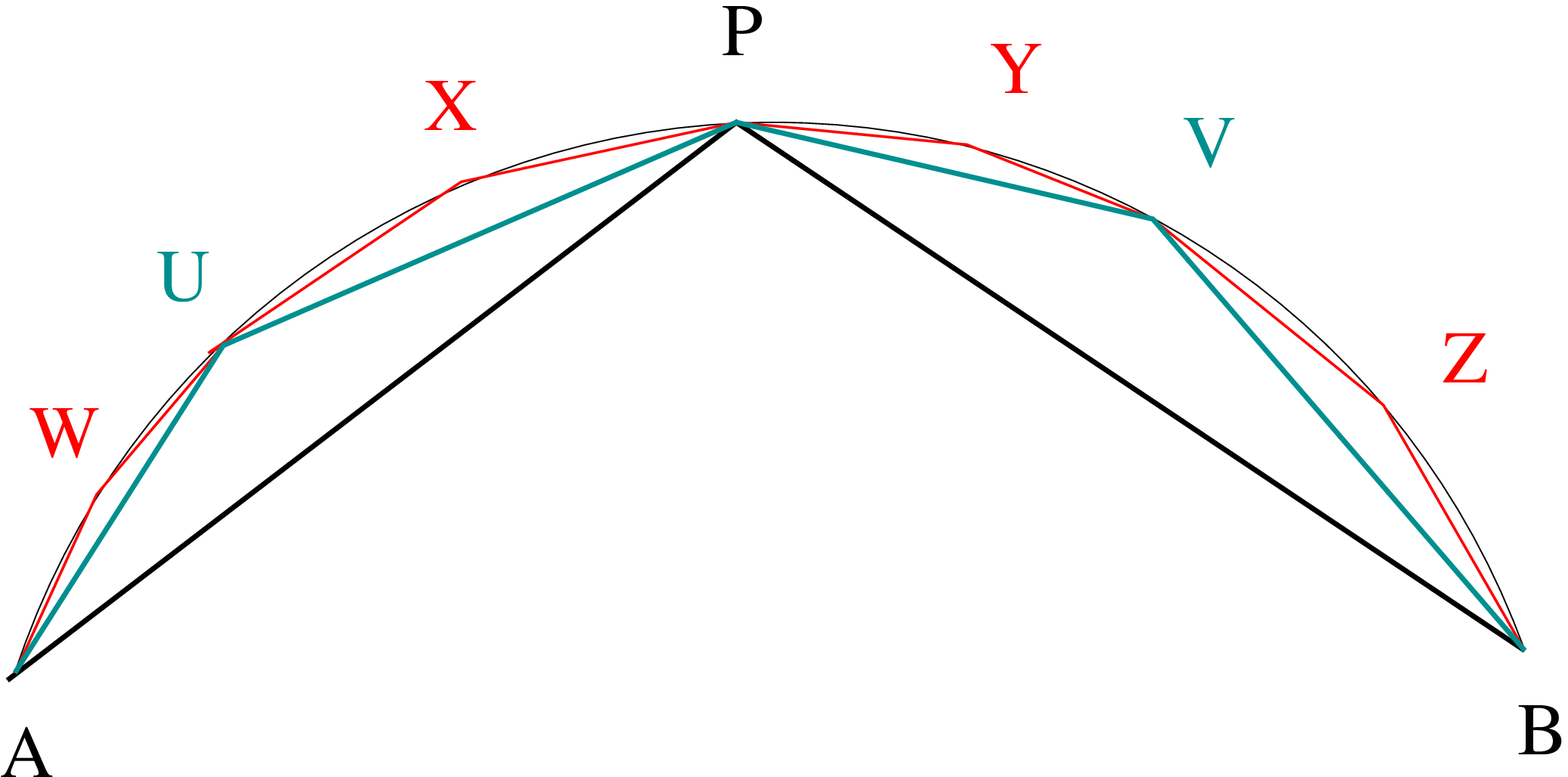,width=8cm}
}}
\caption[]{Two successive refinements on the polygon $[A,P,B]$, symbolized by the cyan lines and the red lines. The area of the triangles defined by the cyan polygon, e.g. $S(\widehat{AUP})$, is roughly 8 times smaller than $S(\widehat{APB})$, while the area of the triangles defined by the red polygon, e.g. $S(\widehat{AWU})$, is roughly 64 times smaller than $S$. This estimate comes from assuming that $\widehat{APB}$ is isosceles as well as triangles on the cyan and red polygons; it also uses the small angle approximation, which is not enforced on the figure, for clarity.}
\label{fig:tworefinements}
\end{figure}
Now, we have to relate equation (\ref{eq:dsdt}) to the variation during a time step of the area of the candidate triangles obtained  from adding refinement points $U$  and $V$ on segments $[A,P]$ and $[P,B]$, respectively, and more importantly, to avoid refining twice, the variations of the area of the next four candidate triangles obtained from adding refinement points $W$, $X$, $Y$ and $Z$ respectively on segments $[A,U]$, $[U,P]$, $[P,V]$ and $[V,B]$ (Fig.~\ref{fig:tworefinements}). To estimate roughly the area of these four candidate triangles, we use the small angle and mid point approximations. With this set of assumptions, we have
\begin{equation}
S(\widehat{AUP}) \simeq S_{(1)} \equiv \frac{1}{16} \frac{d_{AP}^3}{R}, 
\end{equation}
and analogously for $S(\widehat{PVB})$. The small angle approximation also reads $d_{AU} \simeq d_{UP} \simeq d_{AP}/2$. Similarly, we have $d_{AW} \simeq d_{AU}/2$, $d_{WU} \simeq d_{AU}/2$, and so on. As a result, the area of the next refinement level triangles, $\widehat{ijk}=\widehat{AWU}$, $\widehat{UXP}$, $\widehat{PYV}$ and $\widehat{VZB}$, verifies
\begin{equation}
S(\widehat{ijk})\simeq S_{(1)}/8 \simeq S_{(2)}\equiv  S_{(0)}/64.
\end{equation}

To avoid triggering twice refinement, one must have, after time step evolution,  $S_{(2)} \leq S_{\rm add}$, i.e.~$S_{(0)}(t+{\rm dt}) \leq 64 S_{\rm add}$, with, trivially, 
\begin{equation}
S_{(0)}(t+{\rm d}t) \simeq S_{(0)}(t)+ ({\rm d}S_{(0)}/{\rm d}t)\ {\rm d}t.
\label{eq:svariate}
\end{equation}
After refinement, but prior to time step evolution, we have by construction $S_{(0)}(t) \la 8 S_{\rm add} \ll 64 S_{\rm add}$, allowing us to neglect the $S_{(0)}(t)$ contribution in equation (\ref{eq:svariate}) to set the following approximate local constraint on the time step 
\begin{equation}
{\rm d}t \la 64 \frac{S_{\rm add}}{|{\rm d}S_{(0)}/{\rm d}t|}.
\end{equation}
In practice, we implement this condition as follows:
\begin{equation}
{\rm d}t \leq  {\rm d}t_{\rm refinement} \equiv 64 \frac{S_{\rm add}}{\max_i |{\rm d}S_i/{\rm d}t|},
\label{eq:refinementdt} 
\end{equation}
where
\begin{eqnarray}
\left| \frac{{\rm d}S_i}{{\rm d}t} \right| &= &\frac{1}{2} \left| (x_i-x_{i-1}) a_{i+1} + (x_{i+1}-x_i) a_{i-1} \right. \nonumber \\
  & & \left. + (x_{i-1}-x_{i+1}) a_i\right|.
\end{eqnarray}

\item {\em Constraint on the time step from segment length variations:}
let us consider two successive points $A$ and $B$ on the polygon and the distance $\ell$ between them. Then 
\begin{eqnarray}
\frac{{\rm d}\ell}{{\rm d}t} &= & \frac{1}{\ell} (v_A-v_B) (x_A-x_B + a_A-a_B) \label{eq:ddisex}\\
 & \simeq & \frac{1}{\ell} (v_A-v_B)(x_A-x_B)
\left(1+\frac{\partial a}{\partial x} \right) \\
&\simeq & \frac{1}{\ell} (v_A-v_B)(x_A-x_B)(1-2\rho), \label{eq:ddisap}
\end{eqnarray} 
where $a$ symbolizes the acceleration. 

Once the system has evolved during a time step, the variation of $\ell$ should not be such that we refine twice, ${\rm d} \ell \leq 2 d_{\rm add}$, this for every single contributing segment of the orientated polygon.  In practice, the implementation of this condition reads 
\begin{equation}
{\rm d}t \leq  {\rm d}t_{\rm distance} \equiv \frac{2 d_{\rm add}}{\max_i |{\rm d}\ell_i/{\rm d}t|},
\label{eq:distdt}
\end{equation}
where
\begin{equation}
\left|\frac{{\rm d}\ell_i}{{\rm d}t} \right|=\frac{1}{\ell_i} |(v_{i-1}-v_i)(x_{i-1}-x_i+a_{i-1}-a_i)|
\end{equation}
is the magnitude of the derivative of the distance $\ell_i=d_{i-1,i}$ between vertices $i$ and $i-1$.
\end{itemize}
%================================
\subsection{Practical implementation and evolution of the time step}
\label{sec:dtplots}
%================================
\begin{figure}
\centerline{
\hbox{\psfig{file=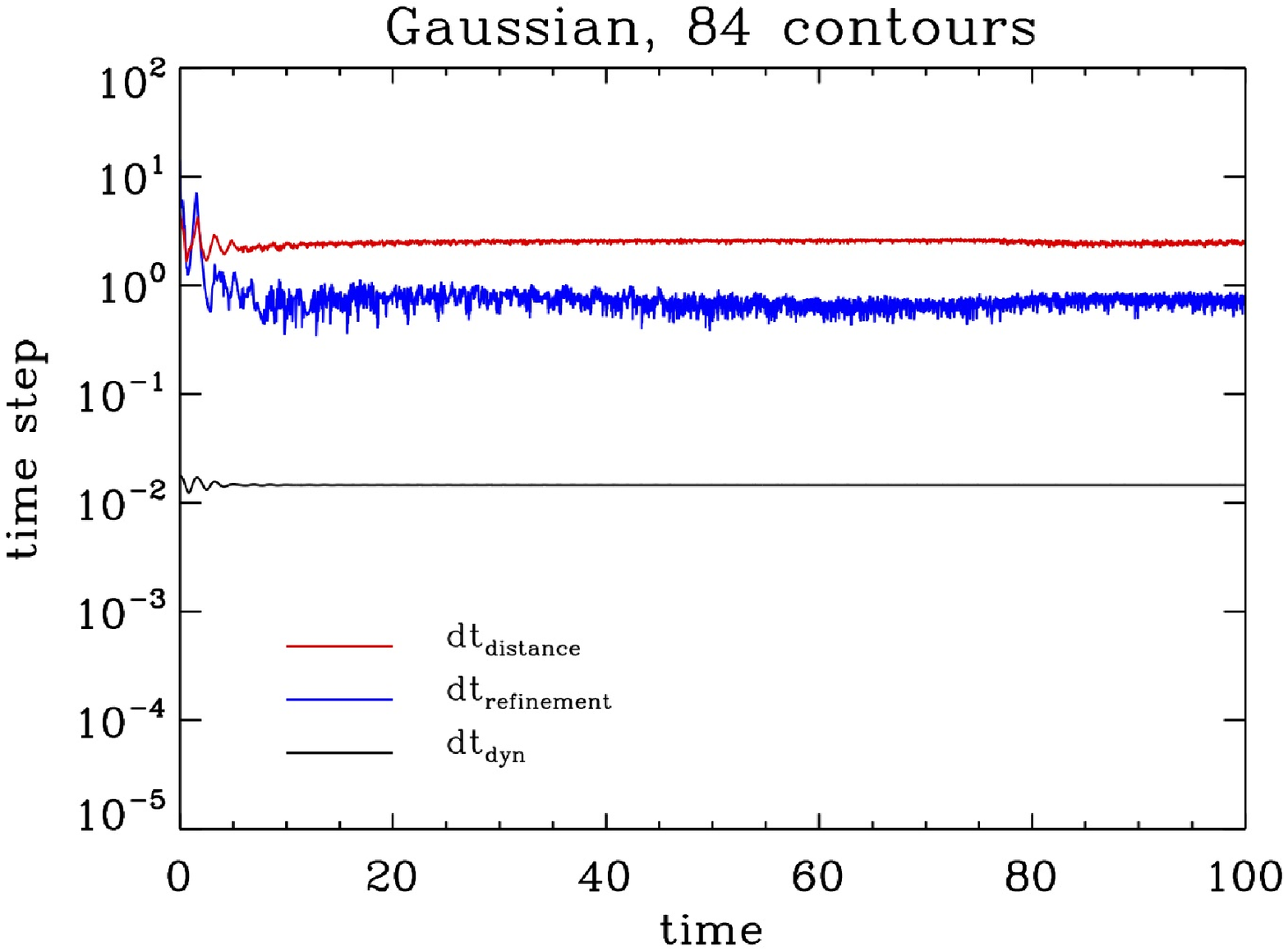,width=8.8cm,bbllx=20pt,bblly=20pt,bburx=575pt,bbury=437pt}}}
\centerline{
\hbox{\psfig{file=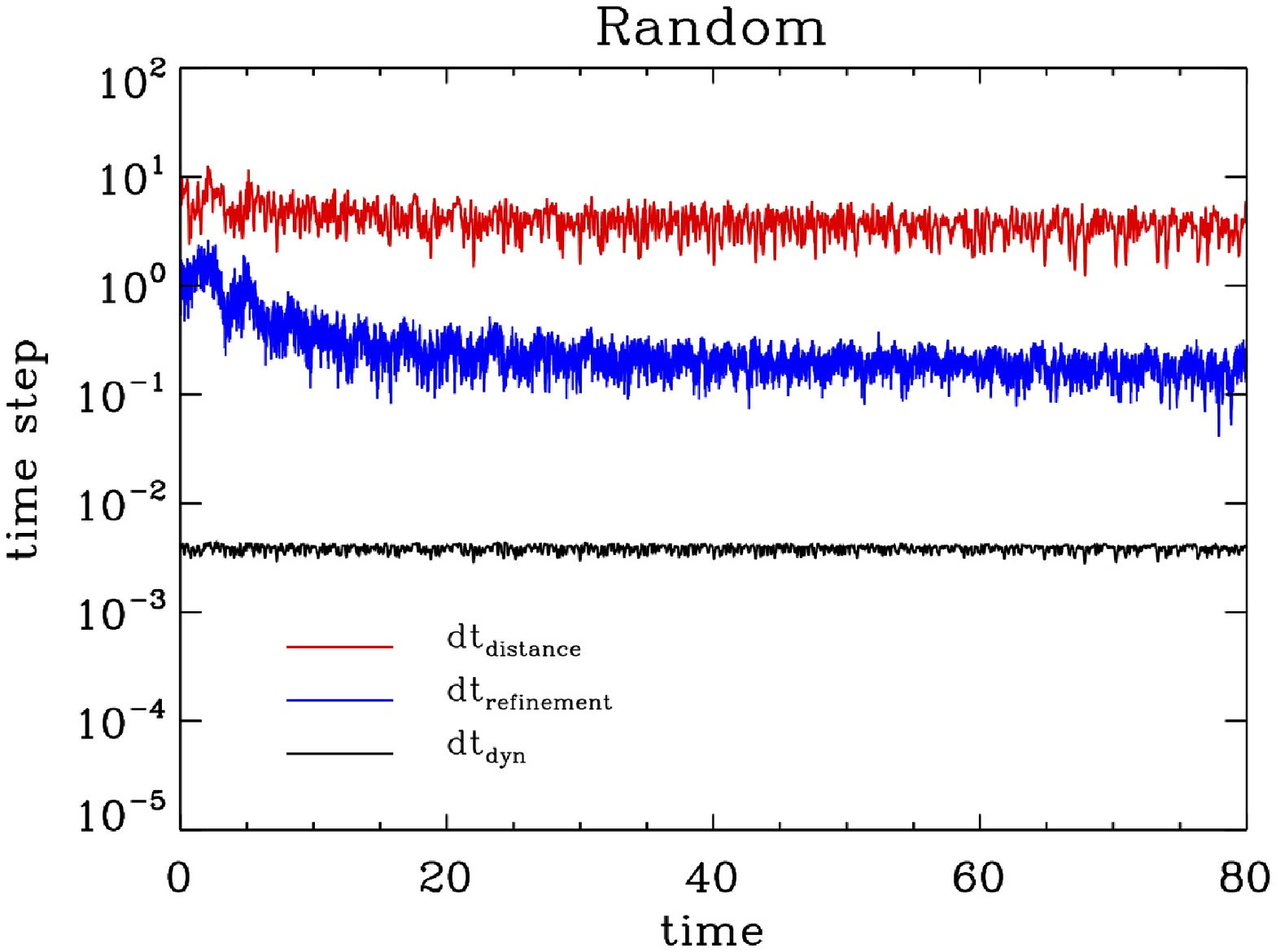,width=8.8cm,bbllx=20pt,bblly=20pt,bburx=575pt,bbury=437pt}}}
\caption[]{The time step as a function of time, obtained from the constraints brought by equation (\ref{eq:courant}), (\ref{eq:refinementdt}) and (\ref{eq:distdt}), for the simulations of Fig.~\ref{fig:gaussianA} and \ref{fig:random}, with $S_{\rm add}=2\times 10^{-8}$ and no unrefinement allowed ({\tt Gaussian} and {\tt Random} in the nomenclature of Table~\ref{tab:simuparam}). Setting stronger constraints on refinement would simply lower down the red and blue curves. For the simulations we performed in this paper, the order ``red $\ga$ blue $>$ black'' was always verified.}
\label{fig:dt_plots}
\end{figure}
 Our final set up for the time step is 
\begin{equation}
{\rm d}t=\min({\rm d}t_{\rm dyn},{\rm d}t_{\rm refinement},{\rm d}t_{\rm distance}),
\end{equation}
where ${\rm d}t_{\rm dyn}$, ${\rm d}t_{\rm refinement}$ and ${\rm d}t_{\rm distance}$ are given by equations (\ref{eq:courant}),  (\ref{eq:refinementdt}) and (\ref{eq:distdt}), respectively. For all the simulations we present in this paper, we used rather small values of $C$ in equation (\ref{eq:courant}) to try to stay on a conservative side. Our dynamical set ups are therefore such that the constraints brought by ${\rm d}t_{\rm distance}$ and ${\rm d}t_{\rm refinement}$ are less restrictive than equation (\ref{eq:courant}), as illustrated by Fig.~\ref{fig:dt_plots} for two of the simulations with Gaussian and random halos initial conditions. It is of course possible to construct settings where it is not the case. 

We also obtain, not surprisingly, ${\rm d}t_{\rm dyn} \simeq$ constant of time, except during the first few dynamical times that correspond to the relaxation phase of the system towards a quiescent state. To preserve even better symplecticity, given our integration scheme, one could also simply use ${\rm d}t={\rm constant}$:\footnote{Note that it is possible to create symplectic integrators with varying time step \citep[see, e.g.,][and references therein]{Richardson2012}.} in this case the time step scheme reduces exactly to leap-frog. However,
using a constant value of ${\rm d}t$ requires, in the framework of equation (\ref{eq:courant}), a prior guess of the maximum projected density over all the run, which is delicate without performing a testbed simulation. Note that the same problem arises in fact for estimating $C$ in equation (\ref{eq:courant}) because the number of dynamical times depends on this maximum projected density. This is particularly relevant for the single waterbag simulations in the close to cold limit, that is with small values of $\Delta p$. As a matter of fact, the values of $C$ in Table~\ref{tab:simuparam} have been chosen in a rather add-hoc way, yet still reasonable. 

As an example, for our coldest waterbag with $\Delta p=0.001$, we used $C=0.001$ which corresponds to an average time step of about $\Delta t \simeq 2.6 \times 10^{-4}$ and a rather large total number of time steps of about $1.9\times 10^5$. 
%==================================================
\section{Tests on energy conservation}
\label{app:enercons}
\begin{figure*}
\centerline{\hbox{\psfig{file=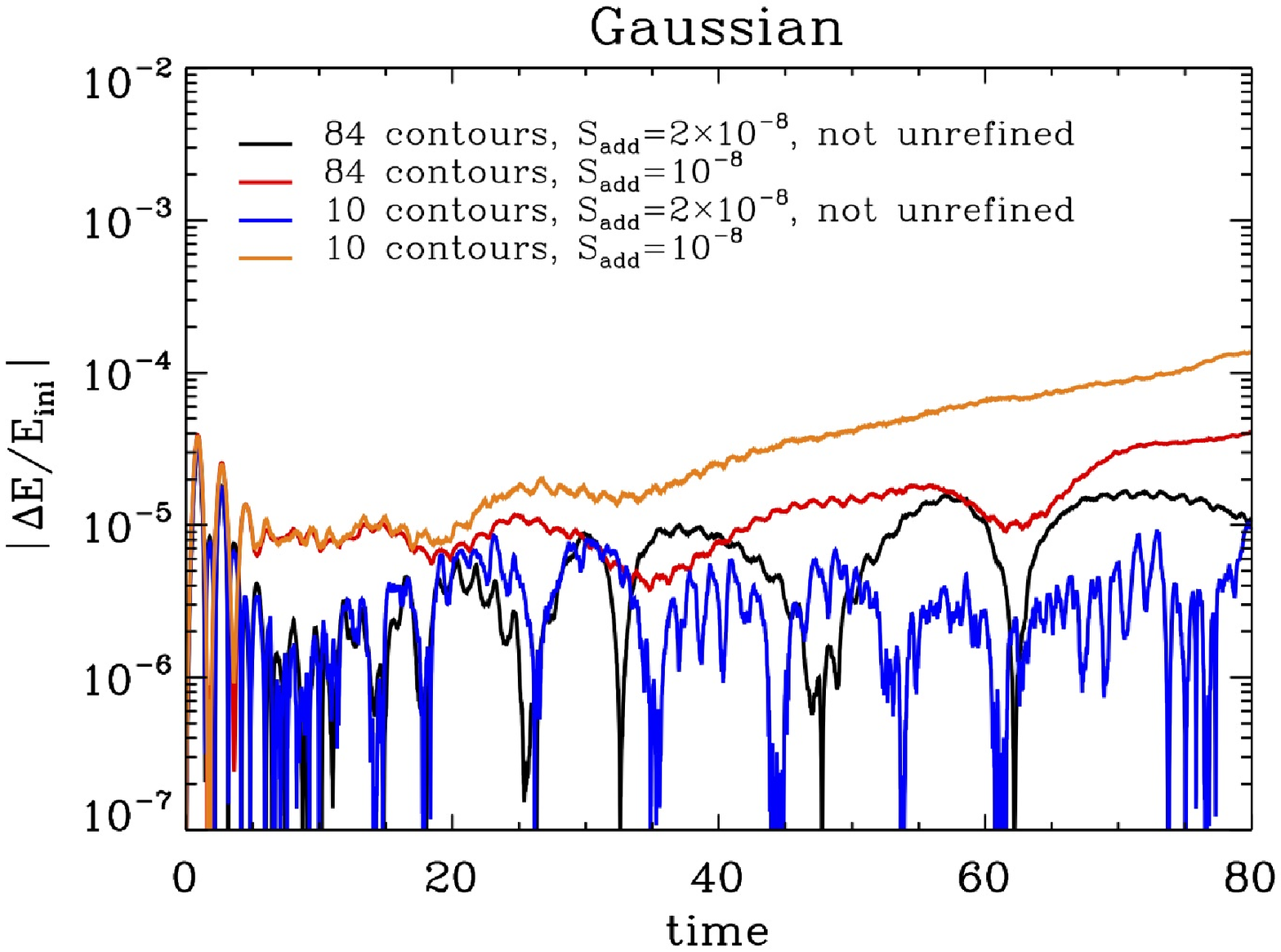,width=8.8cm}
\psfig{file=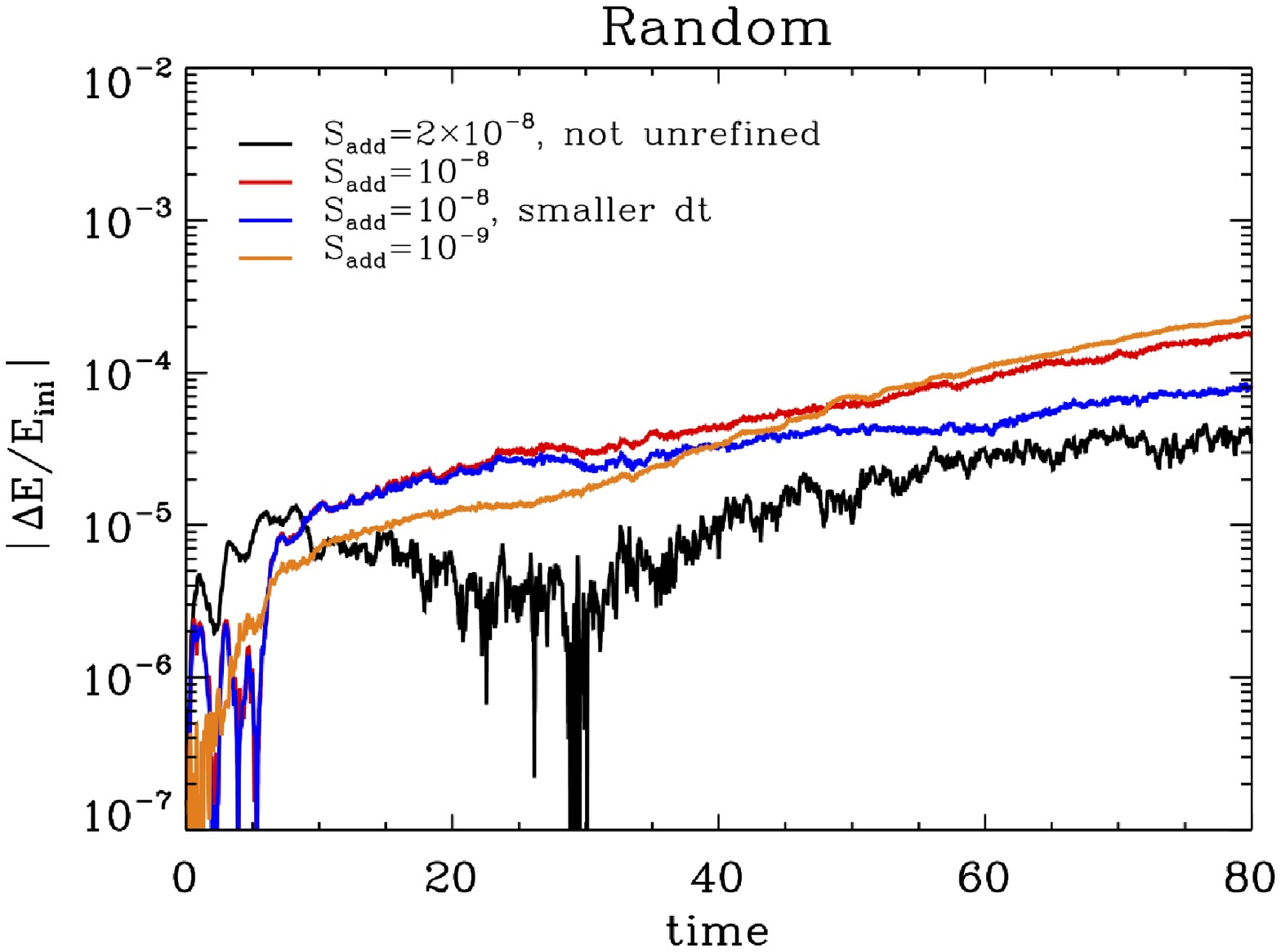,width=8.8cm}}}
\centerline{\hbox{
\psfig{file=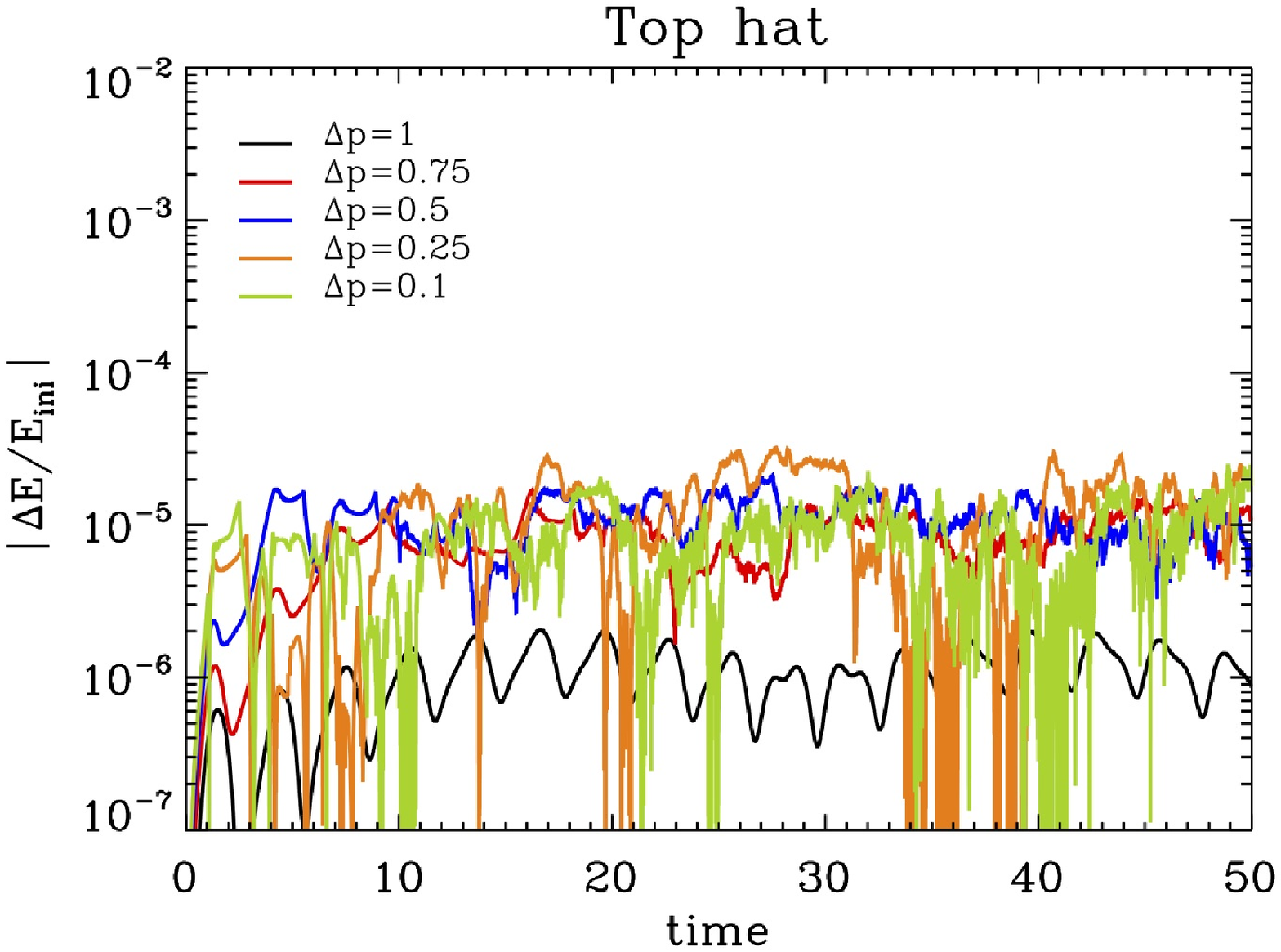,width=8.8cm}
\psfig{file=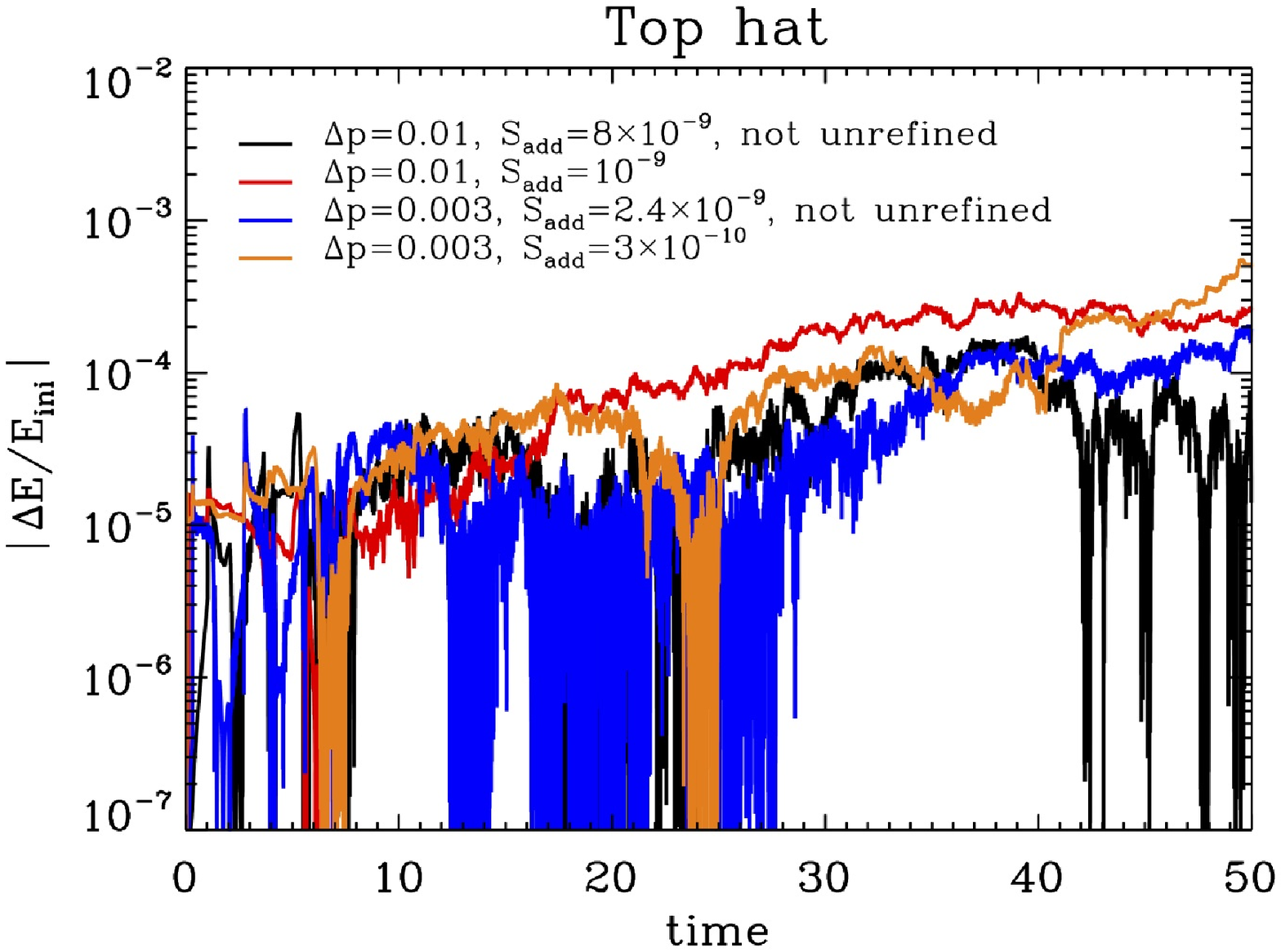,width=8.8cm}
}}
\centerline{\hbox{
\psfig{file=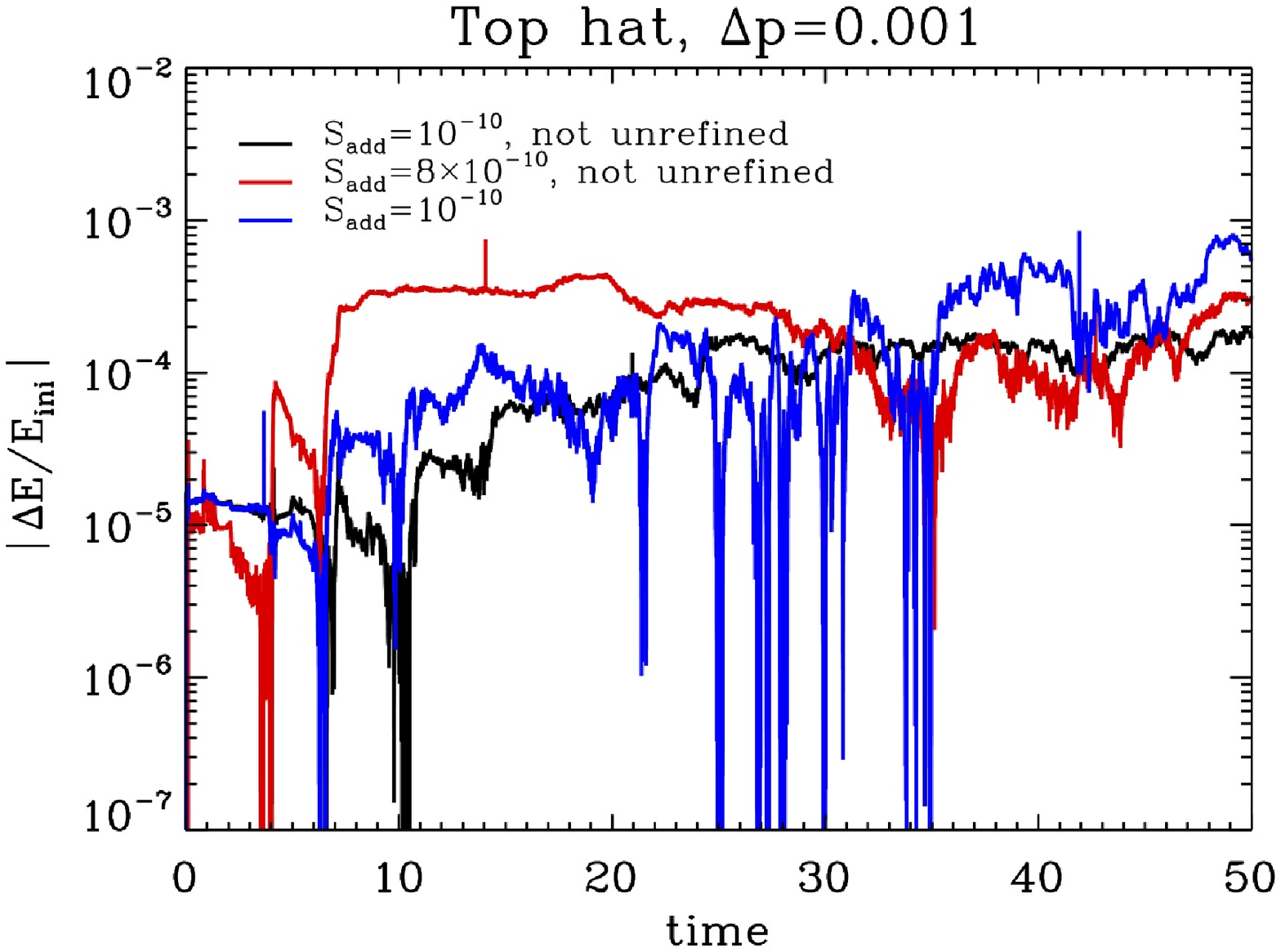,width=8.8cm}
\psfig{file=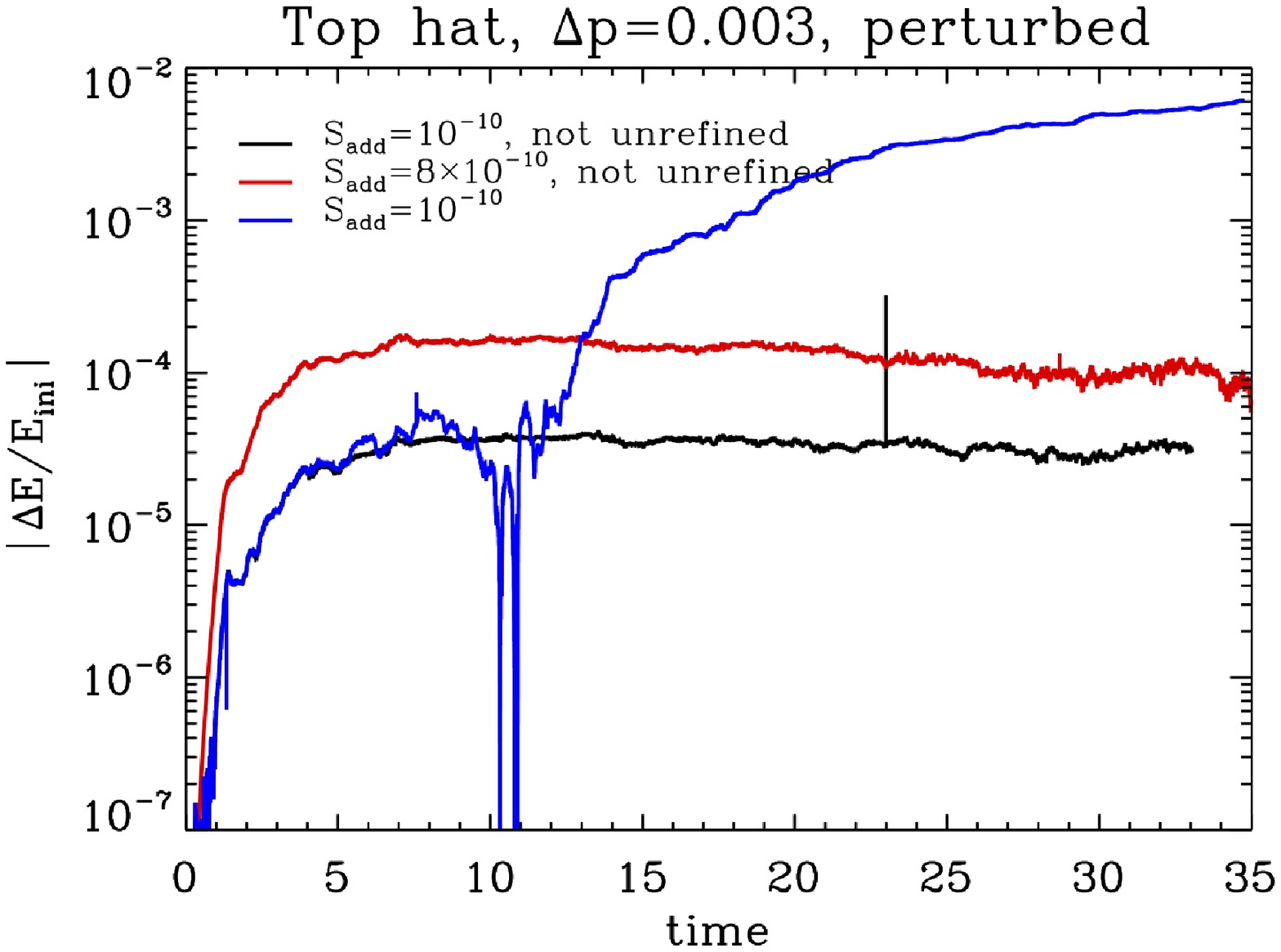,width=8.8cm}
}}
\caption[]{Violation to conservation of total energy in the simulations performed for this article. The ratio $|\Delta E/E_{\rm ini}|=|(E-E_{\rm ini})/E_{\rm ini}|$ is shown as a function of time, where $E_{\rm ini}$ is the total initial energy and $E$ is the total energy measured at time $t$. The top-left and top-right panels correspond respectively to the initially Gaussian distribution function and to the random set of halos, while the four bottom ones treat the single waterbag simulations. The important parameters of the simulations are indicated on each panel. For the warm single waterbags ($\Delta p \geq 1$, middle-left panel), energy conservation is clearly excellent for the parameters of choice (with unrefinement allowed) so we did not bother to investigate other refinement strategies.}
\label{fig:enercons}
\end{figure*}

%==================================================
%
Figure~\ref{fig:enercons} shows the relative deviation from energy conservation as a function of time for all the simulations we performed.\footnote{On the two bottom panels, one can distinguish spikes on some curves, which are mere numerical artifacts due to some defects in the design of the subroutine of our code calculating total energy.}  In practice, energy conservation remains excellent for all the simulations, better than $\sim 2\times 10^{-4}$ in warm cases and than $\sim 10^{-3}$ in colder configurations, except for one of the randomly perturbed waterbag simulations that we discuss below. In fact, with the proper choice of time step and refinement strategy, we can see that energy can be conserved at levels as good as $\sim 5 \times 10^{-5}$ and $\sim 2 \times 10^{-4}$ respectively for the warm and cold configurations studied in this work. 

When it comes to refinement strategy, all our measurements show that it is more optimal to inhibit unrefinement than to allow for it while keeping the number of vertices approximately the same: one just need to examine top-left, top-right, middle right and bottom panels of Fig.~\ref{fig:enercons} to be convinced of this state of facts. Indeed, as foreseen in Appendix~\ref{app:naddrem}, while the numerical noise introduced by unrefinement does not affect too much the dynamical properties of the system at the early stages of the simulations, it becomes increasingly significant with time. This effect is particularly dramatic for the randomly perturbed single waterbag (bottom-right panel of Fig.~\ref{fig:enercons}), which is the most challenging to simulate: energy conservation violation, first nearly as small as for the case without unrefinement up to $t \simeq 10$, suddenly augments dramatically and gets close to the percent level, while it is constrained at about the $10^{-4}$ level without unrefining and a eight times larger value of refinement parameter $S_{\rm add}$. 

Still, this energy conservation analysis shows that unrefinement, although suboptimal, gives, in general, perfectly acceptable results with perhaps the very exception of this perturbed waterbag simulation. Unrefinement might become a must in spherical symmetry or in higher number of dimensions, where the gravitational force variations are much more dramatic. 

As a final note, on top-right panel of Fig.~\ref{fig:enercons}, we test, for the set of random halos, the effect of reducing the time step by a factor two while keeping all the other parameters unchanged. As expected, energy conservation is improved (blue curve with respect to the red one). Note interestingly that the black curve, corresponding to a twice larger time step but a different refinement criterion shows better energy conservation than the blue one.  Clearly, time stepping and refinement affect the dynamics of the system in totally different ways and have to be both carefully checked for.
%==================================================
\section{Circulation along the orientated polygon}
\label{app:circu}
%==================================================
%==================================================
\subsection{Calculation of the force}
\label{sec:orientated}
%==================================================
When performed on the orientated polygon, integral (\ref{eq:circu2}) reads
\begin{equation}
M_{\rm left}(x)=\sum_{i=1}^N \delta f_i\ U(x_{i-1},x_i,v_{i-1},v_i,x).
\label{eq:sumfina}
\end{equation}
In this equation, $(x_i,v_i)$, $i \in [0,\cdots,N]$, are the vertex coordinates. The quantity $\delta f_i$ is given by
\begin{equation}
\delta f_i \equiv f^{\rm right}_i-f^{\rm left}_i,
\end{equation}
where $f^{\rm right}_i$ and $f^{\rm left}_i$ are the values of the phase-space distribution function respectively at the right and at the left of the segment $[i-1,i]$ while following the direction of circulation (increasing $i$). Finally, function $U(x_a,x_b,v_a,v_b,x)$ is defined by
\begin{eqnarray}
U&=& \frac{1}{2}(x_b-x_a)(v_a+v_b) \ {\rm if } \ \max(x_a,x_b) \leq x, \label{eq:U1} \\
&=&\frac{1}{2}(x_b-x)\left[ 2 v_b+\frac{v_a-v_b}{x_a-x_b}(x-x_b)\right]\ {\rm if}\ x_b \leq x < x_a, \nonumber \\
& & \label{eq:U2} \\
&=& \frac{1}{2} (x-x_a) \left[ 2 v_a +\frac{v_b-v_a}{x_b-x_a}(x-x_a) \right] \ {\rm if }\ x_a \leq x < x_b, \nonumber \\
& & \label{eq:U3}
\end{eqnarray}
and $U=0$ otherwise.  Computing function $M_{\rm left}(x)$ quickly for any value of $x$ remains an issue, as the sum (\ref{eq:sumfina}), if performed each time naively, is a slow process. 

The most convenient way to improve the speed of the calculation of function $M_{\rm left}(x)$ is to perform a preliminary sort of the position array,  $\{x_i\}$, $i={0,\cdots,N}$. In our implementation of the algorithm, we did not bother to optimize the sorting procedure that we arbitrarily chose to be Quicksort \citep{1992nrfa.book.....P}.  To perform such an optimization,  one would have to take into account of the fact that, during a time-step, the vertices of the polygon are not expected to move much and thus stay ranked in approximately the right order \citep[see][for an interesting investigation on this matter in the $N$-body case]{2003JCoPh.186..697N}.  However, in our algorithm, sorting is not, in practice, the costliest part of the calculation of the acceleration, for which we give now the final algorithmic details. 

At the end of the sorting procedure, we have a ranked array ${\hat x}_j$ which has been contracted to have strictly ${\hat x}_j < {\hat x}_{j+1}$, along with the hash table  $p(i)$ such that
\begin{equation}
{\hat x}_{p(i)} \equiv x_i.
\label{eq:sortarr}
\end{equation}
In practice, the acceleration is needed for the sampling points of the polygon, so we focus for now on the calculation of $M_{\rm left}({\hat x}_j)$, but we show in Appendix~\ref{app:A} how to access quickly to $M_{\rm left}(x)$ for any value of $x$. 

\begin{figure}
\centerline{\hbox{
\psfig{file=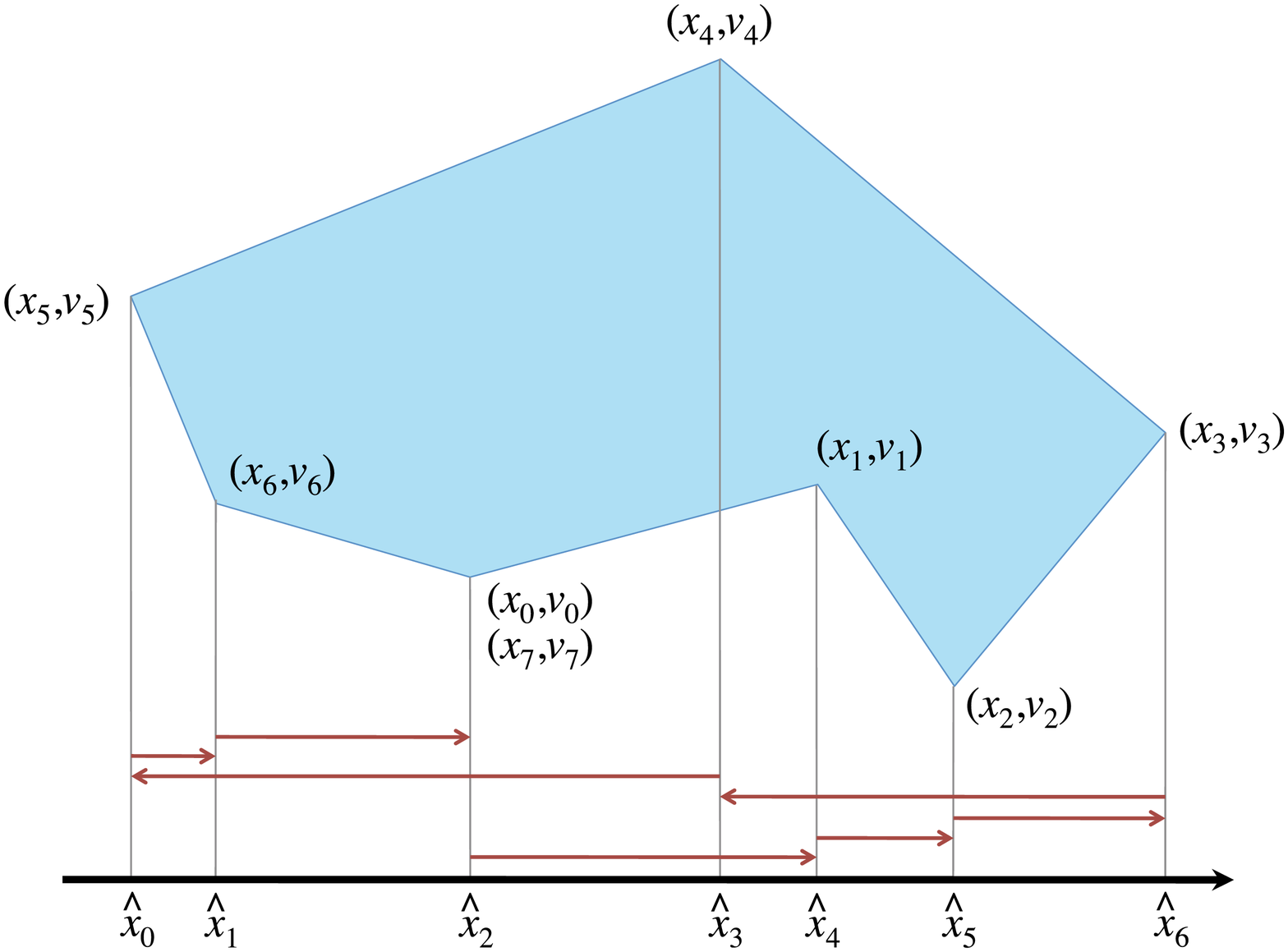,width=8cm}
}}
\caption[]{Algorithm for calculating the acceleration on each point of the orientated polygon.
In this example, we consider a polygon describing the boundary of a unique waterbag. It is
composed of eight points, with $(x_7,v_7)=(x_0,x_0)$ to close the contour. After projection
on the $x$ axis and sorting, we obtain the ranked array ${\hat x}_0,\cdots,{\hat x}_6$. Each segment 
$[x_{i-1},x_{i}]$ contributes to the calculation of the mass in the segments
$[{\hat x}_{j-1},{\hat x}_j]$ such that $\min[p(i-1),p(i)] < j \leq \max[p(i-1),p(i)]$, where $p(i)$ is the
index transform such that ${\hat x}_{p(i)}=x_i$. Here we have
${p}(0,1,2,3,4,5,6,7)=2,4,5,6,3,0,1,2$. For instance, segment $[x_4,x_5]$ contributes
to $[{\hat x}_{j-1},{\hat x}_j]$ with $j=1,2,3$. 

One can understand equations~(\ref{eq:sumfina}), (\ref{eq:U1}), (\ref{eq:U2}) and (\ref{eq:U3}) or
equations~(\ref{eq:U4}), (\ref{eq:summation}) and (\ref{eq:defEj}) by noticing that the area of the polygon is
the difference between two integrals, one on the function $v(x)$ defined by the upper part
of the polygon, the other one on the function defined by the lower part of the polygon.
Furthermore, we see that these integrals are themselves the sum of surfaces of elementary
polygons composed of 4 points, $(x_i,v_i)$, $(x_{i-1},v_{i-1})$, $(x_i,0), (v_i,0)$.
The area of such polygons is given by $|(x_{i}-x_{i-1})(v_i+v_{i-1})/2|$, which with
the correct sign handling reduces to the expressions obtained in the main text.

Note finally that the main part of the calculation of the force is not spent in 
sorting the array ${x_i}$ to obtain the arrays ${\hat x}_j$ and ${\hat v}(i)$.
Indeed, the summation of all the elementary parts of the contour integral
on subsegments $[{\hat x}_{j-1},{\hat x}_j]$ takes, in general, most of the
time spent in the calculation, due to the large overlapping between the segments
$[x_{i-1},x_i]$ as symbolized by the horizontal arrows above the ${\hat x}$ axis.}
\label{fig:projection}
\end{figure}

The next step is now to perform a walk on the orientated polygon. Let us set
\begin{equation}
\delta M_j \equiv M_{\rm left}({\hat x}_j)-M_{\rm left}({\hat x}_{j-1}),
\end{equation}
the amount of mass in segment $[{\hat x}_{j-1},{\hat x}_j]$. Each segment $[x_{i-1},x_{i}]$ of a contour contributes in the sum (\ref{eq:sumfina}) to $\delta M_j$ for $j \in ]\min\{p(i-1),p(i)\},\max\{p(i-1),p(i)\}]$ with the amount
\begin{eqnarray}
{\hat U}_{i,j} & \equiv &  
\left[ v_{i-1} + \frac{v_{i}-v_{i-1}}{x_{i}-x_{i-1}}\left(\frac{{\hat x}_j+{\hat x}_{j-1}}{2}- x_{i-1} \right) \right] \nonumber \\
& \times & ({\hat x}_j-{\hat x}_{j-1})\ \delta f_i \ {\rm sgn}(x_{i}-x_{i-1}) , \label{eq:U4}
\end{eqnarray}
where ${\rm sgn}(x_i-x_{i-1})$ is needed to take into account the direction of circulation with respect to the ordered array ${\hat x}_j$. Hence,
\begin{eqnarray}
\delta M_j &=&\sum_{i \in E_j} {\hat U}_{i,j}, 
\label{eq:summation}\\
E_j &\equiv & \left\{ i \in [1,\cdots,N] / \right. \nonumber \\
       & & \left. \min[p(i-1),p(i)]  <  j  \leq \max[p(i-1),p(i)]  \right\}. \nonumber \\
\label{eq:defEj}
\end{eqnarray}
This gives us the algorithm for computing the acceleration (Fig.~\ref{fig:projection}), given the fact that this sum is performed simultaneously on all the $\delta M_j$'s while scanning all the segments $[x_{i-1},x_{i}]$, $i={1,\cdots,N}$ and simply that $M_{\rm left}(x_i)  = {\hat M}_{\rm left}[{\hat x}_{p(i)}]$ with
${\hat M}_{\rm left}({\hat x}_j) \equiv  \sum_{k \leq j} \delta M_k$. 

Unfortunately, many subsegments $[{\hat x}_{j-1},{\hat x}_j]$ can be contained in the segment $[x_{i-1},x_{i}]$. As a result, the cost of the calculation of $\delta M_j$ can become significant and is in general much more expensive than sorting, because it takes into account the connected nature of the orientated polygon. This cost is reduced to zero for a pure $N$-body approach, where sorting represents the main part of the calculation of the acceleration.

Notice finally that the acceleration $a(x)$ can be expressed as an ensemble of piecewise second order polynomials (see equation \ref{eq:mleftapp} in Appendix~\ref{app:A}). Its is smooth up to its first derivative but its second derivative is discontinuous. The magnitude of the discontinuities is most significant at positions where the border of a waterbag is locally parallel to the velocity axis in phase-space, due to the stepwise nature of the representation of the phase-space distribution function. These discontinuities might trigger long term numerical instabilities and can be reduced only by augmenting the waterbag sampling. This is well illustrated by Fig.~\ref{fig:gaussianA}.  %==================================================
\subsection{Calculation of various profiles}
\label{app:A}
%==================================================
The algorithm just discussed above can in fact be generalized to compute analytically any quantity of the form
\begin{eqnarray}
g(x) &=& \int h(x,v')\ f(x,v')\ {\rm d}v' \\
 &=& \sum_{k=1}^{N_{\rm patch}} f_k \int_{(x,v') \in P_k} h(x,v')\ {\rm d}v'
\end{eqnarray}
as well as any integral 
\begin{equation}
G(x)=\int_{x' \leq x} g(x')\ {\rm d}x',\quad {\cal G}(x)=\int_{x' \leq x} G(x')\ {\rm d}x'
\end{equation}
for any bivariate polynomial 
\begin{eqnarray}
h(x,v) &=& \sum_{l=0}^{n_x}\sum_{m=0}^{n_v}  \alpha_{l,m} h_{l,m}(x,v), \\
h_{l,m}(x,v) &= &x^l v^m.
\end{eqnarray}
First consider 
\begin{equation}
g_{l,m}(x)=\int h_{l,m}(x,v')\ f(x,v')\ {\rm d}v'.
\end{equation}
Using the same sorted array ${\hat x}_i$ as in \S~\ref{sec:orientated} (equation~\ref{eq:sortarr}), 
one can realize that in interval $[{\hat x}_{j-1},{\hat x}_j]$, function $g_{l,m}(x)$ reads
\begin{equation}
g_{l,m}(x) =\frac{x^l}{m+1}  \sum_{q=0}^{m+1} \left(\begin{array}{c} m+1 \\ q \end{array}\right) 
\beta_{m,j,q} \ (x- {\hat x}_{j-1})^q, \label{eq:glm} 
\end{equation}
with
\begin{equation}
\left(\begin{array}{c} m \\ q \end{array}\right) \equiv \frac{m!}{q!\ (m-q)!}
\end{equation}
and where
\begin{eqnarray}
\beta_{m,j,q} &= &\sum_{i \in E_j} U_{i,m,j,q},  \\
U_{i,m,j,q} &=& \left[ v_{i-1} +\frac{v_{i}-v_{i-1}}{x_{i}-x_{i-1}}({\hat x}_{j-1}-x_{i-1}) \right]^{m+1-q} \nonumber \\
& \times & \left[ \frac{v_i-v_{i-1}}{x_i-x_{i-1}} \right]^q\delta f_i\ {\rm sgn}(x_{i}-x_{i-1}),
\end{eqnarray}
$E_j$ being defined by equation (\ref{eq:defEj}).
The way $\beta_{m,j,q}$ is calculated is thus the same as described in \S~\ref{sec:orientated} for $\delta M_j$.

It then becomes a simple algebraic procedure to compute an integral over variable $x$ in equation (\ref{eq:glm}) or even two or more successive integrals. We just perform the explicit calculation  for $l=0$ that we only need here. Similarly as for the cumulative mass function we can write
\begin{eqnarray}
\delta G_{j,m}(x) & \equiv & \int_{{\hat x}_{j-1}}^{x} g_{0,l}(x')\ {\rm d}x', \\
                & = & \frac{1}{m+1} \sum_{q=0}^{m+1} \left(\begin{array}{c} m+1 \\ q \end{array}\right) \frac{ \beta_{m,j,q}}{q+1}  \nonumber \\
                & & \times  (x-{\hat x}_{j-1})^{q+1},
\end{eqnarray}
so for $x \in [{\hat x}_{j-1},{\hat x}_j]$,
\begin{eqnarray}
G_{m}(x) & \equiv & \int_{x' \leq x} g_{0,m}(x')\ {\rm d}x', \\
              & = & \sum_{j' \leq j-1} \delta G_{j',m}({\hat x}_{j'})+\delta G_{j,m}(x),
\end{eqnarray}
and, setting
\begin{eqnarray}
\delta {\cal G}_{j,m}(x) & \equiv & \int_{{\hat x}_{j-1}}^{x} G_{m}(x')\ {\rm d}x', \\
   &=& G_m({\hat x}_{j-1}) (x-{\hat x}_{j-1}) \nonumber \\
   & + & \frac{1}{m+1} \sum_{q=0}^{m+1} \left(\begin{array}{c} m+1 \\ q \end{array}\right) \frac{ \beta_{m,j,q} }{(q+1)(q+2)} \nonumber \\
& & \times (x-{\hat x}_{j-1})^{q+2},
\end{eqnarray}
\begin{eqnarray}
{\cal G}_m(x) & \equiv & \int_{x' \leq x} G_m(x)\ {\rm d}x', \\
 & = & \sum_{j' \leq j-1} \delta {\cal G}_{j',m}({\hat x}_{j'})+\delta {\cal G}_{j,m}(x).
\end{eqnarray}
These preliminary calculations set up the framework for computing various quantities as functions of $x$:
\begin{itemize}
\item {\it The projected density profile} reads
\begin{equation}
\rho(x)  \equiv  \int_{v'} f(x,v')\ {\rm d}v',
\end{equation}
hence $h(x,v)=1$ and $\rho=g_{0,0}$:
\begin{equation}
\rho(x) =  \beta_{0,j,0}+\beta_{0,j,1}\ \delta x_{j},
\end{equation}
with
\begin{equation}
\delta x_{j} \equiv x- {\hat x}_{j-1}.
\end{equation}
\item {\it The mass profile}, $M_{\rm left}(x)=G_0(x)$, reads:
\begin{eqnarray}
M_{\rm left}(x)  & = & \sum_{j' \leq j-1} \delta M_{j'}+\beta_{0,j,0}\ \delta x_j + \frac{1}{2}\beta_{0,j,1}\ (\delta x_{j})^2, \nonumber \\
& & \label{eq:mleftapp}
\end{eqnarray}
with
\begin{eqnarray}
\delta M_j  & = & \beta_{0,j,0}({\hat x}_j-{\hat x}_{j-1}) + \frac{1}{2} \beta_{0,j,1} ({\hat x}_j-{\hat x}_{j-1})^2. \nonumber \\
\label{eq:deltamja}
\end{eqnarray}
Note, hence, that the total mass reads
\begin{equation}
M_{\rm tot}=\sum_j \delta M_j.
\label{eq:totalm}
\end{equation}
\item {\it The gravitational potential} can be defined as follows
\begin{equation}
\phi(x) =\int |x-x'|\ \rho(x')\ {\rm d}x', 
\label{eq:phiexp}
\end{equation}
which can be conveniently rewritten
\begin{eqnarray}
\phi(x)&= &M_{\rm tot}(x_{\rm max}-x) - \int_{x_{\rm min}}^{x_{\rm max}}M_{\rm left}(x')\ {\rm d}x' \nonumber \\
           &+& 2 \int_{x_{\rm min}}^x M_{\rm left}(x')\ {\rm d}x',
\label{eq:phidex}
\end{eqnarray}
where $[x_{\rm min},x_{\rm max}]$ represents the extension of the system in coordinate space. 
Hence, we have 
\begin{eqnarray}
\phi(x) &=& M_{\rm tot}(x_{\rm max}-x) +{\cal G}_0(x_{\rm min})-{\cal G}_0(x_{\rm max}) \nonumber \\
   & & +2[{\cal G}_0(x)-{\cal G}_0(x_{\rm min})],
\end{eqnarray} 
and finally
\begin{eqnarray}
\phi(x) & = & \sum_{j' \leq j-1} \delta \phi_{j'}  + 2  M_{\rm left}({\hat x}_{j-1}) \ \delta x_j  \nonumber \\
& + & \beta_{0,j,0}\ (\delta x_j)^2 +\frac{1}{3} \beta_{0,j,1}\ (\delta x_j)^3  \nonumber \\ 
& - & \frac{1}{2} \sum_{j'} \delta \phi_{j'}+M_{\rm tot}(x_{\rm max}-x),
\label{eq:phical}
\end{eqnarray}
with
\begin{eqnarray}
\delta \phi_j & = & 2 M_{\rm left}({\hat x}_{j-1})( {\hat x}_{j}-{\hat x}_{j-1})  \nonumber \\
                     &+  & \beta_{0,j,0}\ ( {\hat x}_{j}-{\hat x}_{j-1})^2+\frac{1}{3} \beta_{0,j,1} ( {\hat x}_{j}-{\hat x}_{j-1})^3. \nonumber \\
                     \label{eq:deltaphij}
\end{eqnarray}
\item {\em The bulk velocity profile} reads
\begin{eqnarray}
{\bar v}(x) &=& \frac{1}{\rho(x)} \int v' f(x,v')\ {\rm d}v', \\
                 &= &\frac{g_{0,1}(x)}{\rho(x)}.
\end{eqnarray}
Hence, 
\begin{eqnarray}
{\bar v}(x) & = &\frac{1}{\rho(x)}\left[ \frac{1}{2} \beta_{1,j,0} +\beta_{1,j,1}\ \delta x_j+\frac{1}{2} \beta_{1,j,2}\ (\delta x_j)^2 \right]. \nonumber \\
\end{eqnarray}
\item {\em The local velocity dispersion} reads
\begin{eqnarray}
\sigma^2_v(x) &=& \frac{1}{\rho(x)} \int v'^2 f(x,v')\ {\rm d}v'- {\bar v}^2(x) \\
                       &=& \frac{g_{0,2}(x)}{\rho(x)}-{\bar v}^2(x),
\end{eqnarray}
so
\begin{eqnarray}
\sigma^2_v(x) & = & \frac{1}{\rho(x)} \left[ \frac{1}{3} \beta_{2,j,0}+\beta_{2,j,1}\ \delta x_j +\beta_{2,j,2}\ (\delta x_j)^2 \right. \nonumber \\
& & \left. +\frac{1}{3}\beta_{2,j,3}\ (\delta x_j)^3\right]-{\bar v}^2(x).
\end{eqnarray}
\end{itemize}
%=======================================================
\subsection{Global quantities: waterbag area, center of mass, total kinetic and potential energy}
\label{sec:kinepot}
%=======================================================
Calculation of integrals of the form $\int x^l v^m f(x,v)\ {\rm d}x\ {\rm d}v$ is simpler than for profiles, because it can be reduced to a simple circulation over the orientated polygon without having to perform sorting. Therefore, computing the total area of each waterbag, the center of mass coordinates and the total kinetic energy do not represent any difficulty:
\begin{itemize}
\item {\em The area of waterbag $k$}, $V_k$, reads, with the notations of introduction,
\begin{equation}
V_k = \oint_{\partial P_k} v(s){\rm d}x(s).
\end{equation}
This integral that can be obtained from circulation on the part of the orientated polygon that coincides with the waterbag border:
\begin{eqnarray}
V_k & = & \frac{1}{2}\sum_{i, I^{\rm right}_i=k} (x_i-x_{i-1}) (v_i+v_{i-1}) \nonumber \\
     & - & \frac{1}{2}\sum_{i,I^{\rm left}_i=k}(x_i-x_{i-1}) (v_i+v_{i-1}), \label{eq:vkint}
\end{eqnarray}
where the integers $I^{\rm left}_i$ and $I^{\rm right}_i$ identify the waterbags respectively at the left and the right of segment $[i-1,i]$ of the orientated polygon.  In practice, the calculation is of course performed simultaneously for all $k$ by circulating once on the orientated polygon. 
\item {\em The center of mass}, $(x_{\rm M},v_{\rm M})$, reads:
\begin{eqnarray}
x_{\rm M} & \equiv & \frac{1}{M_{\rm tot}} \int x f(x,v)\ {\rm d}x\ {\rm d}v, \\
             & = & \frac{1}{6 M_{\rm tot}} \sum_i \delta f_i (v_{i-1}-v_{i}) \nonumber \\
             &    & \quad \quad \quad \quad \quad \quad \times (x_i^2+x_i x_{i-1} + x_{i-1}^2), \label{eq:sumxm} \\
v_{\rm M} & \equiv & \frac{1}{M_{\rm tot}} \int v f(x,v)\ {\rm d}x\ {\rm d}v, \\
             & = & \frac{1}{6 M_{\rm tot}} \sum_i \delta f_i  (x_i-x_{i-1}) \nonumber \\
             & & \quad \quad \quad \quad \quad \quad \times (v_i^2+v_i v_{i-1}+v_{i-1}^2). 
\label{eq:sumvm}
\end{eqnarray}
Note that we used the interesting property that, since waterbags are bounded by closed contours, $\sum_i g(x_i) h(v_i) \delta f_i=\sum_i g(x_{i-1}) h(v_{i-1}) \delta f_i$, for any function $g$ and $h$.  
\item {\em The total kinetic energy}, $E_{\rm k}$, reads: 
\begin{eqnarray}
E_{\rm k} & = & \frac{1}{2} \int v^2 f(x,v)\ {\rm d}x\ {\rm d}v, \label{eq:kinect} \\
            & = &\frac{1}{24} \sum_i \delta f_i (x_i-x_{i-1}) \nonumber \\
            &  &  \quad \quad \quad \times (v_i^3+v_i^2 v_{i-1}+v_i v_{i-1}^2+v_{i-1}^3).
\end{eqnarray}
\item {\em  The total potential energy}, on the other hand is given by
\begin{equation}
E_{\rm p} = \frac{1}{2} \int \rho(x) \phi(x) {\rm d}x. \label{eq:energiepot}
\end{equation}
This integral seems difficult to compute without sorting the orientated polygon vertices positions. We find convenient to rewrite it as follows:
\begin{equation}
E_{\rm p}=M_{\rm tot} \int  M_{\rm left}(x)\ {\rm d}x - \int M_{\rm left}^2(x)\ {\rm d}x,
\end{equation}
where it has to be reminded that the integral must be performed over the minimum possible interval $[x_{\rm min},x_{\rm max}]$ containing the regions where $\rho(x) > 0$. Then,
\begin{eqnarray}
\int M_{\rm left}(x)\ {\rm d}x & = & \frac{1}{2}\sum_j \delta \phi_j, \\
\int M^2_{\rm left}(x)\ {\rm d}x & = & \sum_j \delta {\tilde M}^2_{{\rm left},j},
\end{eqnarray}
with $\delta \phi_j$ given by equation (\ref{eq:deltaphij}) and
\begin{eqnarray}
\delta {\tilde M}^2_{{\rm left},j} & \equiv & [M_{\rm left}({\hat x}_{j-1})]^2  {\hat x}_{j-1,j} \nonumber \\
 & & + M_{\rm left}({\hat x}_{j-1}) \left[ \beta_{0,j,0} {\hat x}_{j-1,j}^2  +\frac{1}{3} \beta_{0,j,1} {\hat x}_{j-1,j}^3 \right] \nonumber \\
& & + \frac{1}{3}\beta_{0,j,0}^2 {\hat x}_{j-1,j}^3+\frac{1}{4} \beta_{0,j,0}\beta_{0,j,1}  {\hat x}_{j-1,j}^4 \nonumber \\
& & +\frac{1}{20} \beta_{0,j,1}^2  {\hat x}_{j-1,j}^5,
\end{eqnarray}
and 
\begin{equation}
{\hat x}_{j-1,j}\equiv {\hat x}_j-{\hat x}_{j-1}.
\end{equation}
\end{itemize}
%=======================================================
\section{Position-velocity to Action-Angle transformation}
\label{app:actionangles}
%=======================================================
An interesting way to analyze the simulations is to use Action-Angle canonical coordinates \citep[see, e.g.][]{BT2008}. For a point of phase-space coordinates $(x,v)$, the Action $\Omega$ is given by $\Omega=\Omega[E(x,v)]$, with
\begin{equation}
\Omega(E) \equiv \frac{1}{2\pi}J(E),
\end{equation}
where $J(E)$ is the area inside a contour of constant energy 
\begin{equation}
J(E)\equiv \oint_{E(x,v)=E} v(s)\ {\rm d}x(s)
\label{eq:adiabinv}
\end{equation}
and $E(x,v) \equiv v^2/2 + \phi(x)$ is the specific energy at point $(x,v)$ assuming the fixed potential $\phi(x)$. 

The Angle $\Theta$, chosen by convention to vary in $[-\pi,\pi[$, is given by
\begin{equation}
\Theta(x,v)=2\pi\frac{\tau(x,v)}{T[E(x,v)]}-\pi,
\end{equation}
where 
\begin{equation}
\tau(x,v) = \oint_{s \leq s(x,v), E(x',v')=E(x,v)} \frac{{\rm d}x'(s)}{v'(s)}
\label{eq:timespent}
\end{equation}
is the time taken by a point initially located at coordinates $(x_0 >0,0)$ with
\begin{equation}
E(x_0,0)=\phi(x_0)\equiv E(x,v) 
\end{equation}
to reach position $(x,v)$ in the {\em fixed} potential $\phi(x)$, while $T(E)$ represents the total time needed to follow an entire orbit in this stationary system. 
 
To estimate in a fast way the line integrals (\ref{eq:adiabinv}) and (\ref{eq:timespent}), we define a polar-energy system of coordinates $(\psi, E)$ of which we pixelate the superior half space. Converting the origin of this coordinate system in phase-space coordinates requires to compute a position (which might not be unique) $x_{\rm G}(t)$ where the potential equates its minimum. To find $x_{\rm G}$, we use a slightly modified of the dichotomous subroutine {\tt RTBIS} of the Numerical Recipes \citep{1992nrfa.book.....P} to solve the equation
\begin{equation}
a(x_{\rm G})=-\frac{\partial \phi}{\partial x_{\rm G}} \equiv 0.
\end{equation}
If the force is not strictly monotonous, the ensemble $S_{\rm G}$ of solutions for $x_{\rm G}$ is an interval. However, the convex nature of the contours $E(x,v)$=constant implies that the intersection of any straight line in phase-space passing through any $(x_{\rm G} \in S_{\rm G}, v_{\rm G}=0)$ and the contour $E(x,v)$=constant is always a set of two points. That means that after converting $(x,v)$ into polar coordinates
\begin{eqnarray}
x & = & {\cal R} \cos \psi+x_{\rm G},\\
v & = & {\cal R} \sin \psi,
\end{eqnarray}
the equation $E(x,v)=H$, $H > \phi(x_{\rm G})$, has a unique solution ${\cal R}$ for each value of $\psi$ in $]-\pi,\pi]$, which can be found again with {\tt RTBIS}. This allows us to set up in an unambiguous way a local system of coordinates $(\psi,E)$ that we pixelate in the half-space $\psi \in [0,\pi]$, $E \in [E_{\rm min}=\phi(x_{\rm G}),E_{\rm max}=\max_i \frac{1}{2} v_i^2+\phi(x_i)]$. Defining 
\begin{eqnarray}
  \psi_{\ell} &=&\delta \psi \ \ell, \quad \ell \in \{0,\cdots,n_{\psi}\} \\
  E_m &=& E_{\rm min}+\delta E \ m, \quad m \in \{0, \cdots, n_E \},
\end{eqnarray}
with
\begin{eqnarray}
\delta \psi & \equiv & \frac{\pi}{n_{\psi}}, \\
\delta E & \equiv & \frac{E_{\rm max}-E_{\rm min}}{n_E},\label{eq:deltaedef}
\end{eqnarray}
we find ${\cal R}\equiv {\cal R}_{\ell,m}$ for each pair $(\ell,m)$, hence $x_{\ell,m}={\cal R}_{\ell,m} \cos \psi_{\ell} +x_{\rm G}$ and $v_{\ell,m}={\cal R}_{\ell,m} \sin \psi_\ell$. In practice, we took $(n_E+1,n_\psi)=(1024,1024)$ to generate Fig.~\ref{fig:action_angle_gaussian}. On the other hand, Fig.~\ref{fig:tophatae} required $(n_E+1,n_\psi)=(10^5,1024)$ to be able probe small values of the Action without introducing distortions. 
Then,
\begin{eqnarray}
{\tilde \Omega}_m & \equiv  & \Omega(E_m) \simeq \frac{1}{2\pi} \sum_{\ell \geq 1} (x_{\ell-1,m}-x_{\ell,m}) \nonumber \\
                            &     & \quad \quad \quad \quad \quad \quad \quad \times (v_{\ell-1,m}+v_{\ell,m}), \label{eq:mapping1}\\
{\tilde \eta}_{\ell,m} & \equiv & \eta(x_{\ell,m},v_{\ell,m}) \simeq 2\sum_{\ell' > \ell}  \frac{x_{\ell'-1,m}-x_{\ell',m}}{v_{\ell'-1,m}+v_{\ell',m}},\label{eq:mapping2}\\
{\tilde T}_m & \equiv & T(E_m) \simeq 2 \eta_{0,m},
\end{eqnarray}
where 
\begin{equation}
\eta(x,v) \equiv \tau(x,v)-T[E(x,v)]/2. 
\label{eq:defeta}
\end{equation}
The quantities $\eta_{\ell,m}$ can be computed quickly by using the trivial recursion $\eta_{\ell-1,m}=\eta_{\ell,m}+2 (x_{\ell-1,m}-x_{\ell,m})/(v_{\ell-1,m}+v_{\ell,m})$. 
To avoid the singularity occurring for $m=0$, where $E_0=E_{\rm min}$ we just temporarily set $E_0=E_{\rm min}+0.1\ \delta E$ to approximate the asymptotic limit $E \rightarrow E_{\rm min}$ in equations (\ref{eq:mapping1}) and (\ref{eq:mapping2}).

We are now ready to convert each coordinate $(x,v)$ of the vertices of the orientated polygon to Action-Angle coordinates $(\Omega,\Theta)$ by simple bilinear interpolation in $(\psi,E)$ space. Setting $E=v^2/2 +\phi(x)$, $\psi={\rm arccos}[(x-x_{\rm G})/\sqrt{(x-x_{\rm G})^2+v^2}]$ and
\begin{eqnarray}
\ell &= & {\rm int}(\psi/\delta \psi), \\
m& = &{\rm int}[(E-E_{\rm min})/\delta E], \\
w_{\psi} & = & \psi/\delta \psi-\ell, \\
w_{E} &= & (E-E_{\rm min})/\delta E-m,
\end{eqnarray}
we obtain
\begin{eqnarray}
\Omega &=&{\tilde\Omega}_m (1-w_E)+{\tilde \Omega}_{m+1} w_E, \\
\Theta &=& 2\pi \frac{\eta}{T},
\end{eqnarray}
with
\begin{eqnarray}
\eta &=& {\rm sgn}(v) \left[ {\tilde \eta}_{\ell,m} (1-w_E)(1-w_\psi) \right. \nonumber \\
       &  & +{\tilde \eta}_{\ell+1,m} (1-w_E) w_\psi +{\tilde \eta}_{\ell,m+1} w_E (1-w_\psi) \nonumber \\
       & & \left.+{\tilde \eta}_{\ell+1,m+1} w_E w_\psi \right], \\
T &= & {\tilde T}_m (1-w_E)+{\tilde T}_{m+1} w_E.
\end{eqnarray}

When we project the orientated polygon in Action-Angle space, we assume that each of its segments $[i,i+1]$ remains, in first approximation a straight line. Some special care has however to be taken in the vicinity of the boundaries $\Theta=-\pi$ and $\Theta=\pi$, which correspond, in phase-space, to the half-line $(x \geq 0, v=0)$. 
The list of segments of the polygon with 
\begin{eqnarray}
\Theta_i \Theta_{i+1}  &\leq & 0,\\
|\Theta_i-\Theta_{i+1}| & > & \pi,
\end{eqnarray}
are selected. Then the point $I$ of intersection of each of these segments $[i,i+1]$ with the line $v=0$ is found. In phase-space, its projected position is given by
\begin{eqnarray}
x_I=\frac{v_{i+1} x_i-v_i x_{i+1}}{v_{i+1}-v_{i}},
\end{eqnarray}
which can be converted in an Action $\Omega_I$ with the interpolation procedure given above. Then the segment $[i,i+1]$ is replaced in Action-Angle space with 5 segments $[i,A]$, $[A,B]$, $[B,C]$, $[C,D]$ and $[D,i+1]$ with
\begin{eqnarray}
A &=& [s \pi,\Omega_I], \label{eq:eqA} \\
B &=& [s \pi,0], \label{eq:eqB} \\
C &=& [-s \pi,0], \label{eq:eqC} \\
D &=& [-s \pi, \Omega_I], \label{eq:eqD}
\end{eqnarray}
which have all the attributes of segment $[i,i+1]$, in particular with respect to $f^{\rm left}$ and $f^{\rm right}$. In this last set of equations, $s=-1$ if $v_{i} \leq  0$ and $v_{i+1} >0$ or $v_{i} < 0$ and $v_{i+1} \geq 0$, and $s=1$ in the opposite case. With this procedure, it is possible to circulate along the orientated polygon or to draw the waterbags in Action-Angle space using the parity algorithm described in Appendix~\ref{app:drawwat}. 
%=======================================================
\section{Phase-space energy distribution function}
\label{app:fofemet}
%=======================================================
To compute the phase-space energy distribution function, $f_E(E)$ (equation \ref{eq:fofedef}), we need to estimate the amount of mass in the interval $[E,E+\delta E]$,
\begin{equation}
\int_{E(x,v) \in [E,E+\delta E]} f(x,v)\ {\rm d}x\ {\rm d}v=M_E(E+\delta E)-M_E(E),
\end{equation}
where $M_E(E)$ is the mass enclosed inside the contour $E(x,v)=E$, as well as the surface between contours of constant energy $E$ ad $E+\delta E$,
\begin{equation}
\int_{E(x,v) \in [ E,E+\delta E ]}\ {\rm d}x\ {\rm d}v=J(E+\delta E)-J(E),
\label{eq:denomfofE}
\end{equation}  
where $J(E)$ is given by equation (\ref{eq:adiabinv}). 

To compute $J(E)$, we use the method developed in \S~\ref{app:actionangles}, with $\delta E$ given by equation (\ref{eq:deltaedef}) and $n_E=1023$.  On the other hand the calculation of $M_E(E)$ requires a special care. To deal with this issue, we employ a similar technique to that used to estimate $M_{\rm left}(x)$. To achieve this, we use a new system of coordinates, $(E,\eta)$, where $E=v^2/2+\phi(x)$ is easily obtained by using equation (\ref{eq:phical}) to compute $\phi(x)$, while $\eta(x,v)$ is given by equation (\ref{eq:defeta}) and is computed numerically for any $(x,v)$ exactly as in \S~\ref{app:actionangles}. The transformation from $(x,v)$ to $(E,\eta)$ has, like the transformation to Action-Angle space, the property of conserving phase-space volume. In this new system of coordinates, where we assume that the edges of the orientated polygon have stayed approximately straight, we can easily compute function $M_E(E)$ with exactly the same algorithm as for $M_{\rm left}(x)$. The only additional difficulty is to take into account periodic boundaries, i.e. to make appropriate modifications when the waterbag border reaches the edge of the computational domain. The method is analogous to that described at the end of Appendix~\ref{app:actionangles} (equations~\ref{eq:eqA} to \ref{eq:eqD}): it consists in adding a new piece to the orientated polygon, starting from this intersection and running along the edge of the computational domain until reaching the intersection at the other side. The frontier of the computational domain is indeed composed of the two curves of coordinates $[E,T(E)/2]$ and $[E,-T(E)/2]$. These two curves are approximated with a piece of polygon, $[{\hat E}_j,\pm T({\hat E}_j)]$, where ${\hat E}$ is the reordered array of increasing values of $E_i=v_i^2/2+\phi(x_i)$. 
%=======================================================
\section{Drawing waterbags}
\label{app:drawwat}
%=======================================================
 Drawing a polygon on a pixelated image is a very standard but in fact non trivial procedure. If we were using pixelated data for analysis purposes, it would be necessary to enforce mass conservation by computing the surface of the actual intersection of each waterbag with each target pixel. But since we perform all the analyses by circulating directly on the orientated polygon, images are generated here only for examination purpose. So instead of computing the actual intersection of the pixels with the waterbags, which is possible but rather involved, we just test if the center of each pixel is inside a waterbag. To do this, we use the classical parity algorithm, which allows us to avoid taking into account of the orientated nature of the polygon, to circumvent otherwise strong artifacts on the image if there is shell-crossing in phase-space. Shell-crossing should, in theory, not happen, but it can take place in small regions of the system, particularly at late times and where mixing is particularly strong. Affecting to each pixel only the waterbag(s)\footnote{If there is shell crossing in phase-space, several waterbags can contain the center of the pixel: this is an artifact.} containing their center can however induces some very strong aliasing effects, particularly when waterbags are thin or small compared to the pixel size: for instance, a pixel can be found to be empty while it is in fact crossed by many very narrow waterbags. To reduce the aliasing effects, we subsequently redraw the waterbag borders with a simplified pixelation technique described below, and our choice for the attribute of the pixel is then the maximum value of $f$ over all the waterbags that have an edge affected to it. We now detail how these two steps, firstly, finding the waterbag where the center of each pixel lie, then, secondly, reducing aliasing by drawing the edges of the waterbags on the picture, are performed. 

\begin{itemize}
\item{\em Finding the waterbags where the center of each pixel lies.}  Each pixel $(\ell,m)$, $\ell\in\{1,\cdots,n_x\}$, $m\in\{1,\cdots,n_v\}$ is assumed to have its center at position $(\ell+1/2,m+1/2)$ in pixel frame units, hence the pixel frame coverage corresponds to intervals  $[1,n_x+1]$ and $[1,n_v+1]$. In all the pictures of phase-space displayed in this article, the resolution was set to
\begin{equation}
n_x=n_v=1024.
\end{equation}
The image is chosen to cover the range $[x_{\rm min,g},x_{\rm max,g}]$ and $[v_{\rm min,g},v_{\rm max,g}]$ in phase-space, that might not entirely contain the computing volume. Positions $(x_i,v_i)$ in phase-space of the vertices of the polygon are converted in pixel frame units which are implicit from now on:
\begin{eqnarray}
x_i & \rightarrow & \frac{x_i-x_{\rm min,g}}{x_{\rm max,g}-x_{\rm min,g}}n_x+1, \label{eq:gridx}\\
v_i & \rightarrow & \frac{v_i-v_{\rm min,g}}{v_{\rm max,g}-v_{\rm min,g}}n_v+1. \label{eq:gridv}
\end{eqnarray}
In order to draw the waterbags, we create for each line $m$ of the image lying on the coordinate $v=m+1/2$ a list of candidate segments, $[i-1,i]$ of the polygon intersecting it, and store for each of them {\em twice} the coordinate ${\tilde x}_p=(m+1/2-v_{i-1}) (x_i-x_{i-1})/(v_i-v_{i-1})+x_{i-1}$, ${\tilde x}_{\rm p+1}={\tilde x}_p$, where $p$ is an incremental count that at the end gives the number of waterbag borders intersecting with the line. We have to store the information twice because the polygon accounts for the border of two adjacent waterbags (one of them can be the ``null'' infinite region where $f=0$). In addition, to be able to proceed further, we also store two integers, ${\tilde I}_p=I^{\rm left}_i$ and ${\tilde I}_{p+1}=I^{\rm right}_i$ corresponding to the waterbags identity (zero for the ``null'' waterbag), as well as ${\tilde f}_{p}=f^{\rm left}_i$ and ${\tilde f}_{p+1}=f^{\rm right}_i$, to be of course able to affect the right value to the pixel once the waterbag containing its center will be identified. The structure is then reordered in terms of increasing ${\tilde x}_p$. Additionally, a preparation for the parity algorithm is performed. A flag ${\rm Left}(p)$ is assigned to each value of $p$ to decide whether the waterbag is at the left of the intersection [${\rm Left}(p)=1$] or at the right of it [${\rm Left}(p)=0$]. For a given waterbag, which is identified with ${\tilde I}_p$, the flag ${\rm Left}$ alternates between 0 and 1 with increasing $p$'s corresponding to the same ${\tilde I}_p$, starting from 0, except for the ``null'' waterbag, where it is needed to start from 1. 

Then we can, for each line of pixels $m$ of the image, apply the parity algorithm to the waterbags crossing the line. Starting from the first pixel, $\ell=1$ of the line, with coordinate $\chi_{\ell}=\ell+1/2$ and the first index $p=p_{\rm reference}=1$ of the intersection of coordinate ${\tilde x}_p$, and assuming a reference value for the distribution function, $f_{\rm reference}=0$, we iterate as follows:
\begin{description}
\item[(i)] We check that ${\tilde x}_{p_{\rm reference}} < \chi_{\ell}$ to proceed further. Indeed, if ${\tilde x}_{p_{\rm reference}} \geq \chi_\ell$, we consider the center of the pixel to still belong to the same waterbag as defined previously, with $f=f_{\rm reference}$, and we proceed to next pixel, $\ell \rightarrow \ell+1$, until ${\tilde x}_{p_{\rm reference}} < \chi_{\ell}$. 
\item[(ii)] When ${\tilde x}_{p_{\rm reference}} < \chi_{\ell}$, we now increase the index $p$ until both ${\tilde x}_p \geq \chi_\ell$ and ${\rm Left}(p)=1$, save ${\tilde I}_{\rm candidate}\equiv {\tilde I}_p$ and set a new value of $p_{\rm reference}$ equal to the present $p$. 
\item[(iii)] The previous step gives our candidate waterbag possibly containing the pixel: we now {\em decrease} the index $p$ until ${\tilde I}_p={\tilde I}_{\rm candidate}$. Then, for this value of $p$, if ${\tilde x}_p < \chi_\ell$ the value affected to the pixel $(\ell,m)$ is ${\tilde f}_p$, which becomes our new value of reference, $f_{\rm reference}={\tilde f}_p$, otherwise, $f_{\rm reference}$ is unchanged and affected to the pixel $(\ell,m)$. 
\item[(iv)] The process is started again from (i) until all the pixels of the line have been examined, $\ell=n_x$, or when all the intersections have been scanned: in the last case, all the pixels that fail the test (i) are obviously empty, with $f=0$.
\end{description}
It is easy to see that this algorithm works even if the image does not entirely contain the computing domain.

\item{\em Reduction of aliasing effects.} From the previous step, we have computed an image with sampled values of the distribution function, $f_{\ell,m}$. We now redraw the waterbag borders inside the pixelated image as follows. Consider a segment $[i-1,i]$ of the orientated polygon in the grid coordinate frame (equations~\ref{eq:gridx} and \ref{eq:gridv}) and set
\begin{eqnarray}
\delta x & = & x_i-x_{i-1}, \\
\delta v & = & v_i-v_{i-1}.
\end{eqnarray}
Then we affect $\max(f_i^{\rm left},f_i^{\rm right},f_{\ell,m})$ to the following list of pixels
\begin{eqnarray}
\ell_p={\rm int}\left( x_i+\frac{\delta x}{p_{\rm max}} p \right),
m_p={\rm int}\left( v_i+\frac{\delta v}{p_{\rm max}} p \right),
\end{eqnarray}
 for $p \in \{0,\cdots,p_{\rm max} \}$, where
\begin{equation}
p_{\rm max}={\rm int}[\max(|\delta x|,|\delta v|)]+1.
\end{equation}
This procedure does not correspond to computing exactly the intersection of the segment with each pixel, but it provides satisfactory results for visual inspection. 
\end{itemize}
%=======================================================
\section{Lagrangian perturbation theory}
\label{app:analcalc}
%=======================================================
In this appendix, we give the details of the calculations of the phase-space energy distribution function that lead to equations (\ref{eq:powe0}) and (\ref{eq:Epred2}) of \S~\ref{sec:fofE}.  To do so, we describe the cold system we aim to follow by a curve, $x(q,t)$  and $v(q,t)$, where
$q$ is a Lagrangian coordinate, with
\begin{equation}
x(q,t=0)=q.
\end{equation}
This curve in phase-space has an initial projected density approximated by equation (\ref{eq:rhoapp}). Defining, for convenience, the new variables
\begin{eqnarray}
\tau &=&\sqrt{{\bar \rho}_0} t, \\
Q&=&\sqrt{a} q, \\
X(Q,\tau)&=&\sqrt{a} x(q,t),\\
V(Q,\tau)&=&\frac{\partial X}{\partial \tau}=\sqrt{\frac{a}{{\bar \rho}_0}} v,
\end{eqnarray}
the equations of motion read
\begin{eqnarray}
\frac{\partial X}{\partial \tau} &=& V, \\
\frac{\partial V}{\partial \tau} &=&-2 (Q-Q^3), 
\end{eqnarray}
corresponding to an initial rescaled projected density profile 
\begin{equation}
\Sigma_0(Q)=1-3 Q^2.
\end{equation}
Prior to collapse time, the solution of these equations follows Zel'dovich dynamics \citep{Zeldovich70}:
\begin{eqnarray}
X(Q,\tau) &=& Q-(Q-Q^3) \tau^2, \label{eq:xofqt} \\
V(Q,\tau) &=& -2 (Q-Q^3) \tau,
\end{eqnarray}
while the projected density reads
\begin{equation}
\rho(X,\tau)=\Sigma_0(Q)\left|\frac{\partial X}{\partial Q} \right|^{-1}=\frac{1-3 Q^2}{1-\tau^2+3 Q^2 \tau^2},
\end{equation}
where $Q$ is obtained by solving equation (\ref{eq:xofqt}).
For estimating the phase-space energy distribution function, ${f}_E({\cal E})$ in our system of coordinates, it is also useful to write the specific energy of a fluid element as a function of $Q$
\begin{eqnarray}
{\cal E}(Q,\tau) &= & \frac{1}{2} V^2 + \int_0^Q 2 (Q-Q^3) \frac{\partial X}{\partial Q} {\rm d}Q +{\cal E}_{\rm min}(\tau), \\
         & = & (1+\tau^2) Q^2-\frac{1}{2} (1+4 \tau^2) Q^4 \nonumber \\
  & & + \tau^2 Q^6 +{\cal E}_{\rm min}(\tau), 
 \end{eqnarray}
where ${\cal E}_{\rm min}(\tau)$ is the minimum value of the (rescaled) energy at time $\tau$. It obviously coincides with the minimum potential energy $\Phi_{\rm min}$. Computing ${\cal E}_{\rm min}(\tau)$ is not a real difficulty but does not serve our purpose. 

The phase-space energy distribution function can be conveniently expressed as follows:
\begin{equation}
{f}_E({\cal E})=2 \Sigma_0(Q) \left[ \frac{\partial {\cal E}}{\partial Q} \right]^{-1} \left[ \frac{{\rm d} J}{{\rm d} {\cal E}} \right]^{-1},
\end{equation}
with $J$ given by equation (\ref{eq:adiabinv}) and where one needs to find a mean to invert the relation ${\cal E}(Q)$ to find $Q$ as a function of ${\cal E}$.  The calculation of $J({\cal E})$ is however involved, but can be performed easily in the interesting limit where the (rescaled) potential $\Phi(X)$ is, within a constant, a power-law:
\begin{equation}
\Phi(X)={\cal E}_{\rm min}+\Phi_0 X^\beta.
\label{eq:powerlawphi}
\end{equation}
Equation (\ref{eq:adiabinv}) can be written 
\begin{equation}
J({\cal E})=4 \int_0^{X_{\rm max}}\sqrt{2[{\cal E}-\Phi(X)]} {\rm d}X,
\end{equation}
where $X_{\rm max}$ is such that ${\cal E}\equiv \Phi(X_{\rm max})$. Using the expression (\ref{eq:powerlawphi}) gives
\begin{equation}
J({\cal E})=\frac{4 \sqrt{2\pi} \beta\ \Gamma(1+1/\beta)\ \Phi_0^{-1/\beta} ({\cal E}-{\cal E}_{\rm min})^{1/2+1/\beta} }{(2+\beta)\ \Gamma(1/2+1/\beta)} \label{eq:comp0}
\end{equation}
\citep[see, e.g.,][]{Schulz2013}.

At initial time, we have simply $X=Q$ so the (rescaled) potential $\Phi$ reads
\begin{eqnarray}
\Phi(X,\tau=0)&= &{\cal E}_{\rm min}+ X^2-\frac{1}{2} X^4,\\
  & \simeq & {\cal E}_{\rm min} + X^2, \quad X \ll 1, \label{eq:phiini}
\end{eqnarray}
hence $\beta=2$ and $\phi_0=1$. After rescaling, one obtains equation (\ref{eq:phini2}).
The energy reads
\begin{equation}
{\cal E}(Q,0) \simeq {\cal E}_{\rm min}+Q^2, \quad Q \ll 1.
\end{equation}
Hence
\begin{equation}
 f_E({\cal E},t=0)=\frac{1}{\sqrt{2}\pi} ({\cal E}-{\cal E}_{\rm min})^{-1/2}.
\end{equation}
However, we yet have to pass back to our coordinate space to find the correct normalization for $f_E(E,t=0)$ and obtain equation (\ref{eq:powe0}).

Collapse time, or first crossing time, corresponds to the occurrence $| \partial X/\partial Q |=0$ that happens first for $Q=0$. In our system of coordinates, collapse time is equal to unity:
\begin{equation}
\tau_{\rm c}=1.
\end{equation}
At this time, we have
\begin{equation}
X(q,\tau_{\rm c})=Q^3,
\end{equation}
and the projected density has the well known singular behavior:
\begin{equation}
\rho(X,\tau_{\rm c})=\frac{1}{3 X^{2/3}}-1.
\end{equation}
The potential is thus of the form
\begin{equation}
\Phi(X,\tau_{\rm c}) \simeq  \frac{3}{2} |X|^{4/3}+{\cal E}_{\rm min},\quad X \ll 1,
\end{equation}
so $\beta=4/3$ and $\Phi_0=3/2$, and one obtains, after rescaling, equation (\ref{eq:phicol2}). The energy reads,
\begin{equation}
{\cal E}(Q,\tau_{\rm c})\simeq {\cal E}_{\rm min}+ 2 Q^2, \quad Q \ll 1.
\end{equation}
Therefore, after rescaling, one obtains equation (\ref{eq:Epred2}). 
\label{lastpage}
\end{document}